\documentclass[12pt]{article}
\usepackage{graphicx}
\usepackage{array}
\usepackage{amssymb}
\usepackage{amsfonts}
\usepackage{bm}
\usepackage{amsmath}
\usepackage{mathrsfs}
\usepackage{color}
\usepackage{booktabs}
\usepackage{threeparttable}
\usepackage{multirow}
\usepackage{subfigure}
\usepackage{times}
\usepackage{epsfig}
\usepackage{threeparttable}
\usepackage{chngpage}
\usepackage{latexsym}
\usepackage{mathptm}
\usepackage{floatrow}
\usepackage{color}
\usepackage{floatrow}
\usepackage{tabularx}
\usepackage[noadjust]{cite}

\advance \topmargin by -\headheight
\advance \topmargin by -\headsep

\evensidemargin \oddsidemargin
\def\pmb#1{\setbox0=\hbox{#1}
\kern.05em\copy0\kern-\wd0 \kern-.025em\raise.0433em\box0 }

\newcounter{cntd}
  {\end{list}}
  {\end{list}}

\usepackage{lipsum}

\let\OLDthebibliography\thebibliography
\renewcommand\thebibliography[1]{
  \OLDthebibliography{#1}
  \setlength{\parskip}{0pt}
  \setlength{\itemsep}{0pt plus 0.3ex}
}

\begin{document}


\begin{titlepage}

\begin{center}
{\Large\bf Data Based Identification and Prediction of Nonlinear and} 
\end{center}
\vspace*{-0.3in}
\begin{center}
{\Large\bf Complex Dynamical Systems}
\end{center}

\begin{center}
{\large Wen-Xu Wang$^{a,b}$, Ying-Cheng Lai$^{c,d,e}$, and Celso Grebogi$^e$}
\end{center}

\begin{enumerate}
\item [a]
School of Systems Science, Beijing Normal University,
Beijing, 100875, China  
\item [b]
Business School, University of Shanghai for Science and 
Technology, Shanghai 200093, China 
\item [c]
School of Electrical, Computer and Energy Engineering,
Arizona State University, Tempe, Arizona 85287, USA 
\item [d]
Department of Physics, Arizona State University, Tempe, Arizona 85287, USA 
\item [e]
Institute for Complex Systems and Mathematical Biology,
King's College, University of Aberdeen, Aberdeen AB24 3UE, UK
\end{enumerate}

\date\today
\end{titlepage}

\begin{center}
{\large\bf Abstract}
\end{center}

The problem of reconstructing nonlinear and complex dynamical systems 
from measured data or time series is central to many scientific disciplines 
including physical, biological, computer, and social sciences, as well as 
engineering and economics. The classic approach to phase-space reconstruction 
through the methodology of delay-coordinate embedding has been practiced for 
more than three decades, but the paradigm is effective mostly for 
low-dimensional dynamical systems. Often, the methodology yields 
only a topological correspondence of the original system. There are 
situations in various fields of science and engineering where the systems 
of interest are complex and high dimensional with many interacting components. 
A complex system typically exhibits a rich variety of collective dynamics, 
and it is of great interest to be able to detect, classify, understand, 
predict, and control the dynamics using data that are becoming increasingly 
accessible due to the advances of modern information technology. To accomplish 
these tasks, especially prediction and control, an accurate reconstruction 
of the original system is required.

Nonlinear and complex systems identification aims at inferring, from data, 
the mathematical equations that govern the dynamical evolution and the 
complex interaction patterns, or topology, among the various components of 
the system. With successful reconstruction of the system equations and the 
connecting topology, it may be possible to address challenging 
and significant problems such as identification of causal relations among 
the interacting components and detection of hidden nodes. The ``inverse'' 
problem thus presents a grand challenge, requiring new paradigms beyond the 
traditional delay-coordinate embedding methodology.

The past fifteen years have witnessed rapid development of contemporary 
complex graph theory with broad applications in interdisciplinary science 
and engineering. The combination of graph, information, and nonlinear 
dynamical systems theories with tools from statistical physics, 
optimization, engineering control, applied mathematics, and scientific 
computing enables the development of a number of paradigms to address 
the problem of nonlinear and complex systems reconstruction. In this 
Review, we review the recent advances in this forefront and rapidly 
evolving field, with a focus on compressive sensing based methods. 
In particular, compressive sensing is a paradigm developed in recent years 
in applied mathematics, electrical engineering, and nonlinear physics to 
reconstruct sparse signals using only limited data. It has broad applications 
ranging from image compression/reconstruction to the analysis of large-scale 
sensor networks, and it has become a powerful technique to obtain high-fidelity 
signals for applications where sufficient observations are not available.
We will describe in detail how compressive sensing can be exploited to 
address a diverse array of problems in data based reconstruction of nonlinear 
and complex networked systems. The problems include identification of 
chaotic systems and prediction of catastrophic bifurcations, forecasting
future attractors of time-varying nonlinear systems, reconstruction of 
complex networks with oscillatory and evolutionary game dynamics,
detection of hidden nodes, identification of chaotic elements in neuronal
networks, and reconstruction of complex geospatial networks and nodal 
positioning. A number of alternative methods, such as those based on 
system response to external driving, synchronization, noise-induced dynamical
correlation, will also be discussed. Due to the high relevance of network
reconstruction to biological sciences, a special Section is devoted to a 
brief survey of the current methods to infer biological networks. Finally, a 
number of open problems including control and controllability of complex
nonlinear dynamical networks are discussed.   

The methods reviewed in this Review are principled on various concepts 
in complexity science and engineering such as phase transitions, 
bifurcations, stabilities, and robustness. The methodologies have the
potential to significantly improve our ability to understand a variety 
of complex dynamical systems ranging from gene regulatory systems to 
social networks towards the ultimate goal of controlling such systems. 


\newpage

\tableofcontents

\newpage

\section{Introduction} 
\label{sec:CH1_intro}

An outstanding problem in interdisciplinary 
science is nonlinear and complex systems identification, prediction, and
control. Given a complex dynamical system, the various types of 
dynamical processes are of great interest. The ultimate goal in the
study of complex systems is to devise practically implementable 
strategies to control the collective dynamics. A great challenge is 
that the network structure and the nodal dynamics are often unknown 
but only limited measured time series are available. To control the 
system dynamics, it is imperative to map out the system 
details from data. Reconstructing complex network structure and dynamics 
from data, the inverse problem, has thus become a central issue in 
contemporary network science and engineering~\cite{GDA:2002,GAR:2002,GDLC:2003,PG:2003,BDLCNB:2004,YRK:2006,BL:2007,Timme:2007,TYK:2007,NS:2008,Sontag:2008,CMN:2008,WCHLH:2009,DZMK:2009,RWLL:2010,CHR:2010,YSWG:2010,LP:2011,HKKN:2011,ST:2011,YP:2011,WLGY:2011,WYLKG:2011,WYLKH:2011,YLG:2012,PYS:2012,WRLL:2012,BHPS:2012,SBSG:2012,SNWL:2012,SWL:2012,HBPS:2013,ZXZXC:2013,TimC:2014,SWL:2014,SLWD:2014,SWFDL:2014,SWWL:2015}. 
There are broad applications of the solutions of the network 
reconstruction problem, due to the ubiquity of complex interacting patterns 
arising from many systems in a variety of disciplines~\cite{AB:2002,Newman:2003,BLMCH:2006,Newman:book}.

\subsection{Existing works on data based reconstruction of 
nonlinear dynamical systems} \label{subsec:Intro_NLD}

The traditional paradigm of nonlinear time series analysis is the 
delay-coordinate embedding method, the mathematical foundation of 
which was laid by Takens more than three decades ago~\cite{Takens:1981}. 
He proved that, under fairly general conditions, the underlying dynamical 
system can be faithfully reconstructed from time series in the sense that 
a one-to-one correspondence can be established between the reconstructed 
and the true but unknown dynamical systems. Based on the reconstruction, 
quantities of importance for understanding the system can be estimated, 
such as the relative weights of deterministicity and stochasticity of 
the underlying system, its dimensionality, the Lyapunov exponents, and 
unstable periodic orbits that constitute the skeleton of the invariant 
set responsible for the observed dynamics.

There exists a large body of literature on the application of the
delay-coordinate embedding technique to nonlinear/chaotic dynamical 
systems~\cite{KS:book,Hegger:book}. A pioneering work in this field
is Ref.~\cite{PCFS:1980}. The problem of determining the proper time
delay was investigated~\cite{Theiler:1986,LS:1989,LPS:1991,BP:1992,
KF:1993,RCdL:1994,LLH:1996,LL:1998}, with a firm theoretical foundation 
established by exploiting the statistics for testing continuity and 
differentiability from chaotic time series~\cite{PCH:1995,PC:1996,
PCH:1997,GPCR:2001,PMNC:2007}. The mathematical foundation for the required
embedding dimension for chaotic attractors was laid in 
Ref.~\cite{SYC:1991}. There were works on the analysis of transient 
chaotic time series~\cite{JFT:1994,JT:1994,DLK:2000,DLK:2001,TBS:2003},
on the reconstruction of dynamical systems with time delay~\cite{TC:2007},
on detecting unstable periodic orbits from time series~\cite{LK:1989,
BBFFHPRS:1994,PM:1995,PM:1996a,SOSKSG:1996,AM:1997}, on computing the
fractal dimensions from chaotic data~\cite{GP:1983,Grass:1986,
Proc:1988,OP:1989,Lorenz:1991,DGOSY:1993,LLH:1996,LL:1998}, and on estimating
the Lyapunov exponents~\cite{WSSV:1985,Sano:1985,ER:1985,EKRC:1986,
BBA:1991,STY:1998,SY:1999}.

There were also works on forecasting nonlinear dynamical 
systems~\cite{Casdagli:1989,SGMCPW:1990,KR:1990,GS:1990,TE:1992,
Longtin:1993,Murray:1993,Sugihara:1994,FK:1995,SSCBS:1996,HKS:1999,
Sello:2001,MNSSH:2001,Smith:2002,Judd:2003,FS:1987,Gouesbet:1991,
BBBB:1992,Sauer:1994,Sauer:2004,Parlitz:1996,Szpiro:1997,TZJ:2007,
WYLKG:2011}. A conventional approach is to approximate a nonlinear
system with a large collection of linear equations in different regions 
of the phase space to reconstruct the Jacobian matrices on a proper grid
\cite{FS:1987,Gouesbet:1991,Sauer:1994} or fit ordinary differential
equations to chaotic data \cite{BBBB:1992}. Approaches based on chaotic
synchronization~\cite{Parlitz:1996} or genetic 
algorithms~\cite{Szpiro:1997,TZJ:2007} to parameter estimation 
were also investigated. In most existing works, short-term predictions 
of a dynamical system can be achieved by employing the classical 
delay-coordinate embedding paradigm~\cite{Takens:1981,KS:book}.
For nonstationary systems, the method of over-embedding was 
introduced~\cite{HKMS:2000} in which the time-varying parameters were 
treated as independent dynamical variables so that the essential aspects 
of determinism of the underlying system can be restored. 
A recently developed framework based on compressive sensing was 
able to predict the {\em exact} forms of both system equations and 
parameter functions based on available time series for 
stationary~\cite{WYLKG:2011} and time-varying dynamical systems~\cite{YLG:2012}.

\subsection{Existing works on data based reconstruction of complex 
networks and dynamical processes} \label{subsec:Intro_CN}

Data based reconstruction of complex networks in general is deemed to be
an important but difficult problem and has attracted continuous interest, 
where the goal is to uncover the full topology of the network based on 
simultaneously measured time series~\cite{GDA:2002,GAR:2002,GDLC:2003,PG:2003,BDLCNB:2004,YRK:2006,BL:2007,Timme:2007,TYK:2007,NS:2008,Sontag:2008,CMN:2008,WCHLH:2009,DZMK:2009,RWLL:2010,CHR:2010,YSWG:2010,LP:2011,HKKN:2011,ST:2011,YP:2011,
WLGY:2011,WYLKG:2011,WYLKH:2011,YLG:2012,PYS:2012,WRLL:2012,BHPS:2012,SBSG:2012,SNWL:2012,SWL:2012,HBPS:2013,ZXZXC:2013,TimC:2014,SWL:2014,SLWD:2014,SWFDL:2014,SWWL:2015}.
There were previous efforts in nonlinear systems identification
and parameter estimation for coupled oscillators and spatiotemporal
systems, such as the auto-synchronization method~\cite{Parlitz:1996}. 
There were also works on revealing the connection patterns of networks. 
For example, methods were proposed to estimate the network topology 
controlled by feedback or delayed feedback~\cite{YRK:2006,TYK:2007,YP:2011}.
Network connectivity can be reconstructed from the collective dynamical 
trajectories using response dynamics~\cite{ST:2011}. The approach of 
random phase resetting was introduced to reconstruct the details of 
the network structure~\cite{LP:2011}. For neuronal systems, there was 
a statistical method to track the structural changes~\cite{BHPS:2012,HBPS:2013}. 
Some earlier methods required more information about the network than just data. 
For example, the following two approaches require {\em complete} 
information about the dynamical processes, e.g., equations governing 
the evolutions of {\em all} nodes on the network. (1) In Ref.~\cite{YRK:2006}, 
the detailed dynamics at each node is assumed to be known. A replica of 
the network, or a computational model of this ``target'' network, can 
then be constructed, with the exception that the interaction strengths
among the nodes are chosen randomly. It has been demonstrated that in 
situations where a Lyapunov function for the network dynamics exists, 
the connectivity of the model network converges to that of the target 
network~\cite{YRK:2006}. (2) In Ref.~\cite{Timme:2007}, a Kuramoto-type 
of phase dynamics~\cite{Kuramoto:book,Strogatz:2000} on the network is 
assumed, where a steady-state solution exists. By linearizing the network 
dynamics about the steady-state solution, the associated Jacobian matrix 
can be obtained, which reflects the network topology and connectivity. 
Besides requiring complete information about the nodal dynamics, 
the amount of computations required tends to increase dramatically with 
the size of the network~\cite{YRK:2006}. For example, suppose the 
nodal dynamics is described by a set of differential equations. For a 
network of size $N$, in order for its structure to be predicted, the 
number of differential equations to be solved typically increases with 
$N$ as $N^2$. 

For nonlinear dynamical networks, there was a method~\cite{NS:2008} 
based on chaotic time-series analysis through estimating the elements
of the Jacobian matrix, which are the mutual partial derivatives of the 
dynamical variables on different nodes in the network. A statistically
significant entry in the matrix implies a connection between the two
nodes specified by the row and the column indices of that entry. Because 
of the mathematical nature of the Jacobian matrix, i.e., it is meaningful 
only for infinitesimal tangent vectors, linearization of the dynamics 
in the neighborhoods of the reconstructed phase-space points is needed, 
for which constrained optimization techniques~\cite{CDS:1998,CRT:2006a,
CRT:2006b,Candes:2006,Donoho:2006,Baraniuk:2007,CW:2008} were found to be
effective~\cite{NS:2008}. Estimating the Jacobian matrices, however, has 
been a challenging problem in nonlinear dynamics~\cite{ER:1985,Sauer:2004} 
and its reliability can be ensured but only for low-dimensional, 
deterministic dynamical systems. The method~\cite{NS:2008} appeared thus 
to be limited to small networks with sparse connections.

While many of the earlier works required complete or partial information 
about the intrinsic dynamics of the nodes and their coupling functions, 
completely data-driven and model-free methods exist. For example,
The global climate network was reconstructed using the mutual information
method, enabling energy and information flow in the network to be
studied~\cite{DZMK:2009}. The sampling bias of DNA sequences in viruses
from different regions can be used to reveal the geospatial topologies
of the influenza networks~\cite{CHR:2010}. Network structure can also be 
obtained by calculating the causal influences among the time series based 
on the Granger causality~\cite{Granger:1969} method~\cite{BDLCNB:2004,
Sommerladeetal:2012,ZXZXC:2013,Rambetal:2013,Gaoetal:2015}, the overarching 
framework~\cite{Schelteretal:2014}, the transfer entropy 
method~\cite{SBSG:2012}, or the method of inner composition
alignment~\cite{HKKN:2011}. However, such causality based methods are
unable to reveal information about the nodal dynamical equations.
In addition, there were regression-based methods~\cite{MPF:2005} for systems
identification based on the least squares approximation through the 
Kronecker-product representation~\cite{YB:2007}, which would require 
large amounts of data. 

In systems biology, reverse engineering of gene regulatory networks
from expression data is a fundamentally important problem, and 
it attracts a tremendous amount of interest~\cite{EricH:Science,Oates:NRG,
Richard:Cell,Timothy:science,Faith:PLoS,JohnJeremy:bioinf,Perkins:PLoS,
Bonneau:GB,Margolin:BMC}. The wide spectrum of methods for modeling 
genetic regulatory networks can be categorized based on the level of 
details with which the genetic interactions and dynamics are 
modeled~\cite{Bonneau:NatureCB}. One of the classical mathematical
formalisms used to model the dynamics of biological processes is
differential equations, which can capture the dynamics of each
component in a system at a detailed level~\cite{Turing,Goodwin:JTBI,
Goodwin:JTBII}. A major limitation of this approach is its overwhelming 
complexity and the resulting computational requirement, which limits 
its applicability to small-scale systems. In contrast, Boolean 
network models assume that the states of components in the system are 
binary and the state transitions are governed by logic 
operations~\cite{Kauffman:Nature69,Kauffman:JTB,Shmulevich:PRL}. Since 
the gene expressions are usually described by their expression
levels and the interaction patterns between genes may not be logic
operations, in some cases the Boolean network models may not be biologically 
appropriate. Another class of methods explicitly model
biological systems as a graph in which the vertices represent basic
units in the system and the edges characterize the relationships
between the units. The graph itself can be constructed either by
directly comparing the measurements for the vertices based on
certain metric, such as the Euclidean distance, mutual information,
or correlation coefficient~\cite{Eisen:PNAS,Atul:PNAS,Faith:PLoS},
or by some probabilistic approaches for Bayesian network 
learning~\cite{Friedman:JCB,NirFriedman:Science,Segal:NatureG}. However,
inferring regulatory interactions based on Bayesian networks is an
intractable problem~\cite{Chickering:NP,Chickering:JMLR}. The linear
regression models for learning regulatory networks assume that
expression level of a gene can be approximated by a linear
combination of the expressions of other genes~\cite{Bonneau:GB,Richard:Cell,
Timothy:science,Mika:TCBB}, and such models form a middle ground between 
the models based on differential equations and Boolean logic.

\subsection{Compressive sensing based reconstruction of nonlinear
and complex dynamical systems} \label{subsec:Intro_CS}

A recent line of research~\cite{WLGY:2011,WYLKG:2011,WYLKH:2011,
SNWL:2012,SWL:2012,SWL:2014,SLWD:2014,SWFDL:2014,SWWL:2015} 
exploited compressive sensing~\cite{CRT:2006a,CRT:2006b,Candes:2006,
Donoho:2006,Baraniuk:2007,CW:2008}. The basic principle is that the 
dynamics of many natural and man-made systems are determined by smooth 
enough functions that can be approximated by finite series expansions. 
The task then becomes that of estimating the coefficients in the series 
representation of the vector field governing the system dynamics. In 
general, the series can contain high order terms, and the total number
of coefficients to be estimated can be quite large. While this is a 
challenging problem, if most coefficients are zero (or negligible), the 
vector constituting all the coefficients will be sparse. In addition,
a generic feature of complex networks in the real world is that they 
are sparse~\cite{Newman:book}. Thus for realistic nonlinear dynamical  
networks, the vectors to be reconstructed are typically sparse, and 
the problem of sparse vector estimation can then be solved by the 
paradigm of compressive sensing~\cite{CRT:2006a,CRT:2006b,Donoho:2006,
Baraniuk:2007,CW:2008} that reconstructs a sparse signal from limited 
observations. Since the observation requirements can be relaxed 
considerably as compared to those associated with conventional signal
reconstruction schemes, compressive sensing has evolved into a powerful 
technique to reconstruct sparse signal from small amounts of observations 
that are much less than those required in conventional approaches.
Compressive sensing has been introduced to the field of network 
reconstruction for discrete time and continuous time nodal 
dynamics~\cite{WYLKG:2011,WYLKH:2011}, for evolutionary game 
dynamics~\cite{WLGY:2011}, for detecting hidden 
nodes~\cite{SWL:2012,SLWD:2014}, for predicting and
controlling synchronization dynamics~\cite{SNWL:2012}, and for
reconstructing spreading dynamics based on binary data~\cite{SWFDL:2014}.
Compressive sensing also finds applications in quantum measurement 
science, e.g., to exponentially reduce the experimental configurations 
required for quantum tomography~\cite{Shabanietal:2011}.

\subsection{Plan of this review} \label{subsec:Intro_Plan}

This Review presents the recent advances in the forefront and rapidly
evolving field of nonlinear and complex dynamical systems identification 
and prediction. Our focus will be on the compressive sensing based
approaches. Alternative approaches will also be discussed, which 
include noised-induced dynamical mapping, perturbations, reverse engineering, 
synchronization, inner composition alignment, global silencing, Granger 
causality, and alternative optimization algorithms. 

In {\bf Sec.~\ref{sec:CS_NDSI}}, we first introduce the principle of 
compressive sensing and discuss nonlinear dynamical systems identification 
and prediction. We next discuss a compressive sensing based approach to 
predicting catastrophes in nonlinear dynamical systems under the 
assumption that the system equations are completely unknown and only 
time series reflecting the evolution of the dynamical variables of the 
system are available. We then turn to time-varying nonlinear dynamical 
systems, motivated by the fact that systems with one or a few parameters 
varying slowly with time are of considerable interest in many areas of science
and engineering. In such a system, the attractors in the future can be
characteristically different from those at the present. To predict the
possible future attractors based on available information at the present
is thus a well-defined and meaningful problem, which is challenging
especially when the system equations are not known but only time-series
measurements are available. We review a compressive-sensing based method 
for time-varying systems. This framework allows us to reconstruct the 
system equations and the time dependence of parameters based on limited 
measurements so that the future attractors of the system can be predicted 
through computation.
 
{\bf Section~\ref{sec:CH3_CS_CN_Reconstruction}} focuses on compressive 
sensing based reconstruction of complex networked systems. The following 
problems will be discussed in detail.

{\em Reconstruction of coupled oscillator networks}.
The basic idea is that the mathematical functions determining the 
dynamical couplings in a physical network can be expressed by 
power-series expansions. The task is then to estimate all the 
nonzero coefficients, which can be accomplished by exploiting 
compressive sensing~\cite{WYLKH:2011}.

{\em Reconstruction of social networks based on
evolutionary-game data via compressive sensing}.
Evolutionary games are a common type interactions in a variety
of complex networked, natural and social systems. Given such a 
system, uncovering the interacting structure of the underlying 
network is key to understanding its collective dynamics.
We discuss a compressive sensing based method to uncover the 
network topology using evolutionary-game data.
In particular, in a typical game, agents use different strategies
in order to gain the maximum payoff. The strategies can be divided 
into two types: cooperation and defection. It was shown~\cite{WLGY:2011} 
that, even when the available information about each agent's
strategy and payoff is limited, the compressive-sensing based
approach can yield precise knowledge of the node-to-node interaction
patterns in a highly efficient manner. In addition to numerical 
validation of the method with model complex networks, we discuss  
an actual social experiment in which participants forming a 
friendship network played a typical game to generate short sequences 
of strategy and payoff data. The high prediction accuracy achieved 
and the unique requirement of extremely small data set suggest that
the method can be appealing to potential applications to reveal ``hidden'' 
networks embedded in various social and economic systems.

{\em Detecting hidden nodes in complex networks from time 
series}. The power of science lies in its ability to infer and predict 
the existence of objects from which no direct information can be 
obtained experimentally or observationally. A well known example is 
to ascertain the existence of black holes of various masses in 
different parts of the universe from indirect evidence, such as
X-ray emissions. In complex networks, the problem of detecting
{\em hidden} nodes can be stated, as follows. Consider a network 
whose topology is completely unknown but whose nodes consist of two 
types: one accessible and another inaccessible from the outside world. 
The accessible nodes can be observed or monitored, and we assume that 
time series are available from each node in this group. The inaccessible 
nodes are shielded from the outside and they are essentially ``hidden.'' 
The question is, can we infer, based solely on the available time series 
from the accessible nodes, the existence and locations of the hidden 
nodes? Since no data from the hidden nodes are available, nor can they 
be observed directly, they act as some sort of ``black box'' from the 
outside world. Solution of the network hidden node detection problem 
has potential applications in different fields of significant current 
interest. For example, to uncover the topology of a terrorist 
organization and especially, various ring leaders of the network is 
a critical task in defense. The leaders may be hidden in the sense that 
no direct information about them can be obtained, yet they may rely on 
a number of couriers to operate, which are often subject to surveillance.
Similar situations arise in epidemiology, where the original carrier 
of a virus may be hidden, or in a biology network where one wishes to 
detect the most influential node from which no direct observation can 
be made. We discuss in detail a compressive sensing based 
method~\cite{SWL:2012,SLWD:2014} to ascertain hidden nodes in complex 
networks and to distinguish them from various noise sources. 

{\em Identifying chaotic elements in complex neuronal networks}.
We discuss a completely data-driven approach~\cite{SWL:2014} to reconstructing 
coupled neuronal networks that contain a small subset of chaotic neurons. Such 
chaotic elements can be the result of parameter shift in their individual 
dynamical systems, and may lead to abnormal functions of the network. 
To accurately identify the chaotic neurons may thus be necessary and 
important, for example, for applying appropriate controls to bring the 
network to a normal state. However, due to couplings among the nodes 
the measured time series even from non-chaotic neurons would appear 
random, rendering inapplicable traditional nonlinear time-series 
analysis, such as the delay-coordinate embedding method, which yields 
information about the global dynamics of the entire network. The 
method to be discussed is based on compressive sensing. In particular, 
identifying chaotic elements can be formulated as a general problem of 
reconstructing the nodal dynamical systems, network connections, and all 
coupling functions as well as their weights. 

{\em Data based reconstruction of complex geospatial networks,
nodal positioning, and detection of hidden node}.
Complex geospatial networks with components distributed in the real 
geophysical space are an important part of the modern infrastructure. 
Given a complex geospatial network with nodes distributed in a
two-dimensional region of the physical space, can the locations
of the nodes be determined and their connection patterns be uncovered
based solely on data? In realistic applications, time series/signals 
can be collected from a single location. A key challenge is that the 
signals collected are necessarily time delayed, due to the varying 
physical distances from the nodes to the data collection center. 
We discuss a compressive sensing based approach~\cite{SWWL:2015} that 
enables reconstruction of the full topology of the underlying geospatial 
network and more importantly, accurate estimate of the time delays. A 
standard triangularization algorithm can then be employed to find the
physical locations of the nodes in the network. A hidden source or 
threat, from which no signal can be obtained, can also be detected
through accurate detection of all its neighboring nodes. As a geospatial 
network has the feature that a node tends to connect with geophysically 
nearby nodes, the localized region that contains the hidden node can 
be identified.

{\em Reconstructing complex spreading networks with natural 
diversity and identifying hidden source}. Among the various types
of collective dynamics on complex networks, propagation or spreading dynamics
is of paramount importance as it is directly relevant to issues of
tremendous interests such as epidemic and disease outbreak in the human
society and virus spreading on computer networks. We discuss a theoretical
and computational framework based on compressive sensing to reconstruct
networks of arbitrary topologies, in which spreading dynamics with
heterogeneous diffusion probabilities take place~\cite{SWFDL:2014}. 
The approach enables identification with high accuracy of the external 
sources that trigger the spreading dynamics. Especially, a full 
reconstruction of the stochastic and inhomogeneous interactions presented 
in the real-world networks can be achieved from small amounts of polarized 
(binary) data, a virtue of compressive sensing. After the outbreak of 
diffusion, hidden sources outside the network, from which no direct routes 
of propagation are available, can be ascertained and located with high 
confidence. This represents essentially a new paradigm for tracing, monitoring, 
and controlling epidemic invasion in complex networked systems, which will 
be of value to defending and preserving the systems against disturbances 
and attacks.

{\bf Section~\ref{sec:CH4_alt_method}} presents a number of alternative methods
for reconstructing complex and nonlinear dynamical networks. These are
methods based on (1) response of the system to external driving, 
(2) synchronization (via system clone), (3) phase-space linearization, 
(4) noise induced dynamical correlation, and (5) automated reverse 
engineering. In particular, for the system response based method (1), the
basic idea was to measure the collective response of the oscillator network 
to an external driving signal. With repeated measurements of the dynamical 
states of the nodes under sufficiently independent driving realizations, the
network topology can be recovered~\cite{Timme:2007}. Since
complex networks are generally sparse, the number of realizations of
external driving can be much smaller than the network size.
For the synchronization based method, the idea was to design a replica 
or a clone system that is sufficiently close to the original network without
requiring knowledge about network structure~\cite{YRK:2006}. From the clone 
system, the connectivities and interactions among nodes can be obtained 
directly, realizing network reconstruction. For the phase-space linearization
method, $L_1$-minimization in the phase space of a networked system is
employed to reconstruct the topology without knowledge of the self-dynamics 
of nodes and without using any external perturbation to the state of 
nodes~\cite{NS:2008}. For the method based on noise-induced dynamical 
correlation, the principle was based on exploiting the rich interplay
between nonlinear dynamics and stochastic 
fluctuations~\cite{WCHLH:2009,RWLL:2010,WRLL:2012}. Under the condition
that the influence of noise on the evolution of infinitesimal tangent
vectors in the phase space of the underlying networked dynamical system
is dominant, it can be argued~\cite{RWLL:2010,WRLL:2012} that the dynamical
correlation matrix that can be computed readily from the available nodal
time series approximates the network adjacency matrix, fully unveiling
the network topology. For the automated reverse engineering method, 
the approach was based on problem solving using partitioning, automated 
probing and snipping~\cite{BongardLipson:2007}, a process that is akin
to genetic algorithm. Each of the five methods will be described in 
reasonable details. 

Network reconstruction was pioneered in biological sciences.  
{\bf Section~\ref{sec:CH5_Bio_Net}} is devoted to a concise survey of the
approaches to reconstructing biological networks. Those include methods 
based on correlation, causality, information-theoretic measures, 
Bayesian inference, regression and resampling, supervision and 
semi-supervision, transfer and joint entropies, and
distinguishing between direct and indirect interactions.

Finally, in {\bf Sec.~\ref{sec:CH6_Discussions}}, we offer a general discussion
of the field of data-based reconstruction of complex networks and speculate
on a number of open problems. These include: localization of diffusion
sources in complex networks, reconstruction of complex networks with 
binary state dynamics, possibility of developing a universal framework
of structural estimator and dynamics approximator for complex networks,
and a framework of control and controllability for complex nonlinear 
dynamical networks.

\clearpage

\section{Compressive sensing based nonlinear dynamical systems 
identification} \label{sec:CS_NDSI}

\subsection{Introduction to the compressive sensing algorithm} 
\label{subsec:CS_intro}

Compressive sensing is a paradigm developed in recent years by applied 
mathematicians and electrical engineers to reconstruct sparse signals 
using only limited data~\cite{CRT:2006a,CRT:2006b,Candes:2006,Donoho:2006,
Baraniuk:2007,CW:2008}. The observations are measured by linear projections 
of the original data on a few predetermined, random vectors. Since
the requirement for the observations is considerably less comparing to 
conventional signal reconstruction schemes, compressive sensing has been 
deemed as a powerful technique to obtain high-fidelity signal 
especially in cases where sufficient observations are not available.
Compressive sensing has broad applications ranging from image 
compression/reconstruction to the analysis of large-scale sensor 
networks~\cite{Baraniuk:2007,CW:2008}. 

Mathematically, the problem of compressive sensing is to
reconstruct a vector $\mathbf{a} \in \mathbb{R}^N$ from linear
measurements $\mathbf{X}$ about $\mathbf{a}$ in the form:
\begin{equation}\label{eq:CS_general}
\mathbf{X} = \mathcal{G} \cdot \mathbf{a} 
\end{equation}
where $\mathbf{X} \in \mathbb{R}^M$ and $\mathcal{G}$ is an $M\times N$ 
matrix. By definition, the number of measurements is much less than that of 
the unknown signal, i.e, $M\ll N$. Accurate recovery of the original 
signal is possible through the solution of the following convex 
optimization problem 
\cite{CRT:2006a,CRT:2006b}:
\begin{equation} \label{eq:CS_1_2}
\min \|\mathbf{a}\|_1 \quad \mbox{subject}\quad \mbox{to} \quad
\mathcal{G} \cdot \mathbf{a} = \mathbf{X}, 
\end{equation} 
where
\begin{equation}
\|\mathbf{a}\|_1=\sum_{i=1}^{N}|\mathbf{a}_i|
\end{equation} 
is the $l_{1}$ norm of vector $\mathbf{a}$.

The general principle of solving the convex optimization problem
Eq.~(\ref{eq:CS_1_2}) can be described briefly, as follows. By introducing a
new variable vector $\mathbf{u}\in \mathbb{R}^N$, Eq.~(\ref{eq:CS_1_2}) 
can be recast into a linear constraint minimization problem
\cite{CRT:2006a,CRT:2006b}:
\begin{equation} \label{eq:CS_2_1}
\min  \sum_{i=1}^{N}u_i \quad \mbox{subject}\quad \mbox{to}\quad
\begin{cases} a_i-u_i\leq 0 \\ - a_i-u_i\leq 0\\ 
\mathcal{G} \cdot \mathbf{a} = \mathbf{X} \end{cases}.
\end{equation} 
Defining $\mathbf{z}=[\mathbf{a}^T,\mathbf{u}^T]^T$, we can rewrite 
Eq.~(\ref{eq:CS_2_1}) as
\begin{equation} \label{eq:CS_2_2}  
\langle c_0, z\rangle \quad
\mbox{subject}\quad \mbox{to}\quad
\begin{cases}f_i(z)\leq 0 \\ f'_i(z)\leq 0\\ 
\mathcal{G}_0\cdot \mathbf{z} = \mathbf{X} \end{cases},
\end{equation} 
where $f_i(z)=\langle \mathbf{c}_i, \mathbf{z}\rangle$ 
and $f'_i(z)=\langle \mathbf{c'}_i, \mathbf{z}\rangle$ 
($\langle\cdot\rangle$ denotes the inner product of the two
vectors). Here, $\mathbf{c}_0$, $\mathbf{c}_i$, 
$\mathbf{c}'_i\in\mathbb{R}^{2N}$, $\mathcal{G}_0$ is a
$M\times 2N$ matrix, $(\mathbf{c}_0)_j=0$ for $j\leq N$ 
and $(\mathbf{c}_0)_j=1$ for $j> N$; $(\mathbf{c}_i)_j=1$ 
for $j=i$, $(\mathbf{c}_i)_j=-1$ for $j=N+i$;
$(\mathbf{c}'_i)_j=1$ for $j=i$, $(\mathbf{c}'_i)_j=-1$ for $j=N+i$;
$\mathcal{G}_0=[\mathcal{0}^{M\times N},\mathcal{G}]$. To solve the 
linear constraint minimization problem in Eq.~(\ref{eq:CS_2_2}), one 
can use the Karush-Kuhn-Tucker conditions~\cite{Ye:1999}, i.e., at the 
optimal point $\mathbf{z}^*$, there exist vectors $\mathbf{v}^*\in\mathbb{R}^M$,
$\mathbf{q}^*\in\mathbb{R}^N$, $\mathbf{q}'^*\in\mathbb{R}^N$,
where $(\mathbf{q}^*)_j \geq 0$ and $(\mathbf{q}^{'*})_j\geq 0$ 
for $j = 1, \ldots, N$, such that the following hold:
\begin{eqnarray} \label{eq:CS_2_3}
\mathbf{c}_0 + \mathcal{G}_0^T \cdot \mathbf{v}^* & + & \sum_i 
q_i^* \mathbf{c}_i+\sum_i q_i^{'*}\mathbf{c}'_i=\bf{0}, \\ \nonumber
q_i^*f_i(z^*) & = & 0,i=1,...,N, \\ \nonumber
q_i^{'*}f'_i(z^*) & = & 0,i=1,...,N, \\ \nonumber
\mathcal{G}_0 \cdot \mathbf{z}^* & = & \mathbf{X}, \\ \nonumber
f_i(z^*)& \leq & 0, \\ \nonumber
f'_i(z^*) & \leq & 0.
\end{eqnarray}
Equation~(\ref{eq:CS_2_3}) can be solved by the classical Newton method 
in the valid solution set determined by the inequality constraints
\begin{equation} \label{eq:CS_valid}
\{q_i\geq0,q_i^{'}\geq 0,f_i(z)\leq 0,f'_i(z)\leq 0\},
\end{equation} 
where a point $(\mathbf{z},\mathbf{v},\mathbf{q},\mathbf{q}')$ in this set is 
called as an interior point. Define a residual vector for all the equality
conditions in Eq.~(\ref{eq:CS_2_3}) as 
$\mathbf{r}=[\mathbf{r}_1^T,\mathbf{r}_2^T,\mathbf{r}_3^T,\mathbf{r}_4^T]^T$, 
with $\mathbf{r}_i$ ($i=1,\ldots,4$) given by
where
\begin{eqnarray} \label{eq:CS_2_4}
\mathbf{r}_1 & = & \mathbf{c}_0 + \mathcal{G}_0^T \cdot \mathbf{v} 
+\sum_i q_ic_i+\sum_i q_i^{'} \mathbf{c}'_i, \\ \nonumber
\mathbf{r}_2 & = & - \mathbf{\lambda} \cdot \mathbf{f}, \\ \nonumber
\mathbf{r}_3 & = & - \mathbf{\lambda}'\cdot \mathbf{f}', \\ \nonumber
\mathbf{r}_4 & = & \mathcal{G}_0 \cdot \mathbf{z} - \mathbf{X},
\end{eqnarray}
where $\mathbf{f}=[f_1(z),f_2(z),...,f_N(z)]^T$,
$\mathbf{f}'=[f'_1(z),f'_2(z),...,f'_N(z)]^T$, $\mathbf{\lambda}$ 
and $\mathbf{\lambda}'$ are diagonal matrices with 
$(\mathbf{\lambda})_{ii}=q_i$ and $(\mathbf{\lambda}')_{ii}=q'_i$. 
To find the solution to Eq.~\eqref{eq:CS_2_3}, one linearizes the 
residual vector $r$ using the Taylor expansion about the point 
$(\mathbf{z},\mathbf{v},\mathbf{q},\mathbf{q}')$, which gives
\begin{equation} \label{eq:CS_2_5}
\mathbf{r}(\mathbf{z}+\Delta \mathbf{z}, \mathbf{v}+\Delta \mathbf{v}, 
\mathbf{q} +\Delta \mathbf{q}, \mathbf{q}'+\Delta \mathbf{q}')  
= \mathbf{r}(\mathbf{z}, \mathbf{v}, \mathbf{q}, \mathbf{q}')
+ \mathcal{J}(\mathbf{z},\mathbf{v},\mathbf{q},\mathbf{q}') \cdot \left(
          \begin{array}{c}
          \Delta \mathbf{z} \\
         \Delta \mathbf{v} \\
         \Delta \mathbf{q} \\
         \Delta \mathbf{q}' \\
        \end{array}
        \right),
\end{equation} 
where $\mathcal{J}$ is the Jacobian matrix of $\mathbf{r}$ given by
\begin{equation}
\mathcal{J} = \left(
\begin{array}{cccc}
\mathbf{0} & \mathcal{G}_0^T & \mathcal{C}^T & \mathcal{C}'^T \\
-\mathbf{\lambda}\cdot\mathcal{C} & \mathbf{0} & -\mathcal{F} & \mathbf{0} \\
-\mathbf{\lambda}'\cdot\mathcal{C}' & \mathbf{0}& \mathbf{0} & -\mathcal{F}' \\
\mathcal{G}_0 & \mathbf{0} & \mathbf{0} & \mathbf{0} \\
    \end{array}
  \right),
\end{equation} 
the $N\times 2N$ matrices $\mathcal{C}$ and $\mathcal{C}'$ have $\mathbf{c}_i$ 
and $\mathbf{c}'_i$ as rows, respectively, $\mathcal{F}$ and $\mathcal{F}'$ 
are diagonal matrices with $(\mathcal{F})_{ii}=f_i(\mathbf{z})$ and 
$(\mathcal{F}')_{ii}=f'_i(\mathbf{z})$. The steepest descent direction 
can be obtained by setting zero the left-hand side of Eq.~(\ref{eq:CS_2_5}),
which gives 
\begin{equation} \label{2_6}
\left( \begin{array}{c}
      \Delta \mathbf{z} \\
   \Delta \mathbf{v} \\
      \Delta \mathbf{q} \\
  \Delta \mathbf{q}' \\
   \end{array}
       \right)=
       -\mathcal{J}^{-1} \cdot \mathbf{r}.
\end{equation} 
With the descent direction so determined, to solve Eq.~\eqref{eq:CS_2_3}, 
one can update the solution by 
$\mathbf{z} = \mathbf{z}+s\Delta\mathbf{z}$,
$\mathbf{v}=\mathbf{v}+s \Delta\mathbf{v}$, 
$\mathbf{q}=\mathbf{q}+s \Delta\mathbf{q}$,
$\mathbf{q}'=\mathbf{q}'+s \Delta\mathbf{q}'$ 
with step length $s$, where $s$ should be chosen to guarantee that 
$(\mathbf{z}+s\Delta\mathbf{z},\mathbf{v}+s\Delta\mathbf{v},
\mathbf{q} + s\Delta\mathbf{q},\mathbf{q}'+s\Delta\mathbf{q}')$ is an 
interior point of the valid solution set Eq.~\eqref{eq:CS_valid}. Iterating 
this procedure gives the reconstructed sparse signal $\mathbf{a}$.

\subsection{Mathematical formulation of systems identification based on 
compressive sensing} \label{subsec:CS_NLD}

The inverse problem of identifying nonlinear dynamical
systems can be cast in the framework of compressive sensing so that
optimal solutions can be obtained even when the number of base
coefficients to be estimated is large and/or the amount of available
data is small. Consider systems described by
\begin{equation} \label{eq:C2_DS_model}
\dot{{\bf x}} = {\bf F}({\bf x}), 
\end{equation}
where ${\bf x} \in R^m$ represents the set of externally accessible
dynamical variables and ${\bf F}$ is a smooth vector function in
$R^m$. The $j$th component of ${\bf F}({\bf x})$ can be represented
as a power series:
\begin{eqnarray} \label{eq:C2_DS_Expansion}
[{\bf F}({\bf x})]_j = \sum_{l_1=0}^{n}\sum_{l_2=0}^{n}
\cdots\sum_{l_m=0}^{n} (a_j)_{l_1,\cdots,l_m} \cdot x_1^{l_1}
x_2^{l_2} \cdots x_m^{l_m},
\end{eqnarray}
where $x_k$ ($k=1,\cdots, m$) is the $k$th component of the
dynamical variable, and the scalar coefficient of each product term
$(a_j)_{l_1,\cdots,l_m}\in R$ is to be determined from measurements.
Note that the terms in Eq.~(\ref{eq:C2_DS_Expansion}) are all 
possible products of different components with different powers, and 
there are $(1+n)^m$ terms in total.

Without loss of generality, one can examine
one dynamical variable of the system. (Procedures for other
variables are similar.) For example, to construct the measurement
vector ${\bf X}$ and the matrix ${\bf G}$ for the case of $m = 3$
(dynamical variables $x$, $y$, and $z$) and $n = 3$, one has the
following explicit equation for the first dynamical variable: 
\begin{displaymath}
[{\bf F}({\bf x})]_1 \equiv (a_1)_{0,0,0}x^0y^0z^0 +
(a_1)_{1,0,0}x^1y^0z^0 + \cdots + (a_1)_{3,3,3}x^3y^3z^3. 
\end{displaymath}
Denote the coefficients of $[{\bf F}({\bf x})]_1$ by ${\bf a}_1 =
[(a_1)_{0,0,0},(a_1)_{1,0,0},\cdots ,(a_1)_{3,3,3}]^T$. Assuming
that measurements of ${\bf x}(t)$ at a set of time
$t_1,t_2,\ldots,t_w$ are available, one can write
\begin{displaymath}
{\bf g}(t) = \big[
x(t)^0y(t)^0z(t)^0, x(t)^0y(t)^0z(t)^1, \cdots, x(t)^3y(t)^3z(t)^3
\big] 
\end{displaymath}
such that $[{\bf F}({\bf x}(t))]_1 = {\bf g}(t) \cdot {\bf a}_1$. From 
the expression of $[{\bf F}({\bf x})]_1$, one can choose the 
measurement vector as 
\begin{displaymath}
\mathbf{X}=\left[\dot{x}(t_1),\dot{x}(t_2), \cdots,\dot{x}(t_w)\right]^T,
\end{displaymath}
which can be calculated from time series. Finally, one obtains the
following equation in the form 
$\mathbf{X} = \mathcal{G} \cdot \mathbf{a}_1$:
\begin{eqnarray}\label{eq:YeqsAX}
 \left ( \begin{array}{cc}
  \dot{x}(t_1)\\
  \dot{x}(t_2)\\
  \vdots \\
  \dot{x}(t_w)\\
\end{array}\right )=
\left(
\begin{array}{c}
  {\bf g}(t_1) \\
  {\bf g}(t_2) \\
  \vdots  \\
  {\bf g}(t_w) \\
\end{array}
\right ) \left(
\begin{array}{c}
   {\bf a}_1
\end{array}
\right).
\end{eqnarray}
To ensure the restricted isometry property~\cite{CRT:2006a,CRT:2006b}, one 
normalizes $\mathcal{G}$ by dividing elements in each column by the $L_2$
norm of that column: $(\mathcal{G}')_{ij} = (\mathcal{G})_{ij}/L_2(j)$ with
$L_2(j) =\sqrt{ \sum_{i=1}^{M} [(\mathcal{G})_{ij}]^2}$, so that
$\mathbf{X} = \mathbf{\mathcal{G}'} \cdot \mathbf{a}_1'$. After the
normalization, ${\bf a_1'}={\bf a_1}\cdot L_2$ can be determined via
some standard compressive-sensing algorithm~\cite{CRT:2006a,CRT:2006b}.
As a result, the coefficients ${\bf a_1}$ are given by ${\bf a_1'}/L_2$.
To determine the set of power-series coefficients corresponding to a
different dynamical variable, say $y$, one simply replaces the
measurement vector by ${\bf X} =\left[\dot{y}(t_1),\dot{y}(t_2),
\cdots,\dot{y}(t_w)\right]^T$ and use the same matrix $\mathcal{G}$.
This way all coefficients ${\bf a_1}$, ${\bf a_2}$, and ${\bf a_3}$
of three dimensions can be estimated.

\subsection{Reconstruction and identification of chaotic systems}
\label{subsec:chaotic_systems}

The problem of predicting dynamical systems based on time series
has been outstanding in nonlinear dynamics because, despite previous 
efforts~\cite{KS:book} in exploiting the delay-coordinate embedding 
method~\cite{Takens:1981,PCFS:1980} to decode the topological properties 
of the underlying system, how to accurately infer the underlying {\em
nonlinear system equations} remains largely an unsolved problem.
In principle, a nonlinear system can be approximated by a large
collection of linear equations in different regions of the phase
space, which can indeed be accomplished through reconstructing the Jacobian
matrices on a proper grid that covers the phase-space region of
interest~\cite{FS:1987,Sauer:2004}. However, the accuracy and
robustness of the procedure are challenging issues, including
the difficulty with the required computations. For example, in order
to be able to predict catastrophic bifurcations, local reconstruction
of a large set of linearized dynamics is not sufficient but
rather, an accurate prediction of the underlying nonlinear
equations themselves is needed.

In 2011 it was proposed~\cite{WYLKG:2011} that compressive sensing 
provides a powerful method for data based nonlinear systems 
identification, based on the principle that it is possible to fully 
reconstruct dynamical systems from time series because the dynamics of 
many natural and man-made systems are determined by functions that can be
approximated by series expansions in a suitable base. The task
is then to estimate the coefficients in the series representation.
In general, the number of coefficients to be estimated can be large
but many of them may be zero (the sparsity condition). According to the
conventional wisdom this would be a difficult problem as a large
amount of data is required and the computations involved can be
extremely demanding. However, compressive sensing~\cite{CRT:2006a,
CRT:2006b,Candes:2006,Donoho:2006,Baraniuk:2007,CW:2008}. 
provides a viable solution to the problem, where the basic principle 
is to reconstruct a sparse signal from small amount of observations,
as measured by linear projections of the original signal on a few 
predetermined vectors. 

\begin{figure}
\centering
\includegraphics[width=\linewidth]{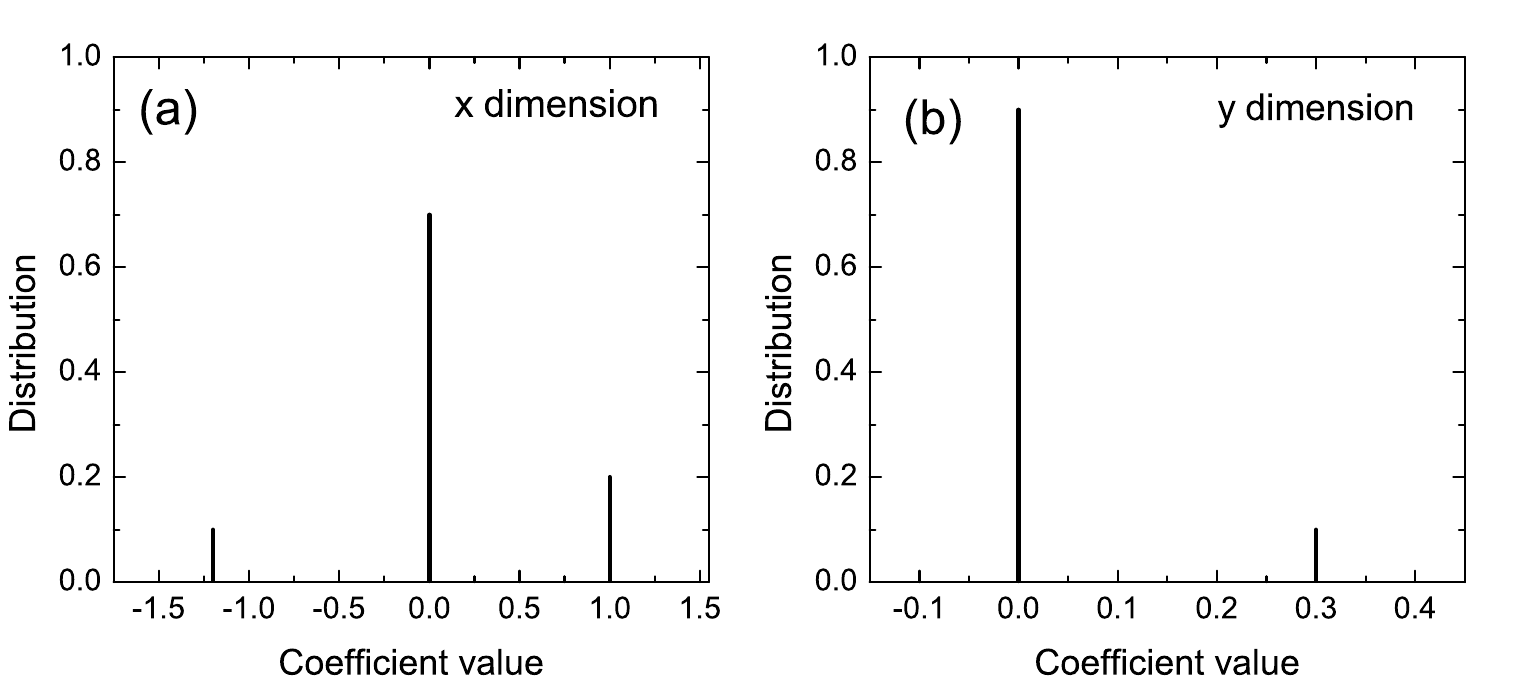}
\caption{\small {\bf An example of chaotic systems identification based
on compressive sensing}. For the H\'{e}non map, in (a) $x$ dimension 
and (b) $y$ dimension, distributions of the predicted values of ten 
power-series coefficients up to order 3: constant, $y$, $y^2$, $y^3$, 
$x$, $xy$, $xy^2$, $x^2$, $x^2y$ and $x^3$. From Ref.~\cite{WYLKG:2011}
with permission.} 
\label{fig:distribution}
\end{figure}

In Ref.~\cite{WYLKG:2011}, a number of classic chaotic systems were
used to demonstrate the compressive sensing based approach. Here we
quote one example, the H\'{e}non map~\cite{Henon:1976}, a classical 
model that has been used to address many fundamental issues in chaotic 
dynamics. The prediction of the map equations resembles that of a vector 
field. The map is given by: $(x_{n+1},y_{n+1}) = (1 - a x_n^2 + y_n, bx_n)$, 
where $a$ and $b$ are parameters. For $b = 0.3$, the map exhibits
periodic and chaotic attractors for $a < a_c \approx 1.42625$, where $a_c$ 
is the critical parameter value for a boundary 
crisis~\cite{GOY:1983a,Ott:book,LT:book}, above which almost all 
trajectories diverge. Assuming power-series expansions up to order $3$ 
in the map equations, the authors were able to identify the map 
coefficients with high accuracy using only a few 
data points. Figure~\ref{fig:distribution} shows the distributions of 
the estimated power-series coefficients, where extremely narrow peaks 
about zero indicate that a large number of the coefficients are effectively 
zero, which correspond to nonexistent terms in the map equations.
Coefficients that are not included in the zero peak correspond then 
to the existent terms and they determine the predicted map equations.
Note that, to predict correctly the map equations, the number of required 
data points is extremely low. Similar results were obtained~\cite{WYLKG:2011} 
for the classic standard map~\cite{CI:1973,Chirikov:1979}, the chaotic Lorenz 
system~\cite{Lorenz:1963}, and the chaotic R\"{o}ssler 
oscillator~\cite{Rossler:1976}. 

\begin{figure}
\centering
\includegraphics[width=\linewidth]{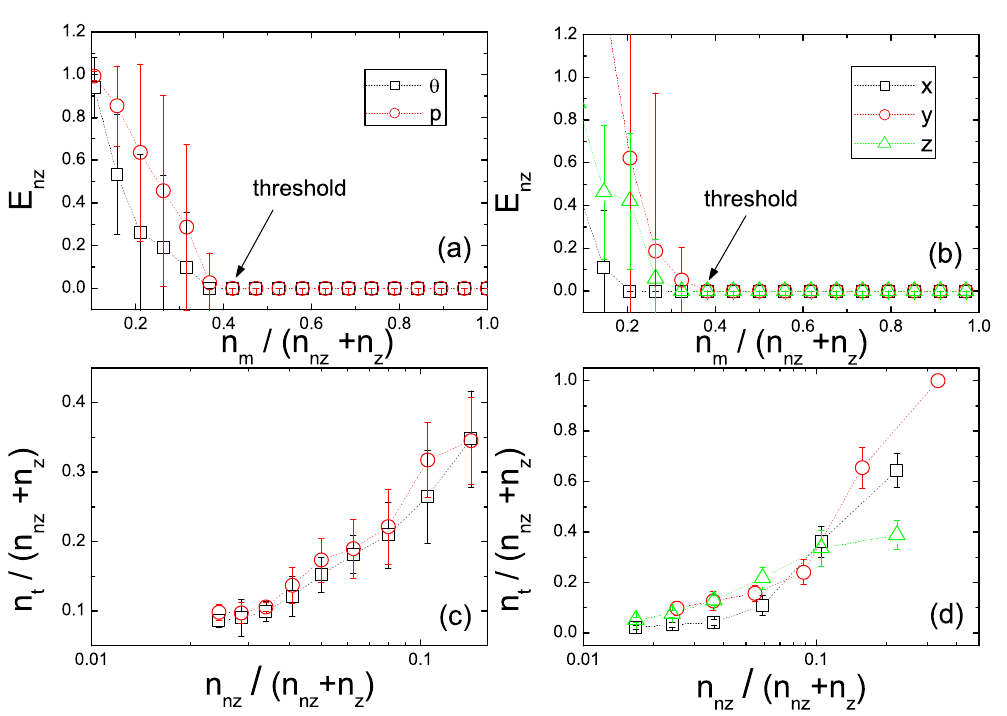}
\caption{\small {\bf Performance characterization of compressive sensing
based chaotic systems identification.} Prediction
errors $E_{nz}$ in dynamical equations as a function of the ratio of
the number $n_m$ of measurements to the total number $n_{nz}+n_z$ of
terms to be predicted for (a) the standard map and (b) the Lorenz
system. The ratio of the threshold $n_t$ to $n_{nz}+n_z$ for
different equations as a function of the ratio $n_{nz}/(n_{nz}+n_z)$
for (c) the standard map and (d) the Lorenz system. In (a) and (b),
$n_{nz}+n_z$ is 20 and 35, respectively. The error bars represent
the standard deviations obtained from 30 independent realizations.
In (c) and (d), $n_{nz}+n_z$ can be adjusted by the order of the power
series. In (c), the points range from order 3 to order 11, and
in (d) from order 2 to order 7, and $E_{nz}$ and $E_z$
exhibit the same threshold. From Ref.~\cite{WYLKG:2011} with permission.}
\label{fig:measurement}
\end{figure}

To quantify the performance of the compressive sensing based systems 
identification method with respect to the amount of required data,  
the prediction errors were calculated~\cite{WYLKG:2011}, which are 
defined separately for nonzero (existing) and zero terms in the
dynamical equations. The relative error of a nonzero term is defined
as the ratio to the true value of the absolute difference between
the predicted and true values. The average over the errors of all
terms in a component is the prediction error $E_{nz}$ of nonzero
terms for the component. In contrast, the absolute error $E_z$ is
used for zero terms. Figures~\ref{fig:measurement}(a) and 
\ref{fig:measurement}(b) show $E_{nz}$ as a function of the ratio 
of the number $n_m$ of measurements to the total number $n_{nz} + n_z$ 
of terms to be predicted, for the standard map and the Lorenz
system, respectively. Note that, for the standard map, it is
necessary to explore alternative bases of expansion so that the
sparsity condition can be satisfied. A practical strategy is that, assuming
that a rough idea about the basic physics of the underlying dynamical
system is available, one can choose a base that is compatible with
the knowledge. In the case of the standard map, a base including 
the trigonometric functions can be chosen. The results in 
Figs.~\ref{fig:measurement}(a) and \ref{fig:measurement}(b)
indicate that, when the number $n_m$ of measurements exceeds a threshold 
$n_t$, $E_{nz}$ becomes effectively zero. For convenience, one can
define $n_t$ by using the small threshold value $E_{nz} = 10^{-3}$
so that $n_t$ is the minimum number of required measurements for an
accurate prediction. Figures~\ref{fig:measurement}(a) and
\ref{fig:measurement}(b) show that $n_t$ is much less than
$n_{nz} + n_z$ if $n_{nz}$, the number of nonzero terms is small.
The performance of the method can thus be quantified by the
threshold with respect to the numbers of measurements and terms to
be predicted. As shown in Figs.~\ref{fig:measurement}(c) and
\ref{fig:measurement}(d) for the standard map and the Lorenz system,
respectively, as the nonzero terms become sparser among all terms to
be predicted (characterized by a decrease in $n_{nz}/(n_{nz}+n_z)$
when $n_{nz}+n_z$ is increased), the ratio of the threshold $n_t$ to
the total number of terms $n_{nz}+n_z$ becomes smaller. These
results demonstrated the advantage of the compressive-sensing based
method to predict dynamical systems, i.e., high accuracy and
extremely low required measurements. In general, to predict a 
nonlinear dynamical system as accurately as possible, many
reasonable terms should be assumed in the expansions, insofar as the
percentage of nonzero terms is small so that the sparsity condition
of compressive sensing is met.

There are situations where the system is high-dimensional or
stochastic. A possible solution is to employ the Bayesian inference 
to determine the system equations. In general the computational 
challenge associated with the approach can be formidable, but the 
power-series or more general expansion based compressive-sensing method 
may present an effective strategy to overcome the difficulty.

\subsection{Predicting catastrophe in nonlinear dynamical systems}
\label{subsec:pred_catastrophe}

\subsubsection{Predicting catastrophic bifurcations based on compressive
sensing} \label{subsubsec:pred_crises}

Nonlinear dynamical systems, in their parameter space, can often exhibit 
catastrophic bifurcations that ruin the desirable or ``normal'' 
state of operation. Consider, for example, the phenomenon of 
crisis~\cite{GOY:1983a} where, as a system parameter is changed, 
a chaotic attractor collides with its own basin boundary and is 
suddenly destroyed. After the crisis, the state of the system is 
completely different from that associated with the attractor 
before the crisis. Suppose that the state before 
the crisis is normal and desirable and the state after the crisis 
is undesirable or destructive. The crisis can thus be regarded as a 
{\em catastrophe} that one strives to avoid at all cost. Catastrophic
events, of course, can occur in different forms in all kinds of
natural and man-made systems. A question of paramount importance
is how to predict catastrophes in advance of their possible occurrences. 
This is especially challenging when the equations of the underlying 
dynamical system are unknown and one must then rely on measured time 
series or data to predict any potential catastrophe.

Compressive sensing based nonlinear systems identification 
provides an approach to forecasting potential catastrophic 
bifurcations~\cite{WYLKG:2011}. Assume that an accurate model 
of the system is not available, i.e., the system equations are 
unknown, but the time evolutions of the key variables of the 
system can be accessed through monitoring or measurements. The 
method~\cite{WYLKG:2011} thus consists of three steps: (i) predicting
the dynamical system based on time series, (ii) identifying the
parameters of the system, and (iii) performing a computational bifurcation
analysis using the predicted system equations to locate potential
catastrophic events in the parameter space so as to determine the
likelihood of system's drifting into a catastrophe regime. In
particular, if the system operates at a parameter setting close to
such a critical bifurcation, catastrophe is imminent as a small
parameter change or a random perturbation can push the system
beyond the bifurcation point. Consider the concrete case of crises
in nonlinear dynamical systems. Once a complete set of system
equations has been predicted and the parameters have been
identified, one proceeds to examine the available parameter space.
It should be noted that, to explore the multi-parameter space of a
dynamical system can be challenging, but this can often
lead to the discovery of new phenomena in dynamics. Examples are
the phenomenon of double crises in systems with two bifurcation 
parameters~\cite{SUGY:1995} and the hierarchical structures in the 
parameter space~\cite{BG:2008,SBU:2010}. 

\begin{figure}
\centering
\includegraphics[width=\linewidth]{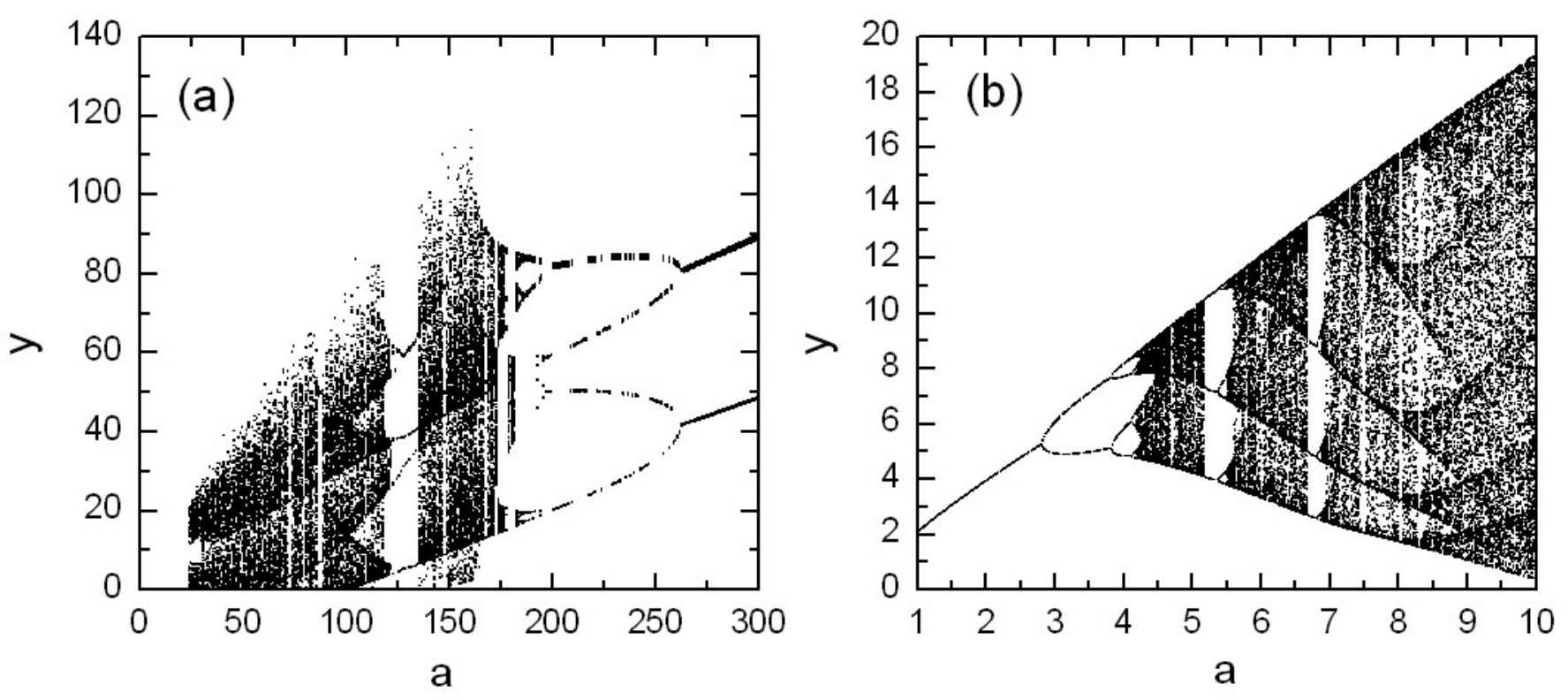}
\caption{\small {\bf Bifurcation diagrams predicted by the compressive
sensing based method of nonlinear systems identification.}
(a) A predicted Lorenz system given by  
$\dot{x}=10.000548307148881\times y - 10.001897147696283\times x$,
$\dot{y} = x(a- 1.000933186801829\times z)-1.000092963203845\times
y$, $\dot{z}=0.999893761636553\times xy - 2.666477325955504\times
z$. (b) A predicted R\"ossler system given by $\dot{x} =
-0.999959701293536\times y - 0.999978902248041\times z$, $\dot{y}
= 1.000004981649221\times x + 0.200005996113158\times y$, $\dot{z}
= 0.199997011085648 + 0.999999156496251\times z( x - a)$. In both 
cases, $n_m = 18$ and $n_{nz} + n_z = 35$. From Ref.~\cite{WYLKG:2011} 
with permission.}
\label{fig:Lorenz_Rossler}
\end{figure}

In Ref.~\cite{WYLKG:2011}, a number of examples were given in which
the bifurcation diagrams computed from the predicted system equations
agree well with those from the original systems, so all possible 
critical bifurcation points can be predicted accurately based on time 
series only. Figures~\ref{fig:Lorenz_Rossler}(a) and 
\ref{fig:Lorenz_Rossler}(b) illustrate a predicted bifurcation 
diagram from the chaotic Lorenz and R\"{o}ssler systems, respectively.

\subsubsection{Using compressive sensing to predict tipping points in 
complex systems} \label{subsubsec:pred_tipping_points}

It is increasingly recognized that a variety of complex dynamical
systems ranging from ecosystems and the climate to economical, social,
and infrastructure systems can exhibit {\em tipping points} at which
irreversible transition from a normal functioning state to a catastrophic
state occurs~\cite{Gladwell:book,Schefferetal:2009}. Examples of such
transitions are blackouts in the power grids, sudden emergence of massive
jamming in urban traffic systems, the shutdown of the thermohaline
circulation in the North Atlantic~\cite{Rahmstorf:2002}, sudden extinction
of species in ecosystems~\cite{DJ:2010,DVKG:2012,TC:2014,LVSB:2014}, and
the occasional switches of shallow lakes from clear to turbid
waters~\cite{Scheffer:2004}. In fact, the seemingly abrupt transitions
are the consequence of {\em gradual} changes in the system which can, for
example, be caused by slow drifts in the external conditions. To understand
the dynamical properties of the system near a tipping point, to predict the
tendency for the system to drift toward the tipping point, to issue early
warnings, and finally to apply control to reverse or to slow down the trend,
are significant but extremely challenging research problems. Compounded with
the difficulty is the fact the complex systems are often {\em interdependent
and non-stationary}. For example, the evolution of an ecosystem depends on
human behaviors, which in turn affects the well being of the human society.
Infrastructure systems such as the power grids and communication networks
are interdependent upon each other~\cite{BPPSH:2010,GBSH:2012}, both being 
affected by human activities (social system). The concept of interdependence 
is prevalent in many disciplines. At the present there is little understanding 
of tipping points in interdependent complex systems in terms of their 
emergence and dynamical properties.

\begin{figure}
\centering
\includegraphics[width=\linewidth]{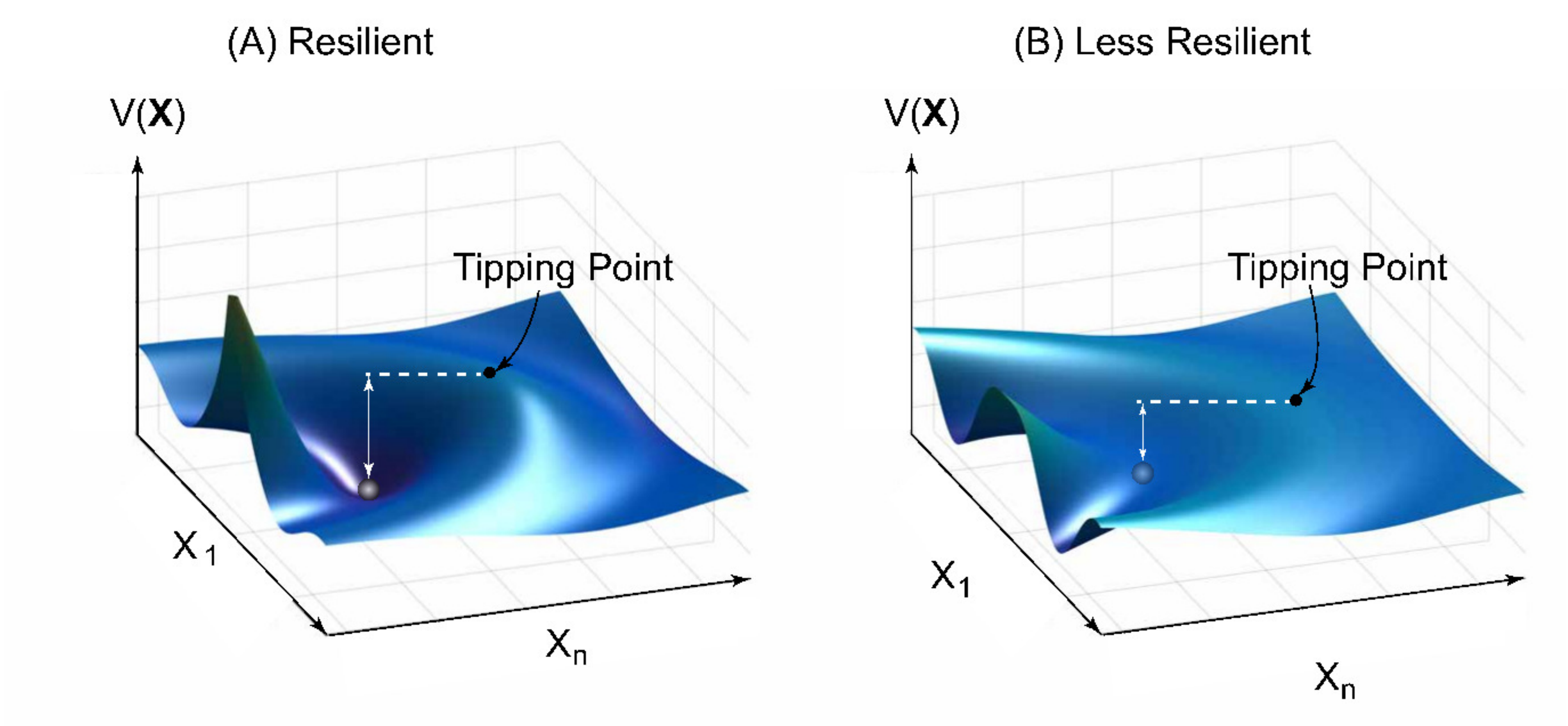}
\caption{\small {\bf Relationship between tipping point and resilience}. 
For a real world complex dynamical system, the potential landscape is 
time-dependent or {\em non-stationary}: there are slow but ultimately large
drifts in the system parameters and equations leading to catastrophic
collapse, and there is fast but small random perturbation. The origins
of these drifts and perturbation are discussed in the text.}
\label{fig:TP_Resilience}
\end{figure}

In a dynamical system, the existence of one or several tipping points is
intimately related to the concept of {\em resilience}~\cite{Scheffer:2010,
Linkovetal:2013,PSRCL:2013,PSR:2013,DCvanNS:2015,GBB:2016}, 
which can be understood
heuristically by resorting to the intuitive picture of a ball moving in
a valley under gravity, as shown in Fig.~\ref{fig:TP_Resilience}. To the
right of the valley there is a hill, or a potential barrier in terms of
classical mechanics. The downhill side to the right of the barrier corresponds
to a catastrophic behavior. Normal functioning of the system is represented
as the confinement of ball's motion within the valley.
If the valley is sufficiently deep (or the height of the barrier
is sufficiently large), as shown in Fig.~\ref{fig:TP_Resilience}(a), there
will be little probability for the ball to move across the top of the
barrier towards catastrophic behavior, implying that the system is more
resilient to random noise or external disturbances. However, if the
barrier height is small, as shown in Fig.~\ref{fig:TP_Resilience}(b), the
system is less resilient as small perturbation can push the ball over
to the left side of the barrier. The top of the barrier thus corresponds
to a tipping point, across which from the left the system will
essentially collapse. In mechanics, the situation can be
formulated using a potential function so that, mathematically,
the motion of the ball can be described by the Hamilton's
equations~\cite{Goldstein:book}. Given a dynamical system, if the potential
landscape can be determined, it will be possible to locate the tipping
points. In systems biology, the potential function is called the Waddington
landscape~\cite{Waddington:book}, which essentially determines the biological
paths for cell development and differentiation~\cite{Huang:2009,MML:2009}
- the {\em landscape metaphor}. In the past few years, a quantitative
approach to mapping out the potential landscape for gene circuits or gene 
regulatory networks has been developed~\cite{WXW:2008,WXWH:2010,WZXW:2011,
ZXZWW:2012}. In nonlinear dynamical systems, a similar concept exists -
{\em quasipotential}~\cite{GT:1984,GHT:1991,TL:2010}.

Because of non-stationarity, for a complex system in the real world, 
Fig.~\ref{fig:TP_Resilience} in fact represents only a ``snapshot'' of
the potential landscape. For complex dynamical systems, the potential
landscape must necessarily be {\em time-dependent or non-stationary},
which completely governs the emergence of tipping points and the
global dynamics of the system. There are two types of time-varying
disturbances to the landscape: (1) slow but eventually large changes
in parameters and system equations and (2) fast but small random
perturbation. The physical origins of these disturbances can be
argued, as follows.

Take, for example, the challenge of adding distributed renewable
energy generation to the power grid. The time scale for changes is
years. Over that time period, significant new renewables can be added, 
and yet the precise timing, location, and amount of distributed 
renewable energy generation is unpredictable, because it is not possible to
know how social decisions (in terms of regulation, business planning,
and consumer choice) will play out. (A similar situation occurs for climate
change: slow, long-term, secular, and nonlinear changes in climatic
averages would occur, which will impact generation via water availability and
temperature, etc. At the same time, weather remains a high frequency pattern
on top of these slow changes.) Clearly the current grid is largely stable
vis-a-vis existing distributions of weather variables, but how will that
change in the future? For example, in the future, charging networks may
comprise mainly slow charging stations~\cite{GW:2009} as the widespread use
of fast charging stations would raise the power demand in the
electrical grid~\cite{MRS:2013}, causing tremendous difficulties for the
managers to operate the grid. There can also be changing regulatory
incentives or management structures. All these can lead to {\em social
non-stationarity}. Mathematically, the social factors can be modeled by
adiabatic changes to the system, which affect the potential landscape on
time scales that are slower than that of the intrinsic dynamics.

At short time scales, precise hourly patterns of electricity generation 
may fluctuate due to changes in wind and cloud activity - a type of
{\em environmental non-stationarity}. There can also be {\em technological
non-stationarity} such as shifting demand patterns associated with
consumer electronics, plug-in hybrid and electric vehicles, etc. These
occur on short time scale, which can be modeled as random perturbation 
or noise to the system.

Referring to Fig.~\ref{fig:TP_Resilience}, the topographic landscape metaphor
of resilience, one can see that, if the system is stationary, there is a fixed
threshold across which the system will collapse. In this case, the system
resilience can simply be characterized by the distance to the threshold.
However, for a non-stationary system, it is not possible to measure the 
threshold distance by establishing the absolute positions of the ball (system)
and the tipping point and then computing the difference between them.
Instead, one must attempt to estimate the differences directly in real time.
Two open questions are: is it possible to determine the non-stationary 
landscape so as to predict the emergence and the locations of the 
time-dependent tipping points? Can human intervention or control strategies 
be developed to prevent or significantly slow down the system's evolution 
toward a catastrophic tipping point? 

At the present there are no answers to these questions of the grand 
challenge nature, but we wish to argue that compressive sensing 
based nonlinear systems identification can provide insights into the 
fundamental issues pertinent to these questions. Consider, 
for example, a complex power grid system, in which the time series 
such as the voltage and power at all generator nodes are available, 
as well as social interaction data that are typically polarized (e.g., 
binary). There were recent efforts in reconstructing complex networks
of nonlinear oscillators based on continuous time series~\cite{WYLKH:2011} 
and in uncovering epidemic propagation networks using binary 
data~\cite{SWFDL:2014} (to be discussed in 
Sec.~\ref{sec:CH3_CS_CN_Reconstruction}).  
In principle, these approaches can be combined to deal with 
nonstationary complex systems. Specifically, large but slowly varying 
physical non-stationarity can be modeled through the appearance of 
additional, concrete mathematical terms involving voltage, phase, 
and current variables, or through the disappearance of certain 
terms. Social non-stationarity can be modeled by functions of Boolean 
variables that generate polarized data or can be reconstructed using 
these data. It would then be possible to establish a 
mathematical framework combining reconstruction methods for continuous 
time series and polarized data. Potentially, this could represent
an innovative and concrete approach to incorporating social data into
a complex physical/technological system and assessing, quantitatively,
the effect of social non-stationarity on the dynamical evolution of
the system.

\subsection{Forecasting future states (attractors) of nonlinear dynamical
systems} \label{subsec:future_states}

A dynamical system in the physical world is constantly subject to 
random disturbance or adiabatic perturbation. Broadly speaking, there 
are two types of perturbations: stochastic or deterministic. Stochastic 
disturbances (or noise) can typically be described by random processes 
and they do not alter the intrinsic structure of the underlying 
equations of the system. Deterministic perturbations, however, can cause
the system equations or parameters to vary with time. Suppose the 
perturbations are adiabatic, i.e., $T_i$, the time scale of the intrinsic 
dynamics of the system, is much smaller than $T_e$, the time scale of the 
external perturbation. In this case, some ``asymptotic states'' or 
``attractors'' of the system can still be approximately defined in a time 
scale that is much larger than $T_i$ but smaller than $T_e$. When the dynamics 
in such a time interval is examined in a long run, the attractor of the 
system will depend on time. Often, one is interested in forecasting the 
``future'' asymptotic states of the system. Take the climate system as an 
example. The system is under random disturbances, but adiabatic perturbations 
are also present, such as CO$_2$ injected into the atmosphere due to human 
activities, the level of which tends to increase with time. The time 
scale for appreciable increase in the CO$_2$ level to occur (e.g., 
months or years) is much larger than the intrinsic time scales of the 
system (e.g., days). The climate system is thus an adiabatically 
time-varying, nonlinear dynamical system. It is of interest to
forecast what the future attractors of the system might be in order to
determine whether it will behave as desired or sustainably. The issue of
sustainability is, of course, critical to many other natural and man-made
systems as well. To be able to forecast the future states of such 
systems is essential to assessing their sustainability.

It was demonstrated that compressive sensing can be exploited for 
predicting the future states (attractors) of adiabatically time
varying dynamical systems~\cite{YLG:2012}. The general problem 
statement is: given a nonlinear dynamical system whose equations 
or parameters vary adiabatically with time, but otherwise are completely 
unknown, can one predict, based solely on measured time series, the 
future asymptotic attractors of the system? To be concrete, consider 
the following dynamical system:
\begin{equation} \label{eq:system_TV}
d{\bf x}/dt = {\bf F}[{\bf x}, {\bf p}(t)],
\end{equation}
where ${\bf x}$ is the dynamical variable of the system in the 
$d$-dimensional phase space and ${\bf p}(t) = [p_1(t),\ldots,p_K(t)]$ 
denotes $K$ independent, time-varying parameters of the system. Assume
that both the form of ${\bf F}$ and ${\bf p}(t)$ are unknown but at time 
$t_M$, the end of the time interval during which measurements are taken, 
the time series ${\bf x}(t)$ for $t_M - T_M \le t \le t_M$ are available, 
where $T_M$ denotes the measurement time window. The idea was to predict,
using compressive sensing, the precise mathematical forms of ${\bf F}$ and 
${\bf p}(t)$ based on the available time series at $t_M$ so that the 
evolution and the likely attractors of the system for $t > t_M$ can be 
computationally assessed and anticipated~\cite{YLG:2012}. The predicted 
forms of ${\bf F}$ and ${\bf p}(t)$ at time $t_M$ would contain errors 
that in general will increase with time. In addition, for $t > t_M$ new 
perturbations can occur to the system so that the forms of ${\bf F}$ and 
${\bf p}(t)$ may be further changed. It is thus necessary to execute
the prediction algorithm frequently using time series available at the
time. In particular, the system could be monitored at all times so that
time series can be collected, and predictions should be carried out at $t_i$'s,
where $\ldots > t_i > \ldots > t_{M+2} > t_{M+1} > t_M$. For any $t_i$,
the prediction algorithm is to be performed based on available time series in
a suitable window prior to $t_i$.

To formulate the problem of predicting time-varying dynamical
systems in the framework of compressive sensing, one can expand all 
components of the time-dependent vector field ${\bf F}[{\bf x}, {\bf p}(t)]$ 
in Eq.~\eqref{eq:system_TV} into a power series in terms of both
dynamical variables ${\bf x}$ and time $t$. The $i$th component
${\bf F}[{\bf x}, {\bf p}(t)]_i$ of the vector field can be written as
\begin{equation} \label{eq:series_TV}
\sum_{l_1,\cdots,l_m=1}^{n} [(\alpha_i)_{l_1,\cdots,l_m}x_1^{l_1} \cdots 
x_m^{l_m}\cdot\sum_{w=0}^{v}(\beta_i)_w t^w]
\equiv \sum_{l_1,\cdots,l_m=1}^{n}\sum_{w=0}^{v}
(c_i)_{l_1,\cdots,l_m;w}x_1^{l_1} \cdots x_m^{l_m}\cdot t^w,
\end{equation}
where $x_k$ ($k=1,\cdots, m$) is the $k$th component of the dynamical 
variable and $c_i$ is the $i$th component of the coefficient vector to be 
determined. Assume that the time evolution of each term can be approximated 
by the power series expansion in time, i.e., 
$\sum_{w=0}^{v}(\beta_i)_w t^w$. The power-series expansion
allows us to cast Eq.~\eqref{eq:system_TV} into the standard
form of compressive sensing, Eq.~\eqref{eq:CS_general}. In principle, 
if every combined scalar coefficient $(c_i)_{l_1,\cdots,l_m;w}$ 
associated with the corresponding term in Eq.~\eqref{eq:series_TV} 
can be determined from time series for $t \le t_M$, the vector field 
component $[{\bf F}({\bf x},{\bf p}(t))]_i$ becomes known.
Repeating the procedure for all components, the entire vector field 
for $t > t_M$ can be found.

\begin{figure}
\centering
\includegraphics[width=\linewidth]{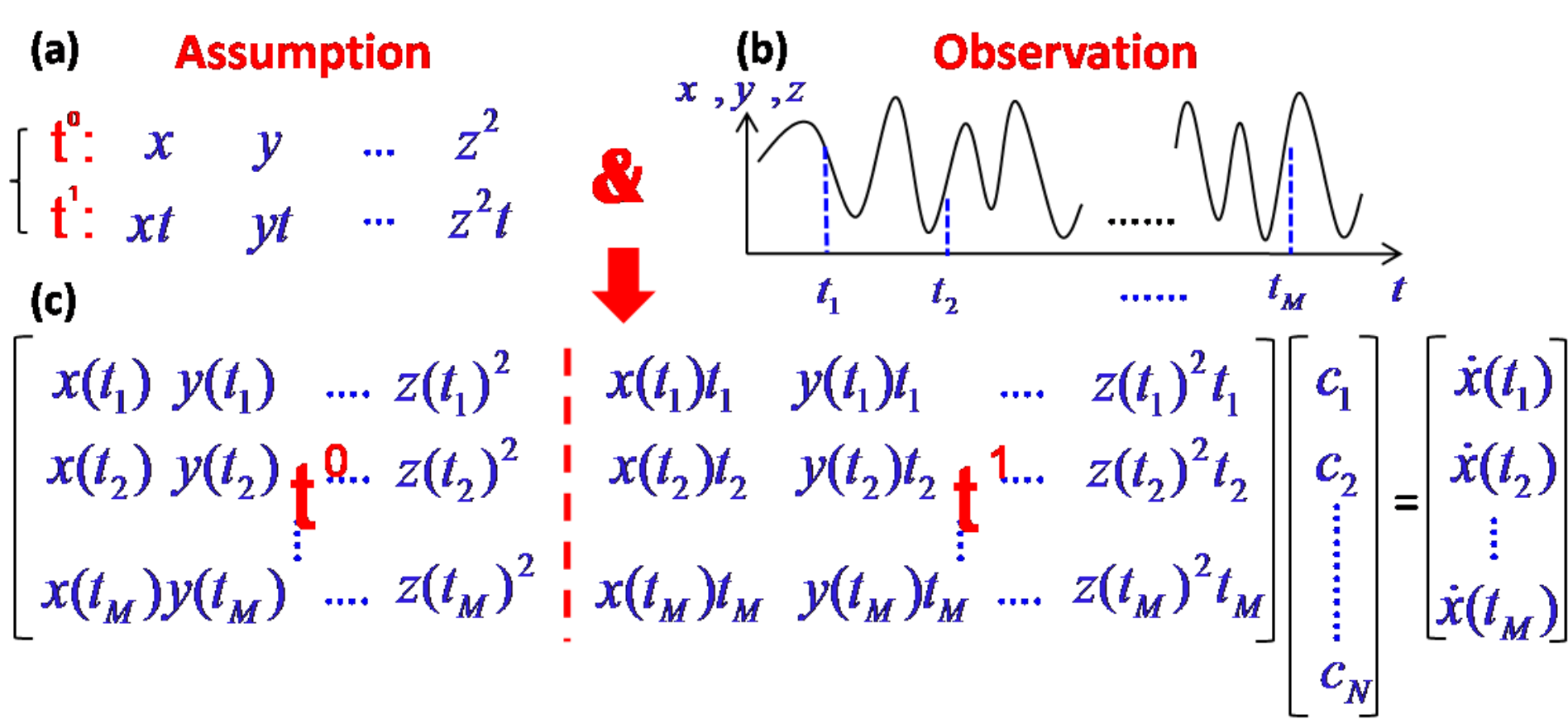}
\caption{\small {\bf Illustration of compressive sensing based scheme for 
predicting future attractors}: the problem of identifying time-varying 
nonlinear dynamical systems can be solved by treating time as an 
independent expansion variable and invoking compressive sensing.
From Ref.~\cite{YLG:2012} with permission.}
\label{fig:schematic_TV}
\end{figure}

To explain the compressive sensing based method in an intuitive way, one can 
consider the special case where the number of components of the dynamical
variables is $m = 3$ ($x$, $y$, and $z$), the order of the power series
is $l_1+l_2+l_3 \leq 2$, and the maximum power of time $t$ in 
Eq.~\eqref{eq:series_TV} is $v=1$, i.e., only the $t^0$ and $t^1$ terms 
are included. Focusing on one dynamical variable, say $x$, the total 
number of terms in the power-series expansion is 20, as specified in 
Fig.~\ref{fig:schematic_TV}(a). Let the measurements 
$x(t)$, $y(t)$, and $z(t)$ be taken at times $t_1, \ t_2, \ \ldots, \ t_M$, 
as shown in Fig.~\ref{fig:schematic_TV}(b). The values of all 
20 power-series terms at these time instants can then be obtained, as 
shown in Fig.~\ref{fig:schematic_TV}(c), where all the terms 
are divided into two blocks according to the distinct powers of the time 
variable $t$: $t^0$ and $t^1$. The projection matrix $\mathcal{G}$ in 
Eq.~\eqref{eq:CS_general} thus consists of these two blocks. (In the
general case where higher powers of the time variable is involved, 
$\mathcal{G}$ would contain a corresponding number of blocks.) The 
components of vector $\mathbf{X}$ in Eq.~\eqref{eq:CS_general} are the 
first derivatives $dx/dt$ evaluated at $t_1, \ t_2, \ \ldots, \ t_M$, 
which can also be approximated by the measured time series $x(t)$ at 
these times. As shown in Fig.~\ref{fig:schematic_TV}(c), 
Eq.~\eqref{eq:series_TV} for this simple example can be 
written in the form of Eq.~\eqref{eq:CS_general}. To ensure sparsity, 
one can assume many terms in the power-series expansion up to some 
high order $n$ so that the total number of terms in 
Eq.~\eqref{eq:series_TV}, $N$, will be quite large. As a
result, $\mathbf{a}$ is high-dimensional but most of its components
are zero. The number of measurements, $M$, needs not be as large as 
$N$: $M \ll N$. Another requirement of compressive sensing is the 
restricted isometric property that can be guaranteed by normalizing the 
matrix $\mathcal{G}$ and by using linear-programming based signal-recovery 
algorithms~\cite{CRT:2006a,CRT:2006b,Candes:2006,Donoho:2006,
Baraniuk:2007,CW:2008}. To determine the set of power-series coefficients 
corresponding to a different dynamical variable, say $y$, one simply 
replaces the measurement vector by
$\mathbf{X} =\left[\dot{y}(t_1),\dot{y}(t_2),\ldots,\dot{y}(t_M)\right]^T$.
The matrix $\mathcal{G}$, however, remains the same. The problem of 
forecasting {\it time-varying} nonlinear dynamical systems then fits 
perfectly into the compressive-sensing paradigm.

\begin{figure}
\centering
\includegraphics[width=\linewidth]{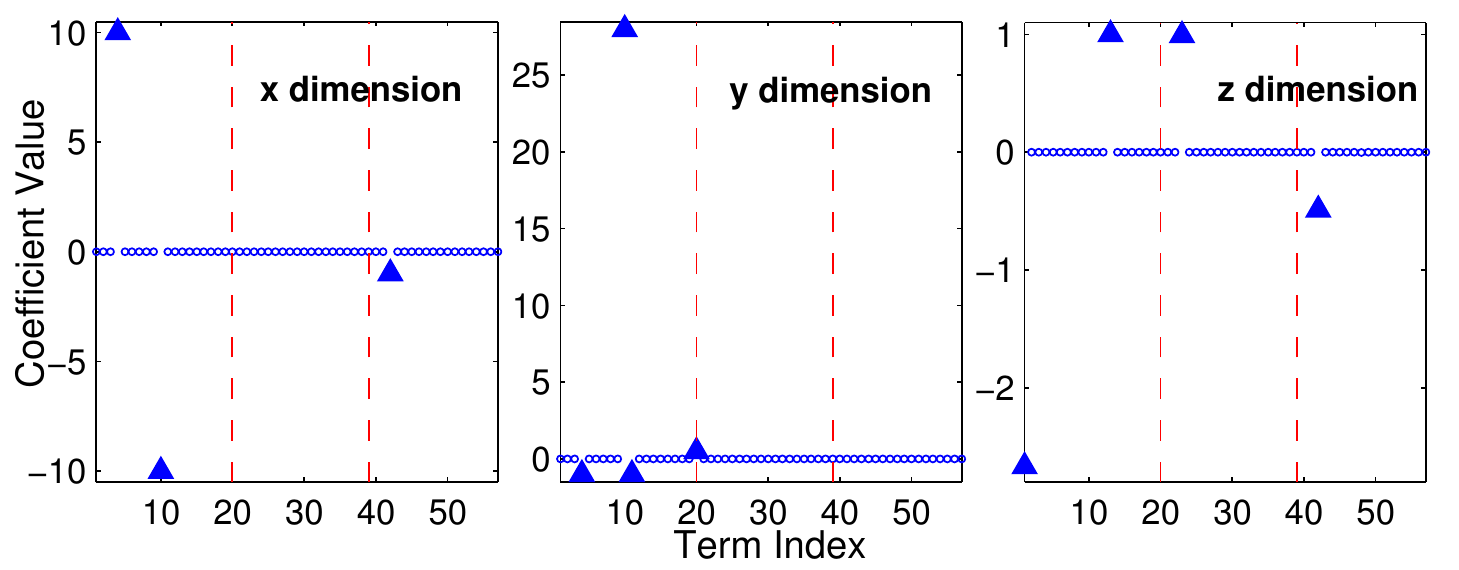}
\caption{\small {\bf An example of predicting time-varying chaotic systems.}
For the time-varying Lorenz chaotic systems Eq.~\eqref{eq:Lorenz_TV}
with $k_1(t) = - t^2$, $k_2(t) = 0.5 t$, $k_3(t) = t$, and 
$k_4(t) = - 0.5 t^2$, predicted values of coefficients of power-series
terms versus the term index for the $x$-, $y$-, and $z$-equations,
where solid triangles and open circles denote nonzero and zero
coefficients, respectively. Note that, the number $M$ of data
points used for prediction is about $50\%$ of the total number $N$
of unknown coefficients in each power-series expansion. From 
Ref.~\cite{YLG:2012} with permission.}
\label{fig:single_detection_TV}
\end{figure}
 
As a proof of principle, the authors of Ref.~\cite{YLG:2012} used the 
classical Lorenz chaotic system \cite{Lorenz:1963} as an example by
incorporating explicit time dependence in a number of additional
terms. The modified Lorenz system is given by
\begin{eqnarray} \label{eq:Lorenz_TV} 
\dot{x} & = &  -10(x - y) + k_1(t)y, \\ \nonumber 
\dot{y} & = & 28x - y - xz + k_2(t)z, \\ \nonumber 
\dot{z} & = & -(8/3)z+xy+[k_3(t) + k_4(t)]y,
\end{eqnarray}
where $k_1(t) = -t^2$, $k_2(t) = 0.5 t$, $k_3(t) = t$, and $k_4(t) =
-0.5 t^2$. Suppose that the system equations are unknown
but only measured time series $x(t)$, $y(t)$, and $z(t)$ in a finite
time interval are available. The number of dynamical variables is $m
= 3$ and we choose the orders of the power-series expansions in the
three variables according to $l_1+l_2+l_3 \leq 3$. The maximum power
in the time dependence is chosen to be $v = 2$ so that explicit
time-dependent terms $t^0$, $t^1$ and $t^2$ are considered. The
total number of coefficients to be predicted is then
$(v+1)\sum_{i=1}^3(i+1)(i+2)/2=57$. (Note that, using low-order
power-series expansions in both the dynamical variables and time is
solely for facilitating explanation and presentation of results,
while the forecasting principle is the same for realistic dynamical
systems where much higher orders may be needed.) 
Figure~\ref{fig:single_detection_TV} shows the predicted coefficient values
versus the term index for all three dynamical variables, where in
each panel, solid triangles and open circles denote the predicted
non-zero and zero coefficients, respectively, and the red dashed
dividing lines indicate the terms associated with different powers
of the time variable, i.e., $t^0$, $t^1$ and $t^2$ (from left to
right). The meaning of these results can be explained by using any
one of the dynamical variables. For example, for the $x$-component
of the vector field, the prediction algorithm gives only 3 nonzero
coefficients. By identifying the corresponding values of the term
index, one can see that they correspond to the two terms without
explicit time dependence: $y$, $x$, and the term that contains
explicit such dependence: $t^2y$, respectively. A comparison of the
predicted nonzero coefficient values with the actual ones in the
original Eq.~\eqref{eq:Lorenz_TV} indicates that the method works
remarkably well. Similar results were obtained for the $y$ and $z$
components of the vector field. 

\begin{figure}
\centering
\includegraphics[width=\linewidth]{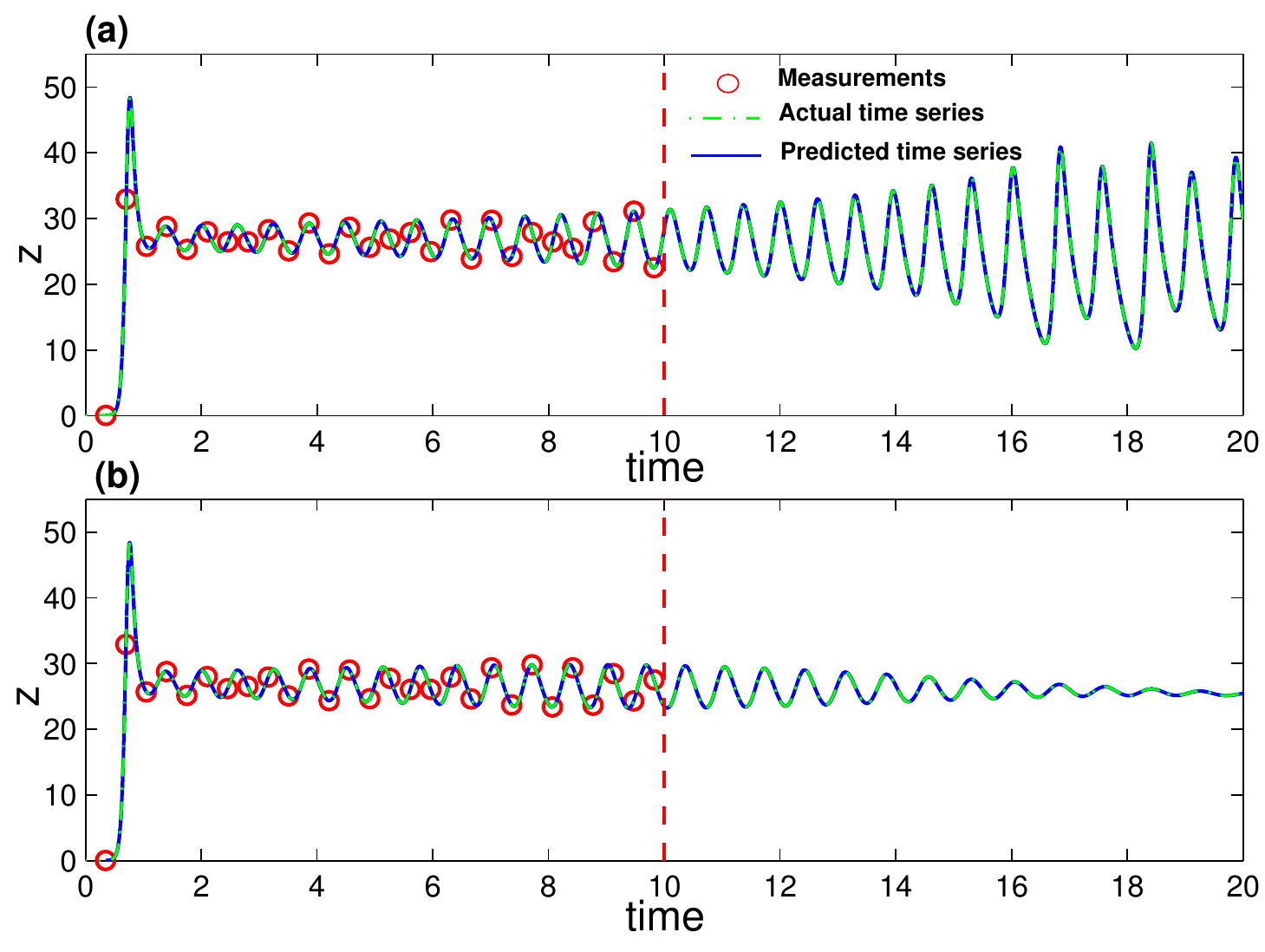}
\caption{\small {\bf Predicting future attractors of time varying dynamical
systems.} For the time-varying Lorenz system Eq.~\eqref{eq:Lorenz_TV},
predicted time series and measured values of the dynamical variable 
$z(t)$ for (a) time-independent case where $k_i(t) = 0$ ($i=1,\ldots,4$) 
and (b) time-dependent case where $k_1(t) = -0.01t^2$, $k_2(t) = 0.01t$, 
$k_3(t) = 0.01t$, and $k_4(t) = -0.01t^2$. Red circles denote the
measurements used for prediction, while the green dash and
blue solid lines represent the actual and predicted time series,
respectively. In both panels, $t = 0$ and $t = 10$ correspond to the
beginning and the end of the measurement window, i.e., $t_1$ and
$t_M$, respectively. From Ref.~\cite{YLG:2012} with permission.}
\label{fig:time_series_TV}
\end{figure}

\begin{figure}
\centering
\includegraphics[width=\linewidth]{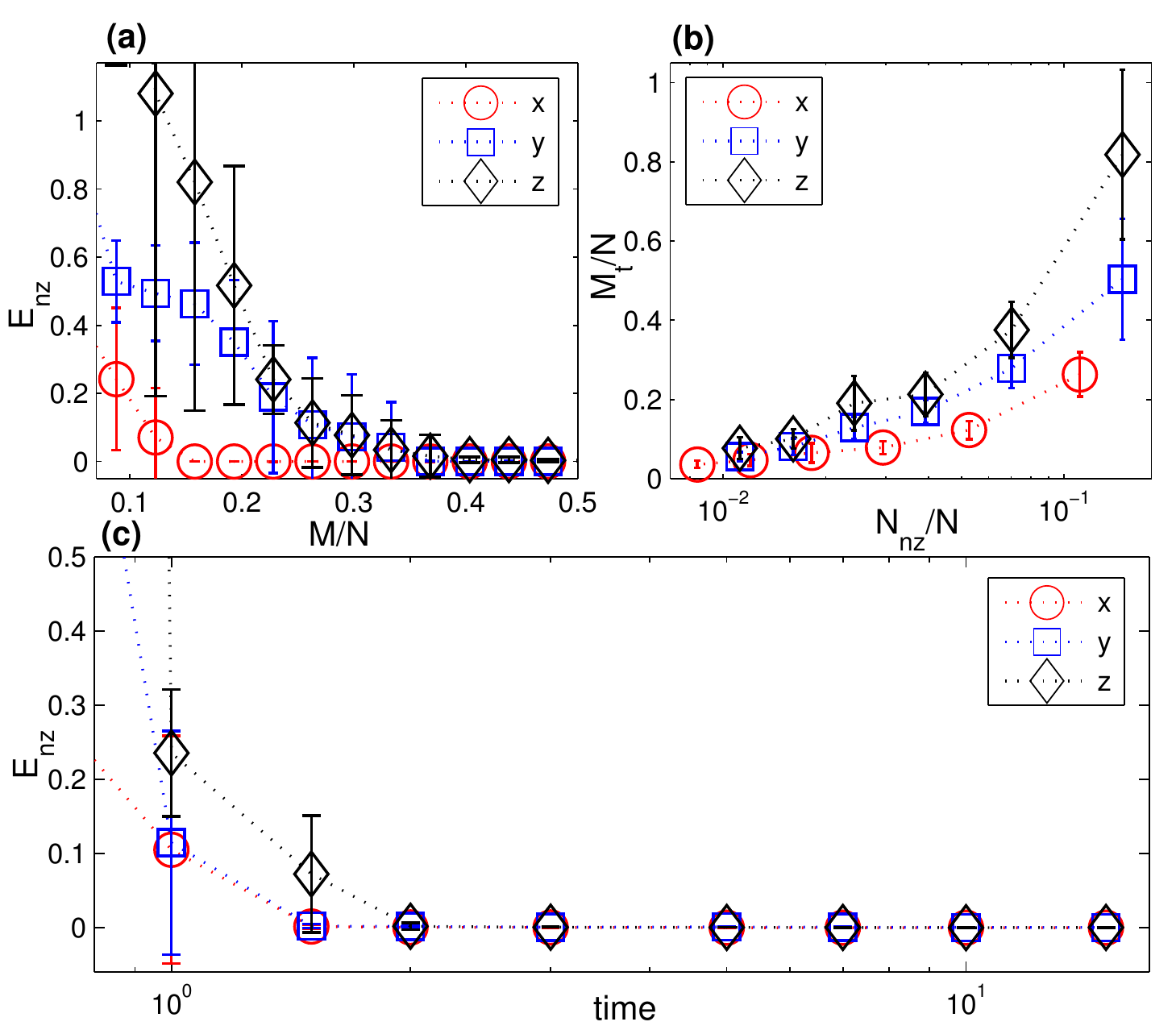}
\caption{\small {\bf Performance analysis of compressive sensing based
identification of time varying nonlinear dynamical systems.}
For the time-varying Lorenz system as in Fig.~\ref{fig:single_detection_TV},
(a) prediction errors $E_{nz}$ as a function of the ratio $M/N$, and (b)
ratio $M_t/N$ as a function of the ratio $N_{nz}/N$, where $N$ can be
increased by using higher-order power-series expansions (e.g., up to 7).
(c) With fixed number of measurement $M$, prediction errors $E_{nz}$ as
a function of the length of measurement window. The error bars are
obtained from 20 independent realizations of the prediction algorithm.
The prediction errors $E_z$ associated with non-existent terms show
similar behaviors. From Ref.~\cite{YLG:2012} with permission.}
\label{fig:error_TV}
\end{figure}

When the vector field ${\bf F}[{\bf x},{\bf p}(t)]$ of the
underlying dynamical system has been predicted, one can solve Eq.
(\ref{eq:system_TV}) numerically to assess the state variables
at any future time and the asymptotic attractors. 
Figures~\ref{fig:time_series_TV}(a,b) present one example, where 
a forecasted time series calculated from the predicted vector field 
is shown, together with the values of the corresponding dynamical 
variable from the actual Lorenz system at a number of time instants. 
The two cases shown are where the parameter functions $k_i(t)$ 
($i=1,\ldots,4$) are all zero and time-varying, respectively. Excellent 
agreement was again obtained, indicating the power of the method to 
predict the future states and attractors of time-varying dynamical 
systems. The interpretation and implication of 
Figs.~\ref{fig:time_series_TV}(a,b) are the following.
Note that $t = 0$ and $t = 10$ correspond to the beginning and the end
of the measurement time window $[t_1,t_M]$, respectively.
For the original classical Lorenz system without time-varying
parameters, the asymptotic attractor is chaotic, as can be seen from
Fig.~\ref{fig:time_series_TV}(a). However, as external perturbations
are turned on at $t = 0$, there are four time-varying parameters
in the system for $t > 0$. In this case, the asymptotic attractor
becomes a fixed-point, as can be seen from the asymptotic behavior
$z \rightarrow \mbox{constant}$ in Fig.~\ref{fig:time_series_TV}(b).
In both cases, by using limited amount of measurements, namely,
available time series in the window $[t_1,t_M]$,
one obtains quite accurate forecasting results. The
result exemplified in Fig.~\ref{fig:time_series_TV}(b) is especially
significant, as it indicates that the future state and attractors of
time-varying dynamical systems can be accurately predicted
based on limited data availability.

An error and performance analysis was carried out in Ref.~\cite{YLG:2012}
by using the indicators $E_{nz}$ and $E_z$, the prediction error for 
existent and non-existent terms, respectively.
Figure~\ref{fig:error_TV}(a) shows, for the time-varying Lorenz 
system, $E_{nz}$ versus the ratio of the number $M$ of measurements to 
the total number $N$ of terms to be predicted. For all dynamical 
variables, one observes that, as $M$ exceeds a threshold value $M_t$, 
$E_{nz}$ becomes effectively zero, where $M_t$ can be defined
quite arbitrarily, e.g., the minimum number of measurements required to
achieve $E_{nz}=10^{-3}$. The data requirement for accurate prediction
can then be assessed by examining how $M_t$ depends on the sparsity of
the coefficient vector to be predicted, which can be defined as the ratio
of the number $N_{nz}$ of the nonzero terms to the total number $N$ of
terms to be predicted. Note that, $N$ or the ratio $N_{nz}/N$ can be
adjusted by varying the order of the assumed power series. From 
Fig.~\ref{fig:error_TV}(b), one can see that, as $N_{nz}/N$ is 
decreased (e.g., by increasing $N$) so that the vector to be predicted 
becomes more sparse, the ratio $M_t/N$ also decreases. In particular, 
for the smallest value of $N_{nz}/N$ examined, where $N = 357$, only 
about $5\%$ of the data points are needed for accurate prediction, 
despite the time-varying nature of the underlying dynamical system.
Figure~\ref{fig:error_TV}(c) shows the prediction errors with respect
to different length of the measurement window for a fixed number of data 
points. It can be seen that, when the length exceeds a certain (small) 
value so that the time series extends to the whole attractor in the 
phase space, $E_{nz}$ approaches zero rapidly.

Dynamical systems are often driven by time-periodic forces, such as 
the classical Duffing system~\cite{HR:1976}. In such a case, it is 
necessary to explore alternative bases of expansion with respect to 
the time variable other than power series to ensure the sparsity 
condition. A realistic strategy to choose a suitable expansion base 
is to make use of the basic physics underlying the dynamical system 
of interest. Insofar as an appropriate base can be chosen so that 
the coefficient vector to be predicted is sparse, the compressive
sensing based methodology is applicable. 

\clearpage

\section{Compressive sensing based reconstruction of complex 
networked systems} \label{sec:CH3_CS_CN_Reconstruction}

Compressive sensing has recently been introduced to the field of network 
reconstruction for continuous time coupled oscillator 
networks~\cite{WYLKG:2011,WYLKH:2011}, for evolutionary game 
dynamics on networks~\cite{WLGY:2011}, for detecting hidden 
nodes~\cite{SWL:2012,SLWD:2014}, for predicting and controlling 
synchronization dynamics~\cite{SNWL:2012,Sommerladeetal:2014}, 
for reconstructing spreading dynamics based on binary data~\cite{SWFDL:2014}, 
and for reconstructing complex geospatial networks through estimating 
the time delays~\cite{SWWL:2015}. In this Section we shall the 
methodologies and the main results.

\subsection{Reconstruction of coupled oscillator networks}
\label{subsec:CS_CON}

We describe a compressive sensing based framework that enables a full
reconstruction of coupled oscillator networks whose vector fields
consist of a limited number of terms in some suitable base of
expansion~\cite{WYLKH:2011}. There are two facts that justify the
use of compressive sensing. First, complex networks in the real world
are typically sparse~\cite{Strogatz:2001,AB:2002,Newman:book}. Second, 
the mathematical functions determining the dynamical couplings in a 
physical network can be expressed by power-series expansions. The task 
is then to estimate all the nonzero coefficients. Since the underlying 
coupling functions are unknown, the power series can contain high-order
terms. The number of coefficients to be estimated can therefore be
quite large. However, the number of nonzero coefficients may be only 
a few so that the vector of coefficients is effectively sparse. Because
the network structure as well as the dynamical and coupling functions
are sparse, compressive sensing stands out as a feasible framework for
full reconstruction of the network topology and dynamics.  

A complex oscillator network can be viewed as a high-dimensional dynamical 
system that generates oscillatory time series at various nodes. The
dynamics of a node can be written as
\begin{equation} \label{eq:maineq_CON}
\dot{{\bf x}}_i = {\bf F}_i({\bf x}_i) + \sum_{j=1,j\neq i}^{N}
\mathcal{C}_{ij} \cdot ({\bf x}_j - {\bf x}_i), \ \ (i=1,\cdots ,N),
\end{equation}
where ${\bf x}_i \in R^m$ represents the set of externally
accessible dynamical variables of node $i$, $N$ is the number of
accessible nodes, and $\mathcal{C}_{ij}$ is the coupling matrix
between the dynamical variables at nodes $i$ and $j$ denoted by
\begin{equation}
\mathcal{C}_{ij} =\left(
  \begin{array}{cccc}
    c_{ij}^{1,1} & c_{ij}^{1,2} & \cdots & c_{ij}^{1,m} \\
    c_{ij}^{2,1} & c_{ij}^{2,2} & \cdots & c_{ij}^{2,m} \\
    \cdots & \cdots  & \cdots  & \cdots  \\
    c_{ij}^{m,1} & c_{ij}^{m,2} & \cdots & c_{ij}^{m,m} \\
  \end{array}
\right).
\end{equation}
In $\mathcal{C}_{ij}$, the superscripts $kl$ ($k,l=1,2,...,m$) stand
for the coupling from the $k$th component of the dynamical
variable at node $i$ to the $l$th component of the dynamical
variable at node $j$. For any two nodes, the number of possible
coupling terms is $m^2$. If there is at least one nonzero element
in the matrix $\mathcal{C}_{ij}$, nodes $i$ and $j$ are coupled and,
as a result, there is a link (or an edge) between them in the
network. Generally, more than one element in $\mathcal{C}_{ij}$ can be
nonzero. Likewise, if all the elements of $\mathcal{C}_{ij}$ are zero,
there is no coupling between nodes $i$ and $j$. The connecting
topology and the interaction strengths among various nodes of the
network can be predicted if we can reconstruct all the coupling matrices
$\mathcal{C}_{ij}$ from time-series measurements.

Generally, the compressive sensing based method consists of the 
following two steps. First, one rewrites Eq.~(\ref{eq:maineq_CON}) as
\begin{equation} \label{eq:4_CON}
\dot{{\bf x}}_i = [{\bf F}_i({\bf x}_i) - \sum_{j=1,j\neq i}^{N}
\mathcal{C}_{ij} \cdot {\bf x}_i ]  + 
\sum_{j=1,j\neq i}^{N} \mathcal{C}_{ij} \cdot {\bf x}_j, 
\end{equation}
where the first term in the right-hand side is exclusively a
function of ${\bf x}_i$, while the second term is a function of
variables of other nodes (couplings). We define the first term to
be ${\bf f}_i({\bf x}_i)$, which is unknown. In general, the
$k$th component of ${\bf f}_i({\bf x}_i)$ can be represented
by a power series of order up to $n$:
\begin{equation} \label{eq:2_1_CON}
\left[ {\bf f}_i({\bf x}_i)\right]_k \equiv \left[{\bf
F}_i({\bf x}_i) - \sum_{j=1,j\neq i}^{N} \mathcal{C}_{ij}\cdot {\bf x}_i
\right]_k = \sum_{l_1=0}^{n}\sum_{l_2=0}^{n}
\cdots\sum_{l_m=0}^{n} [({\bf \alpha}_i)_k]_{l_1,\cdots,l_m}
[({\bf x}_i)_1]^{l_1} [({\bf x}_i)_2]^{l_2} \cdots
[({\bf x}_i)_m]^{l_m},
\end{equation}
where $({\bf x}_i)_{k}$ ($k=1,\cdots, m$) is the $k$th component
of the dynamical variable at node $i$, the total number of products is 
$(1+n)^m$, and $[({\bf \alpha}_i)_k]_{l_1,\cdots,l_m}\in R^m$ is the 
coefficient scalar of each product term, which is to be determined from 
measurements as well. Note that the terms in Eq.~(\ref{eq:2_1_CON}) are all 
possible products of different components with different power of
exponents. As an example, for $m=2$ (the components are $x$ and
$y$) and $n=2$, the power series expansion is $\alpha_{0,0} +
\alpha_{1,0}x + \alpha_{0,1}y + \alpha_{2,0}x^2 + \alpha_{0,2}y^2
+ \alpha_{1,1}xy +\alpha_{2,1}x^2y + \alpha_{1,2} xy^2 +
\alpha_{2,2}x^2y^2$.

Second, one rewrites Eq.~(\ref{eq:4_CON}) as
\begin{eqnarray} \label{eq:6_CON}
\dot{{\bf x}}_i = {\bf f}_i({\bf x}_i) + \mathcal{C}_{i1} \cdot {\bf
x}_1 + \mathcal{C}_{i2} \cdot {\bf x}_2 + \cdots + \mathcal{C}_{iN} \cdot {\bf x}_N
\end{eqnarray}
The goal is to estimate the various coupling matrices ${\bf
C}_{ij}(j=1,\cdots, i-1,i+1,\cdots,N)$ and the coefficients of
${\bf f}_i({\bf x}_i)$ from sparse time-series measurements.
According to the compressive sensing theory, to reconstruct the
coefficients of Eq.~(\ref{eq:6_CON}) from a small number of
measurements, most coefficients should be zero - the sparse
signal requirement. To include as many coupling forms as possible, 
one expands each term $\mathcal{C}_{ij} {\bf x}_j$ in Eq.~(\ref{eq:6_CON}) 
as a power series in the same form of ${\bf f}_i({\bf x}_i)$ but 
with different coefficients:
\begin{equation} \label{eq:7_CON}
\dot{{\bf x}}_i = {\bf f}_1({\bf x}_1) + {\bf f}_2({\bf
x}_2)+ \cdots + {\bf f}_N({\bf x}_N).
\end{equation}
This setting not only includes many possible coupling forms but also 
ensures that the sparsity condition is satisfied so that the prediction 
problem can be formulated in the compressive-sensing framework. For an
arbitrary node $i$, information about node-to-node coupling, or
about the network connectivity, is contained completely in 
${\bf f}_j(j\neq i)$. For example, if in the equation of $i$, a
term in ${\bf f}_j(j\neq i)$ is not zero, there then exists
coupling between $i$ and $j$ with the strength given by the
coefficient of the term. Subtracting the coupling terms
$-\sum_{j=1,j\neq i}^{N} \mathcal{C}_{ij}\cdot {\bf x}_i$ from 
${\bf f}_i$ in Eq.~(\ref{eq:2_1_CON}), which is the sum of coupling
coefficients of all ${\bf f}_j$ $(j\neq i)$, the nodal dynamics
${\bf F}_i({\bf x}_i)$ can be obtained. That is, once the
coefficients of Eq.~(\ref{eq:7_CON}) have been determined, the node
dynamics and couplings among the nodes are all known.

The formulation of the method can be understood in a more detailed and 
concrete manner by focusing on one component of the dynamical variable 
at all nodes in the network, say component 1. (Procedures for other 
components are similar.) For each node, one first expands the corresponding
component of the vector field into a power series up to power $n$.
For a given node, due to the interaction between this component
and other $(m-1)$ components of the vector field, there are
$(n+1)^m$ terms in the power series. The number of coefficients to
be determined for each individual nodal dynamics is thus $(n+1)^m$.
Now consider a specific node, say node $i$. For every other node
in the network, possible couplings from node $i$ indicates the
need to estimate another set of $(n+1)^m$ power-series
coefficients in the functions of $\mathbf{f}_j(\mathbf{x}_j)$. 
There are in total $N(n+1)^m$ coefficients that need to be determined. 
The vector $\mathbf{a}$ to be determined in the compressive sensing 
framework contains then $N(n+1)^m$ components. For example, to construct 
the measurement vector $\mathbf{X}$ and the matrix $\mathcal{G}$ for
the case of $m = 3$ (dynamical variables $x$, $y$, and $z$) and $n
= 3$, one obtains the following explicit dynamical equation for the
first component of the dynamical variable of node $i$:
\begin{eqnarray}\label{eq:x_i1}
\Gamma_i(x_i) & = & (a_i)_{000}\cdot x_i^0y_i^0z_i^0 + \cdots +
(a_i)_{003}\cdot x_i^0y_i^0z_i^3   \\ \nonumber &+ &
(a_i)_{010}\cdot x_i^0y_i^1z_i^0 + \cdots +  (a_i)_{100}\cdot
x_i^1y_i^0z_i^0 + \cdots  +  (a_i)_{333}\cdot x_i^3y_i^3z_i^3.
\end{eqnarray}
We can denote the coefficients of $\Gamma_i(x_i)$ by ${\bf a}_i =
[(a_i)_{000},(a_i)_{001},\cdots ,(a_i)_{333}]^T$. Assuming that
measurements ${\bf x}_i(t)$ ($i=1,\ldots,N$) at a set of time
$t_1,t_2,\ldots,t_M$ are available, one denotes
\begin{equation}
{\bf g}_i(t) = \big[ x_i(t)^0y_i(t)^0z_i(t)^0,
x_i(t)^0y_i(t)^0z_i(t)^1, \cdots, x_i(t)^3y_i(t)^3z_i(t)^3 \big],
\end{equation}
such that $\Gamma_i[x_i(t)]= {\bf g}_i(t) \cdot {\bf a}_i$.
According to Eq.~(\ref{eq:x_i1}), the measurement vector can be
chosen as $\mathbf{X} =\left[\dot{x}_i(t_1),\dot{x}_i(t_2),
\cdots,\dot{x}_i(t_M)\right]^T$, which can be calculated from time
series. Finally, one obtains the following equation in the form
$\mathbf{X} = \mathcal{G} \cdot \mathbf{a}$:
\begin{equation} \label{eq:YeqsAX_CON}
\nonumber \left ( \begin{array}{cc} %
  \dot{x}_i(t_1)\\
  \dot{x}_i(t_2)\\
  \vdots \\
  \dot{x}_i(t_M)\\
\end{array}\right ) = 
\left(
\begin{array}{cccc}
  {\bf g}_1(t_1) & {\bf g}_2(t_1) & \cdots & {\bf g}_N(t_1) \\
  {\bf g}_1(t_2) & {\bf g}_2(t_2) & \cdots & {\bf g}_N(t_2) \\
  \vdots & \vdots & \vdots & \vdots \\
  {\bf g}_1(t_M) & {\bf g}_2(t_M) & \cdots & {\bf g}_N(t_M) \\
\end{array}
\right ) \cdot \left( \begin{array}{c}
  {\bf a}_1 \\
  {\bf a}_2 \\
  \vdots \\
  {\bf a}_N \\
\end{array}
\right).
\end{equation}
To ensure the restricted isometry property~\cite{CRT:2006a,CRT:2006b,
Candes:2006,Donoho:2006,Baraniuk:2007,CW:2008}, one can normalize the
coefficient vector by dividing the elements in each column by the $L_2$
norm of that column: $(\mathcal{G})_{ij} =
(\mathcal{G})_{ij}/L_2(j)$ with $L_2(j) =\sqrt{ \sum_{i=1}^{M}
[(\mathcal{G})_{ij}]^2}$. After $\mathbf{a}$ is determined via some
standard compressive-sensing algorithm, the coefficients are given
by $\mathbf{a}/L_2$. To determine the set of power-series
coefficients corresponding to a different component of the
dynamical variable, say component 2, one simply replaces the
measurement vector by $\mathbf{X} = \left[\dot{y_i}(t_1),\dot{y_i}(t_2),
\cdots,\dot{y_i}(t_M)\right]^T$ and use the same matrix
$\mathcal{G}$. This way all coefficients can be estimated. After
the equations of all components of $i$ are determined, one can
repeat this process for all other nodes to reconstruct the whole system.

\begin{figure*}
\centering
\includegraphics[width=\linewidth]{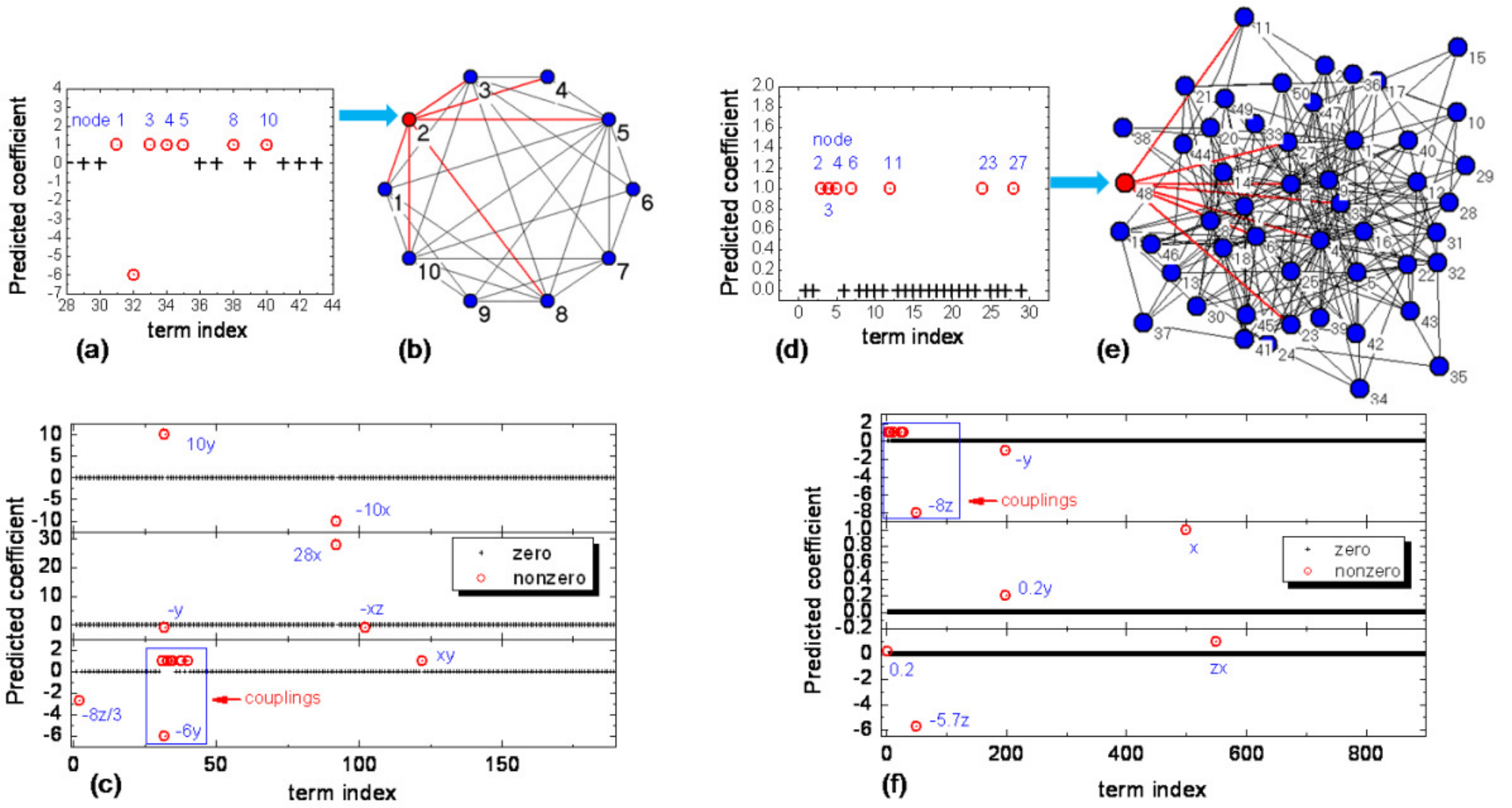}
\caption{\small {\bf Examples of compressive sensing based network 
reconstruction: networks of coupled Lorenz and R\"{o}ssler chaotic 
oscillators.} (a) Predicted coupling terms in the $z$ variable for 
node $\#2$ in a random network (b) of 10 coupled chaotic Lorenz 
oscillators. (c) Predicted terms in the dynamics of node $\#2$ and couplings
between node $\#2$ and other nodes. (d) Predicted coupling terms
in the $x$-variable for node $\#48$ in a scale-free network (e)
of 50 coupled chaotic R\"ossler oscillators. (f) Predicted terms
in the dynamics of node $\#48$ and couplings between node $\#48$
and other nodes. In (c) and (f), top to bottom panels: predicted
terms with coefficients in the $x$, $y$ and $z$ variables, where
the corresponding true values of the existent terms are marked. In
(a) and (d), the node numbers corresponding to the existing terms
are marked and the coupling forms are $c^{3,2}$ and $c^{1,3}$,
respectively. Here, ``term index'' refers to the order of to be
predicted coefficients appearing in Eq.~(\ref{eq:7_CON}) and in the
vector $\mathbf{a}$ of Eq.~(\ref{eq:YeqsAX_CON}). The average degrees
$\langle k\rangle$ for the random and scale-free networks are 6
and 10, respectively. The number of data points used for prediction is 
140 and the time interval for data collection is $\Delta t =1$. From
Ref.~\cite{WYLKH:2011} with permission.} 
\label{fig:CS_1_CON}
\end{figure*}

The working of the method was illustrated by using networks of 
coupled chaotic Lorenz and R\"ossler oscillators as 
examples~\cite{WYLKH:2011}. The classical Lorenz and
R\"ossler systems are given by $[\dot{x},\dot{y},\dot{z}] = 10
(y-x), x(28 - z)-y, xy - (8/3)z]$ and $[\dot{x},\dot{y},\dot{z}] =
-y-z, x + 0.2y, 0.2 + z(x - 5.7)]$, respectively. Since $m = 3$,
the power series of $x$, $y$ and $z$ can be chosen such that $l_1 +
l_2 + l_3 \le 3$. The total number of the coefficients to be
estimated is then $N\sum_{i=1}^3 (i+1)(i+2)/2+1=19N+1$, where
$i=l_1 + l_2 + l_3$ ranges from $1$ to $3$. 
Random and scale-free network topologies were studied~\cite{WYLKH:2011}. 
In particular, the Lorenz oscillator network was chosen to be a
Erd\H{o}s-R\'{e}nyi type of homogeneous random network~\cite{ER:1959},
generated by assuming a small probability of link
for any pair of nodes. The coupling between nodal dynamics was 
assumed to occur between the $y$ and the $z$ variables in the
Lorenz equations, leading to the following coupling matrix:
$c_{ij}^{3,2} = 1$ if nodes $i$ and $j$ are connected and
$c_{ij}^{3,2} = 0$ otherwise. The R\"ossler oscillator network was 
assumed to be a Barab\'asi-Albert type of scale-free network~\cite{BA:1999} 
with a heterogeneous degree distribution. The
coupling scheme is $c_{ij}^{1,3} = 1$ for link between $i$ and
$j$. Both types of network structures are illustrated
schematically in Fig. \ref{fig:CS_1_CON}. Time series were generated by
integrating the whole networked system with time step $h =
10^{-4}$ for $6\times 10^6$ steps. However, the number of
``measured'' data points required for the method to be successful
can be orders of magnitude less than $6\times 10^6$, {\em a
fundamental advantage of compressive-sensing method}.
Specifically, random measurements were collected from the integrated
time series and the number of elements in each row of the matrix
$\mathbf{G}$ is given by $N(n+1)^m$.

Figure~\ref{fig:CS_1_CON} shows some representative results. For the
random Lorenz network, the inferred coefficients were shown of node
$\#$2 associated with both the couplings with other nodes
[Fig.~\ref{fig:CS_1_CON}(a)] and those with its own dynamics
[Fig.~\ref{fig:CS_1_CON}(a)]. The term index is arranged from low to
high values, corresponding to the order from low to high node
number. The predicted coupling strengths between node $\#$2 and
others are shown in Fig.~\ref{fig:CS_1_CON}(a), where each term
according to its index corresponds to a specific node. Nonzero
terms belonging to nodes other than node $\#$2 indicate inter-node
couplings. The predicted interactions with nonzero coefficients
(the value is essentially unity) are in agreement with the
neighbors of node $\#$2 in the sample random network in
Fig.~\ref{fig:CS_1_CON}(b). The term 32 related to $-6y$ is the
coupling strength from node $\#$2, which equals the sum of the
coupling strengths from the other connected nodes. Figure
\ref{fig:CS_1_CON}(c) displays the inferred coefficients for both nodal
dynamics and coupling terms in the three components $x$, $y$ and
$z$. All predicted terms with nonzero coefficients are in
agreement with those in the equations of the dynamics of node
$\#$2, together with the inter-node coupling terms $c^{3,2}$.

Figure~\ref{fig:CS_1_CON}(d) shows the predicted links between node
$\#48$ and others in a R\"ossler oscillator network with a
scale-free structure. All existing couplings have been accurately
inferred, as compared to the structure presented in
Fig.~\ref{fig:CS_1_CON}(e), even though the interaction patterns among
nodes are heterogeneous. Both the detected local dynamical and
coupling terms associated with node $\#$48 are indicated in
Fig.~\ref{fig:CS_1_CON}(f), where in the $x$ component, the term $-8z$
is the combination of the local-dynamical term $-z$ and the
coupling of node $\#$48 with 7 neighboring nodes. Since all the
couplings have been successfully detected, the local-dynamical
term $-z$ in the $x$ component can be separated from the
combination so that all terms of node $\#$48 are predicted. Similar
results were obtained for all other nodes, leading to a complete and 
accurate reconstruction of the underlying complex networked 
system~\cite{WYLKH:2011}.

A performance analysis was carried out~\cite{WYLKH:2011}, with the
result that the number of required data points is much smaller than 
the number of terms in the power series
function, a main advantage of the compressive-sensing technique.
Insofar as the number of data points exceeds a critical value, the
prediction errors are effectively zero, indicating the robustness
of the reconstruction. One empirical observation was that, if the 
sampling frequency is high, the number of data points is not able to 
cover the dynamics in the whole phase space. In order to obtain a faithful
prediction of the whole system, the sampling frequency must be sufficiently 
low. Another important question was how the structural properties of the
network affect the prediction precision. A calculation~\cite{WYLKH:2011} 
of the dependence of the prediction error on the average degree 
$\langle k\rangle$ and the network size $N$ indicated that,
regardless of the network size, insofar as the network connections
are sparse, the prediction errors remain to be quite small, providing 
further support for the robustness of the compressive-sensing based method.

\begin{figure}
\centering
\includegraphics[width=\linewidth]{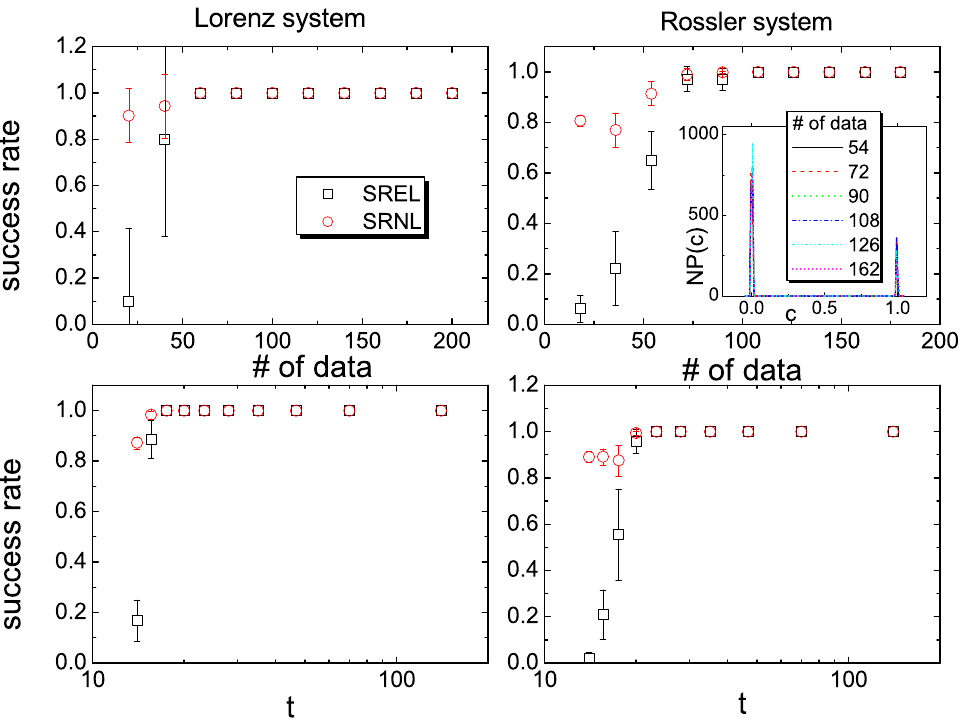}
\caption{\small {\bf Reconstruction accuracy of full network topology}. 
Success rates of existing links SREL and nonexistent links SRNL as
a function of the number of data points and length $t$ of time
series for random Lorenz and R\"ossler networks. The inset in the
upper-right panel shows the distribution of coupling strength in
the R\"ossler network for different numbers of data points. From
Ref.~\cite{WYLKH:2011} with permission.}
\label{fig:CS_4_CON}
\end{figure}

With respect to reconstruction of the network topology, computations 
demonstrated~\cite{WYLKH:2011} that, despite the small
prediction errors, all existing links in the original network can
be predicted extremely reliably. The performance of the method for 
predicting network structures can be quantified through the
success rates for existing links (SREL) and
nonexistent links (SRNL), defined to be the ratio between the
number of successfully predicted links and total number of links
and the ratio of the number of correctly predicted nonexistent
links to the total number of nonexistent links, respectively.
Figure~\ref{fig:CS_4_CON} shows the success rates versus the
number of data points and $t$ for both random Lorenz and R\"ossler
oscillator networks. It can be seen that, when the number of data
points and $t$ are sufficiently large, both SREL and SRNL reach
$100\%$. The inset in the upper-right panel shows the distribution
of the coupling strengths in the R\"ossler network. For all tested
numbers of data points ($>54$), there exist two sharp peaks
centered at $c=0$ and $c=1$, corresponding to the absence of
coupling and the existing coupling of strength $1.0$,
respectively. The narrowness of the two peaks in the distribution
makes it feasible to distinguish existing links (with nonzero
coupling strength) from nonexistent links (effectively with zero
coupling). This provides an explanation for the $100\%$ success
rates shown in Fig.~\ref{fig:CS_4_CON}.

The method was also tested~\cite{WYLKH:2011} on a number of real-world 
networks, ranging from social to biological and technological networks,
where the nodal dynamics were assumed to be of the Lorenz and
R\"ossler types. For five real-world networks, the prediction
errors are shown in Table~\ref{table:data1_CON}. It can be seen that all
errors are small. 

\begin{table} 
\caption{Prediction errors $E_{nz}$ and $E_z$ for five real-world 
networks. From Ref.~\cite{WYLKH:2011} with permission.}
\begin{center}
\begin{tabular}{|c|c|c|c|c|}
  \hline
Nodal dynamics & \multicolumn{2}{c}{Lorenz }  & \multicolumn{2}{c|}{R\"ossler} \\
\hline
Errors  & $E_{nz}$ & $E_{z}$ & $E_{nz}$ & $E_{z}$ \\ \hline
Dolphin social network~\cite{LSBHSD:2003}  & $3.6\times 10^{-3}$ & $7.4\times 10^{-6}$ & $2.8\times 10^{-3}$ & $2.2 \times 10^{-7}$\\
  \hline
  Friendship network of karate club~\cite{Zachary:1977} & $3.3\times 10^{-3}$ & $3.6\times 10^{-6}$ & $1.9\times 10^{-3}$ & $2.7\times 10^{-7}$\\
  \hline
Network of political book purchases~\cite{WYLKH:2011}  & $3.8\times 10^{-3}$ & $8.3\times 10^{-5}$ & $2.8\times 10^{-3}$ & $2.1\times 10^{-7}$\\
  \hline
Electric circuit networks~\cite{MIKLSASA:2004} & $4.4\time 10^{-3}$ & $5.8\times 10^{-6}$ & $2.9\times 10^{-3}$ & $2.7\times 10^{-7}$\\
  \hline
The neural network of C.~Elegans~\cite{WS:1998} & $1.1\times 10^{-3}$ & $1.9\times 10^{-5}$ & $2.8\times 10^{-3}$ & $2.2\times 10^{-7}$\\
  \hline
\end{tabular}
\label{table:data1_CON}
\end{center}
\end{table} 

\subsection{Reconstruction of complex networks with evolutionary-game
dynamics} \label{subsec:CS_Game}

Many complex dynamical systems in biology, social science, and economics
can be mathematically modeled by evolutionary games~\cite{Smith:book,
Weibull:book,HS:book,Nowak:book,SF:2007}. For example, in recent years
evolutionary game models played an important role in addressing the 
biodiversity problem through microscopic modeling and the mechanism of
species competition and coexistence at the level of individual
interactions~\cite{FA:2001,TCH:2005,RMF:2007a,RMF:2007b,PA:2008,
SS:2008,BRSF:2009,SWYL:2010,NYWLG:2010,YWLG:2010,HMT:2010,WLG:2010,Frey:2010,
NWLG:2010,VP:2010,WNLG:2011,JSM:2012,LEGN:2012,ABLMO:2012,ABLM:2012,
JWLN:2012,JSM:2013,KKWF:2013,HDL:2013,PDHL:2013,KPWH:2013,DARF:2014,
ZYHPDL:2014}. It was demonstrated~\cite{WLGY:2011} that compressive sensing
can be exploited to reconstruct the full topology of the underlying 
network based on evolutionary game data. In particular, in a typical 
game, agents use different strategies in order to gain the maximum 
payoff. Generally, the strategies can be divided into two types: 
cooperation and defection. It was shown~\cite{WLGY:2011} that, even when 
the available information about each agent's strategy and payoff is 
limited, the compressive-sensing based method can yield precise knowledge 
about the node-to-node interaction patterns in a highly efficient manner.
The basic principle was further demonstrated by using an actual social
experiment in which participants forming a friendship network played a 
typical game to generate short sequences of strategy and payoff data.

In an evolutionary game, at any time a player can choose one of
two strategies $S$: cooperation (C) or defection (D), which can be
expressed as $\mathbf{S}(C) = (1, 0)^T$ and $\mathbf{S}(D) = (0,
1)^T$. The payoffs of the two players in a game are determined by their
strategies and the payoff matrix of the specific game. For
example, for the prisoner's dilemma game (PDG)~\cite{NM:1992} and
the snowdrift games (SG)~\cite{HD:2004}, the payoff matrices are
\begin{eqnarray}
\mathcal{P}_{PDG} = \left (%
\begin{array}{cc}
  1 & 0 \\
  b & 0 \\
\end{array}
\right) \ \ \hbox{or} \ \
\mathcal{P}_{SG} = \left (%
\begin{array}{cc}
  1 & 1-r \\
  1+r & 0 \\
\end{array}
\right),
\end{eqnarray}
respectively, where $b$ ($1<b<2$) and $r$ ($0<r<1$) are parameters
characterizing the temptation to defect. When a defector
encounters a cooperator, the defector gains payoff $b$ in the PDG
and payoff $1+r$ in the SG, but the cooperator gains the sucker
payoff 0 in the PDG and payoff $1-r$ in the SG. At each time step,
all agents play the game with their neighbors and gain payoffs.
For agent $i$, the payoff is
\begin{equation}
P_i = \sum_{j \in \Gamma_i}\mathbf{S}_i^T \cdot \mathcal{P} \cdot \mathbf{S}_j,
\end{equation}
where $\mathbf{S}_i$ and $\mathbf{S}_j$ denote the strategies of
agents $i$ and $j$ at the time and the sum is over the neighboring
set $\Gamma_i$ of $i$. After obtaining its payoff, an agent
updates its strategy according to its own and neighbors' payoffs,
attempting to maximize its payoff at the next round. Possible
mathematical rules to capture an agent's decision making process
include the best-take-over rule~\cite{NM:1992}, the Fermi 
equation~\cite{ST:1998}, and payoff-difference-determined updating
probability \cite{SSP:2008}. In the computational study of evolutionary
game dynamics, the Fermi rule has been commonly used, 
which is defined, as follows. After a player
$i$ randomly chooses a neighbor $j$, $i$ adopts $j$'s status
$\mathbf{S}_j$ with the probability~\cite{ST:1998}:
\begin{equation}
W(\mathbf{S}_i\leftarrow \mathbf{S}_j)=\frac{1}{1+\exp{[(P_i-P_j)/\kappa]}},
\end{equation}
where $\kappa$ characterizes the stochastic uncertainties in the
game dynamics. For example, $\kappa=0$ corresponds to absolute
rationality where the probability is zero if $P_j < P_i$ and one
if $P_i < P_j$, and $\kappa \rightarrow \infty$ corresponds to
completely random decision. The probability $W$ thus characterizes
the bounded rationality of agents in society and the natural
selection based on relative fitness in evolution.

The key to solving the network-reconstruction problem lies in the
relationship between agents' payoffs and strategies. The
interactions among agents in the network can be characterized by
an $N \times N$ adjacency matrix $\mathcal{A}$ with elements
$a_{ij}=1$ if agents $i$ and $j$ are connected and $a_{ij}=0$
otherwise. The payoff of agent $x$ can be expressed by
\begin{eqnarray} \label{eq:main_Game}
P_x(t) & = & a_{x1}\mathbf{S}_x^T(t) \cdot \mathcal{P} \cdot
\mathbf{S}_1(t) + \cdots  + a_{x,x-1}\mathbf{S}_x^T(t)\cdot
\mathcal{P} \cdot \mathbf{S}_{x-1}(t) \\ \nonumber
& & + a_{x,x+1}\mathbf{S}_x^T(t)
\cdot \mathcal{P} \cdot \mathbf{S}_{x+1}(t) 
+ \cdots + a_{xN}\mathbf{S}_x^T(t) \cdot \mathcal{P} \cdot \mathbf{S}_N(t),
\end{eqnarray}
where $a_{xi}$ ($i=1, \cdots, x-1, x+1, \cdots, N$) represents a
possible connection between agent $x$ and its neighbor $i$,
$a_{xi}\mathbf{S}_x^T(t) \cdot \mathcal{P} \cdot \mathbf{S}_i(t)$
($i=1, \cdots, x-1, x+1, \cdots, N$) stands for the possible
payoff of agent $x$ from the game with $i$ (if there is no
connection between $x$ and $i$, the payoff is zero because
$a_{xi}=0$), and $t=1, \cdots, m$ is the number of round that all
agents play the game with their neighbors. This relation provides
a base to construct the vector $\mathbf{P}_x$ and matrix
$\mathcal{G}_x$ in a proper compressive-sensing framework to
obtain solution of the neighboring vector $\mathbf{A}_x$ of agent
$x$. In particular, one can write
\begin{eqnarray}
\mathbf{P}_x &=& (P_x(t_1),P_x(t_2),\cdots,P_x(t_m))^T, \nonumber \\
\mathbf{A}_x &=& (a_{x1},\cdots , a_{x,x-1},a_{x,x+1}, \cdots , a_{xN})^T,
\end{eqnarray}
and $\mathcal{G}_x = $
\begin{eqnarray}
\left(
\begin{array}{cccccc}
 F_{x1}(t_1)& \cdots & F_{x,x-1}(t_1) & F_{x,x+1}(t_1) &\cdots & F_{xN}(t_1) \\
 F_{x1}(t_2)& \cdots & F_{x,x-1}(t_2) & F_{x,x+1}(t_2) &\cdots & F_{xN}(t_2) \\
 \vdots & \cdots & \vdots & \vdots & \vdots & \vdots \\
 F_{x1}(t_m)& \cdots & F_{x,x-1}(t_m) & F_{x,x+1}(t_m) &\cdots & F_{xN}(t_m) \\
\end{array}
\right), \nonumber
\end{eqnarray}
where $F_{xy}(t_i) = \mathbf{S}_x^T(t_i) \cdot \mathcal{P} \cdot
\mathbf{S}_y(t_i)$. The vectors $\mathbf{P}_x$, $\mathbf{A}_x$ and
matrix $\mathcal{G}_x$ satisfy
\begin{equation}
\mathbf{P}_x = \mathcal{G}_x \cdot \mathbf{A}_x,
\end{equation}
where $\mathbf{A}_x$ is sparse due to the sparsity of the
underlying complex network, making the compressive-sensing
framework applicable. Since $\mathbf{S}_x^T(t_i)$ and
$\mathbf{S}_y(t_i)$ in $F_{xy}(t_i)$ come from data and
$\mathcal{P}$ is known, the vector $\mathbf{P}_x$ can be obtained
directly while the matrix $\mathcal{G}_x$ can be calculated from
the strategy and payoff data. The vector $\mathbf{A}_x$ can thus
be predicted based solely on the time series. Since the
self-interaction term $a_{xx}$ is not included in the vector
$\mathbf{A}_x$ and the self-column $[F_{xx}(t_1),\cdots ,
F_{xx}(t_m)]^T$ is excluded from the matrix $\mathcal{G}_x$, the
computation required for compressive sensing can be reduced. In a
similar fashion, the neighboring vectors of all other agents can
be predicted, yielding the network adjacency matrix $\mathcal{A} =
(\mathbf{A}_{1},\mathbf{A}_{2},\cdots , \mathbf{A}_{N})$.

The method was first tested~\cite{WLGY:2011} by implementing PDG 
and SG on three types of standard complex networks: random~\cite{ER:1959}, 
small-world~\cite{WS:1998} and scale-free~\cite{BA:1999}. In particular,
time series of strategies and payoffs were recorded during the system's 
evolution towards the steady state, which were used for uncovering the 
interaction network topology. The performance of the method can be 
quantified in terms of the amount of required measurements for 
different game types and network structures through the success rates 
of existent links (SREL) and non-existent links (SRNL). If the 
predicted value of an element of the adjacency matrix $\mathcal{A}$ is close 
to 1, the corresponding link can be deemed to exist. If the value 
is close to zero, the prediction is that there is no link. In practice, 
a small threshold can be assigned, e.g., 0.1, so that the ranges of 
the existent and non-existent links are $1\pm 0.1$ and $0\pm 0.1$, 
respectively. Any value outside the two intervals is regarded as a 
prediction failure. For a single player, SREL is defined as the 
ratio of the number of successfully predicted neighboring links to 
the number of actual neighbors, and SRNL is similarly defined. Averaging 
over all nodes leads to the values of SREL and SRNL for the entire
network. The reason for treating the success rates for existent
and non-existent links separately lies in the sparsity of the
underlying complex network, where the number of non-existent links
is usually much larger than the number of existent links. The
choice of the threshold does not affect the values of the success
rates, insofar as it is not too close to one, nor too close to zero.

\begin{figure}
\centering
\includegraphics[width=\linewidth]{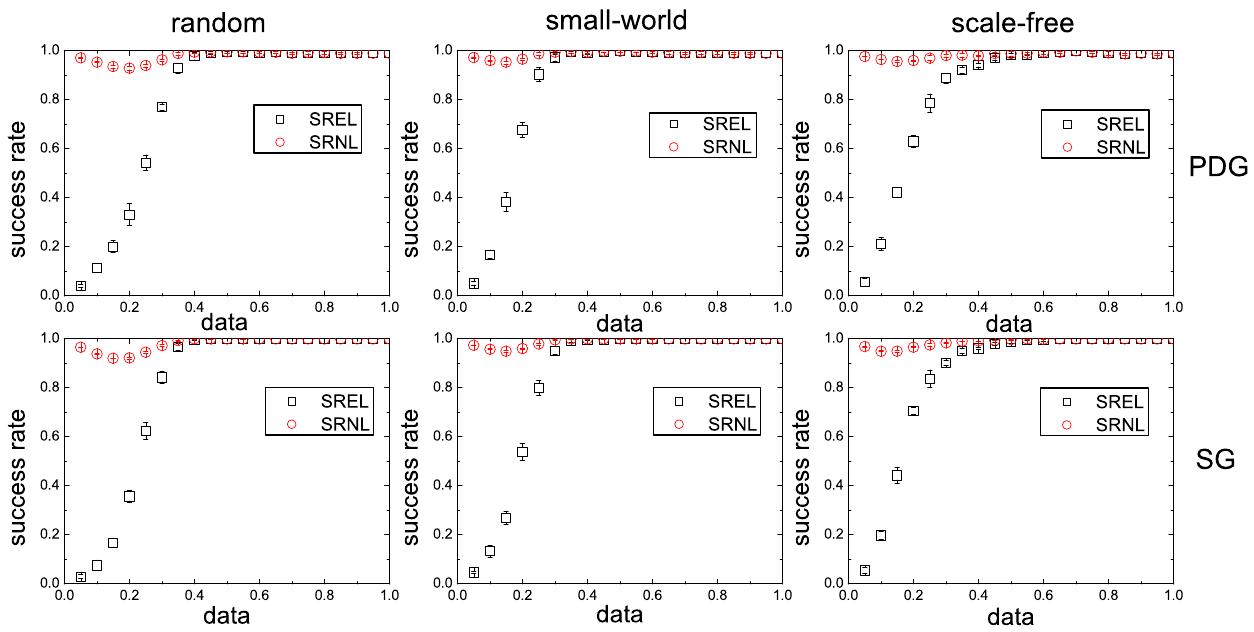}
\caption{\small {\bf Reconstructing the full topology of complex networks
hosting evolutionary game dynamics}. 
Success rates of inferring three types of networks:
random, small-world and scale-free, with PDG and SG dynamics. The
network size $N$ is 100. Each data point is obtained by averaging
over 10 network realizations. For each realization, measurements
are randomly picked from time series of temporary evolution. The
error bars denote the standard deviations. The payoff parameters
for the PDG and the SG are $b=1.2$ and $r=0.7$, respectively. 
Other values of $b$ and $r$ were also tested, yielding 
similar success rates. The average node degrees of all
used networks are fixed to $6$ and the noise parameter $\kappa=0.1$.
From Ref.~\cite{WLGY:2011} with permission.}
\label{fig:rate_simu_Game}
\end{figure}

The success rates of prediction for two types of games and three
types of network topologies are shown in Fig.~\ref{fig:rate_simu_Game}.
The length of the time series is represented by the number of
measurements collected during the temporal evolution normalized by
the number $N$ of agents, e.g., the value of one means that the
number of used measurements equals $N$. For all combinations of
game dynamics and network topologies examined, perfect success
rate can be achieved with extremely low amount of data. For
example, for random and small-world networks, the length of data
required for achieving $100\%$ success rate is between 0.3 and
0.4. This value is slightly larger (about 0.5) for scale-free
networks, due to the presence of hubs whose connections are much
denser than most nodes, although their neighboring vectors are
still sparse. Figure~\ref{fig:rate_simu_Game} thus demonstrates that
the method is both accurate and efficient. The exceptionally low
data requirement is particularly important for situations where
only rare information is available. From this standpoint,
evolutionary games are suitable to simulate such situations as
meaningful data can be collected only during the transient phase
before the system reaches its steady state, and game dynamics are
typically fast convergent so that the transients are short. In
addition, the robustness of the method was tested in situations where 
the time series are contaminated by noise~\cite{WLGY:2011}. For
example, the case was studied where random noise of amplitude
up to $30\%$ is added to the payoffs of PDG. When the size of
the data exceeds 0.4, the success rate approaches $100\%$ for
random networks. Similar performance was achieved for
small-world and scale-free networks. The noise immunity embedded
in the method is not surprising, as compressive sensing represents
an optimization scheme that is fundamentally resilient to noise.
In contrast, another type of noise, noise $\kappa$ in the
strategy updating process, plays a positive role in network
reconstruction, because of the fact that this kind of noise
can increase the relaxation time towards one of the absorbing
states (all C or all D), thereby providing more information for
successful reconstruction.

\begin{figure}
\centering
\includegraphics[width=\linewidth]{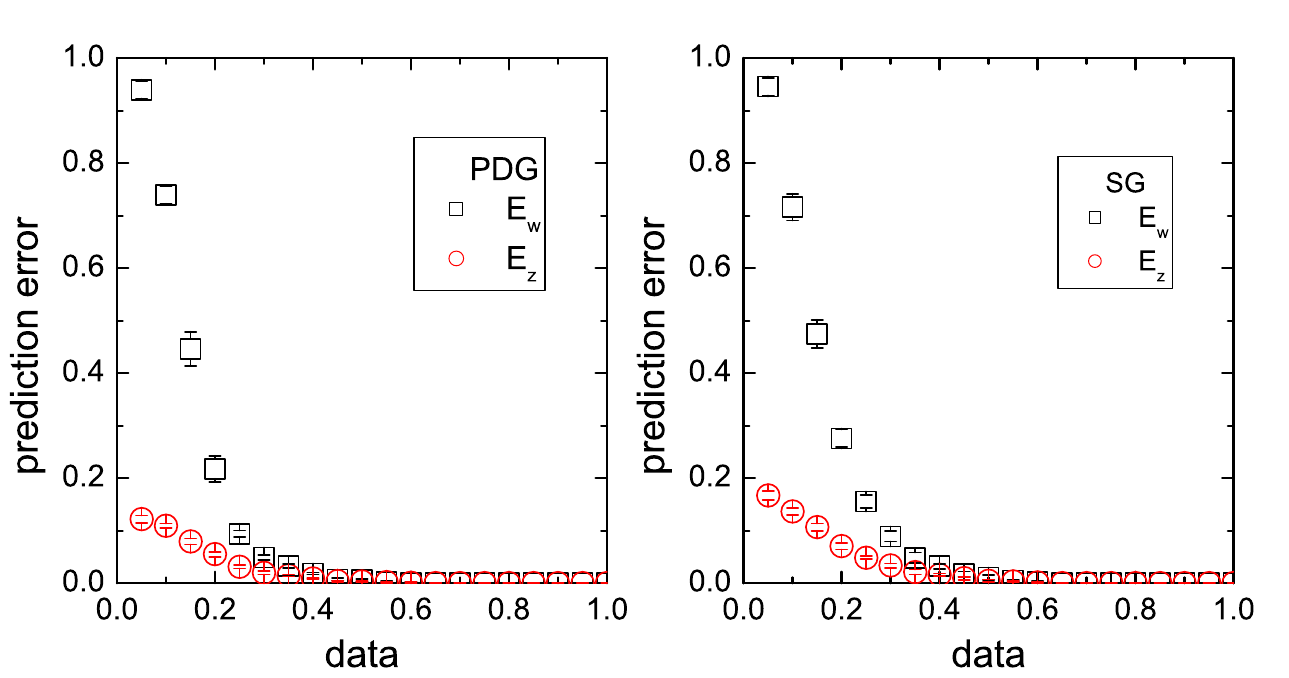}
\caption{\small {\bf Reconstructing the full topology of weighted 
complex networks hosting evolutionary game dynamics}.
Prediction errors $E_w$ in link weights and $E_z$ in
non-existent links for PDG on weighted scale-free networks. The
network size is 100 and the weights follow a uniform distribution,
ranging from 1.0 to 6.0. Each value of the prediction error is
obtained using 10 independent network realizations. Other
parameters are the same as Fig.~\ref{fig:rate_simu_Game}.
From Ref.~\cite{WLGY:2011} with permission.}
\label{fig:weighted_rate_Game}
\end{figure}

The method can be generalized straightforwardly to weighted networks 
with inhomogeneous node-to-node interactions. Using weights to 
characterize various interaction strengths, the elements of the weighted 
adjacency matrix $\mathcal{W}$ can be defined as
\begin{eqnarray}
W_{ij}=\left\{%
\begin{array}{ll}
    w_0 > 1, & \hbox{if $i$ connects to $j$} \\
    0,       & \hbox{otherwise.} \\
\end{array}%
\right.
\end{eqnarray}
In the context of evolutionary games on networks, the weight $W_{ij}$
characterizes the situation of aggregate investment. In
particular, for both players, more investments in general will
lead to more payoffs. Given the link weights, the weighted payoff
$P_{i}^w$ of an arbitrary individual is given by
\begin{equation}
P_i^w = \sum_{j \in \Gamma_i}w_{ij}\mathbf{S}_i^T \cdot \mathcal{P}
\cdot \mathbf{S}_j,
\end{equation}
where $\Gamma_i$ denotes the neighboring set of $i$. With the
evolutionary-game dynamics, the weighted network structure is
taken into account by the weighted payoff $P_{i}^w$. To uncover
such a network from data, the weighted payoff vector
$\mathbf{P}_x^w$, matrix $\mathcal{G}_x$, and the weighted
neighboring vector $\mathbf{W}_x$ for an arbitrary individual $x$
are needed. The vectors $\mathbf{P}_x^w$ and $\mathbf{W}_x$ are given by
\begin{eqnarray}
\mathbf{P}_x^w & = & (P_x^w(t_1),P_x^w(t_2),\cdots,P_x^w(t_m))^T,
\nonumber \\
\mathbf{W}_x & = & (W_{x1},\cdots,W_{x,x-1},W_{x,x+1},\cdots,W_{xN})^T.
\end{eqnarray}
Similar to unweighted networks, one has
\begin{equation}
\mathbf{P}_x^w = \mathcal{G}_x \cdot \mathbf{W}_x,
\end{equation}
where $\mathbf{W}_x$ can be calculated from the strategy and
payoff data. The prediction accuracy can be conveniently
characterized by various prediction errors, which are defined
separately for link weights and non-existent links with zero
weight. In particular, the relative error of a link weight is
defined as the ratio of the absolute difference between the
predicted weight and the true weight to the latter. The average
error over all link weights is the prediction error $E_w$.
However, a relative error for a zero (non-existent) weighted link
cannot be defined, so it is necessary to use the absolute error $E_z$. 
Figure~\ref{fig:weighted_rate_Game} shows the prediction errors for PDG
dynamics on a scale-free network with random link weights chosen
uniformly from the interval $[1.0,6.0]$. It can be seen that the
prediction errors decrease fast as the number of measurements is
increased. As the relative data size exceeds about 0.4, the two types of
prediction errors approach essentially zero, indicating that all
link weights have been successfully predicted without failure and
redundancy, despite that the link weights are random. Random and 
small-world networks were also tested with the finding that, to
achieve the same level of accuracy, the requirement for data can
be somewhat relaxed as compared with scale-free networks.

\begin{figure}
\centering
\includegraphics[width=0.8\linewidth]{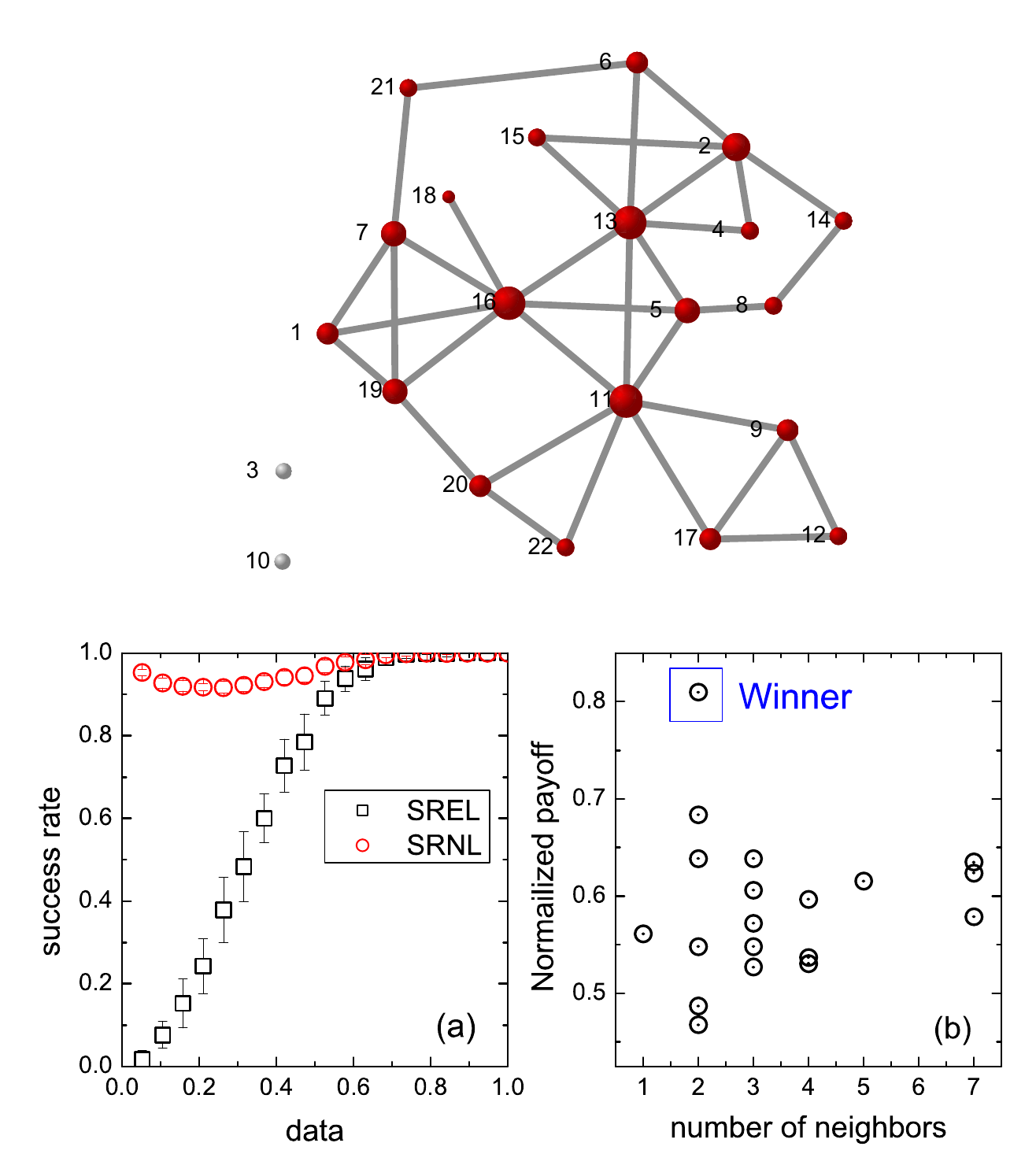}
\caption{\small {\bf Reconstructing an experimental social network using
evolutionary game dynamics}.
(a) Structure of experimental social network. (b) Success
rates of uncovering the network topology and (c) normalized payoff
of player as a function of node degrees. The 10 independent
realizations used in calculating the average success rates were
randomly chosen from the data base of 31 rounds of games.
From Ref.~\cite{WLGY:2011} with permission.}
\label{fig:social_rate_Game}
\end{figure}

The compressive sensing based reconstruction framework can be applied
to real social networks. Especially, an experimental was reported in
Ref.~\cite{WLGY:2011}, where 22 participants from Arizona State University
played PDG together iteratively and, at each round each player was
allowed to change his/her strategies to optimize the payoff. The
payoff parameter is set (arbitrarily) to be $b = 1.2$. The player
who had the highest normalized payoff (original payoff divided by
the number of neighbors) summed over time was the winner and
rewarded. During the experiment, each player was allowed to
communicate only with his/her direct neighbors for strategy
updating. Prior to experiment, there was a social tie (link)
between two players if they had already been acquainted to each
other; otherwise there was no link. Among the 22 players, two
withdrew before the experiment was completed, so they were treated
as isolated nodes. The network structure is illustrated in
Fig.~\ref{fig:social_rate_Game}(a). It exhibits typical features of
social networks, such as the appearance of a large density of
triangles and a core consisting of 4 players (nodes 5, 11, 13, and
16), which is fully connected within and has more links than other
nodes in the network. The core essentially consists of players who
were responsible for recruiting other players to participate in
the experiment. Each of the 20 players who completed the
experiment played 31 rounds of games, and he/she recorded his/her
own strategy and payoff at each time, which represented the
available data base for prediction. The data used for each
prediction run was randomly picked from this data base. The
pre-existed friendship ties among the participants tend to favor
cooperation and preclude the system from being trapped in the
social dilemma, due to the relatively short data streams. However,
for a long run, a full defection state may occur. In this sense,
the recorded data were taken during the transient dynamical phase
and were thus suitable for network reconstruction. The results are
shown in Fig.~\ref{fig:social_rate_Game}(b). It can be seen that the 
social network was successfully uncovered, despite the complicated
process of individual's decision making during the experiment.
Compared to the simulation results, larger data set (about 0.6) is
needed for a perfect prediction of social ties. This can be
attributed to the relative smaller size and denser connections in
the social network than in model networks.

An interesting phenomenon is that the winner picked in terms of
the normalized payoff had only two neighbors, in contrast to the
players with the largest node degree, whose normalized payoffs were
approximately at the average level, as shown in
Fig.~\ref{fig:social_rate_Game}(c). In addition, the payoffs of players
of smaller degrees were highly non-uniform, while those of higher
degrees showed smaller difference. This suggests that players of
high degree may not act as leaders due to their average normalized
payoffs. This experimental finding was in striking agreement with
numerical predictions in literature about the relationship between
individuals' normalized payoffs and their node degrees~\cite{SPD:2008}. 
It was also observed from experimental data
that a typical player with a large number of neighbors failed to
stimulate their neighbors to follow his/her strategies, suggesting
that hubs may not be as influential in social networks. However,
this finding should not be interpreted as a counter-example to the
leader's role in evolutionary games \cite{SPL:2006,SSP:2008},
since the network based on friendship may violate the absolute
selfish assumption of players who tend to be reciprocal with each
other.

The method, besides being fully applicable to complex networks governed
by evolutionary-game type of interactions, can be applied to other contexts
where the dynamical processes are discrete in time and the amount of
available data is small. For example, inferring gene regulatory networks
from sparse experimental data is a problem of paramount importance in
systems biology~\cite{GBLC:2003,BBAB:2007,GTF:2007,HLTSG:2009}.
For such an application, Eq.~(\ref{eq:main_Game}) should 
be replaced by the Hill equation, which models generic interactions among 
genes. In an expansion using base functions specifically suited for gene 
regulatory interactions, a compressive-sensing framework may be 
established. The underlying reverse-engineering problem can then be 
solved. A challenge that must be overcome is to represent the Hill 
function by an appropriate mathematical expansion so that the sparsity 
requirement for compressive sensing can be met.

\subsection{Detection of hidden nodes in complex networks} 
\label{subsec:CD_HN}

\subsubsection{Principle of detecting hidden nodes based on compressive
sensing} \label{subsubsec:HN_principle}

When dealing with an unknown complex networked system that has a large 
number of interacting components organized hierarchically, curiosity demands
that we ask the following question: are there hidden objects that are
not accessible from the external world? The problem of inferring the
existence of hidden objects from observations is quite challenging but
it has significant applications in many disciplines of science and
engineering. Here the meaning of ``hidden'' is that no direct observation of
or information about the object is available so that it appears to the
outside world as a black box. However, due to the interactions between
the hidden object and other observable components in the system, it
may be possible to utilize ``indirect'' information to infer the
existence of the hidden object and to locate its position with respect
to objects that can be observed. The difficulty to develop effective
solutions is compounded by the fact that the indirect information on
which any method of detecting hidden objects relies can be subtle and
sensitive to changes in the system or in the environment. In
particular, in realistic situations noise and random disturbances are
present. It is conceivable that the ``indirect'' information can be
mixed up with that due to noise or be severely contaminated. The
presence of noise thus poses a serious challenge to detecting hidden
nodes, and some effective ``noise-mitigation'' method must be
developed.

One approach to addressing the problem of detecting a hidden 
node~\cite{SWL:2012,SLWD:2014} was based on compressive 
sensing. The basic principle principle is that the existence 
of a hidden node typically leads to ``anomalies'' 
in the quantities that can be calculated or deduced from observation. 
Simultaneously, noise, especially local random disturbances applied at 
the nodal level, can also lead to large variance in these quantities.
This is so because, a hidden node is typically connected to a few
nodes in the network that are accessible to the external world,
and a noise source acting on a particular node in the network may
also be regarded as some kind of hidden object. Thus, the key to any
detection methodology is to identify and {\em distinguish} the effects
of hidden nodes on detection measure from those due to {\em local}
noise sources.

\begin{figure}
\centering
\includegraphics[width=0.8\linewidth]{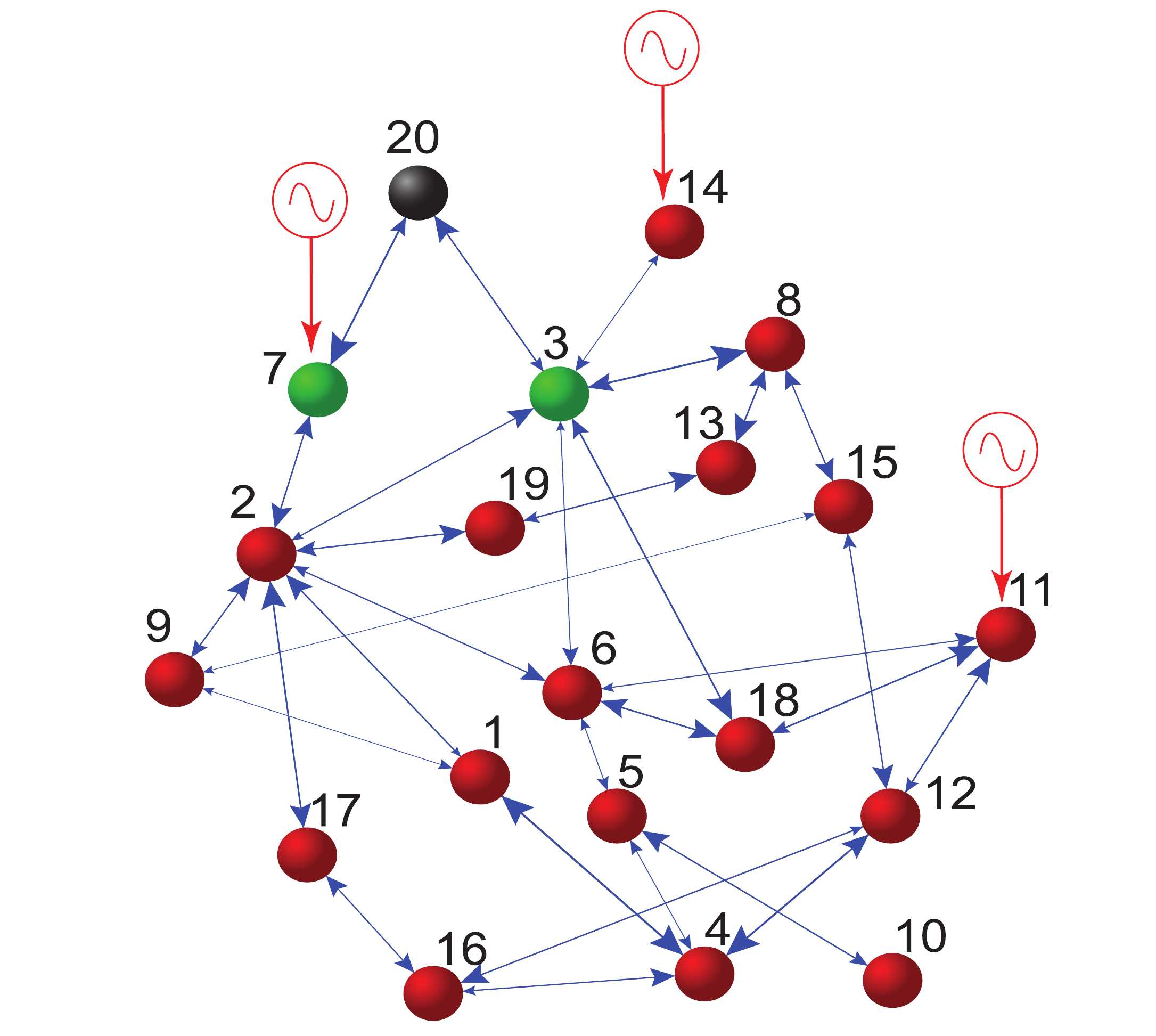}
\caption{\small {\bf Illustration of the principle of hidden node detection 
in a concrete setting}. An example of a complex network with a hidden node. 
Time series from all nodes except hidden node $\#$20 can be measured, which
can be detected when its immediate neighbors, nodes $\#$3 and $\#$7
are unambiguously identified. Nodes $\#$7, $\#$11, and $\#$14 are
driven by local noise sources. From Ref.~\cite{SLWD:2014} with permission.}
\label{fig:HD_Ros_HN}
\end{figure}

The key issues associated with detection of hidden nodes can be explained
in a concrete manner by using the network setting shown
schematically in Fig.~\ref{fig:HD_Ros_HN}, where there are 20 nodes, the
couplings among the nodes are weighted, and the entire network is in a
noisy environment, but a number of nodes also receive relatively
strong random driving, e.g., nodes 7, 11, and 14. 
Assume an oscillator network so that the
nodal dynamics are described by nonlinear differential equations, and
that time series can be measured simultaneously from all nodes in the
network except one, labeled as $\#$20, which is a hidden node. The
task of ascertaining the presence and locating the position of the
hidden node are equivalent to identifying its immediate neighbors,
which are nodes $\#$3 and $\#$7 in Fig.~\ref{fig:HD_Ros_HN}. Note that,
in order to be able to detect the hidden node based on information
from its neighboring nodes, the interactions between the hidden node
and its neighbors must be directional from the former to the latter or
be bidirectional. Otherwise, if the coupling is solely from the
neighbors to the hidden node, the dynamics of the neighboring nodes
will not be affected by the hidden node and, consequently, time series
from the neighboring nodes will contain absolutely no information
about the hidden node, rendering it undetectable. The action of
local noise source on a node is naturally directional, i.e., from the
source to the node.

In Ref.~\cite{SWL:2012}, it was demonstrated that, when the
compressive-sensing paradigm is applied to uncovering the network
topology~\cite{WYLKH:2011}, the predicted linkages associated with
nodes $\#$3 and $\#$7 are typically anomalously dense, and this piece
of information is basically what is needed to identify them as the
neighboring nodes of the hidden node. In addition, when different 
segments of the measurement data are used to reconstruct the coupling 
weights for these two nodes, the reconstructed 
exhibit significantly larger variances than those 
associated with other nodes. However, the predicted linkages associated 
with the nodes driven by local noise sources can exhibit behaviors 
similar to those due to the hidden nodes, leading to uncertainty in 
the detection of the hidden node. This issue is critical to developing 
algorithms for real-world applications. One possible solution~\cite{SLWD:2014}
was to exploit the principle of {\em differential signal} to investigate 
the behavior of the predicted link weights as a function of the data 
used in the reconstruction. Due to the advantage of compressive sensing, 
the required data amount can be quite small and, hence, even if the 
method requires systematic increase in the data amount, it would still 
be reasonably small. It was argued that demonstrated~\cite{SLWD:2014} 
that, when the various ratios of the predicted weights associated with 
all pairs of links between the possible neighboring nodes and the hidden 
node are examined, those associated with the hidden nodes and nodes under 
strong local noise show characteristically distinct behaviors, rendering 
unambiguous identification of the neighboring nodes of the hidden node. 
Any such ratio is essentially a kind of differential signal, because 
it is defined with respect to a pair of edges.

\subsubsection{Mathematical formulation of compressive sensing based
detection of a hidden node}
\label{subsubsec:HN_Math}

\paragraph*{Compressive-sensing based method to uncover weighted 
network dynamics and topology.}
Consider the typical setting of a complex network of $N$ coupled
oscillators in a noisy environment. The dynamics of each individual
node, when it is isolated from other nodes, can be described as
$\dot{\mathbf{x}}_{i}=\mathbf{F}_{i}(\mathbf{x}_{i}) + \xi \mathbf{\eta}_{i}$,
where $\mathbf{x}_{i} \in \mathbb{R}^{m}$ is the vector of state variables,
${\mathbf \eta}_{i}$ are an $m$-dimensional vector whose entries are
independent Gaussian random variables of zero mean and unit variance,
and $\xi$ denotes the noise amplitude. A weighted network can be described by
\begin{equation} \label{Eq:coupling_HN}
\dot{\mathbf{x}}_{i}=\mathbf{F}_{i}(\mathbf{x}_{i})+
\sum_{j=1,j\neq i}^{N}\mathcal{W}_{ij} \cdot [\mathbf{H}(\mathbf{x}_{j})
-\mathbf{H}(\mathbf{x}_{i})] + \xi \eta_{i},
\end{equation}
where $\mathcal{W}_{ij} \in \mathbb{R}^{m \times m}$ is the coupling matrix
between node $i$ and node $j$, and $\mathbf{H}$ is the coupling function. 
Defining
\begin{displaymath}
\mathbf{F}_{i}^{\prime}(\mathbf{x}_{i}) \equiv
\mathbf{F}_{i}(\mathbf{x}_{i})- \mathbf{H}(\mathbf{x}_{i}) \cdot
\sum_{j=1,j\neq i}^{N}\mathcal{W}_{ij},
\end{displaymath}
one has
\begin{equation} \label{Eq:Nolabel1_HN}
\dot{\mathbf{x}}_{i}=\mathbf{F}_{i}^{\prime}(\mathbf{x}_{i})+
\sum_{j=1,j\neq i}^{N}\mathcal{W}_{ij} \cdot \mathbf{H}(\mathbf{x}_{j})
+\xi \eta_{i},
\end{equation}
i.e., all terms directly associated with node $i$ have been grouped
into $\mathbf{F}_{i}^{\prime}(\mathbf{x}_{i})$. One can then expand
$\mathbf{F}^{\prime}(\mathbf{x}_{i})$ into the following form:
\begin{equation} \label{eq:expansion_general_HN}
\mathbf{F}_{i}^{\prime}(\mathbf{x}_{i})=\sum_{\gamma}
\tilde{\mathbf{a}}_{i}^{(\gamma)}
\cdot\tilde{\mathbf{g}}_{i}^{(\gamma)}(\mathbf{x}_{i}),
\end{equation}
where $\tilde{\mathbf{g}}_{i}^{(\gamma)}(\mathbf{x}_{i})$
are a set of orthogonal and complete base functions chosen
such that the coefficients $\tilde{\mathbf{a}}_{i}^{(\gamma)}$
are sparse. While the coupling function ${\mathbf H}({\mathbf x}_i)$ can
be expanded in a similar manner, for simplicity we assume that they are 
linear: $\mathbf{H}(\mathbf{x}_{i})=\mathbf{x}_{i}$. This leads to
\begin{equation} \label{Eq:expandEq_HN}
\mathbf{\dot{x}}_{i}=\sum_{\gamma}\tilde{\mathbf{a}}_{i}^{(\gamma)}
\cdot\tilde{\mathbf{g}}_{i}^{(\gamma)}(\mathbf{x}_{i})+
\sum_{j=1,j\neq i}^{N}\mathcal{W}_{ij}\cdot\mathbf{x}_{j}+\xi \eta_{i},
\end{equation}
where all the coefficients $\tilde{\mathbf{a}}_{i}^{(\gamma)}$ and weights
$\mathcal{W}_{ij}$ need to be determined from time series $\mathbf{x}_{i}$.
In particular, the coefficient vector $\tilde{\mathbf{a}}_{i}^{(\gamma)}$
determines the nodal dynamics and the weighted matrices $\mathcal{W}_{ij}$'s
give the full topology and coupling strength of the entire network.

Suppose simultaneous measurements are available of all state variables
$\mathbf{x}_{i}(t)$ and $\mathbf{x}_{i}(t+\delta t)$ at $M$ different
uniform instants of time at interval $\Delta t$ apart, where 
$\delta t \ll \Delta t$ so that the derivative vector $\mathbf{\dot{x}}_{i}$ 
can be estimated at each time instant. Equation~\eqref{Eq:expandEq_HN} 
for all $M$ time instants can then be written in a matrix form with the 
following measurement matrix:
\begin{equation} \label{Eq:matG_HN}
\mathcal{G}_{i}=\left(
\begin{array}{cccccc}
\tilde{\mathbf{g}}_{i}(t_{1}) & \mathbf{x_{1}}(t_{1}) & \cdots & \mathbf{x}_{k}(t_{1}) & \cdots & \mathbf{x}_{N}(t_{1})\\
\tilde{\mathbf{g}}_{i}(t_{2}) & \mathbf{x_{1}}(t_{2}) & \cdots & \mathbf{x}_{k}(t_{2}) & \cdots & \mathbf{x}_{N}(t_{2})\\
\vdots & \vdots & \cdots & \vdots & \cdots & \vdots\\
\tilde{\mathbf{g}}_{i}(t_{M}) & \mathbf{x_{1}}(t_{M}) & \cdots & \mathbf{x}_{k}(t_{M}) & \cdots & \mathbf{x}_{N}(t_{M})
\end{array}
\right),
\end{equation}
where the index $k$ in ${\mathbf x}_k(t)$ runs from 1 to $N$, $k \neq i$, and
each row of the matrix is determined by the available time series at one
instant of time. The derivatives at different time can be written in a
vector form as
${\mathbf X}_i= [\dot{\mathbf x}_i(t_1), \cdots, \dot{\mathbf x}_i(t_M)]^T$,
and the coefficients from the functional expansion and the weights associated
with all links in the network can be combined concisely into a vector 
$\mathbf{a}_{i}$ as
\begin{equation} \label{eq:Nolabel2_HN}
\mathbf{a}_{i}=[\tilde{\mathbf{a}}_{i},\mathbf{W}_{1i},\cdots,
\mathbf{W}_{i-1,i},\mathbf{W}_{i+1,i},\cdots,\mathbf{W}_{N,i}]^{T},
\end{equation}
where $[\cdot]^{T}$ denotes the transpose. For a properly chosen expansion 
base and a general complex network whose connections are sparse, 
the vector ${\mathbf a}_i$ to be determined is sparse as well. Finally,
Eq.~\eqref{Eq:expandEq_HN} can be written as
\begin{equation} \label{Eq:matEq_HN}
\mathbf{X}_{i} = \mathcal{G}_{i}\cdot\mathbf{a}_{i} + \xi\eta_i.
\end{equation}
In the absence of noise or if the noise amplitude is negligibly
small, Eq.~\eqref{Eq:matEq_HN} represents a linear equation but the
dimension of the unknown coefficient vector $\mathbf{a}_{i}$ can be much
larger than that of $\mathbf{X}_{i}$, and the measurement matrix
will have many more columns than rows. In order to fully recover the
network of $N$ nodes with each node having $m$ components, it is
necessary to solve $N \times m$ such equations. Since $\mathbf{a}_{i}$ is
sparse, insofar as its number of non-zero coefficients is smaller than the
dimension of $\mathbf{X}_{i}$, the vector $\mathbf{a}_{i}$ can be uniquely
and efficiently determined through compressive sensing.

\paragraph*{Detection of hidden node.}
A meaningful solution of Eq.~(\ref{Eq:matEq_HN}) based on compressive sensing 
requires that the derivative vector $\mathbf{X}_{i}$ and the measurement 
matrix $\mathcal{G}_{i}$ be entirely known which, in turn, requires time
series from all nodes. In this case, the information available for 
reconstruction of the complex networked system is deemed to be
{\em complete}~\cite{SWL:2012,SLWD:2014}. In the presence of a hidden 
node, for its immediate neighbors, the available information
will not be complete in the sense that some entries of the vector
$\mathbf{X}_{i}$ and the matrix $\mathcal{G}_{i}$ are unknown.
Let $h$ denote the hidden node. For any neighboring node of $h$,
the vector $\mathbf{X}_{i}$ and the matrix $\mathcal{G}_{i}$ in
Eq.~(\ref{Eq:matEq_HN}) now contain unknown entries at the locations
specified by the index $h$. For any other node not in the
immediate neighborhood of $h$, Eq.~\eqref{Eq:matEq_HN} is unaffected.
When compressive-sensing algorithm is used to solve Eq.~\eqref{Eq:matEq_HN},
there will then be large errors in the solution of the coefficient
vector $\mathbf{a}_{i}$ associated the neighboring nodes of $h$,
regardless of the amount of data used. In general, the so-obtained
coefficient vector $\mathbf{a}_{i}$ will not appear sparse. Instead,
most of its entries will not be zero, a manifestation of which is
that the node would appear to have links with almost every other
node in the network. In contrast, for nodes not in the neighborhood
of $h$, the corresponding errors will be small and can be reduced
by increasing the data amount, and the corresponding coefficient
vector will be sparse. It is this observation which makes identification
of the neighboring nodes of the hidden node possible in a noiseless
or weak-noise situation~\cite{SWL:2012}.

The need and the importance to distinguish the effects of hidden node 
from these of noise can be better seen by separating the term associated
with $h$ in Eq.~(\ref{Eq:expandEq_HN}) from those with other accessible
nodes in the network. Letting $l$ denote a node in the immediate
neighborhood of the hidden node $h$, we have
\begin{equation} \label{Eq:HiddCS_HN}
\mathbf{X}_{l} = \mathcal{G}^{\prime}_{l}\cdot \mathbf{a}^{\prime}_{l}
+ (\mathcal{W}_{lh} \cdot {\mathbf x}_h +  \xi \eta_l),
\end{equation}
where $\mathcal{G}^{\prime}_{l}$ is the new measurement matrix that
can be constructed from all available time series. While background
noise may be weak, the term $\mathcal{W}_{lh} \cdot {\mathbf x}_h$
can in general be large in the sense that it is comparable in magnitude
with other similar terms in Eq.~(\ref{Eq:expandEq_HN}). Thus, when the
network is under strong noise, especially for those nodes that are
connected to the neighboring nodes of the hidden node, the effects
of hidden node on the solution can be entangled with those due to noise.
In addition, if the coupling strength from the hidden node is weak,
it would be harder to identify the neighboring nodes. For example, a 
hidden node in a network with Gaussian weight distribution will be harder
to detect, due to the finite probability of the occurrence of near
zero weights. When the coupling strength is comparable or smaller than
the background noise amplitude, the corresponding link cannot be
detected. 

\paragraph*{Method to distinguish hidden nodes from local noise sources.} 

The basic idea to distinguish the effects of hidden node and of local 
noise sources~\cite{SLWD:2014} is based on the following consideration.
Take two neighboring nodes of the hidden node, labeled as $i$ and $j$.
Because the hidden node is a common neighbor of nodes $i$ and $j$,
the couplings from the hidden node should be approximately proportional
to each other, with the proportional constant determined by the ratio
of their link weights with the hidden node. When the dynamical equations
of nodes $i$ and $j$ are properly normalized, the terms due to the
hidden node tend to cancel each other, leaving the normalization constant
as a single unknown parameter that can be estimated subsequently. 
This parameter is the {\em cancellation ratio}, denoted as $\Omega_{ij}$.
As the data amount is increased, $\Omega_{ij}$ tends to its true value.
Practically one then expects to observe a systematic change in the 
estimated value of the ratio as data used in the compressive-sensing 
algorithm is increased from some small to relatively large amount. If
only local noise sources are present, the ratio should show no
systematic change with the data amount. Thus the distinct behaviors of
$\Omega_{ij}$ as the amount of data is increased provides a way
to distinguish the hidden node from noise and, at the same time, to
ascertain the existence of the hidden node. 

For simplicity, assume that all coupled oscillators share the same local 
coupling configuration and that each oscillator is coupled to any of its 
neighbors through one component of the state vector only. Thus, each 
row in the coupling matrix $\mathcal{W}_{ih}$ associated with a link 
between node $i$ and $h$ has only one non-zero element. Let $p$ denote 
the component of the hidden node coupled to the first component of node 
$i$, the dynamical equation of which can then be written as
\begin{equation} \label{Eq:withHid_HN}
[{\mathbf {\dot x}}_{i}]_{1} =[\sum_{\gamma}\tilde{\mathbf{a}}_{i}^{(\gamma)}
\cdot\tilde{\mathbf{g}}_{i}^{(\gamma)}(\mathbf{x}_{i})]_{1} +
[\sum_{k\neq i, h}^{N}\mathcal{W}_{ij}\cdot\mathbf{x}_{j}]_{1} 
+ w^{1p}_{ih} \cdot [{\mathbf x}_h]_{p} + \xi \eta_{i},
\end{equation}
where $[{\mathbf x}_h]_{p}$ denotes the time series of the $p$th component
of the hidden node, which is unavailable, and $w^{1p}_{ih}$ is the coupling
strength between the hidden node and node $i$. The dynamical equation of the
first component of neighboring node $j$ of the hidden node has a similar form.
Letting
\begin{equation} \label{Eq:omega_HN}
\Omega_{ij}=w_{ih}^{1p}/w_{jh}^{1p},
\end{equation}
be the cancellation ratio, multiplying $\Omega_{ij}$ to the equation of 
node $j$, and subtracting from it the equation for node $i$, one obtains
\begin{eqnarray} \label{Eq:hiddenNei_HN}
[{\mathbf{\dot x}}_{i}]_1 &=&\Omega_{ij}[{\mathbf {\dot x}}_{j}]_{1}
 + \sum_{\gamma}\tilde{{\mathbf a}}_{i}^{(\gamma)} \cdot
 \tilde{{\mathbf g}}_{i}^{(\gamma)}({\mathbf x}_i)
 +\sum_{k\neq i, h} w_{ik}^{1p} [{\mathbf x}_k]_{p} 
- \Omega_{ij}\sum_{\gamma}\tilde{{\mathbf a}}_{j}^{(\gamma)} \cdot
 \tilde{{\mathbf g}}_{j}^{(\gamma)}({\mathbf x}_j) \\ \nonumber
&-& \Omega_{ij} \sum_{k\neq j, h} w_{jk}^{1p} [{\mathbf x}_k]_{p} 
+(w_{ih}^{1p} -\Omega_{ij}w_{jh}^{1p}) \cdot [{\mathbf x}_h]_{p}
+\xi \eta_{i} - \Omega_{ij} \xi \eta_{j}.
\end{eqnarray}
It can be seen that the terms associate with $[{\mathbf x}_h]_{p}$ vanish 
and all deterministic terms on the left-hand side of 
Eq.~(\ref{Eq:hiddenNei_HN}) are known, which can then be
solved by the compressive-sensing method. From the coefficient vector 
so estimated, one can identify the coupling of nodes $i$ and $j$ to 
other nodes, except for the coupling between themselves since such terms 
have been absorbed into the nodal dynamics, and the couplings to their 
common neighborhood are degenerate in Eq.~(\ref{Eq:hiddenNei_HN}) and 
cannot be separated from each other. Effectively, one has combined 
the two nodes together by introducing the cancellation ratio $\Omega_{ij}$.

As a concrete example, consider the situation where each oscillator has 
three independent dynamical variables, named as $x$, $y$ and $z$. For 
the nodal and coupling dynamics polynomial expansions of order up to 
$n$ can be chosen. The $x$ component of the nodal dynamics 
$[\mathbf{F}_{i}^{\prime}(\mathbf{x}_{i})]_{x}$ for node $i$ is:
\begin{equation}
\nonumber
[\mathbf{F}_{i}^{\prime}(\mathbf{x}_{i})]_{x}=
\sum_{l_{x}=0}^{n}\sum_{l_{y}=0}^{n}\sum_{l_{z}=0}^{n}
[a_{l_{x}l_{y}l_{z}}]_{x}\cdot x_{i}^{l_{x}}y_{i}^{l_{y}}z_{i}^{l_{z}},
\end{equation}
and the coupling from other node $k$ to the $x$ component can be written as
\begin{displaymath}
C_{ik}^{x}=w_{ik}^{xx}\cdot x_{k}+w_{ik}^{xy}\cdot y_{k}
+w_{ik}^{xz}\cdot z_{k},
\end{displaymath}
where $w_{ik}^{xy}$ denotes the coupling weight from the $y$ component 
of node $k$ to the $x$ component of node $i$, and so on. The nodal 
dynamical terms in the matrix $\mathcal{G}_{i}$ are
\begin{displaymath}
[\mathbf{\tilde{g}}_{i}]_{x}=[x^0_iy^0_iz^0_i,x^1_iy^0_iz^0_i,
\cdots,x^n_iy^n_iz^n_i],
\end{displaymath}
and the corresponding coefficients are $[a_{l_{x}l_{y}l_{z}}]_{x}$. The 
vector of coupling weights is
\begin{displaymath} 
[\mathcal{W}_{ij}]_{x}=[w_{ij}^{xx},w_{ij}^{xy},w_{ij}^{xz}].
\end{displaymath}
Equation~(\ref{Eq:hiddenNei_HN}) becomes
\begin{displaymath}
\left(\begin{array}{c}
\dot{x}_{i}(t_{1})\\
\dot{x}_{i}(t_{2})\\
\vdots\\
\dot{x_{i}(t_{M})}
\end{array}\right)
\approx
\left(\begin{array}{ccccccc}
{\dot x}_j(t_1) & 1     & [{\tilde {\mathbf g}}_i(t_1)]_x & [{\tilde {\mathbf g}}_j(t_1)]_x     & x_{1}(t_{1}) & \cdots & z_{N}(t_{1})\\
{\dot x}_j(t_2) & 1     & [{\tilde {\mathbf g}}_i(t_2)]_x & [{\tilde {\mathbf g}}_j(t_2)]_x     & x_{1}(t_{2}) & \cdots & z_{N}(t_{2})\\
\vdots          &\vdots&\vdots                        & \vdots                         & \vdots      & \cdots & \vdots    \\
{\dot x}_j(t_M) & 1     & [{\tilde {\mathbf g}}_i(t_M)]_x & [{\tilde {\mathbf g}}_j(t_M)]_x     & x_{1}(t_{M}) & \cdots & z_{N}(t_{M})\\
\end{array}\right)
\cdot
\begin{pmatrix}
\Omega_{ij}\\
c           \\
{\mathbf {\tilde a}}^{\prime}_i\\
{-\Omega_{ij}\cdot \mathbf {\tilde a}}^{\prime}_j\\
w_{i1}^{xx} - \Omega_{ij} w_{j1}^{xx} \\
\vdots\\
w_{iN}^{xz} - \Omega_{ij} w_{jN}^{xz}
\end{pmatrix},
\end{displaymath}
where $c$ is the sum of constant terms from the dynamical equations 
of nodes $i$ and $j$, and ${\mathbf {\tilde a}}^{\prime}_i$ is the 
coefficient vector to be estimated which excludes all the constants. 
Using compressive sensing to solve this equation, one can recover 
the cancellation ratio $\Omega_{ij}$ and the equations of node $i$.
When $\Omega_{ij}$ is known the dynamics of node $j$ can be recovered from
the coefficient vector ${-\Omega_{ij}\cdot \mathbf {\tilde a}}^{\prime}_j$.

In Ref.~\cite{SLWD:2014}, an analysis and discussions were provided about
the possible extension of the method to systems of characteristically
different nodal dynamics and/or with multiple hidden nodes. In particular,
it was shown that the method can be readily adopted to network systems whose
nodal dynamics are not described by continuous-time differential
equations but by discrete-time processes such as evolutionary-game dynamics.
In such a case, the derivatives used for continuous-time systems can be
replaced by the agent payoffs. The cancellation factors can then be
calculated from data to differentiate the hidden nodes from local noise
sources. It was also shown that, under certain conditions with respect to
the coupling patterns between the hidden nodes and their neighboring
nodes, the cancellation factors can be estimated even when there are
multiple, entangled hidden nodes in the network.
 
\subsubsection{Examples of hidden node detection in the presence of
noise} \label{subsubsec:HN_Results}

The methodology of hidden node detection in the presence of noise can be 
illustrated using coupled oscillator networks. (Results from 
evolutionary-game dynamical networks can be found in Ref.~\cite{SLWD:2014}.)
As discussed in Sec.~\ref{subsec:CS_CON}, given
such a networked system, one can use compressive sensing to uncover 
all the nodal dynamical equations and coupling functions~\cite{WYLKH:2011}. 
The expansion base needs to be chosen properly so that the number of 
non-zero coefficients is small as compared with the total number $N_t$ 
of unknown coefficients. All $N_t$ coefficients constitute a coefficient 
vector to be estimated. The amount of data used can be conveniently 
characterized by $R_m$, the ratio of the number $M$ of data points used 
in the reconstruction, to $N_t$. 

\begin{figure}
\centering
\includegraphics[width=0.6\linewidth]{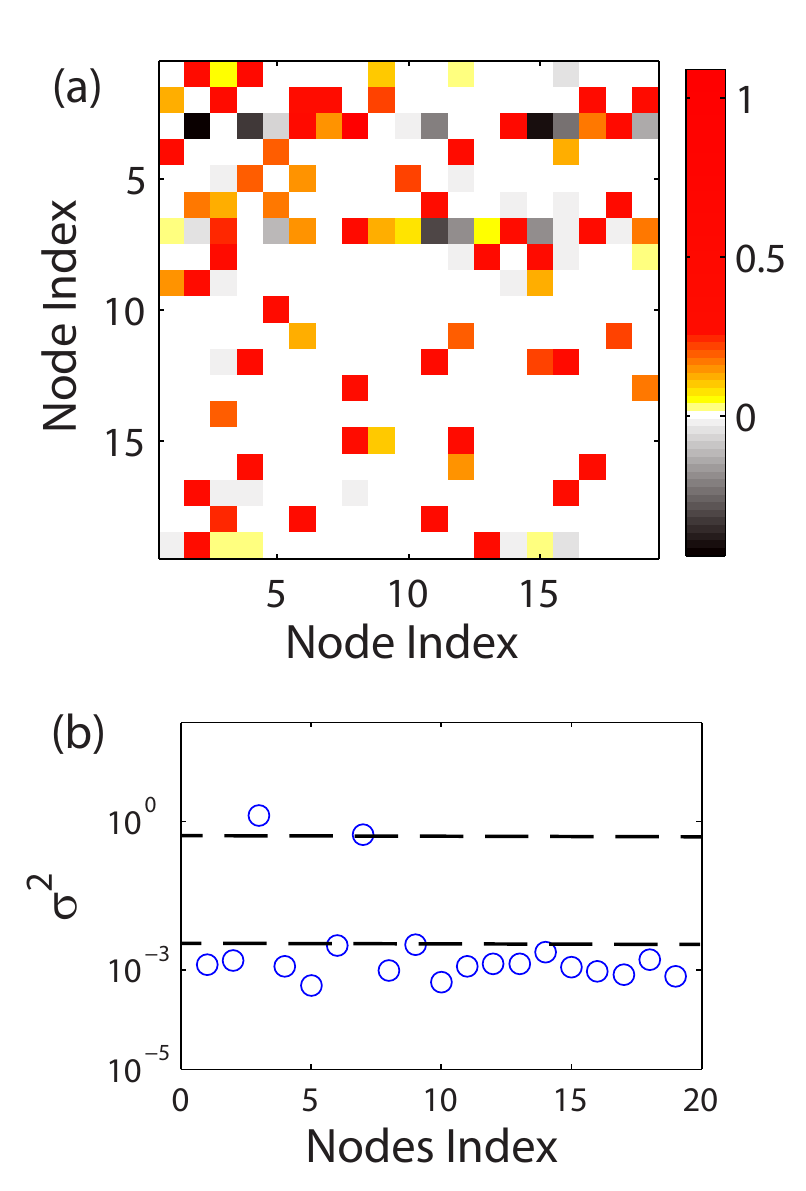}
\caption{\small {\bf Example of hidden node detection without
local noise sources}. 
For the network in Fig.~(\ref{fig:HD_Ros_HN}), (a) predicted
coupling matrix for all nodes except node $\#$20. Time series from
nodes $\#$1 to $\#$19 are available, while node $\#$20 is hidden.
The predicted weights are indicated by color coding and the amount
of data used is $R_{m}=0.7$. The abnormally dense patterns in the 3rd
and 7th rows suggest that nodes $\#$3 and $\#$7 are the immediate
neighbors of the hidden node. (b) Variance $\sigma^2$ of the predicted
coefficients for all accessible nodes, which is calculated using
20 independent reconstructions based on different segments of the
data. The variances associated with nodes $\#$3 and $\#$7 are
apparently much larger than those with the other nodes, confirming that
these are the neighboring nodes of the hidden node. There is a
definite gap between the values of the variance associated with
neighboring and non-neighboring nodes of the hidden node, as
indicated by the two horizontal dashed lines in (b). 
From Ref.~\cite{SLWD:2014} with permission.}
\label{fig:Weights_HN}
\end{figure}

The method to differentiate hidden nodes and noise was 
tested~\cite{SLWD:2014} using random networks of nonlinear/chaotic 
oscillators, where the nodal dynamics were chosen to be those from 
the R\"{o}ssler oscillator~\cite{Rossler:1976},
\begin{displaymath}
[\dot x_i,\dot y_i,\dot z_i]=[ -y_i-z_i, x_i + 0.2y_i,0.2+z_i(x_i-5.7)],
\end{displaymath}
which exhibits a chaotic attractor. The size of the network was 
varied from 20 to 100, and the probability of connection between any 
two nodes is 0.04. The network link weights are equally distributed in 
$[0.1, 0.5]$ (arbitrary). Background noise of amplitude $\xi$ was 
applied (independently) to every oscillator in the network, with 
amplitude varying from $10^{-4}$ to $5\times10^{-3}$. The noise 
amplitude is thus smaller than the average coupling strength
of the network. The tolerance parameter $\varepsilon$ in the 
compressive sensing algorithm can be adjusted in accordance with the 
noise amplitude~\cite{SLWD:2014}. Time series are generated by using
the standard Heun's algorithm~\cite{Gardiner:book} to integrate the 
stochastic differential equations. To approximate the velocity field, 
a third-order polynomial expansion was used in the compressive-sensing 
formulation. (In Ref.~\cite{SLWD:2014}, more examples can be found 
using network systems of varying sizes, different weight distributions 
and topologies, and alternative nodal dynamics.)

\paragraph*{Illustration of hidden node detection.}
Consider the network in 
Fig.~\ref{fig:HD_Ros_HN}, where only background noise is present and 
there are no local noise sources. Linear coupling between any pair of
connected nodes is from the $z$-component to the $x$-component in the
R\"{o}ssler system. From the available time series (nodes $\#$1-19), 
the coefficient vector can be solved using compressive sensing.
In particular, for node $i$, the terms associated with couplings from
the $z$-components of other nodes appear in the $i$th row of the
coupling matrix. As shown in Fig.~\ref{fig:Weights_HN}(a), when the data
amount is $R_{m}=0.7$, the network's coupling matrix can be predicted.
The predicted links and the associated weights are sparse for all
nodes except for nodes $\#$3 and $\#$7, the neighbors of the hidden node.
While there are small errors in the predicted weights due to
background noise, the predicted couplings for the two neighbors of
the hidden node, which correspond to the 3rd and the 7th row in the
coupling matrix, appear to be from almost all other nodes in the
network and some coupling strength is even negative. Such anomalies
associated with the predicted coupling patterns of the neighboring
nodes of the hidden node cannot be removed by increasing the data
amount. Nonetheless, it is precisely these anomalies which hint at
the likelihood that these two ``abnormal'' nodes are connected
with a hidden node.

While abnormally high connectivity predicted for a node is likely 
indication that it belongs to the neighborhood of the hidden node, 
in complex networks there are hub nodes with abnormally large degrees, 
especially for scale-free networks~\cite{BA:1999}. In order to 
distinguish a hidden node's neighboring node from some hub node, one
can exploit the variance of the predicted coupling constants, which
can be calculated from different segments of the available data sets. 
Due to the intrinsically low-data requirement associated with 
compressive sensing, the calculation of the variance is feasible 
because any reasonable time series can be broken into a number of 
segments, and prediction can be made from each data segment. For nodes 
not in the neighborhood of the hidden node, the variances are small 
as the predicted results hardly change when different segments of the 
time series are used. However, for the neighboring nodes of the hidden 
node, due to lack of complete information needed to construct the
measurement matrix, the variance values can be much larger.
Figure~\ref{fig:Weights_HN}(b) shows the variance $\sigma^2$ in the 
predicted coupling strength for all 19 accessible nodes. It can be 
seen that the values of the variance for the neighboring nodes of the 
hidden node, nodes $\#$3 and $\#$7, are at or above the upper dashed line 
and are significantly larger than those associated with all 
other nodes that all fall below the lower dashed line. This indicates 
strongly that they are indeed the neighboring nodes of the hidden node. 
The gap between the two dashed lines can be taken as a quantitative 
measure of the detectability of the hidden node. The larger the gap, 
the more reliable it is to distinguish the neighbors of the hidden node 
from the nodes that not in the neighborhood. The results in 
Fig.~\ref{fig:Weights_HN} thus indicate that the location of the hidden 
node in the network can be reliably inferred in the presence 
of weak background noise. The size of the gap, or the hidden-node
detectability depends on the system details. A systematic analysis of the
detectability measure was done~\cite{SLWD:2014}, where it was found 
that the variance due to the hidden node is mainly determined by the 
strength of its coupling with the accessible nodes in the network. 
It was also found that system size and network topology have little 
effect on the hidden-node detectability. It is worth emphasizing
that the detectability relies also on successful reconstruction of all nodes
that are not in the neighborhood of the hidden node, which determine the
lower dashed line in Fig.~\ref{fig:Weights_HN}.

\begin{figure}
\centering
\includegraphics[width=0.8\linewidth]{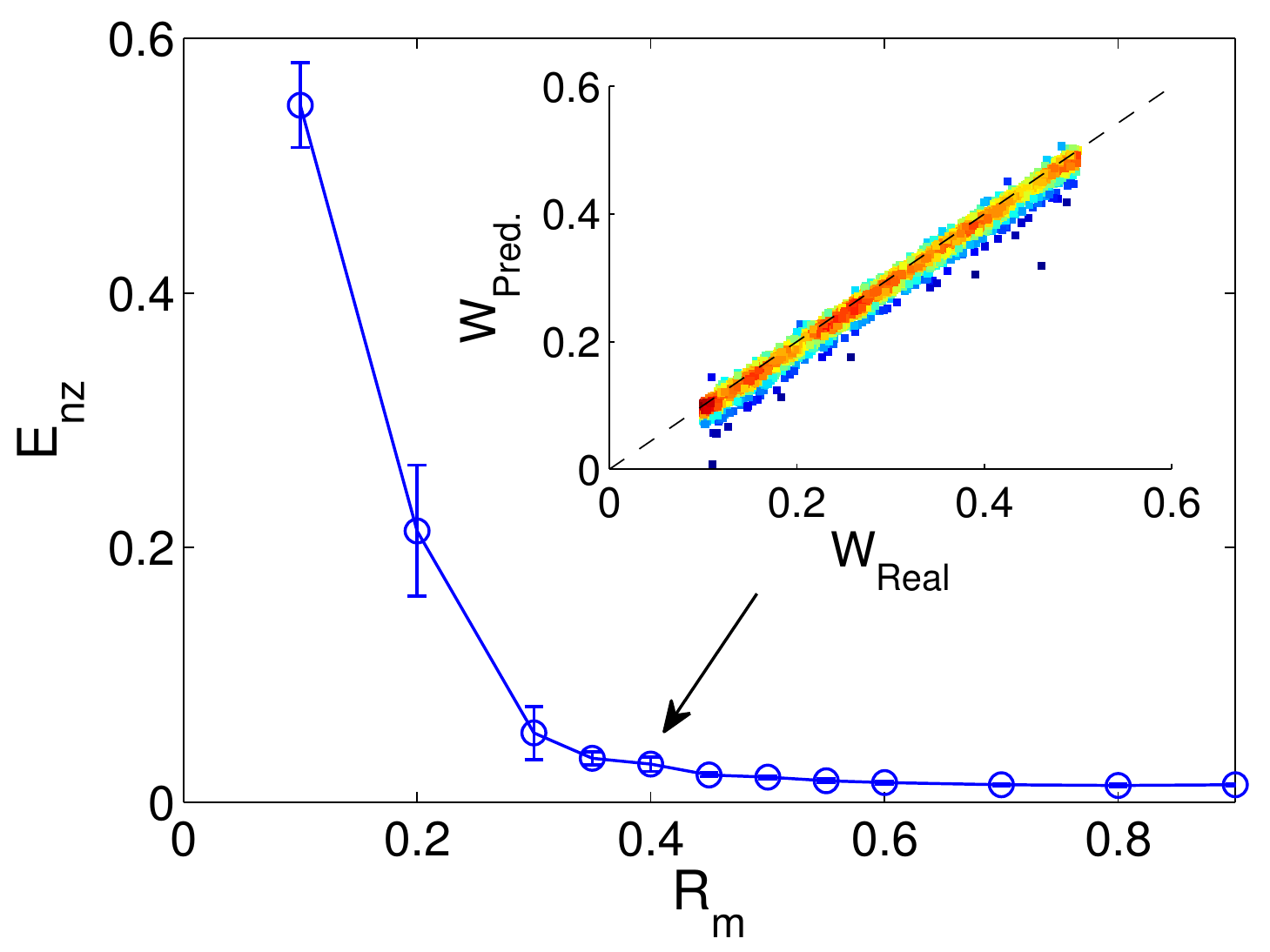}
\caption{\small {\bf Error analysis of hidden node detection in the presence of
weak background noise}. For random networks of size $N = 100$ with uniform 
weight distribution in $[0.1, 0.5]$, prediction error $E_{nz}$ associated 
with nonzero coefficients of the dynamical equations of all nodes except 
for the neighboring nodes of the hidden node, as a function of normalized
data amount $R_{m}$. The background noise amplitude is $\xi=10^{-3}$ for
all nodes. All data points are obtained from 10 independent realizations.
Inset is a comparison of the predicted and actual weights for all
existent links. Each dot represents one such link, and its $x$-value is
the actual weight while the $y$-value is the corresponding predicted
result. The color for each dot is determined by the dot density around it,
while bright color signifies high density. The arrow indicates the value
of $R_{m}$ used in the comparison study. The tolerance of the
compressive-sensing algorithm is set to be $\varepsilon=0.5$.
From Ref.~\cite{SLWD:2014} with permission.}
\label{fig:Enz_HN}
\end{figure}

The reliability of the reconstruction results can be quantified by
investigating how the prediction errors in the link weights of all 
accessible nodes, except the predicted neighbors of the hidden node, 
change with the data amount. For an existent link, one can use the 
normalized absolute error $E_{nz}$, the error in the estimated weight 
with respect to the true one, normalized by the value of the true link 
weight. Figure~\ref{fig:Enz_HN} shows the results for $N = 100$. The 
link weights are uniformly distributed in the interval $[0.1, 0.5]$ and 
the background noise amplitude is $\xi=10^{-3}$. The tolerance parameter 
in the compressive-sensing algorithm was set to be $\varepsilon=0.5$, 
which is optimal for this noise amplitude. (Details of determining the 
optimal tolerance parameter for different values of the background noise 
amplitude can be found in Ref.~\cite{SLWD:2014}.) It can be seen that 
for $R_{m}>0.4$, $E_{nz}$ decreases to the small value of about $0.01$, 
which is determined by the background noise level. As $R_{m}$ is increased 
further, the error is bounded by a small value determined by the noise 
amplitude, indicating that the reconstruction is robust. Although the 
value of $E_{nz}$ does not decrease further toward zero due to noise, 
the prediction results are reliable in the sense that the predicted 
weights and the real values agree with each other, as shown in the
inset of Fig.~\ref{fig:Enz_HN}, a comparison of the actual and the
predicted weights for all existent links. All the predicted results 
are in the vicinities of the corresponding actual values, as indicated 
by a heavy concentration of the dots along the diagonal line. The central 
region in the dot distribution has brighter color than the marginal regions,
confirming that vast majority of the predicted results are accurate. 
In Ref.~\cite{SLWD:2014}, it was further shown that robust reconstruction 
can be achieved regardless of the network size, connection topology and 
weight distributions, insofar as sufficient data are available.

The error measure $E_{nz}$ to characterize the accuracy of the
reconstruction is similar to $z$-scores, or the standard score in
statistics, with the minor difference being that the z-scores use the standard
derivatives of the distribution to normalize the raw scores, while 
the exact values were used~\cite{SLWD:2014} in the examples. In realistic 
applications the exact values are usually not available, so it is necessary 
to use the $z$-score measure.

\begin{figure}
\centering
\includegraphics[width=0.6\linewidth]{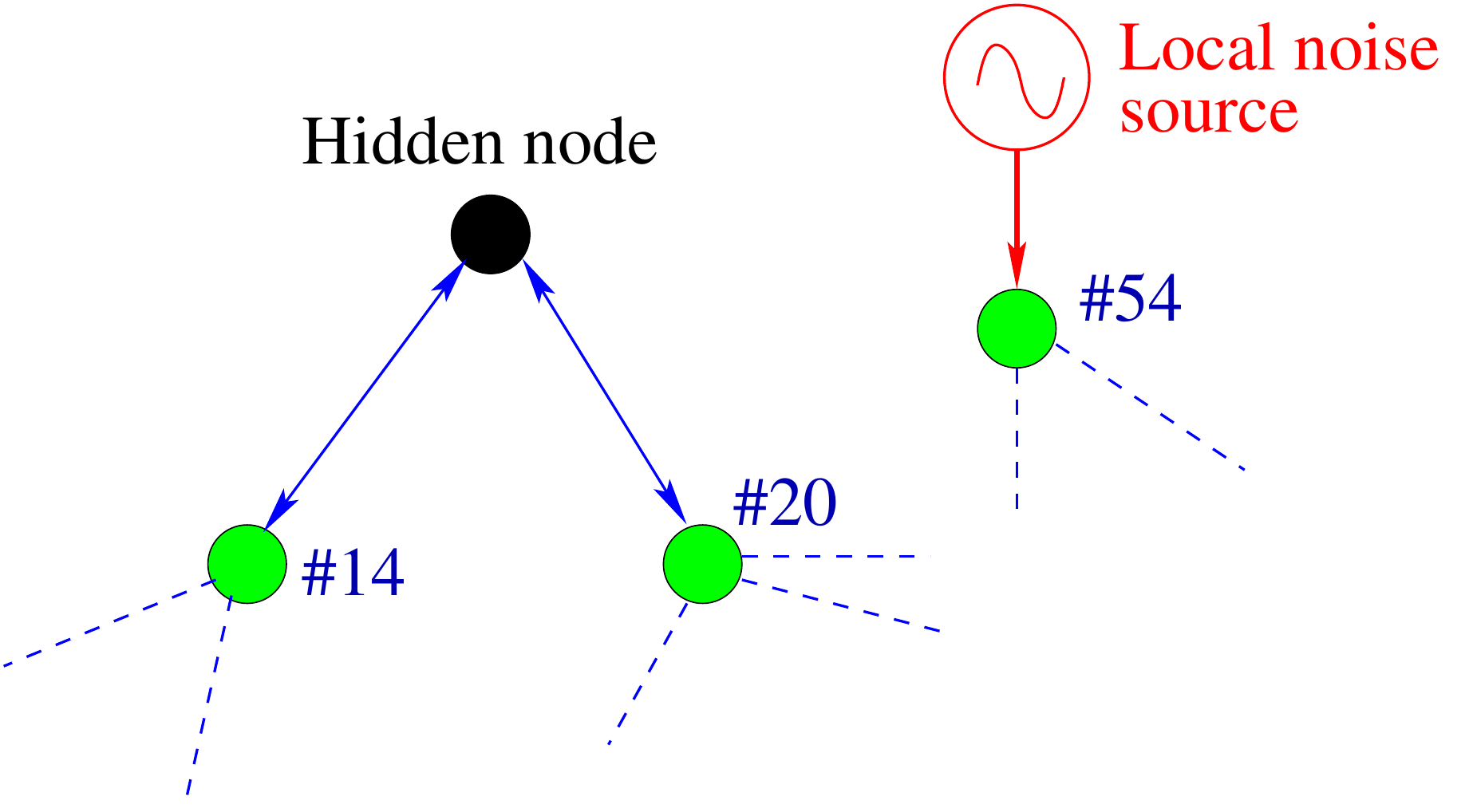}
\caption{\small {\bf Typical local structure of a hidden node}.  
Schematic illustration of a hidden node and its coupling configuration 
with two neighbors in a random network of $N = 61$ nodes, where 60 are 
accessible. A strong noise source is present at node $\#$54.
From Ref.~\cite{SLWD:2014} with permission.}
\label{fig:HN_SN_Schematic_HN}
\end{figure}

It should be emphasized that there are two types of ``dense'' connections: 
one from reconstruction and another intrinsic to the network. In particular,
in the two-dimensional representation of the reconstruction results [e.g., 
Fig.~\ref{fig:Weights_HN}(A)], the neighboring nodes of the hidden node
typically appear densely linked to many other nodes in the network. These 
can be a result of lack of incomplete information (i.e., time series) due
to the hidden node (in this case, there is indeed a hidden node), or
the intrinsic dense connection pattern associated with, for example,
a hub node in a scale-free network. The purpose of examining the variances 
of the reconstructed connections from independent data segments is for 
distinguishing these two possibilities. Extensive 
computations~\cite{SLWD:2014} indicated that a combination of the dense 
connection and large variance can ascertain the existence of hidden node 
with confidence.

\paragraph*{Differentiating hidden node from local noise sources.}
When strong noise sources are present at certain nodes, the predicted 
coupling patterns of the neighboring nodes of these nodes will show 
anomalies. (Here the meaning of the term ``strong'' is that the 
amplitudes of the random disturbances are order-of-magnitude larger 
than that of background noise.) The method based on the cancellation 
ratio was demonstrated~\cite{SLWD:2014} to be effective at distinguishing 
hidden nodes from local noise sources, insofar as the hidden node
has at least two neighboring nodes not subject to such disturbances. To
be concrete, consider a network of $N = 61$ coupled chaotic R\"{o}ssler 
oscillators, which has 60 accessible nodes and one hidden node ($\#$61) 
that is coupled to two neighbors: nodes $\#$14 and $\#$20, as illustrated 
schematically in Fig.~\ref{fig:HN_SN_Schematic_HN}. Assume a strong 
noise source is present at node $\#$54. It was found~\cite{SLWD:2014} 
that the reconstructed weights match their true values to high accuracy.
It was also found that the reconstructed coefficients including the
ratio $\Omega_{ij}$ are all constant and invariant with respect to 
different data segments, a strong indication that the pair of nodes are the
neighboring nodes of the same hidden node, thereby confirming its existence.

\begin{figure}
\centering
\includegraphics[width=0.5\linewidth]{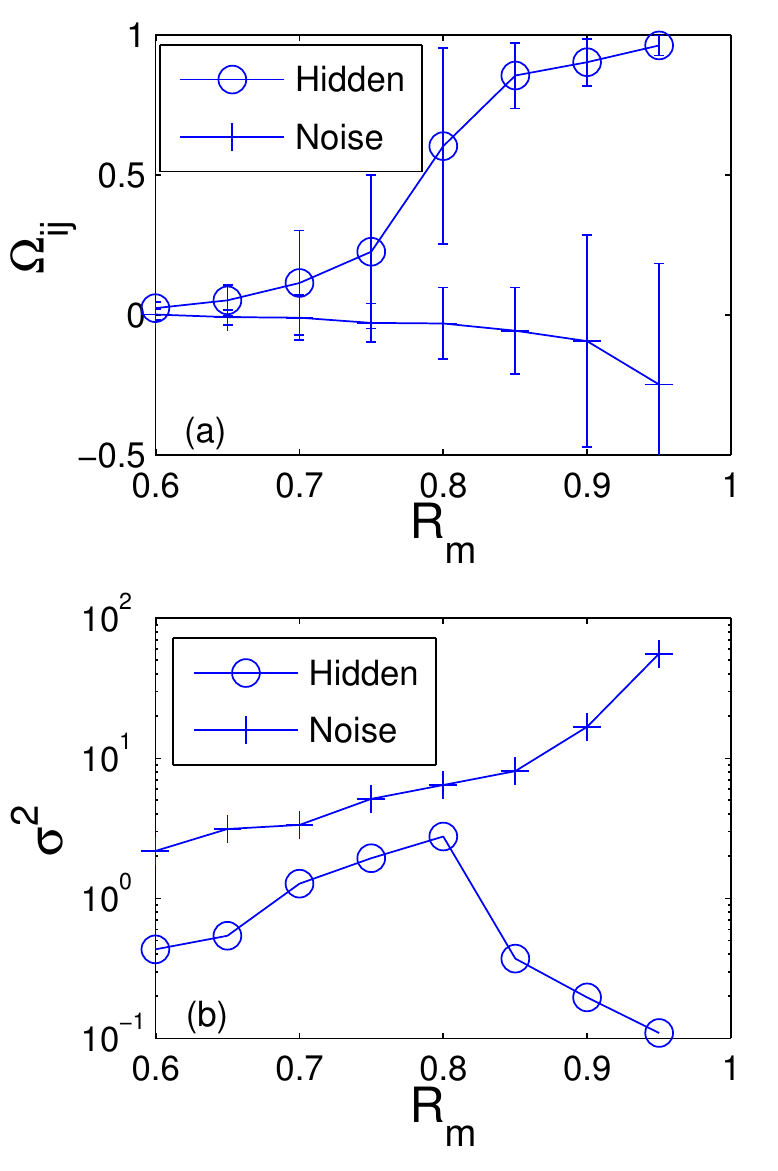}
\caption{\small {\bf Distinguishing hidden nodes from local noise sources}.
For the network described in Fig. \ref{fig:HN_SN_Schematic_HN},
(a) Predicted values of the cancellation ratio $\Omega_{ij}$ obtained 
from the differential signal of two neighboring nodes of the hidden 
node ($\#$14 and $\#$20, indicated by circles) and from the differential 
signal of nodes $\#$14 and $\#$54, where the latter is driven by noise 
of amplitude $\xi = 10^{-2}$ (crosses). (b) Average variance values 
of the predicted local coefficient vectors for the two combinations. 
The background noise amplitude is $\xi=10^{-5}$. The results were 
obtained from 20 independent realizations. From Ref.~\cite{SLWD:2014} 
with permission.}
\label{fig:Comp_HN}
\end{figure}

When there are at least two accessible nodes in the neighborhood of
the hidden node that are not subject to strong noisy disturbance,
such as nodes $\#$14 and $\#$20, as the data amount $R_m$ is increased 
towards $100\%$, the cancellation ratio should also increase and approach 
unity. This behavior is shown as the open circles in 
Fig.~\ref{fig:Comp_HN}(a). However, when a node is driven by a local noise 
source, regardless of whether it is in the neighborhood of the hidden node, 
the cancellation ratio calculated from this node and any other accessible 
node in the network exhibits a characteristically different behavior. 
Consider, for example, nodes $\#$14 and $\#$54. The reconstructed 
connection patterns of these two nodes both show anomalies, as they 
appear to be coupled with all other nodes in the network. In contrast 
to the case where the pair of nodes are influenced by the hidden node 
only, here the cancellation ratio does not exhibit any appreciable increase 
as the data amount is increased, as shown with the crosses in 
Fig.~\ref{fig:Comp_HN}(a). In addition, the average variance values of 
the predicted coefficient vectors of the two nodes exhibit 
characteristically different behaviors, depending on whether any one 
node in the pair is driven by strong noise or not. In particular, for 
the node pair $\#$14 and $\#$20, since neither is under strong noise, 
the average variance will decrease toward zero as $R_m$ approaches unity, 
as shown in Fig.~\ref{fig:Comp_HN}(b) (open circles). In contrast, for 
the node pair $\#$14 and $\#$54, the average variance will increase with 
$R_m$, as shown in Fig.~\ref{fig:Comp_HN}(b) (crosses). This is because, 
when one node is under strong random driving, the input to the 
compressive-sensing algorithm will be noisy, so its performance will 
deteriorate. However, compressive sensing can perform reliably when the 
input data are ``clean,'' even when they are sparse. Increasing the data 
amount beyond a threshold is not necessarily helpful, but longer and 
noisier data sets can degrade significantly the performance. The results 
in Figs.~\ref{fig:Comp_HN}(a,b) thus demonstrate that the cancellation
ratio between a pair of nodes, in combination with the average variance
of the predicted coefficient vectors associated with the two nodes, can
effectively distinguish a hidden node from a local noise source.
If there are more than one hidden node or there is a cluster of hidden
nodes, the procedure to estimate the cancellation factors is similar but
requires additional information about the neighboring nodes of the hidden
nodes. The cancellation-factor based method can be extended to network
systems with nodal dynamics not of the continuous-time type, such as
evolutionary-game dynamics~\cite{SLWD:2014}.

\begin{figure}
\centering
\includegraphics[width=0.8\linewidth]{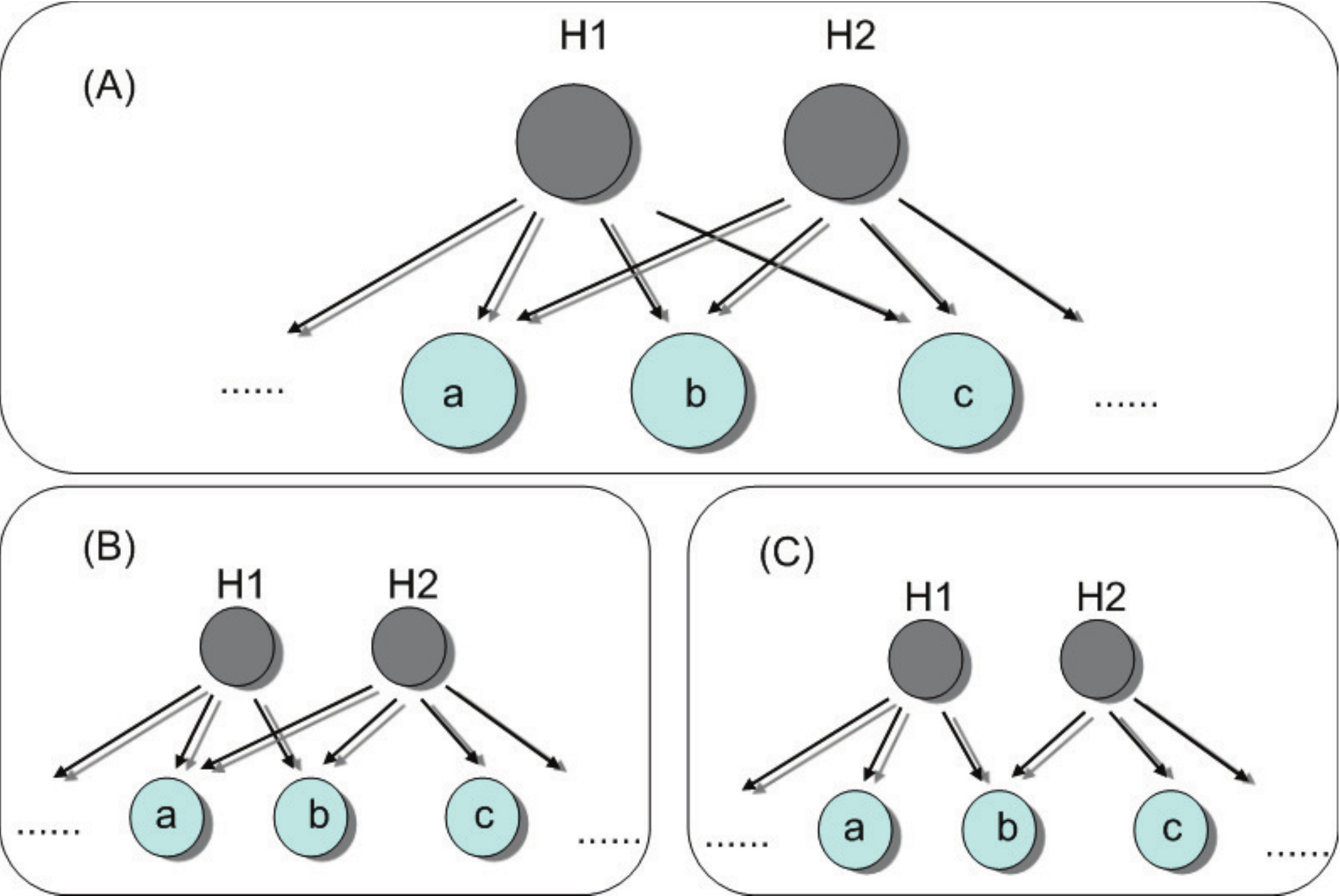}
\caption{\small {\bf Schematic illustration of configurations of two entangled
hidden nodes}. The dark nodes, H1 and H2 represent the two hidden nodes,
and the light blue nodes, a, b and c are accessible nodes. In all
panels, each hidden node has at least two neighboring accessible nodes.
In panel (A), all three accessible nodes are jointly coupled to the two
hidden nodes. In panel (B), two accessible nodes are shared by the hidden
nodes, while in panel (C) only one accessible node is coupled to both
hidden nodes. From Ref.~\cite{SLWD:2014} with permission.}
\label{Fig:MultiNode_HN}
\end{figure}

\paragraph*{Multiple entangled hidden nodes.}
When there are multiple hidden nodes in a network, there is a possibility that
an identified node is connected to more than one hidden node. For example,
node $b$ in all three panels of Fig.~\ref{Fig:MultiNode_HN} is affected by
hidden nodes $H1$ and $H2$. The cancellation factor(s) can still be estimated
if each hidden node in the network has at least two nodes (otherwise it
can be treated as a local noise source).

For simplicity, first consider the situation of two entangled hidden nodes
in that they have overlapping neighborhoods. Some possible
coupling patterns are shown in Fig.~\ref{Fig:MultiNode_HN}. In panel (A), the
two hidden nodes share three NH nodes but with different coupling strength.
In panels (B) and (C), the hidden nodes share two or one common node(s).
If the two hidden nodes do not share any nodes, the cancellation factors
can be estimated independently using the same method as for the situation of
one hidden node. Anther extreme case is that the two hidden nodes have and
only have two identical nodes, which is equivalent to the case of one
hidden node.

A procedure was developed~\cite{SLWD:2014} to estimate the cancellation 
factors for the situation in Fig.~\ref{Fig:MultiNode_HN}(A). The 
procedure can be extended to the other two cases in a straightforward 
manner by setting zero the weights from the hidden nodes to nodes $a$ 
and (or) $b$. For any of the three neighboring nodes $i \in [a, b, c]$, 
its dynamical equation can be expanded as
\begin{equation} \label{Eq:with2Hid_HN}
[{\mathbf {\dot x}}_{i}]_{1} =[\sum_{\gamma}\tilde{\mathbf{a}}_{i}^{(\gamma)}
\cdot\tilde{\mathbf{g}}_{i}^{(\gamma)}(\mathbf{x}_{i})]_{1} +
[\sum_{k\neq i, H_1, H_2}^{N}\mathcal{W}_{ij}\cdot\mathbf{x}_{j}]_{1}
+ w^{1p}_{i,H_1} \cdot [{\mathbf x}_{H_1}]_{p} + w^{1p}_{i,H_2} \cdot 
[{\mathbf x}_{H_2}]_{p} + \xi \eta_{i}.
\end{equation}
To cancel the hidden-node effect in one node, e.g., node $b$, one 
needs the time series of the two other non-hidden (NH) nodes, $a$ 
and $c$, so as to cancel the coupling terms from the two hidden nodes. 
Let the corresponding cancellation factors be $\Omega_{ba}$ and 
$\Omega_{bc}$. A new dynamical equation without the interferences from 
the hidden nodes can then be obtained:
\begin{eqnarray} \label{Eq:Cancel_HN}
[{\mathbf {\dot x}}_{a}]_{1} &-& \Omega_{ba} [{\mathbf {\dot x}}_{b}]_{1} - \Omega_{ca}[{\mathbf {\dot x}}_{c}]_{1} \nonumber \\
        &=& [\sum_{\gamma}\tilde{\mathbf{a}}_{a}^{(\gamma)} \cdot\tilde{\mathbf{g}}_{a}^{(\gamma)}(\mathbf{x}_{a})]_{1}
        - \Omega_{ba}\cdot[\sum_{\gamma}\tilde{\mathbf{a}}_{b}^{(\gamma)} \cdot\tilde{\mathbf{g}}_{b}^{(\gamma)}(\mathbf{x}_{b})]_{1} -
        \Omega_{ca}\cdot[\sum_{\gamma}\tilde{\mathbf{a}}_{c}^{(\gamma)} \cdot\tilde{\mathbf{g}}_{c}^{(\gamma)}(\mathbf{x}_{c})]_{1} \nonumber \\
        &+& [ w^{1p}_{a,H_1} - \Omega_{ba} w^{1p}_{b,H_1} - \Omega_{ca} w^{1p}_{c,H_1}] \cdot [{\mathbf x}_{H_1}]_{p} +
                [ w^{1p}_{a,H_2} - \Omega_{ba} w^{1p}_{b,H_2} - \Omega_{ca} w^{1p}_{c,H_2}] \cdot [{\mathbf x}_{H_2}]_{p} \nonumber \\
        &+& [\sum_{j\neq i, H_1, H_2}^{N} { (\mathcal{W}_{aj} - \Omega_{ba} \mathcal{W}_{bj} -\Omega_{ca} \mathcal{W}_{cj} )\cdot\mathbf{x}_{j}  }]_{1} +
            \xi (\eta_{a} -\Omega_{ba} \eta_{b} - \Omega_{ca}\eta_{c}),
\end{eqnarray}
where it is assumed that the coefficients associated with the coupling 
terms from the hidden nodes are zero. This can be achieved when the equation
\begin{equation}
\left(\begin{array}{cc}
w_{b,H_{1}} & w_{c,H_{2}}\\
w_{b,H_{2}} & w_{c,H_{2}}
\end{array}
\right) \cdot
\left(\begin{array}{c}
\Omega_{ba}\\
\Omega_{ca}
\end{array}\right)
\equiv \mathcal{M}_{w}\cdot
\left(\begin{array}{c}
\Omega_{ba}\\
\Omega_{ca}
\end{array}\right)
= \left(\begin{array}{c}
w_{a,H_{1}}\\
w_{a,H_{2}}
\end{array}\right) 
\end{equation}
holds and has only one trivial solution. The couplings to the hidden 
nodes should thus satisfy the condition $\mbox{det}(\mathcal{M}_{w})=2$. 
One can then estimate the cancellation factors using Eq.~(\ref{Eq:Cancel_HN}),
which is free of influence from the hidden nodes.

\begin{figure}
\centering
\includegraphics[width=0.7\linewidth]{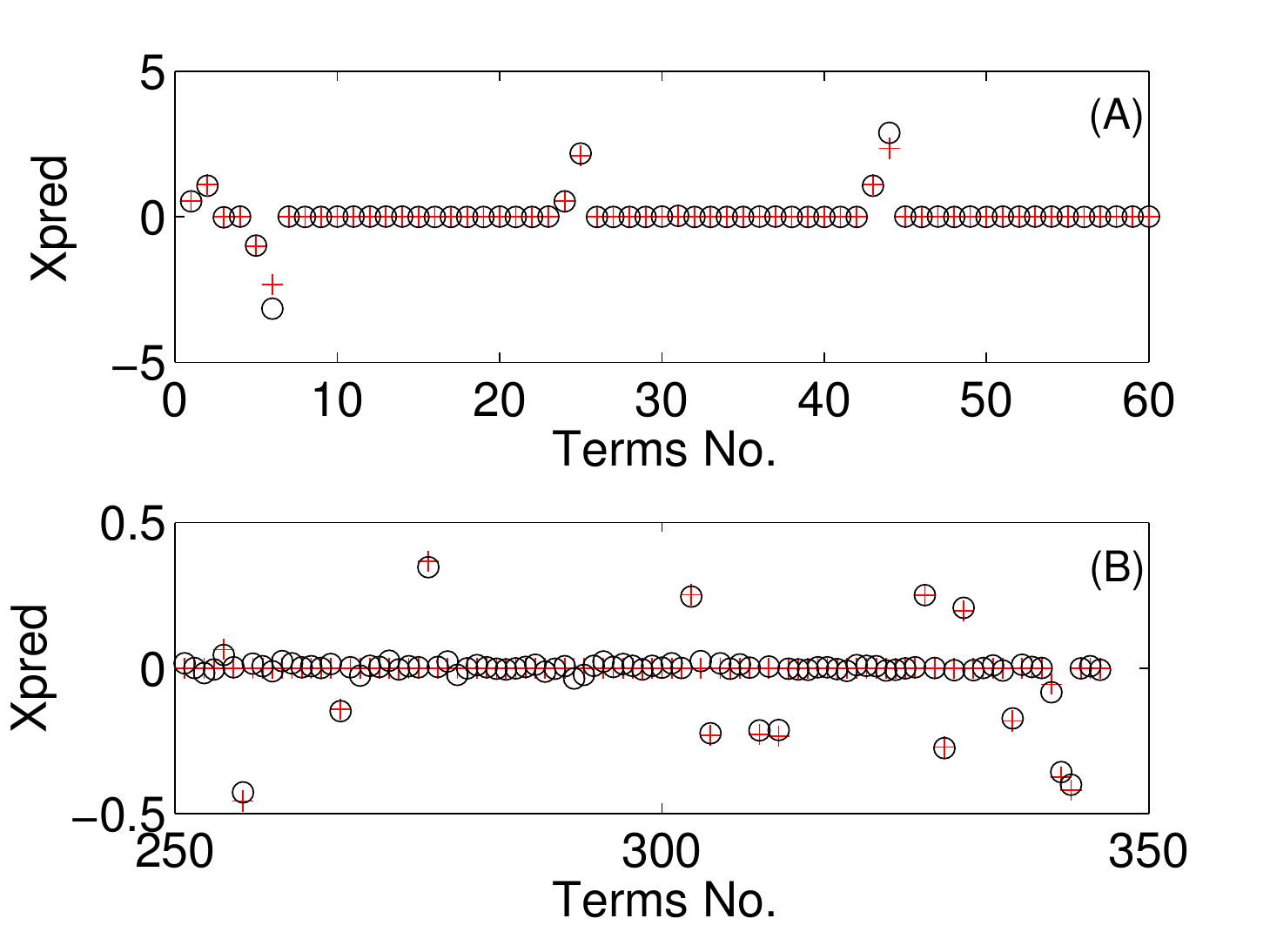}
\caption{\small {\bf Demonstration of detection of multiple entangled hidden
nodes.} For a small oscillator network with two entangled hidden nodes,
comparison of the predicted coefficients (black circles) and the actual
ones (red crosses) obtained by applying the method of cancellation factor.
In panel (A), coefficients from No.~1 to No.~60 correspond to these
for the dynamical equations. Panel (B) specifies the predicted results
for the coupling functions. From Ref.~\cite{SLWD:2014} with permission.}
\label{Fig:Xpred_HN}
\end{figure}

To demonstrate the procedure, a small
chaotic R\"ossler oscillator network of five nodes was tested, as 
illustrated in Fig.~\ref{Fig:MultiNode_HN}(A). A reconstruction 
procedure was carried out for \#b, utilizing the time series
from all three NH nodes. Since the unknown coefficients are highly
correlated and dense in this small system, the least squares method
can be chosen for reconstruction with relative data amount $R_{m}=1.2$. 
The predicted and the actual results are shown in the Fig.~\ref{Fig:Xpred_HN}. 
In panel (A), terms \#1 and \#2 denote the two cancellation factors. The 
nodal dynamics for node \#b, \#a and node \#c are listed in order. 
Panel (B) shows the entangled couplings in the network. All predicted 
terms match well with the actual ones.

When an NH node is coupled with $K$ ($K\geq3$) hidden nodes, their successful
detection requires that every hidden node be connected with two or more
NH nodes. Then $K$ cancellation factors can be estimated when the coupling
weights satisfy the condition $\mbox{det}(\mathcal{M}_{w})=K$, where the
elements $m_{ih}$ in $\mathcal{M}_{w}$ correspond to the coupling terms from
the $h$th hidden node to the $i$th NH node.

\subsection{Identifying chaotic elements in neuronal networks}
\label{subsec:CS_neuron}

Compressive sensing can be exploited to identify a subset of chaotic 
elements embedded in a network of nonlinear oscillators from time 
series. The oscillators, when isolated, are not identical in that their 
parameters are different, so dynamically they can be in distinct regimes. 
For example, all oscillators can be described by differential equations 
of the same mathematical form, but with different parameters. Consider the
situation where only a small subset of the oscillators are chaotic, and 
the remaining oscillators are in dynamical regimes of regular oscillations. 
Due to the mutual couplings among the oscillators, the measured time series 
from most oscillators would appear random. The challenge is to identify 
the small subset of originally (``truly'') chaotic oscillators.

The problem of identifying chaotic elements from a network of
coupled oscillators arises in biological systems and biomedical
applications. For example, for a network of coupled neurons
that exhibit regular oscillations in a normal state, 
the parameters of each isolated neuron are in regular regime. Under
external perturbation or slow environmental influences the parameters
of some neurons can drift into a chaotic regime. When this occurs the
whole network would appear to behave chaotically, which may correspond
to certain disease. The coupling and nonlinearity stipulate that
irregular oscillations at the network level can emerge even if only
a few oscillators have gone ``bad.'' It is thus desirable to be able
to pin down the origin of the ill-behaved oscillators - the few
chaotic neurons among a large number of healthy ones.

One might attempt to use the traditional approach of time-delayed
coordinate embedding to reconstruct the phase space of the underlying
dynamical system~\cite{Takens:1981,PCFS:1980,KS:book} and then to
compute the Lyapunov exponents~\cite{WSSV:1985,EKRC:1986}. However,
for a network of nonlinear oscillators, the
phase-space dimension is high and an estimate of the largest Lyapunov
exponent would only indicate if the {\em whole} coupled system is
chaotic or nonchaotic, depending on the sign of the estimated exponent.
In principle, using time series from any specific oscillator(s) would
give qualitatively the same result. Thus, the traditional approach
cannot give an answer as to which oscillators are chaotic when isolated.

Recently, a compressive sensing based method was developed~\cite{SWL:2014}
to address the problem of identifying a subset of ill-behaved chaotic 
elements from a network of nonlinear oscillators, majority of them 
being regular. In particular, for a network of coupled, mixed nonchaotic 
and chaotic neurons, it was demonstrated that, by formulating the 
reconstruction task as a compressive sensing problem, the system 
equations and the coupling functions as well as all the parameters can 
be obtained accurately from sparse time series. Using the reconstructed 
system equations and parameters for each and every neuron in the network 
and setting all the coupling parameters to zero, a routine calculation 
of the largest Lyapunov exponent can unequivocally distinguish the 
chaotic neurons from the nonchaotic ones.

\begin{figure}
\centering
\includegraphics[width=0.8\linewidth]{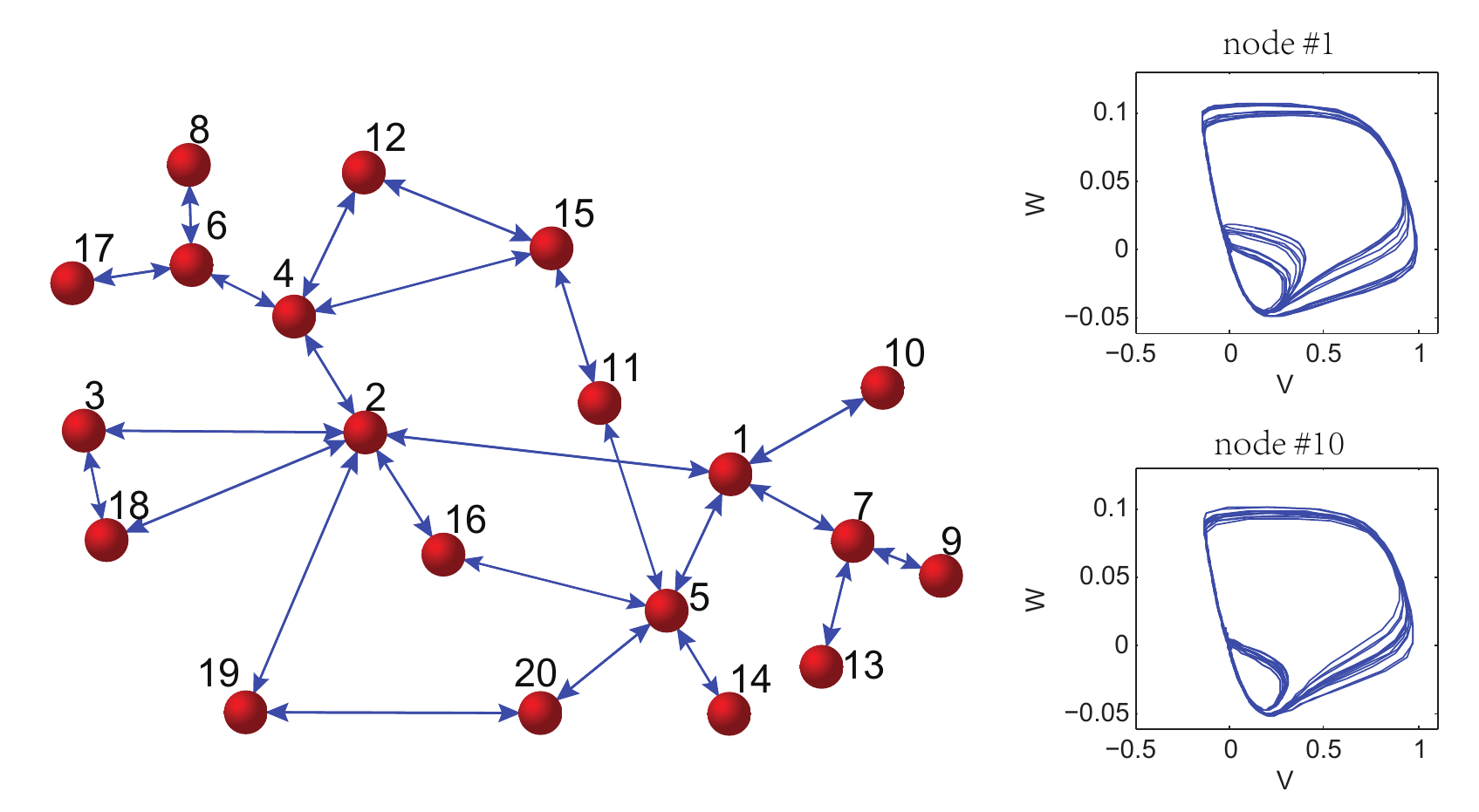}
\caption{\small {\bf A small neuronal network in which most dynamical 
units are regular and only a few are chaotic.} (a) Schematic 
illustration of the network, where the dynamics of each neuron 
is mathematically described by the Fitzhugh-Nagumo (FHN) equations. 
(b,c) Dynamical trajectories of two neurons from the coupled system, 
one being chaotic when isolated and another regular, respectively.
The trajectories give little hint as to which one is originally 
chaotic and which one is regular, due to the coupling. Specifically, 
neuron \#1 is originally chaotic (by setting parameter $a = 0.42$ 
in the FHN equation), while all other neurons are regular (their 
values of the corresponding parameter in the FHN equation are chosen
uniformly from the interval $[0.43,0.45]$). From Ref.~\cite{SWL:2014} 
with permission.}
\label{fig:Illustration_neuron}
\end{figure}

\subsubsection{Basic procedure of identifying chaotic neurons}
\label{subsubsec:CS_neuron_procedure}

Figure~\ref{fig:Illustration_neuron}(a) shows schematically a 
representative coupled neuronal network. Consider a pair of neurons, 
one chaotic and another nonchaotic when isolated (say \#1 and \#10, 
respectively). When they are embedded in a network, due to coupling, the 
time series collected from both will appear random and qualitatively
similar, as shown in Figs.~\ref{fig:Illustration_neuron}(b) and 
\ref{fig:Illustration_neuron}(c). It is visually quite difficult to 
distinguish the time series and to ascertain which node is originally 
chaotic and which is regular. The difficulty is compounded by the fact
that the detailed coupling scheme is not known {\it a priori}. Suppose that 
the chaotic behavior leads to undesirable function of the network and 
is to be suppressed. A viable and efficient method is to apply small 
pinning controls~\cite{WC:2002,LWC:2004,SBGC:2007,YCL:2009} to the 
relatively few chaotic neurons to drive them into some regular regime. 
(An implicit assumption is that, when all neurons are regular, the 
collective dynamics is regular. That is, the uncommon but not unlikely 
situation that a network systems of coupled regular oscillators would 
exhibit chaotic behaviors is excluded.) Accurate identification of the 
chaotic neurons is thus key to implementing the pinning control strategy.

Given a neuronal network, the task is thus to locate all neurons that are
{\em originally} chaotic and neurons that are {\em potentially} likely to
enter into a chaotic regime when they are isolated from the other neurons or
when the couplings among the neurons are weakened. The compressive sensing
based approach~\cite{SWL:2014} consists of two steps. Firstly, the 
framework is employed to estimate, from measured time series only, 
the parameters in the FHN equation for each neuron, as well as the 
network topology and various coupling functions and weights. This can 
be done by expanding the nodal dynamical equations and the coupling
functions into some suitable mathematical base as determined by the 
specific knowledge about the actual neuronal dynamical system, and then 
casting the problem into that of determining the sparse coefficients 
associated with various terms in the expansion. The nonlinear systems 
identification problem can then be solved using the standard compressive
sensing algorithm. Secondly, all coupling parameters are set to zero
so that the dynamical behaviors of each and every {\em individual}
neuron can be analyzed by calculating the Lyapunov exponents. These with 
a positive largest exponent are identified as chaotic.

A typical time series from a neuronal network consists of a sequence of
spikes in the time evolution of the cell membrane potential. It was
demonstrated~\cite{SWL:2014} that the compressive sensing based 
reconstruction method works well even for such spiky time series. The 
dependence of reconstruction accuracy on data amount was analyzed 
to verify that only limited data are required to achieve high accuracy 
in reconstruction.

\subsubsection{Example: identifying chaotic neurons in the 
FitzHugh-Nagumo (FHN) network} \label{subsubsec:CS_neuron_FHN}

The FHN model, a simplified version of the biophysically detailed
Hodgkin-Huxley model~\cite{HH:1952}, is a mathematical paradigm for gaining
significant insights into a variety of dynamical behaviors in real neuronal
systems~\cite{FH:1961,NAY:1962}. For a single, isolated neuron, the 
corresponding dynamical system is described by the following 
two-dimensional, nonlinear ordinary differential equations:
\begin{eqnarray} \label{eq:Single_neuron}
\nonumber
\frac{d V}{dt} &=& \frac{1}{\delta} [ V( V - a)(1 - V) - W], \\
\frac{d W}{dt} &=& V - W - b + S(t),
\end{eqnarray}
where $V$ is the membrane potential, $W$ is the recover variable, $S(t)$ is
the driving signal (e.g., periodic signal), $a$, $b$, and $\delta$ are
parameters. The parameter $\delta$ is chosen to be infinitesimal
so that $V(t)$ and $W(t)$ are ``fast'' and ``slow'' variables,
respectively. Because of the explicitly time-dependent driving signal
$S(t)$, Eq.~(\ref{eq:Single_neuron}) is effectively a three-dimensional 
dynamical system, in which chaos can arise~\cite{Ott:book}. For a 
network of FHN neurons, the equations are
\begin{eqnarray} \label{eq:FHN_network_neuron}
\nonumber
\frac{d V_i}{dt} &=& \frac{1}{\delta} [ V_i ( V_i -a) (1- V_i) - W_i]
                      + \sum_{i=1}^{N}{ c_{ij}(V_j - V_i) } \\
\frac{d W_i}{dt} &=& V_i - W_i -b + S(t),
\end{eqnarray}
where $c_{ij}$ is the coupling strength (weight) between the $i$th and 
the $j$th neurons (nodes). For $c_{ij} = c_{ji}$, the interactions between any
pair of neurons are symmetric, leading to a symmetric adjacency matrix
for the network. For $c_{ij} \ne c_{ji}$, the network is asymmetrically
weighted.

\begin{figure}
\centering
\includegraphics[width=\linewidth]{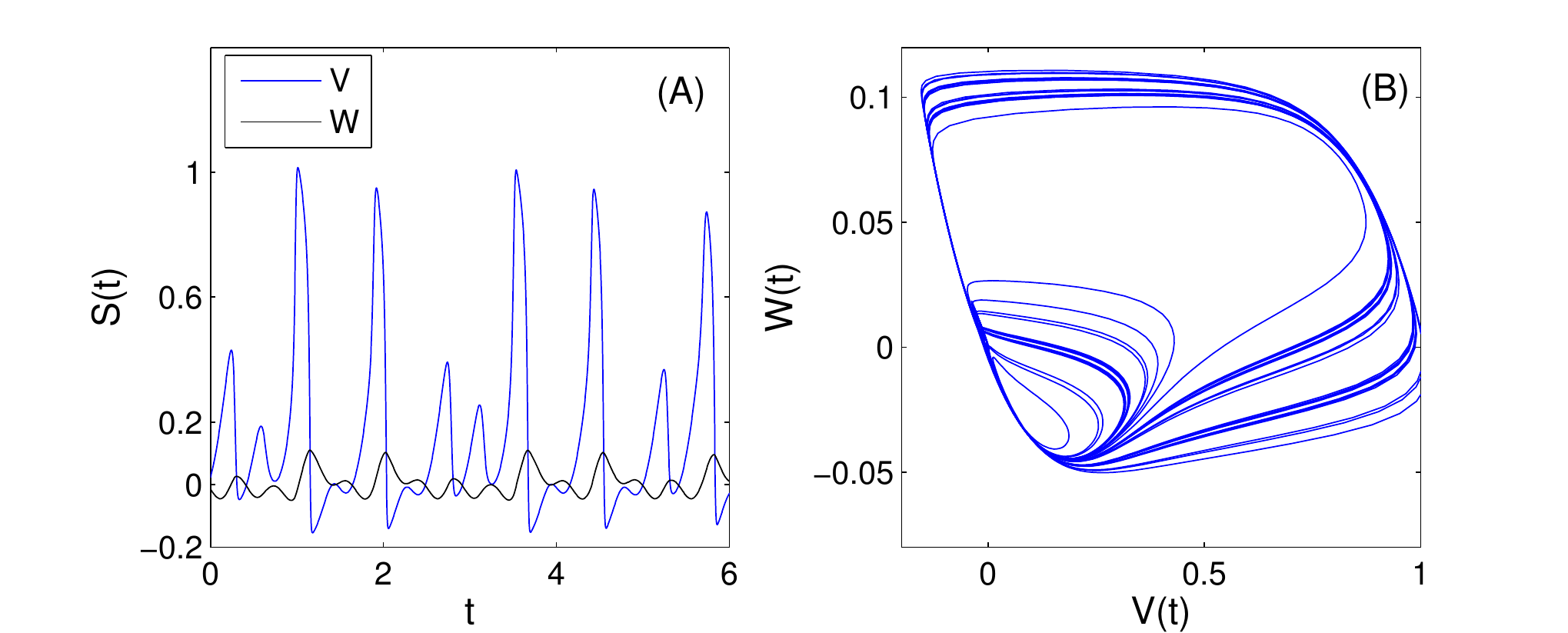}
\caption{\small {\bf Dynamical behavior of a chaotic neuron.} 
(a) Chaotic time series of the membrane potential $V$ and the recovery 
variable $W$ from a single neuron for $a = 0.42$, and (b) the 
corresponding dynamical trajectory. From Ref.~\cite{SWL:2014}
with permission.}
\label{fig:TimeS_neuron}
\end{figure}

\begin{figure}
\centering
\includegraphics[width=\linewidth]{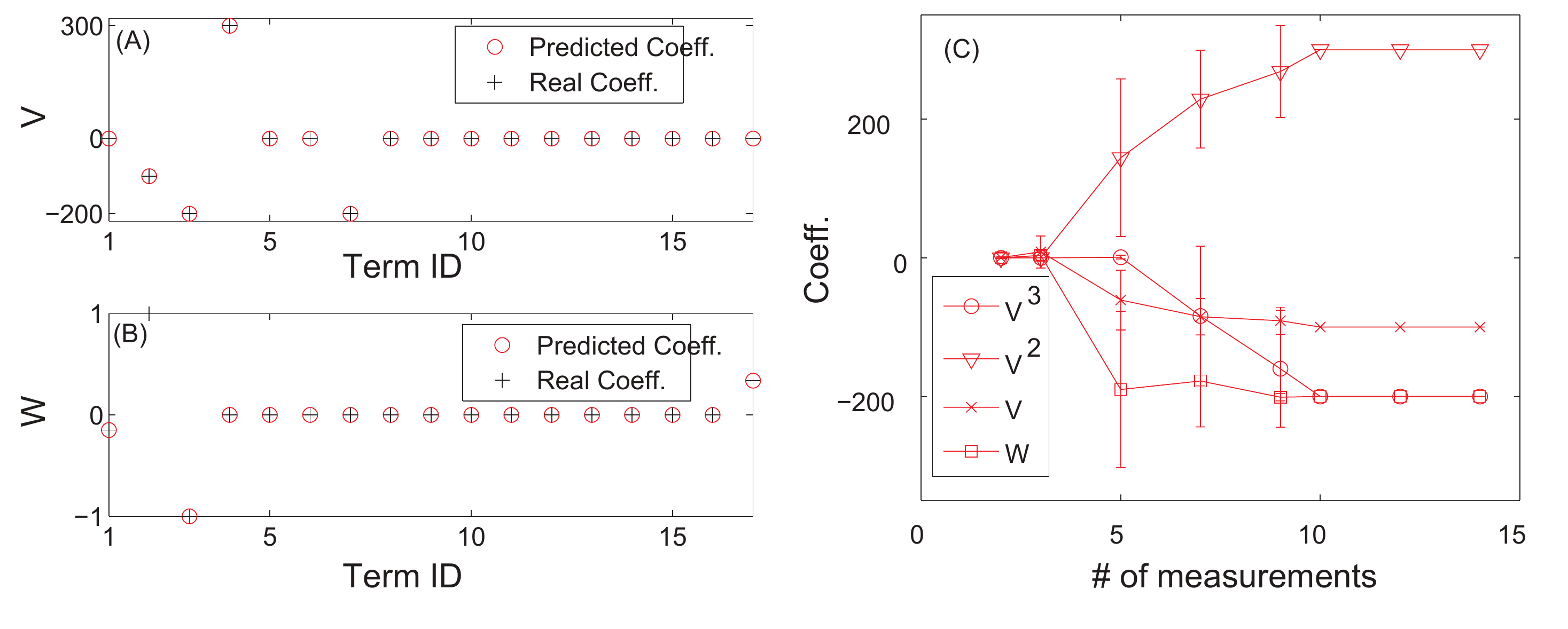}
\caption{\small {\bf Identification of parameters of a single FHN neuron 
using compressive sensing}.
(a,b) Predicted coefficients and comparison with the actual 
parameter values in the dynamical equations of variables $V$ and $W$. 
The number of data points used was 12. (c) Predicted parameters for
a single neuron as the number of data points is increased. The sampling 
interval is $\Delta t = 0.05$. All results are averaged over 10 
independent time series. From Ref.~\cite{SWL:2014} with permission.}
\label{fig:Parameter_neuron}
\end{figure}

Consider the FHN model with sinusoidal driving:
$S(t) = r \sin \omega_0 t$. The model parameters are $r=0.32$, 
$\omega_0=15.0$, $\delta = 0.005$, and $b=0.15$. For $a=0.42$, an 
individual neuron exhibits chaos. Representative chaotic time series and 
the corresponding dynamical trajectory are shown in Fig.~\ref{fig:TimeS_neuron}.
Reconstruction of an isolated neuron can be done by setting zero all 
coupling terms in network. For this purpose power series of order 4 
can be used as the expansion base~\cite{SWL:2014} so that there are 
17 unknown coefficients to be determined. Three consecutive measurements are 
sampled at time interval $\tau=0.05$ apart and a standard two-point formula 
can be used to extrapolate the derivatives. From a random starting point,
12 data points were generated. Results of reconstruction are shown in 
Figs.~\ref{fig:Parameter_neuron}(a) and \ref{fig:Parameter_neuron}(b) 
for variables $V$ and $W$, respectively. The last two coefficients 
associated with each variable represent the strength of the driving 
signal. Since only the variable $W$ receives a sinusoidal input, the last 
coefficient in $W$ is nonzero. By comparing the positions of the nonzero 
terms and the previously assumed vector form $\mathbf{g}_{i}(t)$, one 
can fully reconstruct the dynamical equations of any isolated neuron. 
In particular, from Figs.~\ref{fig:Parameter_neuron}(a) and 
\ref{fig:Parameter_neuron}(b) it can be seen that all estimated 
coefficients agree with their respective true values. 
Figure~\ref{fig:Parameter_neuron}(c) shows how the estimated
coefficients converge to the true values as the number of data points is
increased. It can be seen that, for over 10 data points, all the 
parameters associated with a single FHN neuron can be faithfully 
identified.

\begin{figure}
\centering
\includegraphics[width=\linewidth]{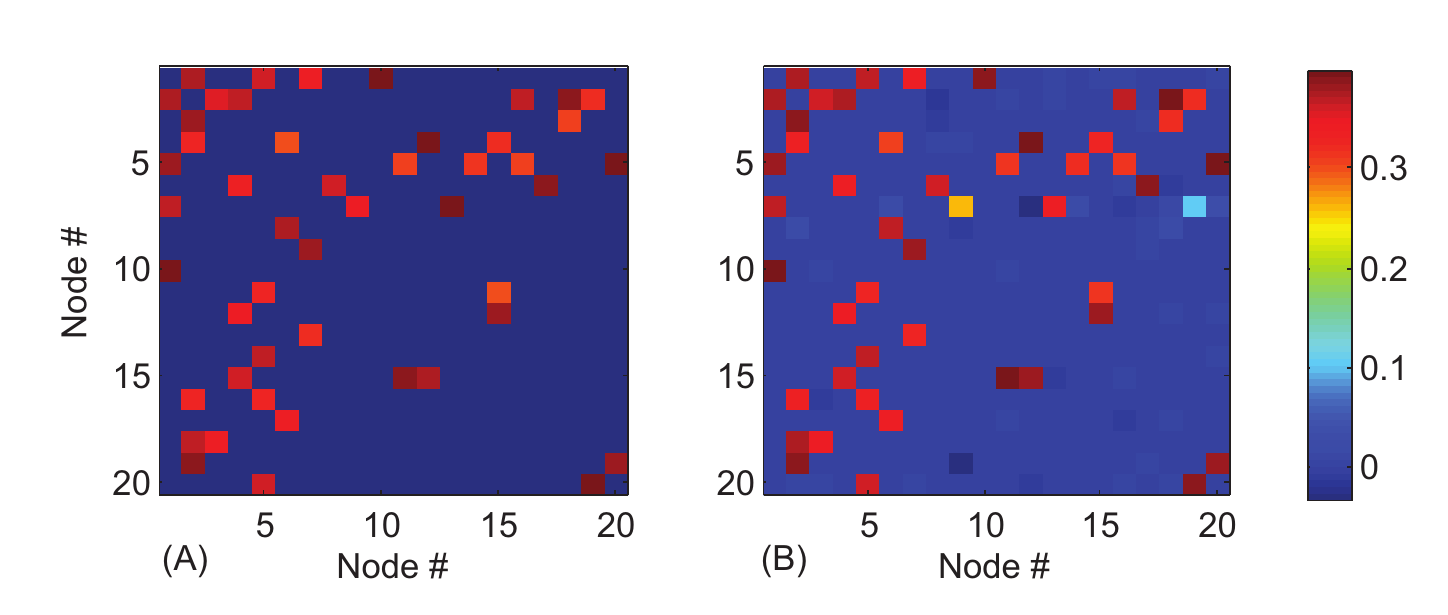}
\caption{\small {\bf Reconstruction of an FHN network.} 
For the network in Fig.~\ref{fig:Illustration_neuron}(a), (a) actual 
and (b) estimated weighted adjacency matrix. The normalized data
amount used in the reconstruction is $R_{m}=0.7$. From Ref.~\cite{SWL:2014} 
with permission.}
\label{fig:Network_neuron}
\end{figure}

Next consider the network of coupled FHN neurons as schematically 
shown in Fig.~\ref{fig:Illustration_neuron}(a), where the coupling 
weights among various pairs of nodes are uniformly distributed in the 
interval $[0.3, 0.4]$. The network topology  is random with connection 
probability $p=0.04$. From time series the compressive sensing matrix for 
each variable of all nodes can be reconstructed. Since the couplings occur 
among the variables $V$ of different neurons, the strengths of all incoming 
links can be found in the unknown coefficients associated with different $V$
variables. Extracting all coupling terms from the estimated coefficients
gives all off-diagonal terms in the weighted adjacency matrix.
Figure~\ref{fig:Network_neuron} shows the reconstructed
adjacency matrix as compared with the real one for $R_m = 0.7$, where
$R_m$ is the relative number of data points normalized by the total
number of unknown coefficients. It can be seen that the compressive
sensing based method can predict all links correctly, in spite of the 
small errors in the predicted weight values. The errors are mainly due 
to the fact that there are large coefficients in the system equations 
but the coupling weights are small.

\begin{figure}
\centering
\includegraphics[width=\linewidth]{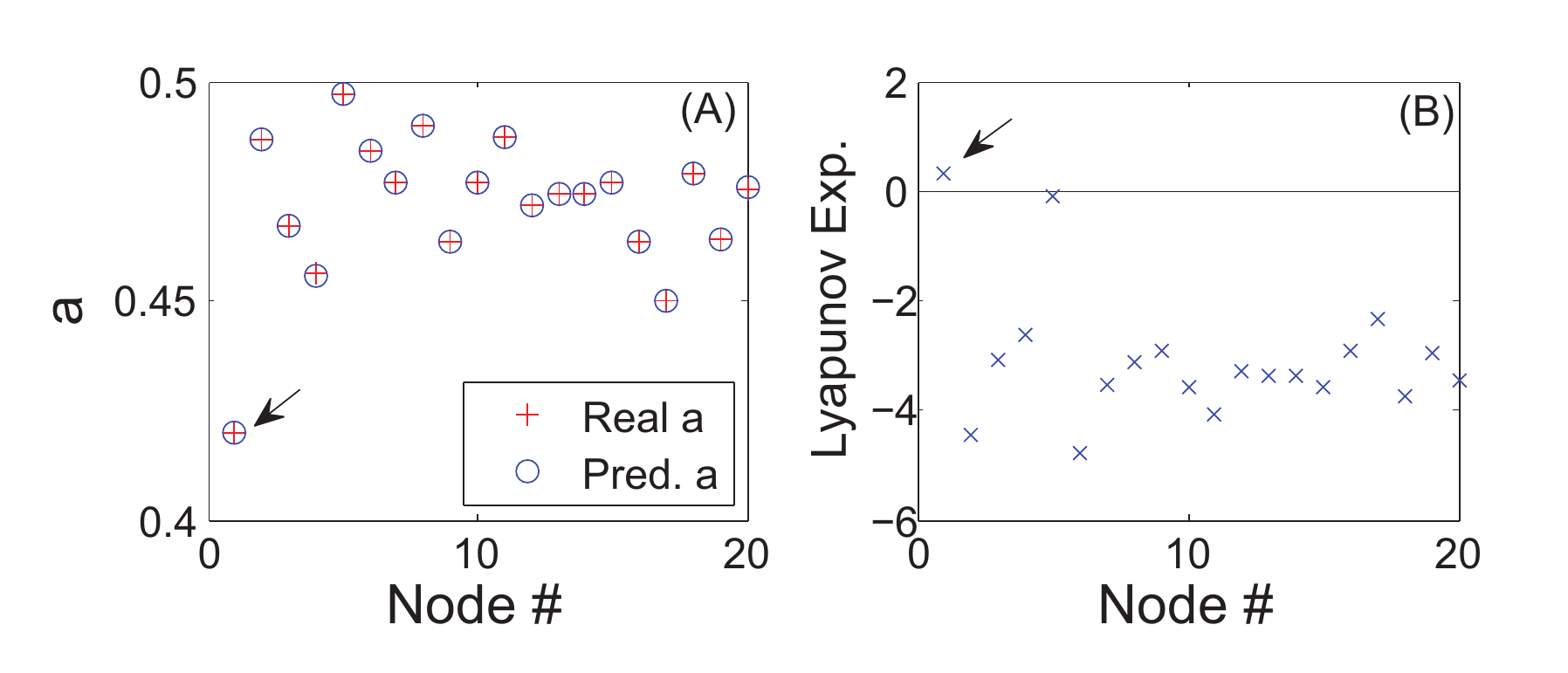}
\caption{\small {\bf Identification of chaotic neurons}.
(a) Estimated values of the parameter $a$ for different neurons (red circles), 
as compared with the actual values (black crosses). The random network 
size is $N=20$ with connection probability $p=0.04$. The normalized data 
amount used in reconstruction is $R_{m}=0.7$. (b) Largest Lyapunov 
exponents calculated from the reconstructed system equations. The 
reference line denotes null value. From Ref.~\cite{SWL:2014} with 
permission.}
\label{fig:Lyp_neuron}
\end{figure}

Using the weighted adjacency matrix, one can identify the coupling terms
in the network vector function so as to extract the terms associated with 
each isolated nodal velocity field. The value of parameter $a$ can then
be identified and the largest Lyapunov exponent can be calculated for 
each individual neuron. The results are shown in 
Figs.~\ref{fig:Lyp_neuron}(a) and \ref{fig:Lyp_neuron}(b). It can be
seen that, for this example, neuron \#1 has a positive largest exponent 
while the largest exponents for all others are negative, so \#1 is 
identified as the only chaotic neuron among all neurons in the network.

\subsection{Data based reconstruction of complex geospatial networks
and nodal positioning} \label{subsec:CS_GeoN}

\begin{figure}
\centering
\includegraphics[width=0.8\linewidth]{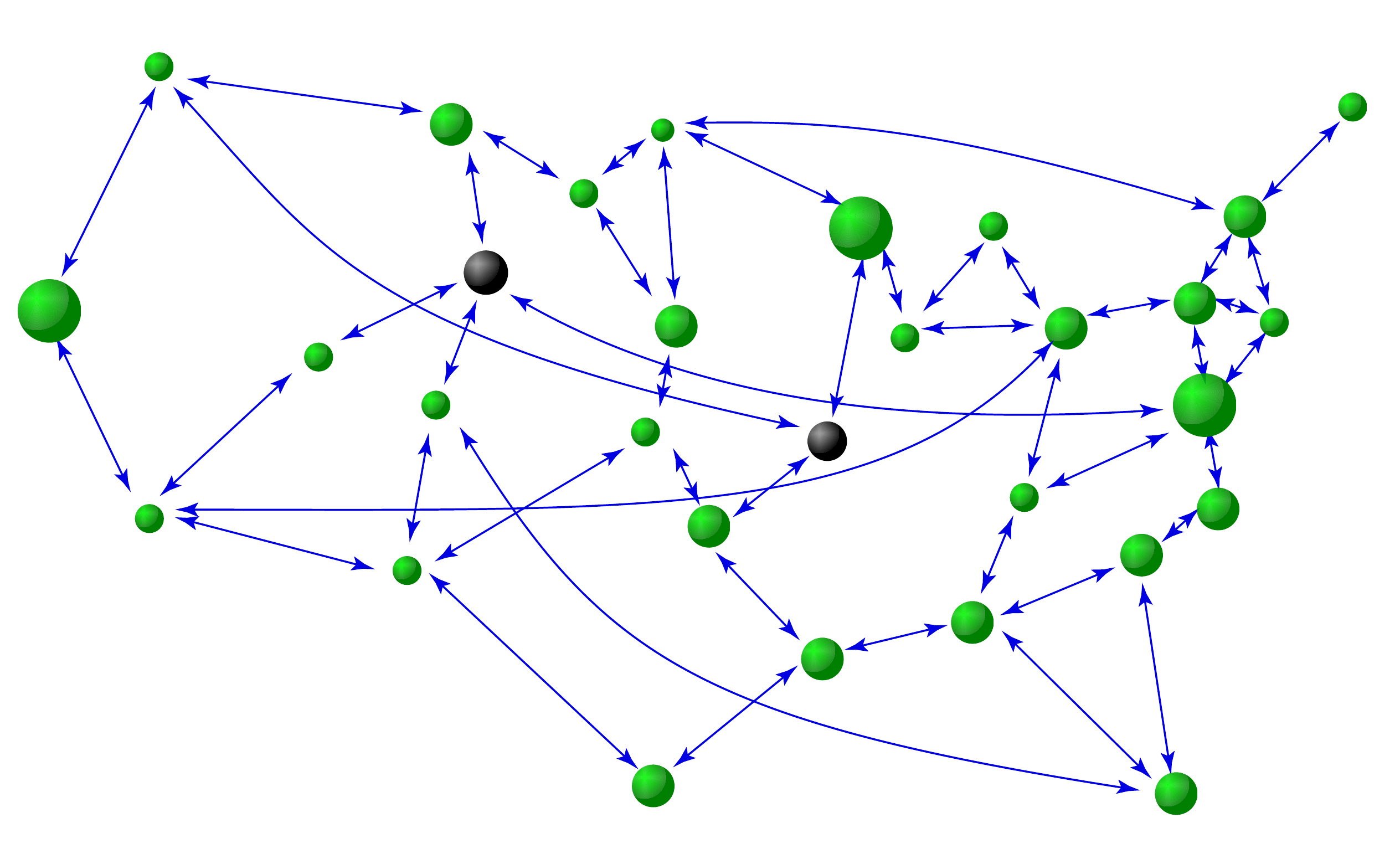}
\caption{\small {\bf A schematic illustration of a complex geospatial network}.
The connection topology, the positions of the nodes in the physical space,
and nodal dynamical equations all are unknown {\it a priori}, but only 
time series from the nodes can be collected at a single node in the 
network (e.g., a data collecting center). The challenges are to 
reconstruct the dynamical network, to locate the precise position of 
each node, and to detect hidden nodes, all based solely on time series 
with inhomogeneous time delays. The green circles denote ``normal'' nodes 
and the dark circles indicate hidden nodes. From Ref.~\cite{SWWL:2015} with
permission.}
\label{fig:scheme_GeoN}
\end{figure}

Complex geospatial networks with components distributed in the real 
geophysical space are an important part of the modern infrastructure. 
Examples include large scale sensor networks and various subnetworks 
embedded in the Internet. For such a network, often the set of active 
nodes depends on time: the network can be regarded as static only in 
relatively short time scale. For example, in response to certain 
breaking news event, a social communication network within the Internet may 
emerge, but the network will dissolve itself after the event and its 
impacts fade away. The connection topologies of such networks are 
usually unknown but in certain applications it is desirable to uncover 
the network topology and to {\em determine the physical locations of 
various nodes} in the network. Suppose time series or signals can be 
collected from the nodes. Due to the distributed physical locations 
of the nodes, the signals are time delayed. An illustrative example
of a complex geospatial network is shown in Fig.~\ref{fig:scheme_GeoN},
where there is a monitoring center that collects data from nodes at 
various locations, but their precise geospatial coordinates are unknown. 
The normal nodes are colored in green. There are also hidden nodes 
that can potentially be the sources of threats (e.g., those represented 
by dark circles). Is it possible to uncover the network topology, 
estimate the time delays embedded in the signals from different nodes, 
and then determine their physical locations? Can the existence of a 
hidden node be ascertained and its actual geophysical location be 
determined? 

A recent work~\cite{SWWL:2015} showed that these questions can be 
addressed by using the compressive sensing based reconstruction 
paradigm. In particular, the time delays of the dynamics at various 
nodes can be estimated using time series collected from {\em a single 
location}. Note that there were previous methods of finding time delays 
in complex dynamical systems, e.g., those based
on synchronization~\cite{WuXQ:2008}, Bayesian estimation~\cite{Husmeier:2003},
and correlation between noisy signals~\cite{WRLL:2012}. The compressive 
sensing based method provides an alternative approach that has the
advantages of generality, high efficiency, low data requirement, and
applicability to large networks. It was demonstrated~\cite{SWWL:2015} 
that the method can yield estimates of the nodal time delays with 
reasonable accuracy. After the time delays are obtained, the actual 
geospatial locations of various nodes can be determined by using, e.g., 
a standard triangular localization method~\cite{STK:2005}, 
given that the locations of a small subset of nodes are known. Hidden 
nodes can also be detected. These results can potentially be useful 
for applications such as locating sensors in wireless networks and 
identifying/detecting/anticipating potential geospatial 
threats~\cite{MC:2007}, an area of importance and broad interest. 

\paragraph*{Reconstruction of time delays using compressive sensing.} 
Consider a continuous-time oscillator network with time delayed 
couplings~\cite{LC:2004,DJD:2004,GRLD:2009}, where for every link, 
the amount of delay is proportional to the physical distance of this link.
The time delayed oscillator network model has been widely used in 
studying neuronal activities~\cite{DJD:2004} and food webs in 
ecology~\cite{GRLD:2009}. Mathematically, the system can be written as 
\begin{equation} \label{Eq:System_GeoN}
\dot{\mathbf x}_i={\mathbf F}_i[{\mathbf x}_i(t)] +
\sum_{j=1, j\neq i}^{N}{\mathcal{W}_{ij} \cdot
[{\mathbf x}_j(t- {\tau}_{ij})- {\mathbf x}_i(t)]},
\end{equation}
for $i = 1, \ldots, N$, where ${\mathbf x}_i\in{\mathbb R}^m$ is the
$m$-dimensional state variable of node $i$ and ${\mathbf F}_i[{\mathbf
x}_i(t)]$ is the vector field for its isolated nonlinear nodal
dynamics. For a link $l_{ij}$ connecting nodes $i$ and $j$, 
the interaction weight is given by the $m\times m$ weight matrix
$\mathcal{W}_{ij} \in {\mathbb R}^{m \times m}$ with its element
$w_{ij}^{p,q}$ representing the coupling from the $q$th component of
node $j$ to the $p$th component of node $i$. For simplicity, we
assume only one component of $w_{ij}^{p,q}$ is non-zero and denotes it
as $w_{ij}$. The associated time delay is denoted as $\tau_{ij}$.
For a modern geospatial network, the speed of signal
propagation is quite high in a proper medium (e.g., optical fiber).
The time delay can thus be assumed to be small and the Taylor expansion 
can be used to express the delay coupling terms 
to the first order, e.g., $x_i(t- \tau_{ji}) \approx
x_i(t) - \tau_{ji} {\dot x}_i$, where ${\dot x}_i$ is the time
derivative. When the coupling function between any pair of nodes
is linear, in a suitable mathematical basis constructed from the time 
series data, $\tilde{\mathbf g}^{(\gamma)} [{\mathbf x}_i (t)]$, 
the coupling and time delayed terms, together with the nodal 
dynamical equations, can be expanded into a series. The task is 
to estimate all the expansion coefficients. Assume linear coupling 
functions and causality so that all $\tau_{ij}$ ($i,j=1,\ldots,N$) 
are positive (for simplicity). All terms directly associated with 
node $i$ can be regrouped into ${\mathbf F}_i^{\prime} [{\mathbf x}_i(t)]$,
where
\begin{equation} \label{eq:general_F_GeoN}
{\mathbf F}_i^{\prime}[{\mathbf x}_i(t)] \equiv
{\mathbf F}_i [{\mathbf x}_i(t)] - {\mathbf x}_i(t)\cdot
\sum_{j=1, j\neq i}^{N} { \mathcal{W}_{ij}},
\end{equation}
and ${\mathbf F}_i^{\prime} [{\mathbf x}_i(t)]$ has been expanded
into the following series form:
\begin{equation} \label{eq:expansion_GeoN}
{\mathbf F}_i^{\prime} [{\mathbf x}_i(t)] =
\sum_{\gamma} { \tilde{\mathbf \alpha}^{(\gamma)} \cdot
\tilde{\mathbf g}^{(\gamma)} [{\mathbf x}_i (t) ]},
\end{equation}
with $\tilde{\mathbf g}^{(\gamma)} [{\mathbf x}_i (t)]$ representing 
a suitably chosen set of orthogonal and complete base functions so 
that the coefficients $\tilde{\mathbf \alpha}^{(\gamma)}$ are sparse.
The time delayed variable ${\mathbf x}_j(t - {\tau}_{ij})$ can be 
expanded as
\begin{equation} \label{eq:tau_Taylor_GeoN}
{\mathbf x}_j( t - {\tau}_{ij})\approx
{\mathbf x}_j(t)-\tau_{ij}\dot{\mathbf x}_j(t).
\end{equation}
All the coupling terms with inhomogeneous time delays associated with 
node $i$ can then be written as
\begin{equation} \label{eq:coupling_term_GeoN}
[\sum_{j=1, j\neq i}^{N}{\mathcal{W}_{ij} {\mathbf x}_j(t - { \tau}_{ij})}]_{p}
\equiv \sum_{j=1, j\neq i}^{N}{[\mathcal{B}_{ij} \cdot {\mathbf x}_j(t)
+ \mathcal{C}_{ij} \cdot{\mathbf x}_j(t)]},
\end{equation}
where $\mathcal{B}_{ij} = \mathcal{W}_{ij}$ and
$\mathcal{C}_{ij}= -\mathcal{W}_{ij} \tau_{ij}$.
In the compressive sensing framework, Eq.~(\ref{Eq:System_GeoN}) can then 
be written in the following compact form:
\begin{equation} \label{Eq:AllCoup_GeoN}
\dot{\mathbf x}_i(t) = \sum_{\gamma} { \tilde{\mathbf \alpha}^{(\gamma)} 
\cdot \tilde{\mathbf g}^{(\gamma)} [{\mathbf x}_i (t) ] }
+ \sum_{j=1, j\neq i}^{N}{[\mathcal{B}_{ij} \cdot {\mathbf x}_j(t)
+ \mathcal{C}_{ij}\cdot{\mathbf x}_j(t)]},
\end{equation}
which is a set of linear equations for data collected at different 
time $t$, where $\tilde{\mathbf \alpha}^{(\gamma)}$, $\mathcal{B}_{ij}$ 
and $\mathcal{C}_{ij}$ are to be determined. If the unknown coefficient 
vectors can be reconstructed accurately, one has complete information 
about the nodal dynamics as represented by 
${\mathbf F}^{\prime}[{\mathbf x}(t)]$, the topology and interacting
weights of the underlying network as represented by $\mathcal{W}_{ij}$,
as well as the time delays associated with the nonzero links because of the 
relations $\mathcal{W}_{ij}= \mathcal{B}_{ij}$. Note that, if the 
coupling form is nonlinear, the relationship between the delay term 
${\mathbf c}_{ij}$ and the time delays ${\tau}_{ij}$ would be hard 
to interpret, especially when the exact coupling form is not known.

\paragraph*{Nodal positioning and reconstruction of geospatial network.} 
After obtaining the time delays, one can proceed to determining the actual
positions of all nodes. If time series are collected simultaneously
from all nodes at the data collecting node, the estimated coupling
delay $\tau_{ij}$ associated with the link $l_{ij}$ is proportional to
the physical distance $d_{ij} = d_{ji}$ of the link. However, in reality
strictly synchronous data collection is not possible. For example, if the
signals are collected, e.g., at a location $s$ outside the network
with varying time delays $\tau_{si}$, the estimated delays associated
with various links in the network are no longer proportional to the
actual distances. The varying delays due to asynchronous data collection 
can be canceled and the distances can still be estimated as 
$d_{ij} = (c/2)( \tau_{ij} + \tau_{ji})$, where $\tau_{ij}$ is the signal 
delay associated with node $j$ from the reconstruction of node $i$, vice 
versa for $\tau_{ji}$, and $c$ is the signal propagation speed.

When the mutual distances between the nodes have been estimated, one 
can determine their actual locations, e.g., by using
the standard triangular localization algorithm~\cite{STK:2005}.
This method requires that the positions of $N_{B}$ reference nodes be
known, the so-called beacon nodes. Starting from the beacon nodes, the
triangulation algorithm makes use of the distances to these reference 
nodes to calculate the Cartesian coordinates of the detected nodes. The 
beacon node set can then be expanded with the newly located nodes. Nodes
that are connected to the new beacon set, each with more than three links 
in the two-dimensional space, can be located. The process continues until 
the locations of all nodes have been determined, or no new nodes can be 
located. The choice of the proper initial beacon set to fully reconstruct 
the network depends on the network topology. For example, for a 
scale-free network, one can choose nodes that were firstly added during 
the process of network generation as the initial beacon set, thereby 
guaranteeing that all nodes can be located using the procedure. An 
empirical rule is then to designate the largest degree nodes as the 
initial beacon node set.

More specifically, given the positions of $k$ reference nodes (or beacon 
nodes) $(x_k, y_k)$, and their distances $d_{i,1}, d_{i,2}, \cdots d_{i,k}$ 
to the target node $i$, one can calculate the position of node $i$ using
the triangular localization method~\cite{STK:2005}, for $k$ larger 
than the space dimension. In general, it is necessary to solve the least 
squares optimization problem $\mathcal{H} \cdot {\mathbf x_i}
={\mathbf b}$, where ${\mathbf x_i}=[x_i, y_i ]^{T}$ is the
position of node $i$, and
$\mathcal{H} = [ {\mathbf x_1},{\mathbf x_2}, \cdots, {\mathbf x_k}]^T$
is the position matrix corresponding to the set of beacon nodes, where
${\mathbf b}=0.5\times[ D_1, D_2, \cdots, D_k]^T$ and
$D_k=d_{ik}^2- y_k^2 + x_k^2$. To locate the positions of all nodes 
in the network, one starts with a small set of beacon nodes whose actual
positions are known. Initially one can locate the nodes that are connected 
to at least three nodes in the set of beacon nodes, insofar as the three 
reference nodes are not located on a straight line. When this is done, 
the newly located nodes can be added into the set of reference nodes 
and the neighboring nodes can be located through the new set of beacon 
nodes. This process can be iterated until the positions of all nodes 
are determined or no more qualified neighboring nodes can be found. 
For a general network, such an initial beacon set may not be easily 
found. A special case is scale-free networks, for which the initial 
beacon set can be chosen as the nodes with the largest degrees. For a 
random network, one can also choose the nodes of the largest degree as the
initial beacon node set, and use a larger beacon set to locate most of 
the nodes in the network. For an arbitrary network topology, the following 
simple method was proposed~\cite{SWWL:2015} to select the set of 
beacon nodes: estimate the distances from one node to all other 
unconnected nodes using the weighted shortest distance and then proceed 
with the triangular localization algorithm. There are alternative 
localization algorithms based on given distances, e.g, the 
multidimensional scaling method~\cite{MDS:2004,MDS:2003}.

\begin{figure}
\centering
\includegraphics[width=\linewidth]{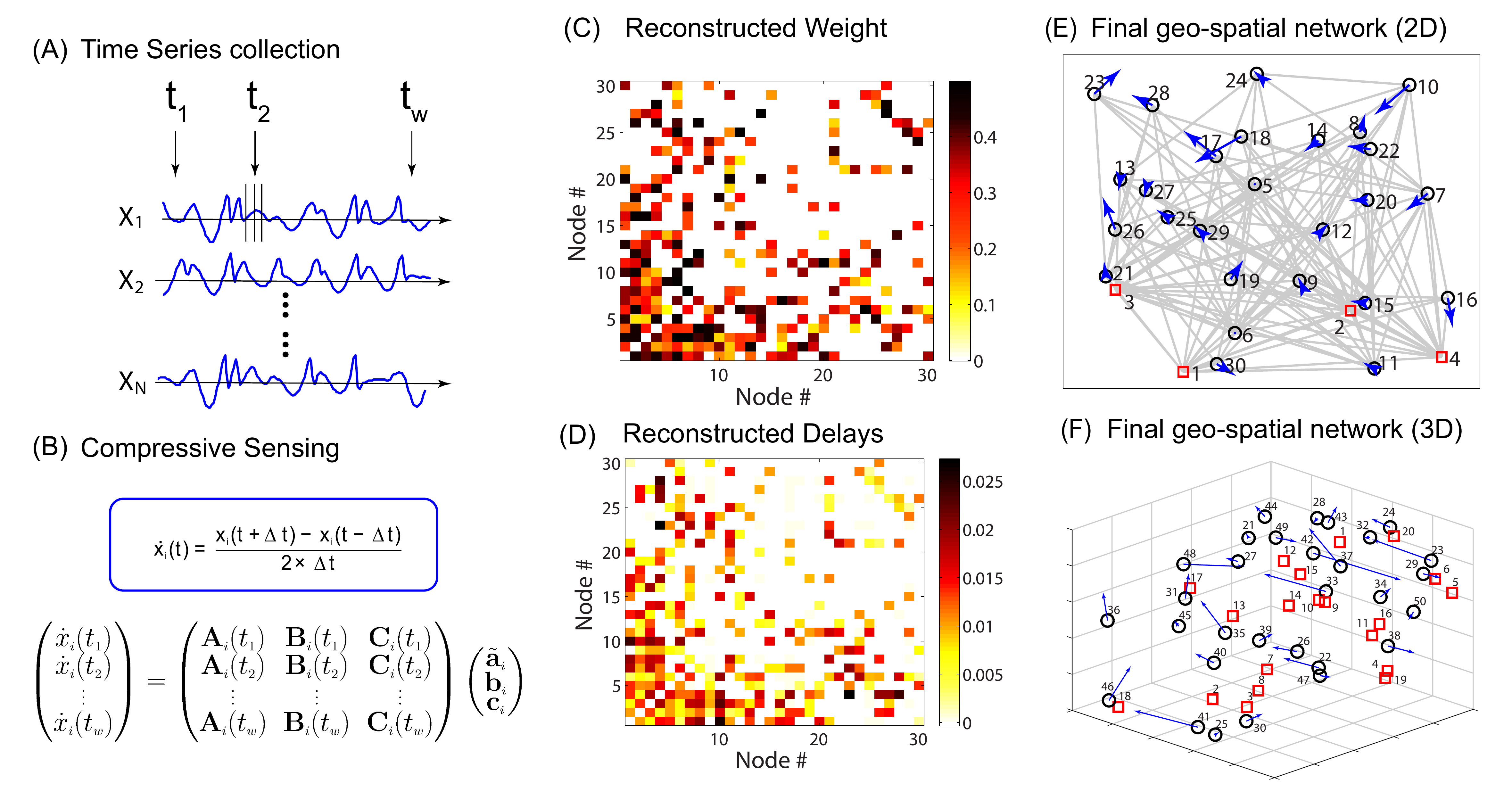}
\caption{\small {\bf Illustration of compressive sensing based method to 
reconstruct a complex geospatial network from time series}. (A) For any 
node, time series of its dynamical variables are collected at $w$ different 
instants of time. (B) The corresponding derivatives are approximated using 
the standard first order Gaussian method, which are needed in constructing
the compressive sensing equations. (C,D) An example of link weights
and time delays obtained from the reconstructed coefficient vectors,
respectively. (E) Given the positions of four beacon nodes (marked as
red rectangles), the locations of the remaining nodes (marked as black
circles) are determined by using the standard triangularization method.
The blue arrows indicate the estimation errors, which point from the
actual to the estimated positions. The various coupling terms are
illustrated using gray lines. There are in total 30 nodes in the
network, connecting with each other via the scale-free topology. The
average outgoing degree is five. The amount of data used is $R_m=0.5$.
(F) Following the same procedure, a geospatial network of 50 nodes which 
are distributed in a three-dimensional cube of unit length is reconstructed. 
There are 20 beacon nodes and the normalized amount of data used is 
$R_m=0.55$. For clarity, the links between these nodes are not shown.
From Ref.~\cite{SWWL:2015} with permission.}
\label{fig:Illustration_GeoN}
\end{figure}

\paragraph*{An example of reconstruction of a complex geospatial network.} 
A numerical example was presented in Ref.~\cite{SWWL:2015} to demonstrate
the reconstruction of complex geospatial networks, in which all nodes
were assumed to be distributed in a two-dimensional square, or a 
three-dimensional cube of unit length. The network topology was 
scale free~\cite{BA:1999} or random~\cite{ER:1959}, and its size 
was varied. A nonlinear oscillator was placed at each node, e.g., 
the R\"ossler oscillator ($[\dot{x},\dot{y},\dot{z}] = 
[-y-z,x+0.2y,0.2x+z(x-0.2)]$). The coupling weights were asymmetric 
and uniformly distributed in the interval [0.1, 0.5]. A small threshold 
was assigned to the estimated weight as $w_0=0.05$ (somewhat arbitrary), 
where if the estimated weight is larger (smaller) than $w_0$, the 
corresponding link is regarded as existent (nonexistent). As a result,
the following holds: $d_{ij}=c \cdot {\tau}_{ij}= c \cdot {\tau}_{ji}$,
and the parameter $c$ was chosen to be $100$ (arbitrarily). Linear 
coupling functions were chosen for any pair of connected nodes, where
the interaction occurs between the $z$-variable of one node and the 
$x$-variable of another. The time series used to reconstruct the 
whole network system were acquired by integrating the coupled delayed 
differential-equation system~\cite{DDEBook} with step size $5\times 10^{-5}$. 
The vector fields of the nodal dynamics were expanded into a power 
series of order $l_x + l_y + l_z \leq 3$. The derivatives required for the
compressive sensing formulation were approximated from time series by
the standard first order Gaussian method. The data requirement was
characterized by $R_{m}$, the ratio of the number of data points used 
to the total number of unknown coefficients to be estimated. The beacon 
nodes were chosen to be those having the largest degrees in the network, 
and their positions were assumed to be known.

Figure~\ref{fig:Illustration_GeoN} summarizes the major steps 
required for reconstructing a complex geospatial network using 
compressive sensing, where $N = 30$ nodes connecting with each 
other in a scale-free manner are randomly distributed in a 
two-dimensional square. Oscillatory time series are collected from 
each node, from which the compressive sensing equations can be obtained, 
as shown in Figs.~\ref{fig:Illustration_GeoN}(A)
and \ref{fig:Illustration_GeoN}(B). The reconstructed coefficients for the
nodal dynamical equations contain the coupling weights $B_{ij}=w_{ij}$ 
and the delay terms $C_{ij}=-w_{ij}\times \tau_{ij}$. The links with 
reconstructed weights larger than the threshold $w_0$ are regarded as 
actual (existent) links, for which the time delays $\tau_{ij}$ can be 
estimated as $\tau_{ij} = - C_{ij}/w_{ij}$. Repeating this procedure 
for all nodes, the weighted adjacency matrix (which defines the network 
topology) and the time delay matrix can be determined. The estimated 
adjacency matrix and the time delays are displayed in 
Figs.~\ref{fig:Illustration_GeoN}(C) and \ref{fig:Illustration_GeoN}(D), 
respectively, which match well with those of the actual network. Note 
that the reconstructed time delays are symmetric with respect to the 
link directions, as shown in Fig.~\ref{fig:Illustration_GeoN}(D), 
which is correct as they depend only on the corresponding physical 
distances. With the estimated time delays, the four largest degree n
odes, node $\#1 \sim \#4$, are chosen as the beacon nodes, so that the 
locations of all remaining nodes can be determined. The fully reconstructed 
geospatial network is shown in Fig.~\ref{fig:Illustration_GeoN}(E), where 
the red rectangles indicate the locations of the beacon nodes. The black 
circles denote the actual locations of the remaining nodes and the heads 
of the blue arrows indicate their estimated positions (shorter arrows 
mean higher estimation accuracy). The amount of data used is relatively 
small: $R_{m}=0.5$.

\begin{figure}[!h]
\centering
\includegraphics[width=0.6\linewidth]{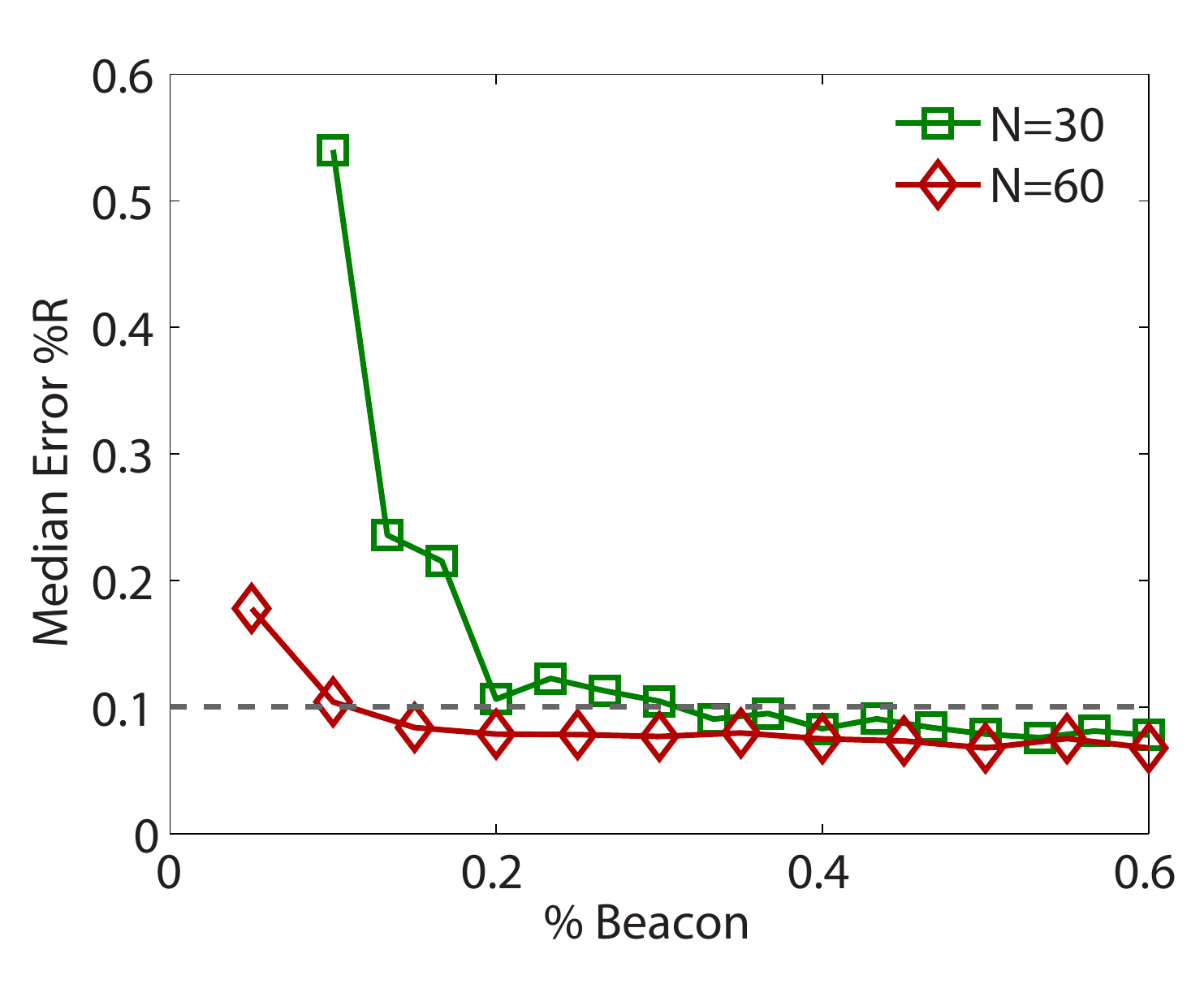}
\caption{\small {\bf Positioning errors.} Normalized positioning error $M_{r}$,
defined as the medium absolute estimated distance error normalized by
the distributed length $L$, as a function of the fraction $R_{B}$ of
the beacon nodes. The networks have the scale-free topology with
the average outgoing degree $k=5$. Two values of the network size are used:
$N = 30$ and $N = 60$. The beacon nodes are chosen as these having
the largest degrees. The time delays are estimated using the data
amount $R_{m}=0.5$, for which the average error is $D_{nz} \approx 0.12$. 
The results were obtained by averaging over 10 independent network 
realizations. From Ref.~\cite{SWWL:2015} with permission.}
\label{fig:locating_GeoN}
\end{figure}

A detailed performance analysis was provided in Ref.~\cite{SWWL:2015}
where the issue of positioning accuracy was addressed. Specifically,
to locate all nodes in a two-dimensional space requires knowledge
of the positions of at least three nodes (minimally four nodes in
the three-dimensional space). Due to noise, the required number of
beacon nodes will generally be larger. Since node positioning is
based on time delays estimated from compressive sensing, which
contain errors, the number of required beacon nodes is larger
than three even in two dimensions. The positioning accuracy can 
be quantified~\cite{SWWL:2015} by using the normalized error $M_r$,
defined as the medium distance error between the estimated
and actual locations for all nodes (except the beacon nodes),
normalized by the distributed length $L$. Figure~\ref{fig:locating_GeoN}
shows $M_r$ versus the fraction $R_B$ of the beacon nodes. The
reconstruction parameters are chosen such that the errors in the
time delay estimation is $D_{nz} \approx 0.12$. For small values
of $R_B$, the positioning errors are large. Reasonable positioning
errors are obtained when $R_B$ exceeds, say, $0.2$.

\begin{figure}[!h]
\centering
\includegraphics[width=\linewidth]{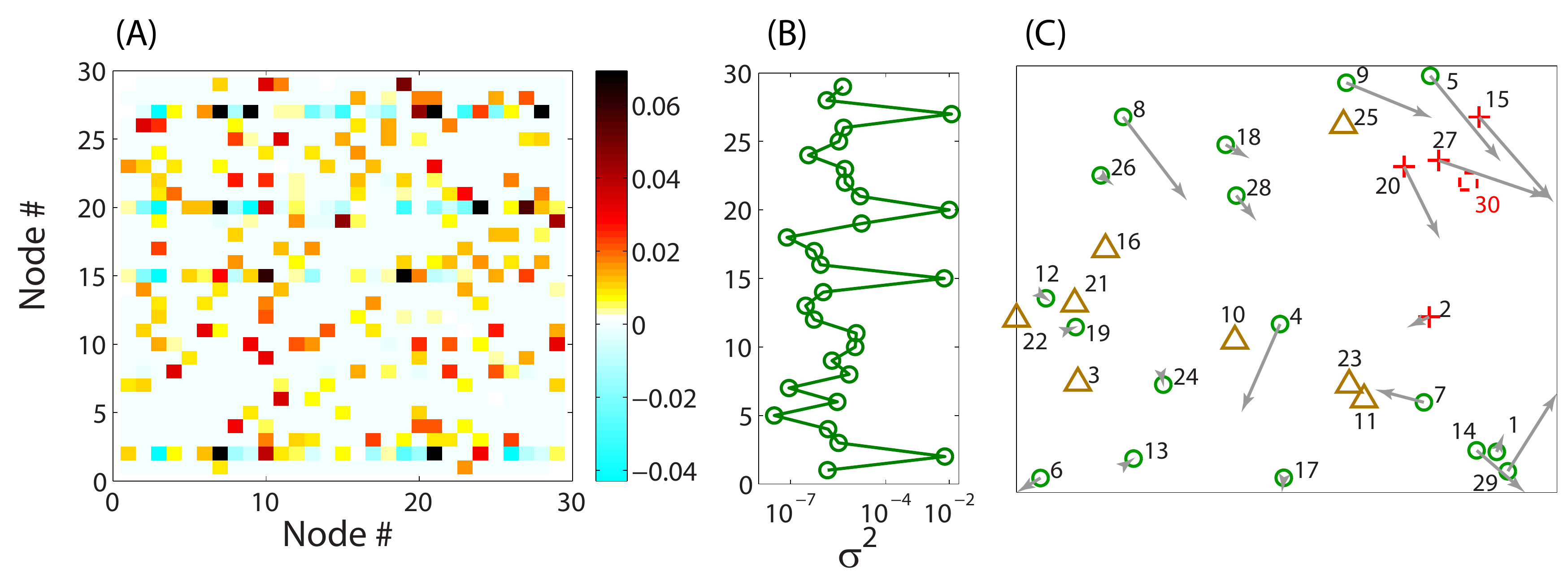}
\caption{\small {\bf Detection of hidden nodes in geospatial networks.}
For a random network of $N = 30$ nodes, illustration of detecting
a hidden node (\#30). (A) Reconstructed time delays using time
series from 29 externally accessible nodes. (B) Average variance
in the reconstructed incoming coupling delays calculated from
different segments of the available time series. (C) Estimated
positions of all accessible nodes in comparison with
the respective actual positions, and the location of the hidden
node. The triangles denote the beacon nodes, whose positions are
known {\it a priori}. The green circles denote ``normal'' nodes
without any hidden node in their immediate neighborhoods, while the
crosses are direct neighbors of the hidden node. The actual position
of the hidden node \#30 is marked as a dashed square. From 
Ref.~\cite{SWWL:2015} with permission.}
\label{fig:Hidden_GeoN}
\end{figure}

The compressive sensing based approach can be used to ascertain the 
existence of a hidden node and to estimate its physical location in 
a complex geospatial network~\cite{SWWL:2015}. 
To detect a hidden node, it is necessary to 
identify its neighboring nodes~\cite{SWL:2012,SLWD:2014}. For an 
externally accessible node, if there is a hidden node in its
neighborhood, the corresponding entry in the reconstructed
adjacency matrix will exhibit an abnormally dense pattern or
contain meaningless values. In addition, the estimated coefficients
for the dynamical and coupling functions of such an abnormal
node typically exhibit much larger variations when different
data segments are used, in comparison with those associated
with normal nodes that do not have hidden nodes in their
neighborhoods. The mathematical formulation of the method to
uncover a hidden node in complex geospatial networks can be found 
in Ref.~\cite{SWWL:2015}. For the network of size $N = 30$ in 
Fig.~\ref{fig:locating_GeoN}, initially, there are only 29 time 
series, one from each of the normal node, and it is not known 
{\em a priori} that there would be a hidden node in the network. 
The network was reconstructed to obtain the estimated weights 
and time delays, as shown in Fig.~\ref{fig:Hidden_GeoN}(A). 
One can see that the connection patterns of some nodes are relatively
dense and the values of the weights and time delays are meaningless
(e.g., negative values), giving the first clue that these nodes may be
the neighboring nodes of a hidden node. To confirm that this
is indeed the case, the available time series are divided into a
number of segments based on the criterion that the data requirement
for reconstruction is satisfied for each segment. As shown in
Fig.~\ref{fig:Hidden_GeoN}(B), extraordinarily large variances 
in the estimated coefficients associated with the abnormal nodes
arise. Combining results from Figs.~\ref{fig:Hidden_GeoN}(A)
and \ref{fig:Hidden_GeoN}(B), one can claim with confidence that the
four nodes are indeed in the immediate neighborhood of a hidden node, 
ascertaining its existence in the network. The method also works if 
there are more than one hidden node, given that they do not share 
common neighboring nodes.

\subsection{Reconstruction of complex spreading networks from binary
data} \label{subsec:CS_SPN}

An important class of collective dynamics is virus 
spreading and information diffusion in social and computer 
networks~\cite{PSV:2001,EK:2002,WMMD:2005,CBBV:2006,GJMP:2008,WGHB:2009,
MA:2010,BV:2011}. We discuss here the problem of reconstructing the 
network hosting the spreading process and identifying the source of 
spreading using limited measurements~\cite{SWFDL:2014}. This is a 
challenging problem due to (1) the difficulty in predicting and monitoring
mutations of deadly virus and (2) absence of epidemic threshold in 
heterogeneous networks~\cite{BPSV:2003,PCH:2010,CPS:2010,Gleeson:2011}. 
The problem is directly relevant to affairs of significant current interest
such as rumor propagation in the online virtual communities, which can cause 
financial loss or even social instabilities (an example being the 2011 
irrational and panicked acquisition of salt in southeast Asian countries 
caused by the nuclear leak in Japan). In such a case, identifying the 
propagation network for controlling the dynamics is of great interest.

A significant challenge in reconstructing a spreading network lies 
in the nature of the available time series: they are polarized, 
despite stochastic spreading among the nodes. Indeed, the link pattern 
and the probability of infection are encrypted in the binary status 
of the individuals, infected or not. 
There were recent efforts in addressing the inverse problem of some
special types of complex propagation networks~\cite{GLK:2010,PTV:2012}.
For example, for diffusion process originated from a single source,
the routes of diffusion from the source constitute a tree-like structure.
If information about the early stage of the spreading dynamics is available,
it would be feasible to decode all branches that reveal the connections
from the source to its neighbors, and then to their neighbors, and so on.
Taking into account the time delays in the diffusion process enables a
straightforward inference of the source in a complex network through
enumerating all possible hierarchical trees~\cite{GLK:2010,PTV:2012}. 
However, if no immediate information about the diffusion process is available, 
the tree-structure based inference method is inapplicable, and the problem 
of network reconstruction and locating the source becomes intractable,
hindering control of diffusion and delivery of immunization. 
The loss of knowledge about the source is common in real situations.
For example, passengers on an international flight can carry a highly
contagious disease, making certain airports the immediate neighbors of the
hidden source, which would be difficult to trace. In another example,
the source could be migratory birds coming from other countries or
continents. A general data-driven approach, applicable in such scenarios,
is an outstanding problem in network science and engineering.

Recently, a compressive sensing based framework was developed
to reconstruct complex spreading networks based on binary 
data~\cite{SWFDL:2014}. Since the dynamics of epidemic propagation 
are typically highly stochastic with binary time series, the standard
power series expansion method to fit the problem into the compressive
sensing paradigm (as discussed in preceding sections) is not applicable,
notwithstanding the methods of alternative sparsity enforcing regularizers 
and convex optimization used in Ref.~\cite{ML:2010} to infer networks.
The idea in Ref.~\cite{SWFDL:2014} was then to develop a scheme to implement 
the highly nontrivial transformation associated with the spreading 
dynamics in the paradigm of compressive sensing. Two prototypical models 
of epidemic spreading on model and real-world (empirical) networks
were studied: the classic 
susceptible-infected-susceptible (SIS) dynamics~\cite{PSV:2001} and
the contact process (CP)~\cite{CPS:2006,VM:2009}. Inhomogeneous infection 
and recovery rates as representative characteristics of the natural 
diversity were incorporated into the diffusion dynamics to better 
mimic the real-world situation.
  
The basic assumption is then that only binary time series can be measured, 
which characterize the status of any node, infected or susceptible, at any 
time after the outbreak of the epidemic. The source that triggers the 
spreading process is assumed to be externally inaccessible (hidden). In 
fact, one may not even realize its existence from the available time series. 
The method developed in Ref.~\cite{SWFDL:2014} enables, based on 
relatively small amounts of data, a full reconstruction of the epidemic 
spreading network with nodal diversity and successful identification 
of the immediate neighboring nodes of the hidden source (thereby 
ascertaining its existence and uniquely specifying its connections to 
nodes in the network). The framework was validated with respect to 
different amounts of data generated from various combinations of the
network structures and dynamical processes. High accuracy, high efficiency
and applicability in a strongly stochastic environment with measurement noise
and missing information are the most striking characteristics of the
framework~\cite{SWFDL:2014}. As a result, broad applications can be 
expected in addressing significant problems such as targeted control 
of virus spreading in computer networks and rumor propagation on social 
networks.

\subsubsection{Mathematical formulation} \label{subsubsec:CS_SPN_Math}

\paragraph*{Spreading processes.}
The SIS model~\cite{PSV:2001} is a classic epidemic model to study a 
variety of spreading behaviors in social and computer networks. 
Each node of the network represents an individual and links are connections 
along which the infection can propagate to others with certain probability.
At each time step, a susceptible node $i$ in state $0$ is infected with 
rate $\lambda_i$ if it is connected to an infected node in state 1. [If 
$i$ connects to more than one infected neighbor, the infection probability 
$P^{01}$ is given by Eq.~(\ref{eq:SIS_infect_SPN}) below.] At the same 
time, infected nodes are continuously recovered to be susceptible at the 
rate $\delta_i$. The CP model~\cite{CPS:2006,VM:2009} describes, e.g., 
the spreading of infection and competition of animals over a territory. 
The main difference between SIS and CP dynamics lies in the influence on 
a node's state from its vicinity. In both SIS and CP dynamics, $\lambda_i$ 
and $\delta_i$ depend on the individuals' immune systems and are selected 
from a Gaussian distribution characterizing the natural diversity. Moreover, 
a hidden source is regarded as infected at all time.

\paragraph*{Mathematical formulation of reconstruction from binary
data based on compressive sensing.} Assume that 
the disease starts to propagate from a fraction of the infected
nodes. The task is to locate any hidden source based solely on binary
time series after the outbreak of infection. The state of an arbitrary node 
$i$ is denoted as $S_i$, where
\begin{eqnarray}
S_i=\left\{
      \begin{array}{ll}
        0, & \hbox{susceptible;} \\
        1, & \hbox{infected.}
      \end{array}
    \right.
\end{eqnarray}
Due to the characteristic difference between the SIS dynamics and CP,
it is useful to treat them separately.

\underline{\em SIS dynamics}. The probability $P_i^{01}(t)$ of an
arbitrary node $i$ being infected by its neighbors at time $t$ is
\begin{equation} \label{eq:SIS_infect_SPN}
P_{i}^{01}(t) =1-(1-\lambda_{i})^{\sum_{j=1,j\ne i}^{N} a_{ij}S_{j}(t)},
\end{equation}
where $\lambda_i$ is the infection rate of $i$, $a_{ij}$ stands for the
elements of the adjacency matrix ($a_{ij}=1$ if $i$ connects to $j$ and
$a_{ij}=0$ otherwise), $S_j(t)$ is the state of node $j$ at $t$, and the 
superscript $01$ denotes the change from the susceptible state (0) to the
infected state (1). At the same time, the recovery probability of $i$ is 
$P_{i}^{10}(t)=\delta_{i}$, where $\delta_i$ is the recovery rate of node 
$i$ and the superscript $10$ denotes the transition from the infected state 
to the susceptible state. Equation~(\ref{eq:SIS_infect_SPN}) can be 
rewritten as
\begin{equation} \label{eq:sis02_SPN}
\ln[1-P_{i}^{01}(t)]=\ln(1-\lambda_{i}) \cdot 
\sum_{j=1,j\ne i}^{N} a_{ij}S_{j}(t).
\end{equation}
Suppose measurements at a sequence of times $t= t_1, t_2, \cdots, t_m$ are
available. Equation~(\ref{eq:sis02_SPN}) leads to the following matrix form
$\mathbf{X}_{m\times 1} = \mathcal{G}_{m\times (N-1)}\cdot
\mathbf{a}_{(N-1)\times 1}$:
\begin{center} 
$\begin{bmatrix}
\ln[1-P_{i}^{01}(t_1)]\\
\ln[1-P_{i}^{01}(t_2)] \\
\vdots\\
\ln[1-P_{i}^{01}(t_m)]
 \end{bmatrix}$
=
$\begin{bmatrix}
S_{1}(t_{1})  &\cdots& S_{i-1}(t_{1}) & S_{i+1}(t_{1}) &\cdots& S_{N}(t_{1}) \\
S_{1}(t_{2})  &\cdots& S_{i-1}(t_{2}) & S_{i+1}(t_{2}) &\cdots& S_{N}(t_{2}) \\
\vdots &\vdots & \vdots &\vdots & \vdots & \vdots\\
S_{1}(t_{m})  &\cdots& S_{i-1}(t_{m}) & S_{i+1}(t_{m}) &\cdots& S_{N}(t_{m})
 \end{bmatrix}
 $      $\begin{bmatrix}
\ln(1-\lambda_i)a_{i1}\\
\vdots\\
\ln(1-\lambda_i)a_{i,i-1}\\
\ln(1-\lambda_i)a_{i,i+1}\\
\vdots\\
\ln(1-\lambda_i)a_{iN}
 \end{bmatrix}$,
\end{center}
between node $i$ and all other nodes, and it is sparse for a general 
complex network. It can be seen that, if the vector $\mathbf{X}_{m\times 1}$ 
and the matrix $\mathcal{G}_{m\times (N-1)}$ can be constructed from time 
series, $\mathbf{a}_{(N-1)\times 1}$ can then be solved by using 
compressive sensing. The main
challenge here is that the infection probabilities $P_i^{01}(t)$ at 
different times are not given directly by the time series of the nodal 
states. A heuristic method to estimate the probabilities can be 
devised~\cite{SWFDL:2014} by assuming that the neighboring set $\Gamma_i$ 
of the node $i$ is known. The number of such neighboring nodes is given 
by $k_i$, the degree of node $i$, and their states at time $t$ can be 
denoted as
\begin{equation}
S_{\Gamma_i}(t) \equiv \{ S_1(t), S_2(t), \cdots, S_{k_i}(t) \}.
\end{equation}
In order to approximate the infection probability, one can use $S_i(t)=0$ 
so that at $t+1$, the node $i$ can be infected with certain probability. 
In contrast, if $S_i(t)=1$, $S_i(t+1)$ is only related with the recovery 
probability $\delta_i$. Hence, it is insightful to focus on the $S_i(t)=0$ 
case to derive $P_i^{01}(t)$. If one can find two time instants: 
$t_{1},t_{2}\in T$ ($T$ is the length of time series), such that 
$S_{i}(t_{1})=0$ and $S_{i}(t_{2})=0$, one can calculate the 
normalized Hamming distance $H[S_{\Gamma_i}(t_{1}),S_{\Gamma_i}(t_{2})]$ 
between $S_{\Gamma_i}(t_{1})$ and $S_{\Gamma_i}(t_{2})$, defined as the
ratio of the number of positions with different symbols between them and 
the length of string. If $H[S_{\Gamma_i}(t_{1}),S_{\Gamma_i}(t_{2})]=0$, 
the states at the next time step, $S_{i}(t_{1}+1)$ and $S_{i}(t_{2}+1)$, 
can be regarded as as i.i.d Bernoulli trials. In this case, using the 
law of large numbers, one has
\begin{equation} \label{eq:neighbor_SPN}
\lim_{l\to\infty} {{1}\over{l}}\sum_{\nu=1}^{l}S_{i}(t_{\nu}+1)
\to P_{i}^{01}(\hat t_{\alpha}), \ \ \forall \ \ t_{\nu}, S_{i}(t_{\nu})=0,\ \
H[S_{\Gamma_i}(\hat t_{\alpha}),S_{\Gamma_i}(t_{\nu})]=0.
\end{equation}
A more intuitive understanding of Eq.~(\ref{eq:neighbor_SPN}) is that, 
if the states of $i$'s neighbors are unchanged, the fraction of times of 
$i$ being infected by its neighbors over the entire time period will 
approach the actual infection probability $P_i^{01}$. Note, however, 
that the neighboring set of $i$ is unknown and to be inferred. A 
strategy is then to artificially enlarge the neighboring set 
$S_{\Gamma_i}(t)$ to include all nodes in the network except $i$. Denote
\begin{equation} \label{eq:string_SPN}
S_{-i}(t)\equiv \{S_{1}(t),S_{2}(t),\dots,S_{i-1}(t),S_{i+1}(t),
\dots,S_{N}(t)\}.
\end{equation}
If $H[S_{-i}(t_{1}),S_{-i}(t_{2})]=0$, the condition
$H[S_{\Gamma_i}(t_{1}),S_{\Gamma_i}(t_{2})]=0$ can be ensured. 
Consequently, due to the nature of the i.i.d Bernoulli trials,  
application of the law of large numbers leads to
\begin{equation*}
\lim_{l\to\infty} {{1}\over{l}}\sum_{\nu=1}^{l}S_{i}(t_{\nu}+1)
\to P_{i}^{01}(\hat t_{\alpha}),
\ \ \forall \ \ t_{\nu}, S_{i}(t_{\nu})=0, \ \ 
H[S_{-i}(\hat t_{\alpha}), S_{-i}(t_{\nu})]=0.
\end{equation*}
Hence, the infection probability $P_{i}^{01}(\hat t_{\alpha})$ of a 
node at $\hat t_{\alpha}$ can be evaluated by averaging over its 
states associated with zero normalized Hamming distance between the
strings of other nodes at some time associated with $\hat t_{\alpha}$. 
In practice, to find two strings with absolute zero normalized Hamming 
distance is unlikely. It is then necessary to set a threshold $\Delta$ 
so as to pick the suitable strings to approximate the law of large 
numbers, that is
\begin{equation} \label{eq:Delta_SIS_SPN}
{{1}\over{l}}\sum_{\nu=1}^{l\gg 1}S_{i}(t_{\nu}+1)\simeq 
{{1}\over{l}}\sum_{\nu=1}^{l\gg 1}P_i^{01}(t_{\nu}),
\ \ \forall \ \ t_{\nu}, S_{i}(t_{\nu})=0, \ \ 
H[S_{-i}(\hat t_{\alpha}), S_{-i}(t_{\nu})]<\Delta,
\end{equation}
where $S_{-i}(\hat t_{\alpha})$ serves as a base for comparison with
$S_{-i}(t)$ at all other times and
$(1/l)\sum_{\nu=1}^{l\gg 1}P_i^{01}(t_{\nu}) \simeq
P_{i}^{01}(\hat t_{\alpha})$.
Since $H[S_{-i}(\hat{t}_\alpha),S_{-i}(t_\nu)]$ is not exactly zero,
there is a small difference between $P_{i}^{01}(\hat{t}_{\alpha})$
and $P_{i}^{01}(t_{\nu})$ ($\nu=1,\cdots,l$). It is thus useful to 
consider the average of $P_{i}^{01}(t_{\nu})$ for all $t_\nu$ to obtain
$P_{i}^{01}(\hat t_{\alpha})$, leading to the right-hand side of
Eq.~(\ref{eq:Delta_SIS_SPN}). Let 
\begin{eqnarray}
\nonumber
\langle S_i(\hat t_{\alpha}+1)\rangle
& = & (1/l)\sum_{\nu=1}^{l\gg 1}S_{i}(t_{\nu}+1) \\ \nonumber
\langle P_{i}^{01}(\hat t_{\alpha}) \rangle & = & 
(1/l)\sum_{\nu=1}^{l\gg 1}P_i^{01}(t_\nu).
\end{eqnarray}
In order to reduce the error in the estimation, one can implement 
the average on $S_{-i}(t)$ over all selected strings using
Eq.~(\ref{eq:Delta_SIS_SPN}). The averaging process is with respect to 
the nodal states $S_{j,j\neq i}(t)$ on the right-hand side of the 
modified dynamical Eq.~(\ref{eq:sis02_SPN}). Specifically, averaging over 
time $t$ restricted by Eq.~(\ref{eq:Delta_SIS_SPN}) on both sides of
Eq.~(\ref{eq:sis02_SPN}) yields
\begin{displaymath}
\langle \ln[1-P_{i}^{01}(t)]\rangle =\ln(1-\lambda_{i})\sum_{j=1,j\ne i}^{N}
a_{ij}\langle S_{j}(t)\rangle.
\end{displaymath}
For $\lambda_i$ small compared with insignificant fluctuations, the 
following approximation holds:
\begin{displaymath} 
\ln[1-\langle P_{i}^{01}(t)\rangle ]\simeq \langle 
\ln[1-P_{i}^{01}(t)]\rangle,
\end{displaymath}
which leads to
\begin{displaymath}
\ln[1-\langle P_{i}^{01}(t)\rangle ]\simeq  \ln(1-\lambda_{i})
\sum_{j=1,j\ne i}^{N} a_{ij}\langle S_{j}(t)\rangle.
\end{displaymath}
Substituting $\langle P_{i}^{01}(\hat{t}_\alpha)\rangle$ by
$\langle  S_{i}(\hat t_{\alpha}+1)\rangle$, one finally gets
\begin{eqnarray} \label{eq:intui_SIS_SPN}
\ln[1-\langle S_{i}(\hat{t}_\alpha+1)\rangle ]\simeq \ln(1-\lambda_{i}) \cdot
\sum_{j=1,j\ne i}^{N} a_{ij}\langle S_{j}(\hat{t}_\alpha)\rangle.
\end{eqnarray}
While the above procedure yields an equation that bridges the links of
an arbitrary node $i$ with the observable states of the nodes, a single
equation does not contain sufficient structural information about the
network. The second step is then to derive a sufficient number of
linearly independent equations required by compressive sensing to reconstruct 
the local connection structure. To achieve this, one can choose a series of
base strings at a number of time instants from a set denoted by $T_{base}$, 
in which each pair of strings satisfy
\begin{equation}
H[S_{-i}(\hat t_{\beta}),S_{-i}(\hat t_{\alpha})]> \Theta,\quad
\forall \hat t_{\alpha},\hat t_{\beta} \in T_{base},
\end{equation}
where $\hat t_{\alpha}$ and $\hat t_{\beta}$ correspond to the time
instants of two base strings in the time series and $\Theta$ is a threshold.
For each string, the process of establishing the relationship between 
the nodal states and connections can be repeated, leading to a set of 
equations at different values of $\hat t_{\alpha}$ in 
Eq.~(\ref{eq:intui_SIS_SPN}). This process finally gives rise to a set 
of reconstruction equations in the matrix form:
\begin{center}\footnotesize
\begin{eqnarray} \label{eq:SIS_matrix_SPN}
\left[
\begin{array}{c}
\ln[1-\langle S_i(\hat t_1+1)\rangle]\\
\ln[1-\langle S_i(\hat t_2+1)\rangle] \\
\vdots\\
\ln[1-\langle S_i(\hat t_m+1)\rangle]
\end{array}
\right]=
\left[
\begin{array}{cccccc}
\langle S_{1}(\hat t_{1})\rangle  &\cdots& \langle S_{i-1}(\hat t_{1})\rangle
& \langle S_{i+1}(\hat t_{1})\rangle &\cdots& \langle S_{N}(\hat t_{1})\rangle \\
\langle S_{1}(\hat t_{2})\rangle  &\cdots& \langle S_{i-1}(\hat t_{2})\rangle
& \langle S_{i+1}(\hat t_{2})\rangle &\cdots& \langle S_{N}(\hat t_{2})\rangle \\
\vdots &\vdots & \vdots &\vdots & \vdots & \vdots\\
\langle S_{1}(\hat t_{m})\rangle  &\cdots& \langle S_{i-1}(\hat t_{m})\rangle
& \langle S_{i+1}(\hat t_{m})\rangle &\cdots& \langle S_{N}(\hat t_{m})\rangle
\end{array}
\right]
\left[
\begin{array}{c}
\ln(1-\lambda_i)a_{i1}\\
\vdots\\
\ln(1-\lambda_i)a_{i,i-1}\\
\ln(1-\lambda_i)a_{i,i+1}\\
\vdots\\
\ln(1-\lambda_i)a_{iN}
\end{array}
\right],
\end{eqnarray}
\end{center}
where $\hat t_1, \hat t_2,\cdots , \hat t_m$ correspond to the time 
associated with $m$ base strings and $\langle \cdot \rangle$ denote 
the average over all satisfied $t$. The vector $\mathbf{X}_{m\times 1}$ 
and the matrix $\mathcal{G}_{m\times (N-1)}$ can then be obtained based 
solely on time series of nodal states and the vector 
$\mathbf{a}_{(N-1)\times 1}$ to be reconstructed is sparse, rendering 
applicable the compressive sensing framework. As
a result, {\em exact reconstruction} of all neighbors of node $i$
from relatively small amounts of observation can be achieved. In a similar 
fashion the neighboring vectors of all other nodes can be uncovered from 
time series, enabling a full reconstruction of the whole network by 
matching the neighboring sets of all nodes.

\begin{figure}[!h]
\centering
\includegraphics[width=\linewidth]{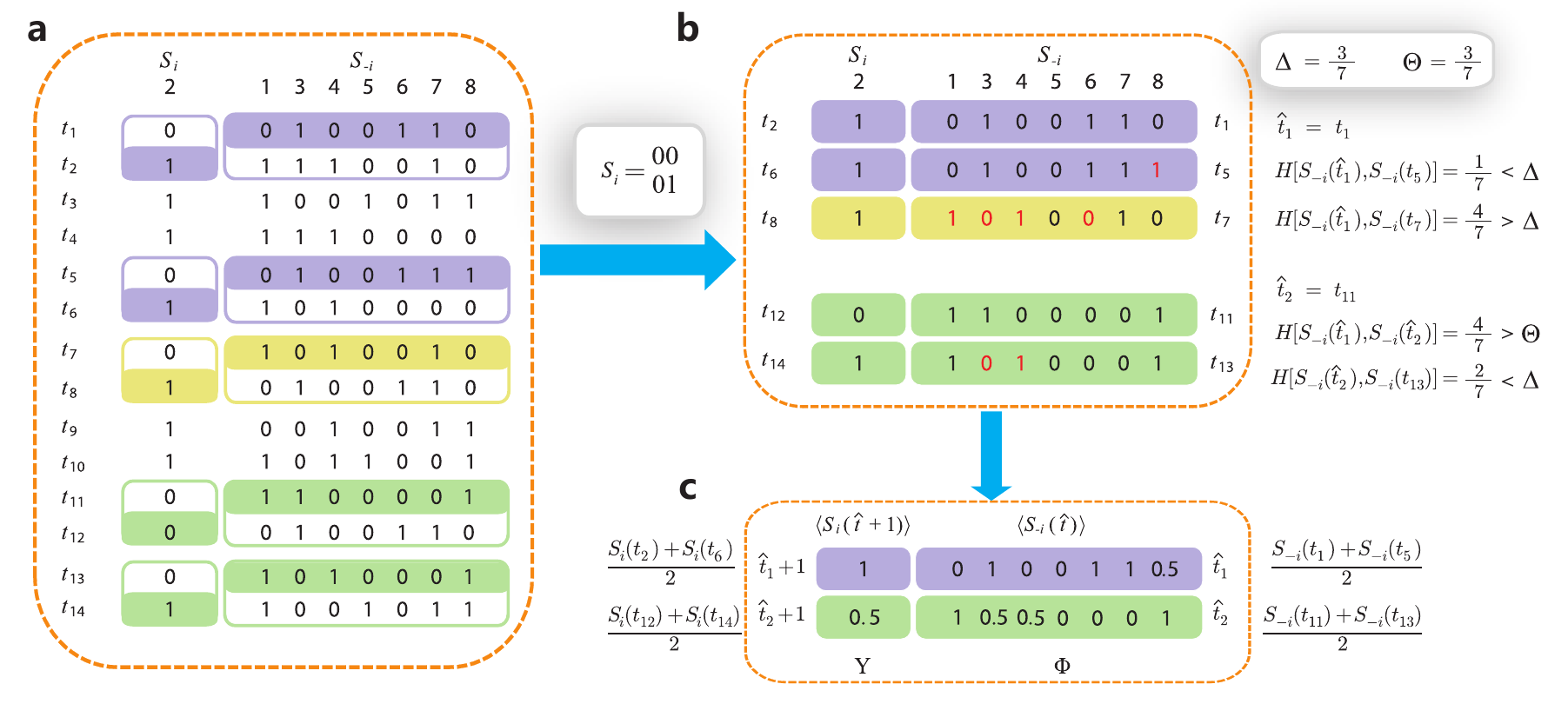}
\caption{\small {\bf Schematic illustration of construction of
$\mathbf{X}$ and $\mathcal{G}$ from binary time series.} 
({\bf a}) 14 snapshots of data at the time instants $t_1$-$t_{14}$ of 8
nodes in a sample network, where $S_i$ is the time series of node 2 and
$S_{-i}$ denotes the strings of other nodes at different times. The
neighborhood of node 2 is to be reconstructed. Only the pairs 00 and 01
in the time series of $S_i$ ($i=2$) and the corresponding $S_{-i}$
contain useful information about the network, as marked by different 
colors. ({\bf b}) Since $S_i(t+1)$ is determined by the neighbors of 
$i$ and $S_{-i}(t)$, one can sort out $S_i(t+1)$ and $S_{-i}(t)$
in the colored sections of the time series in ({\bf a}). With 
the threshold parameters $\Delta=3/7$ and $\Theta = 3/7$, 
the normalized Hamming distance between each pair of strings $S_{-i}(t)$
can be calculated, leading to two base strings at $\hat{t}_1 = t_1$ and 
$\hat{t}_2 = t_{11}$ with $H[S_{-i}(\hat{t}_1),S_{-i}(\hat{t}_2)]>\Theta$.
The colored strings can be separated into two groups that are led by the 
two base strings, respectively. In each group, the normalized Hamming 
distances $H[S_{-i}(\hat{t}_\alpha),S_{-i}(t_\nu)]$ between the base 
string and other strings are calculated and the difference from 
$S_{-i}(\hat{t}_\alpha)$ in each string is marked by red. Using parameter 
$\Delta$, in the group led by $S_{-i}(\hat{t}_1)$, $S_{-i}(t_5)$ and 
$S_i(t_6)$ are preserved, because of $H[S_{-i}(\hat{t}_1),S_{-i}(t_5)]<\Delta$. In contrast, $S_{-i}(t_7)$ and $S_i(t_8)$ are disregarded because
$H[S_{-i}(\hat{t}_1),S_{-i}(t_7)]>\Delta$. In the group led by 
$S_{-i}(\hat{t}_2)$, due to $H[S_{-i}(\hat{t}_2),S_{-i}(t_{13})]<\Delta$, 
the string is preserved. The two sets of remaining strings marked by
purple and green can be used to yield the quantities required by the
reconstruction formula. (Note that different base strings are allowed to
share some strings, but for simplicity, this situation is not illustrated
here.) ({\bf c}) The average values $\langle S_i(\hat{t}_{\alpha}+1) \rangle$
and $\langle S_{-i}(\hat{t}_{\alpha})\rangle$ used to extract the vector 
$\mathbf{X}$ and the matrix $\mathcal{G}$ in the reconstruction formula, where
$\langle S_{-i}(\hat{t}_1)\rangle = [S_{-i}(t_1) + S_{-i}(t_5)]/2$,
$\langle S_{-i}(\hat{t}_2)\rangle = [S_{-i}(t_{11}) + S_{-i}(t_{13})]/2$,
$\langle S_i(\hat{t}_1+1)\rangle = [S_i(t_2) + S_i(t_6)]/2$,
and $\langle S_i(\hat{t}_2+1)\rangle = [S_i(t_{12}) + S_i(t_{14})]/2$
based on the remaining strings marked in different colors. Compressive
sensing can be used to reconstruct the neighboring vector $\mathbf{a}$ of 
node 2 from $\mathbf{X}$ and $\mathcal{G}$ from 
$\mathbf{X}$= $\mathcal{G} \cdot \mathbf{a}$. 
From Ref.~\cite{SWFDL:2014} with permission.}
\label{fig:Hamming_SPN}
\end{figure}

\underline{\em CP dynamics}. The infection probability of an arbitrary
node $i$ is given by
\begin{eqnarray} \label{eq:cp01_SPN}
P_{i}^{01}(t)=\lambda_{i}\sum_{j=1,j\ne i}^{N} a_{ij}S_{j}(t)/k_{i},
\end{eqnarray}
where $k_i$ is the degree of the node $i$, and the recovery probability 
is $P_{i}^{10}(t)=\delta_{i}$. In close analogy to the SIS dynamics, 
one has 
\begin{equation} \label{eq:cp02_SPN}
\langle  S_{i}(\hat t_{\alpha}+1)\rangle  
\simeq\langle  P_{i}^{01}(\hat t_{\alpha})\rangle
=\frac{\lambda_{i}\sum a_{ij}\langle  S_{j}(\hat t_{\alpha})\rangle}{ k_{i}}.
\end{equation}
One can then choose a series of base strings using a proper threshold 
$\Theta$ to establish a set of equations, expressed in the matrix form
$\mathbf{X}_{m\times 1} = \mathcal{G}_{m\times (N-1)} \cdot
\mathbf{a}_{(N-1)\times 1}$, where $\mathcal{G}$ has the same form as in 
Eq.~(\ref{eq:SIS_matrix_SPN}), but $\mathbf{X}$ and $\mathbf{a}$ are 
given by
\begin{eqnarray}
\mathbf{X} &=& \left[\langle S_i(\hat{t}_1 +1)\rangle, \langle S_i(\hat{t}_2 +1)\rangle,
\cdots, \langle S_i(\hat{t}_m +1)\rangle  \right]^\text{T},  \nonumber \\
\mathbf{a} &=& \left[ \frac{\lambda_i}{k_i}a_{i1}, \cdots,
\frac{\lambda_i}{k_i}a_{i,i-1}, \frac{\lambda_i}{k_i}a_{i,i+1}, \cdots,
\frac{\lambda_i}{k_i}a_{iN}  \right]^\text{T}.
\end{eqnarray}
The reconstruction framework based on building up the vector 
$\mathbf{X}$ and the matrix $\mathcal{G}$ is schematically illustrated in 
Fig.~\ref{fig:Hamming_SPN}. It is noteworthy that the framework can 
be extended to directed networks in a straightforward fashion due to 
the feature that the neighboring set of each node can be independently 
reconstructed. For instance, the neighboring vector $\mathbf{a}$ can 
be defined to represent a unique link direction, e.g., incoming links. 
Inference of the directed links of all nodes yields the full topology 
of the entire directed network.

\paragraph*{Inferring inhomogeneous infection rates.}
The values of the infection rate $\lambda_i$ of nodes can be inferred
after the neighborhood of each node has been successfully reconstructed.
The idea roots in the fact that the infection probability of a node
approximated by the frequency of being infected calculated from time
series is determined both by its infection rate and by the number of
infected nodes in its neighborhood. An intuitive picture can be obtained
by considering the following simple scenario in which the number of infected
neighbors of node $i$ does not change with time. In this case, the
probability of $i$ being infected at each time step is fixed. One can
thus count the frequency of the $01$ and $00$ pairs embedded in the
time series of $i$. The ratio of the number of $01$ pairs over the
total number of $01$ and $00$ pairs gives approximately the infection
probability. The infection rate can then be calculated by using
Eqs.~(\ref{eq:SIS_infect_SPN}) and (\ref{eq:cp01_SPN}) for the SIS and 
CP dynamics, respectively. In a real-world situation, however, the number 
of infected neighbors varies with time. The time-varying factor can be 
taken into account by sorting out the time instants corresponding to 
different numbers of the infected neighbors, and the infection probability 
can be obtained at the corresponding time instants, leading to a set of
values for the infection rate whose average represents an accurate
estimate of the true infection rate for each node.

To be concrete, considering all the time instants $t_{\nu}$ associated 
with $k_I$ infected neighbors, one can denote 
\begin{eqnarray}
\nonumber
S_{i}^{(k_I)} & = & (1/l)\sum_{\nu=1}^{l}S_{i} (t_{\nu}+1), \ \mbox{for all}
\ t_{\nu}, \\ \nonumber
\sum_{j\in\Gamma_{i}} S_{j}(t_{\nu}) & = & k_I, \\ \nonumber
S_i(t_\nu) & = &0,
\end{eqnarray}
where $\Gamma_i$ is the neighboring set of node $i$, $k_I$ is the 
number of infected neighbors, and $S_{i}^{(k_I)}$ represents the 
average infected fraction of node $i$ with $k_I$ infected neighbors. 
Given $S_{i}^{(k_I)}$, one can rewrite Eq.~(\ref{eq:SIS_infect_SPN})
by substituting $S_{i}^{(k_I)}$ for $P_{i}^{01}(t)$ and $\lambda_i^{(k_I)}$ 
for $\lambda_i$, which yields 
\begin{displaymath}
\lambda_{i}^{(k_{I})} = 1-\exp\left[\ln \big( 1
- S_{i}^{(k_I)}\big) /k_I\right].
\end{displaymath}
The estimation error can be reduced by averaging $\lambda_{i}^{(k_I)}$ 
with respect to different values of $k_I$, as follows:
\begin{equation} \label{eq:lambda_SIS_SPN}
\lambda_i^{\text{true}}(\text{SIS}) \approx \langle \lambda_i^{(k_I)}  \rangle
= \frac{1}{N_{\lambda_{i}}}\sum_{k_I\in \lambda_{i}}\lambda_{i}^{(k_I)},
\end{equation}
where $\lambda_{i}$ denotes the set of all possible infected neighbors 
during the epidemic process and $N_{\lambda_{i}}$ denotes the number of 
different values of $k_I$ in the set. Analogously, for CP, one can evaluate
$\lambda_i^{\text{true}}$ from Eq.~(\ref{eq:cp01_SPN}) as 
\begin{equation} \label{eq:lambda_CP_SPN}
\lambda_i^{\text{true}}(\text{CP}) \approx \langle \lambda_i^{(k_I)} \rangle
= \frac{1}{N_{\lambda_{i}}}\sum_{k_I\in \lambda_i}
\frac{S_i^{(k_I)} k_{i}}{k_{I}}
\end{equation}
where $k_i = \sum_{j=1}^{N} a_{ij}$ is the node degree of $i$. After
all the links of $i$ have been successfully reconstructed, $S_i^{(k_I)}$ 
can be obtained from the time series in terms of the satisfied 
$S_i(t_\nu +1)$, allowing one to infer $\lambda_i^\text{true}$ via 
Eqs.~(\ref{eq:lambda_SIS_SPN}) and (\ref{eq:lambda_CP_SPN}).

The method so described for estimating the infection rates is 
applicable to any type of networks insofar as the network structure 
has been successfully reconstructed~\cite{SWFDL:2014}.

\subsubsection{Reconstructing complex spreading networks: examples}
\label{subsubsec:CS_SPN_examples}

\paragraph*{Reconstructing networks and inhomogeneous infection and 
recovery rates.}
A key performance indicator of the binary data based reconstruction 
framework is the number of base strings (equations) for a variety 
of diffusion dynamics and network structures. It is necessary to 
calculate the success rates for existent links (SREL) and null
connections (SRNC), corresponding to non-zero and zero element values in
the adjacency matrix, respectively, in terms of the number of base
strings. The binary nature of the network dynamical process and data 
requires that the strict criterion be imposed~\cite{SWFDL:2014}, i.e., 
a network is regarded to have been fully reconstructed if and only if
both success rates reach 100$\%$. The sparsity of links makes it 
necessary to define SREL and SRNC separately. Since the reconstruction 
method is implemented for each node in the network, SREL and SRNC 
can be defined with respect to each individual node and, the two 
success rates for the entire network are the respective averaged 
values over all nodes. The issue of trade-off can also be considered
in terms of the true positive rate (TPR - for correctly inferred links) 
and the false positive rate (FPR - for incorrectly inferred links).

\begin{figure}[h]
\centering
\includegraphics[width=0.6\linewidth]{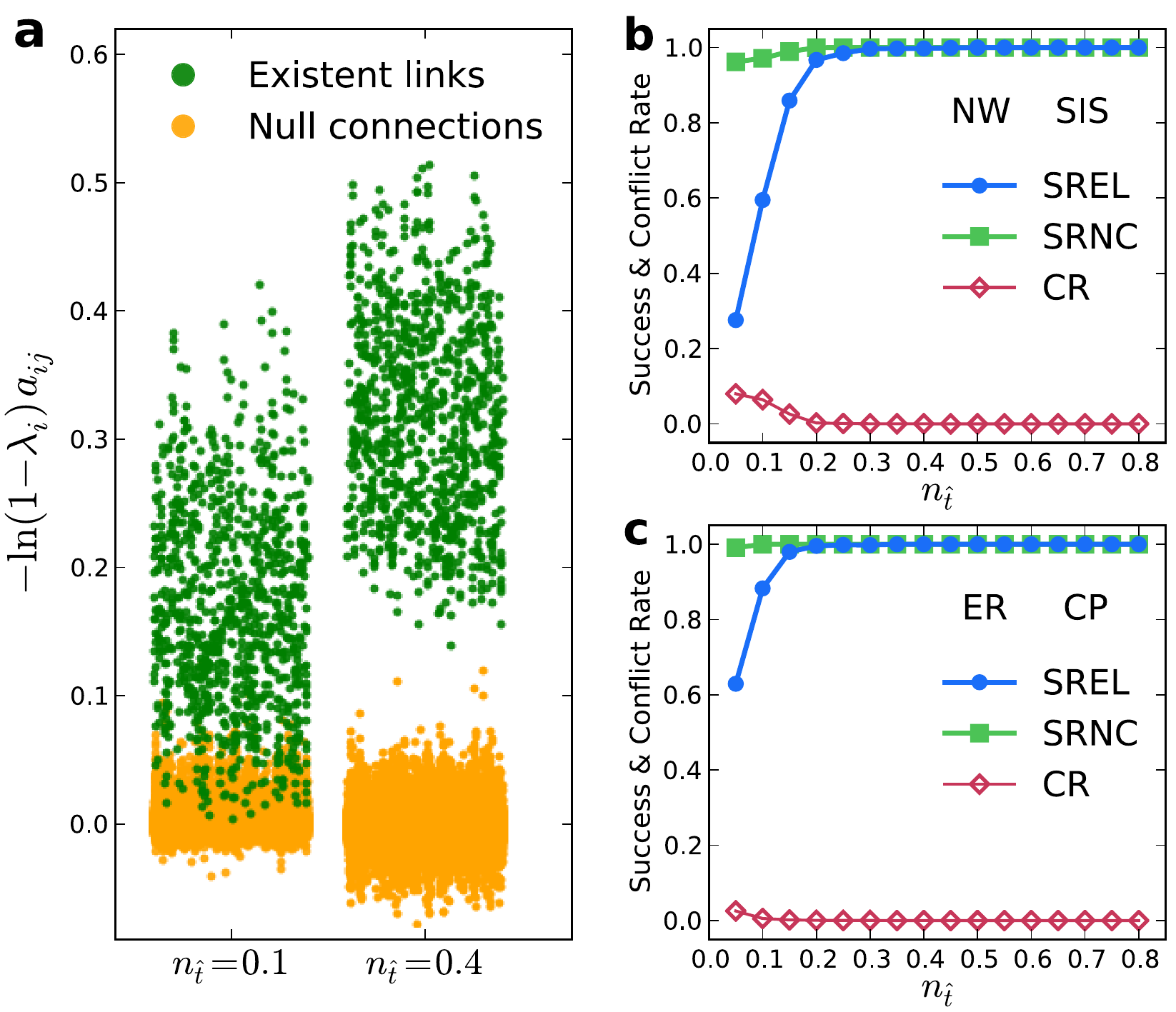}
\caption{\small {\bf Performance of reconstructing complex spreading networks
from binary data.} ({\bf a}) Element values $\ln(1-\lambda_i)a_{ij}$ of 
the vector $\mathbf{X}$ times -1 for different fraction $n_{\hat{t}}$ of 
the base strings for SIS dynamics. ({\bf b}-{\bf c}), success rate (SREL 
and SRNC) and conflict rate (CR) of reconstruction as a function of 
$n_{\hat{t}}$ for SIS dynamics on Newman-Watts (NW) small-world networks
({\bf b}) and CP dynamics on Erd\"os-R\'enyi (ER) random networks ({\bf c}),
respectively. For the SIS dynamics, the parameters were $\Theta=0.25$, 
$\Delta=0.45$, the infection and recovery rates $\lambda_i$ and 
$\delta_i$ were randomly distributed in the ranges $(0.2,0.4)$ and 
$(0.4,0.6)$, respectively. For the CP dynamics, the parameters were 
$\Theta=0.35$, $\Delta=0.45$, $\lambda_i$ and $\delta_i$ were randomly 
distributed in the ranges $(0.7,0.9)$ and $(0.2,0.4)$, respectively. The 
network size $N$ is 200 with average node degree $\langle k\rangle =4$. 
The results were obtained by ensemble averaging over 10 independent 
realizations. The success rate was determined by setting a proper 
threshold. From Ref.~\cite{SWFDL:2014} with permission.}
\label{fig:gap_SPN}
\end{figure}

In Ref.~\cite{SWFDL:2014}, a large number of examples of reconstruction
were presented. Take an example where there is no hidden source, and 
binary time series can be obtained by initiating the spreading process 
from a fraction of infected nodes. Figure~\ref{fig:gap_SPN}(a) shows 
the reconstructed values of the components of the neighboring vector 
$\mathbf{X}$ of all nodes. Let $n_{\hat{t}}$ be the number of base 
strings normalized by the total number of strings. For small values 
of $n_{\hat{t}}$, e.g., $n_{\hat{t}}=0.1$, the values of elements 
associated with links and those associated with null connections
(actual zeros in the adjacency matrix) overlap, leading to ambiguities 
in the identification of links. In contrast, for larger values of 
$n_{\hat{t}}$, e.g., $n_{\hat{t}}=0.4$, an explicit gap emerges between 
the two groups of element values, enabling correct identification of 
all links by simply setting a threshold within the gap. The success 
rates (SREL and SRNC) as a function of $n_{\hat{t}}$ for SIS and CP 
on both homogeneous and heterogeneous networks are shown in 
Figs.~\ref{fig:gap_SPN}(b,c), where nearly perfect reconstruction 
of links are obtained, insofar as $n_{\hat{t}}$ exceeds a relatively 
small value - an advantage of compressed sensing. The exact reconstruction 
is robust in the sense that a wide range of $n_{\hat{t}}$ values can 
yield nearly $100\%$ success rates. The reconstruction method is 
effective for tackling real networks in the absence of any
{\em a priori} knowledge about its topology.

Since the network is to be reconstructed through the union of all 
neighborhoods, one may encounter ``conflicts'' with respect to the
presence/absence of a link between two nodes as generated by the
reconstruction results centered at the two nodes, respectively. Such conflicts 
will reduce the accuracy in the reconstruction of the entire network. 
The effects of edge conflicts can be characterized by analyzing the
consistency of mutual assessment of the presence or absence of link 
between each pair of nodes, as shown in Figs.~\ref{fig:gap_SPN}(b,c). 
It can be seen that inconsistency arises for small values of $n_{\hat{t}}$ 
but vanishes {\em completely} when the success rates reach $100\%$, 
indicating perfect consistency among the mutual inferences of nodes 
and consequently guaranteeing accurate reconstruction of the entire 
network.

\begin{figure}[!h]
\centering
\includegraphics[width=0.8\linewidth]{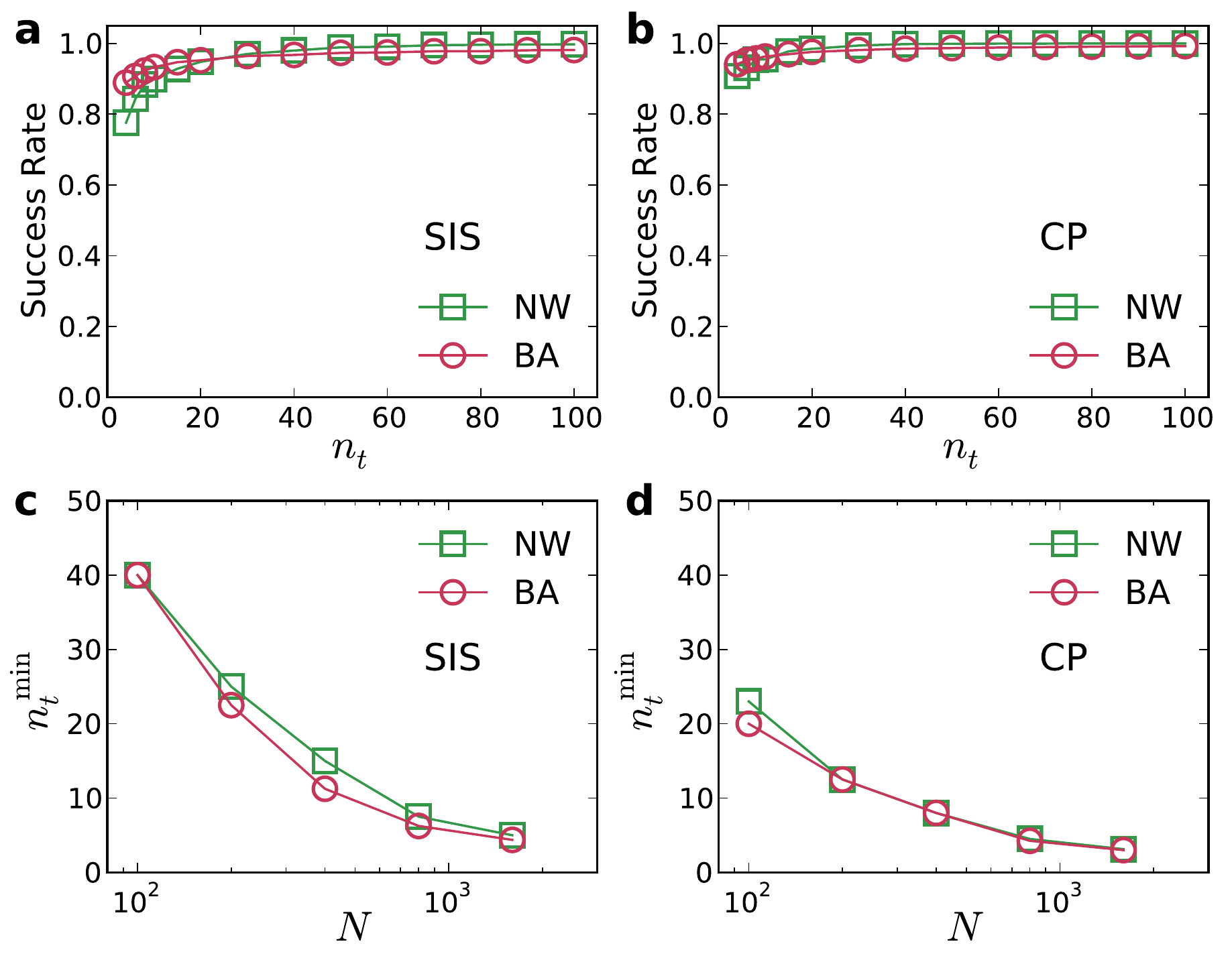}
\caption{\small {\bf Effect of the length of the binary time series 
and network size on reconstruction.}
({\bf a}-{\bf b}) Success rate as a function of the relative length 
$n_t$ of time series for SIS ({\bf a}) and CP ({\bf b}) in combination 
with NW and BA scale free networks. ({\bf c}-{\bf d}) the minimum 
relative length $n_t^{\min}$ that assures at least $95\%$ success 
rate as a function of the network size $N$ for SIS ({\bf c}) and CP 
({\bf d}) dynamics on NW and BA networks, where the success rate is 
taken as the geometric average over SREL and SRNC. For the SIS dynamics, 
$\lambda_i \in (0.1,0.3)$ and $\delta \in (0.2,0.4)$. For the CP dynamics, 
$\lambda_i \in (0.7,0.9)$ and $\delta \in (0.3,0.5)$. In ({\bf a}) and 
({\bf b}), the network size $N$ is 500. The other parameters are the 
same as in Fig.~\ref{fig:gap_SPN}. Note that $n_t \equiv t/N$, where 
$t$ is the absolute length of time series, and 
$n_t^{\min} \equiv t_{\min}/N$, where $t_{\min}$ is the minimum absolute
length of time series required for at least $95\%$ success rate.
From Ref.~\cite{SWFDL:2014} with permission.}
\label{fig:length_SPN}
\end{figure}

While the number of base strings is relatively small compared with
the network size, it is necessary to have a set of strings at 
different time with respect to a base string to formulate the 
mathematical reconstruction framework. How the length of the
time series affects the accuracy of reconstruction was 
studied~\cite{SWFDL:2014}. Figures~\ref{fig:length_SPN}(a,b) show the
success rates as a function of the relative length $n_t$ of time series
for SIS and CP dynamics on both homogeneous and heterogeneous networks,
respectively, where $n_t$ is the total length of time series from the 
beginning of the spreading process divided by the network size $N$.
The results demonstrate that, even for very small values of $n_t$, most
links can already be identified, as reflected by the high values of the
success rate shown. Figures~\ref{fig:length_SPN}(c,d) show the minimum 
length $n_t^\text{min}$ required to achieve at least $95\%$ success 
rate for different network size. For both SIS and CP dynamics on 
networks, $n_t^\text{min}$ decreases considerably as $N$ 
is increased. This seemingly counterintuitive result is due to the 
fact that different base strings can share strings at different times 
to enable reconstruction. In general, as $N$ is increased, $n_{\hat{t}}$ 
will increase accordingly. However, a particular string can belong to 
different base strings with respect to the threshold $\Delta$, accounting 
for the slight increase in the absolute length of the time series and 
the reduction in $n_t^\text{min}$. The dependence of the success rate 
on the average node degree $\langle k\rangle$ for SIS and CP on different 
networks was investigated as well~\cite{SWFDL:2014}. The results in 
Figs.~\ref{fig:gap_SPN} and \ref{fig:length_SPN} demonstrate the high 
accuracy and efficiency of the compressive sensing based reconstruction
method based on small amounts of binary data.

\begin{figure}[!h]
\centering
\includegraphics[width=0.8\linewidth]{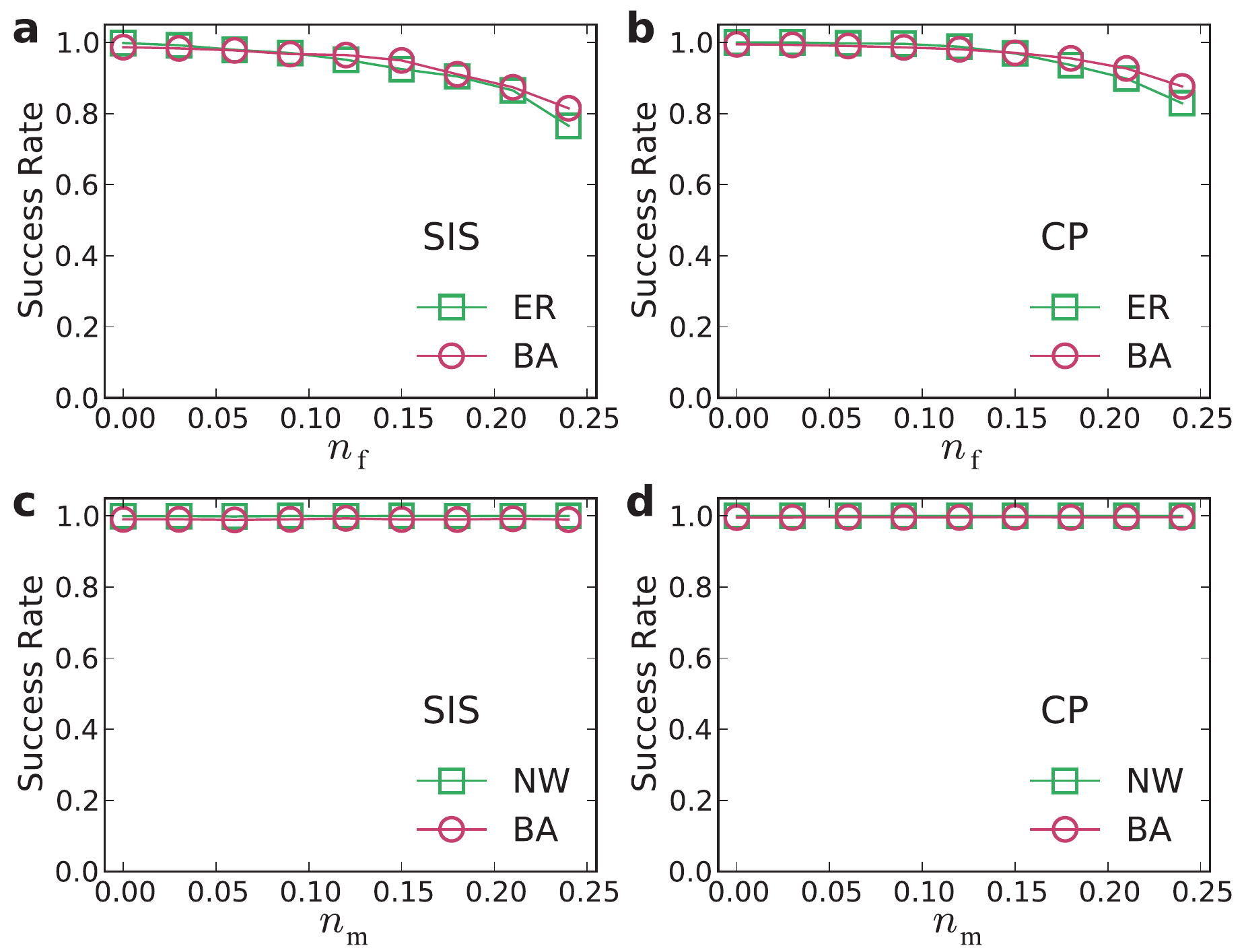}
\caption{\small {\bf Reconstruction against noise and inaccessible nodes.}
({\bf a}-{\bf b}) Success rate as a function of the fraction 
$n_\text{f}$ of the flipped states induced by noise in time series 
for SIS ({\bf a}) and CP ({\bf b}) dynamics on ER and BA networks. 
({\bf c}-{\bf d}) success rate as a function of the fraction 
$n_\text{m}$ of nodes that are externally inaccessible for SIS 
({\bf c}) and CP ({\bf d}) dynamics on NW and BA networks. The 
network size $N$ is 500 and $\langle k\rangle=4$. The other parameters 
are the same as in Fig.~\ref{fig:gap_SPN}. From Ref.~\cite{SWFDL:2014}
with permission.}
\label{fig:noise_SPN}
\end{figure}

In practice, noise is present and it is also common for time series 
from certain nodes to be missing, so it is necessary to test the 
applicability of the method under these circumstances. 
Figures~\ref{fig:noise_SPN}(a,b) show the dependence of the success 
rate on the fraction $n_\text{f}$ of states in the time series that 
flip due to noise for SIS and CP dynamics on two types of networks, 
respectively. It can be seen that the success rates are hardly 
affected, providing strong support for the applicability of the 
reconstruction method. For example, even when $25\%$ of the nodal 
states flip, about $80\%$ success rates can still be achieved for 
both dynamical processes and different network topologies. 
Figures~\ref{fig:noise_SPN}(c,d) present the success rate versus the 
fraction $n_\text{m}$ of unobservable nodes, the states of which
are externally inaccessible. It can be seen that the high success 
rate remains mostly unchanged as $n_\text{m}$ is increased from 
zero to $25\%$, a somewhat counterintuitive but striking result. It
was found that~\cite{SWFDL:2014}, in general, missing information 
can affect the reconstruction of the neighboring vector, as reflected 
by the reduction in the gap between the success rates associated with 
the actual links and null connections. However, even for high values 
of $n_\text{f}$, e.g., $n_\text{f}=0.3$, there is still a clear gap, 
indicating that a full recovery of all links is achievable. Taken 
together, the high accuracy, efficiency and robustness against noise
and missing information provide strong credence for the validity and 
power of the framework for binary time-series based network reconstruction.

Having reconstructed the network structure, one can estimate the infection
and recovery rates of individuals to uncover their diversity in immunity.
This is an essential step to implement target vaccination strategy in a
population or on a computer network to effectively suppress/prevent the
spreading of virus at low cost, as a large body of literature indicates
that knowledge about the network structure and individual characteristics
is sufficient for controlling the spreading dynamics~\cite{CHbA:2003,
FG:2007,KLG:2011,KOGNG:2012}. An effective method was 
proposed~\cite{SWFDL:2014} to infer the individuals' infection rates
$\lambda_i$ based solely on the binary time series of the nodal states 
after an outbreak of contamination. In particular, after all links 
have been successfully predicted, $\lambda_i$ can be deduced from 
the infection probabilities that can be approximated by the corresponding 
infection frequencies. These probabilities depend on both $\lambda_i$
and the number of infected neighbors. The reproduced infection
rates $\lambda_i$ of individuals for both SIS and CP dynamics on
different networks were in quite good agreement with the true values
with small prediction errors. An error analysis revealed uniformly high 
accuracy of the method~\cite{SWFDL:2014}. The inhomogeneous recovery rates 
$\delta_i$ of nodes can be predicted from the binary time series in a 
more straightforward way, because $\delta_i$'s do not depend on the 
nodal connections. Thus the framework is capable of predicting 
characteristics of nodal diversity in terms of degrees and infection 
and recovery rates based solely on binary time series of nodal states.

\paragraph*{Locating the source of spreading.}
Assume that a hidden source exists outside the network but there are
connections between it and some nodes in the network. In practice, the
source can be modeled as a special node that is always infected. Starting
from the neighborhood of the source, the infection originates from the
source and spreads all over the network. One first collects a set of 
time series of the nodal states except the hidden source. 
As discussed in detail in Sec.~\ref{subsec:CD_HN}, the basic
idea of ascertaining and locating the hidden source is based on 
missing information from the hidden source when attempting to 
reconstruct the network~\cite{SWL:2012,SLWD:2014}. In particular,
in order to reconstruct the connections belonging to the immediate
neighborhood of the source accurately, time series from the source are
needed to generate the matrix $\mathcal{G}$ and the vector $\mathbf{X}$.
But since the source is hidden, no time series from it are available, 
leading to reconstruction inaccuracy and, consequently, anomalies in 
the predicted link patterns of the neighboring nodes. It is then 
possible to detect the neighborhood of the hidden source by identifying 
any abnormal connection patterns~\cite{SWL:2012,SLWD:2014}, which can 
be accomplished by using different data segments. If the inferred links 
of a node are stable with respect to different data segments, the 
node can be deemed to have no connection with the hidden source; 
otherwise, if the result of inferring a node's links varies 
significantly with respect to different data segments, the node is
likely to be connected to the hidden source. The standard deviation 
of the predicted results with respect to different data segments can 
be used as a quantitative criterion for the anomaly. Once the 
neighboring set of the source is determined, the source can be 
precisely located topologically.

\begin{figure}[!h]
\centering
\includegraphics[width=\linewidth]{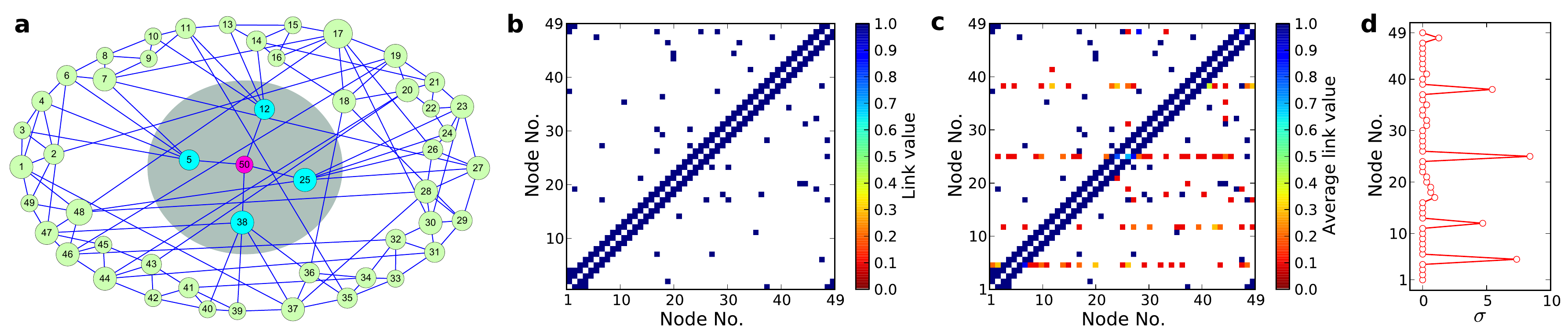}
\caption{\small {\bf An example of locating an external hidden source from time 
series.} ({\bf a}) Hidden source treated as a special node (in red) is 
connected to four nodes in the network (blue). The time series of other 
nodes except the source (No. 50) after the outbreak of an epidemic were 
assumed to be available. ({\bf b}) True adjacency matrix of the NW 
network with identical link weights to facilitate a comparison with 
the reconstructed adjacency matrix. ({\bf c}) Reconstructed adjacency 
matrix from a number of segments in time series. The four neighboring 
nodes of the source are predicted to be densely linked to other nodes, 
as indicated by the average value of the elements in the four rows 
corresponding to these nodes. ({\bf d}) Structural variance $\sigma$ of 
each node. The four neighboring nodes of the source exhibit much larger 
values of $\sigma$ than those from the other nodes, providing unequivocal 
evidence that they belong to the immediate neighborhood of the hidden 
source. From Ref.~\cite{SWFDL:2014} with permission.}
\label{fig:hidden_SPN}
\end{figure}

An example~\cite{SWFDL:2014} is shown in Fig.~\ref{fig:hidden_SPN}, 
where a hidden source is connected with four nodes in the network 
[Fig.~\ref{fig:hidden_SPN}(a)], as can be seen from the network adjacency 
matrix [Fig.~\ref{fig:hidden_SPN}(b)]. The reconstruction framework was
implemented on each accessible node by using different sets of data 
in the time series. For each data set, the neighbors of all nodes 
were predicted, generating the underlying adjacency matrix. Averaging
over the elements corresponding to each location in all the reconstructed
adjacency matrices leads to Fig.~\ref{fig:hidden_SPN}(c), in which 
each row corresponds to the mean number of links in a node's 
neighborhood. The inferred links of the immediate neighbors of the 
hidden source exhibit anomalies. To quantify the anomalies, the 
structural standard deviation $\sigma$ was calculated from different 
data segments, where $\sigma$ associated with node $i$ is defined 
through the $i$th row in the adjacency matrix as 
\begin{equation} \label{eq:struc_var_SPN}
\sigma_i = \frac{1}{N}\sum_{j=1}^{N}\sqrt{\frac{1}{g}
\sum_{k=1}^{g}\big(a_{ij}^{(k)} - \langle a_{ij}\rangle\big)^2},
\end{equation}
where $j$ denotes the column, $a_{ij}^{(k)}$ represents the element 
value in the adjacency matrix inferred from the $k$th group of the data,
$\langle a_{ij}\rangle = (1/g)\sum_{k=1}^{g} a_{ij}^k$ is the mean
value of $a_{ij}$, and $g$ is the number of data segments. Applying
Eq.~(\ref{eq:struc_var_SPN}) to the reconstructed adjacency matrices 
gives the results in Fig.~\ref{fig:hidden_SPN}(d), where the values 
of $\sigma$ associated with the immediate neighboring nodes of the 
hidden source are much larger than those from others (which are 
essentially zero). A threshold can be set in the distribution of 
$\sigma_i$ to identify the immediate neighbors of the hidden source.

\clearpage

\section{Alternative methods for reconstructing complex, nonlinear dynamical
networks} \label{sec:CH4_alt_method}

\subsection{Reconstructing complex networks from response dynamics} 
\label{subsec:CH4_response_dyn}

Probing into a dynamical system in terms of its response to external driving 
signals is commonly practiced in biological sciences. This approach is 
particularly useful for studying systems with complex interactions and dynamical
behaviors. The basic idea of response dynamics was previously exploited for 
reconstructing complex networks of coupled phase oscillators~\cite{Timme:2007}. 
Through measuring the collective response of the oscillator network to 
an external driving signal, the network topology can be recovered through
repeated measurement of the dynamical states of the nodes, provided that the 
driving realizations are sufficiently independent of each other. Since 
complex networks are generally sparse, the number of realizations of
external driving can be much smaller than the network size.

A network of coupled phase oscillators with a complex interacting topology is 
described by
\begin{equation} \label{eq:CH4_CPOs}
\dot{\phi}_i=\omega_i + \sum_{j=1}^{N} J_{ij}f_{ij}(\phi_j - \phi_i) + I_{i,m},
\end{equation}
where $\dot{\phi}_i(t)$ is the phase of oscillator $i$ at time $t$, $\omega_i$ 
is its natural frequency, $J_{ij}$ is the coupling strength from oscillator $j$ 
to oscillator $i$ (weighted adjacency matrix, where $J_{ij}=0$ indicates the 
absence of a link from $j$ to $i$), and $f_{ij}$'s are the pairwise 
coupling functions among the connected oscillators. The nodes on which the
external driving signals are imposed are specified as $I_{i,m}$, where $m=0$ 
indicates no driving signal. When driving is present, the collective frequency 
is
\begin{equation} \label{eq:CH4_coll_freq}
\Omega_m = \omega_i + \sum_{j=1}^{N} J_{ij}f_{ij}(\phi_{j,m} 
- \phi_{i,m}) + I_{i,m}.
\end{equation}
The frequency difference between the driven and non-driven systems is
\begin{equation}
D_{i,m}= \sum_{j=1}^{N} J_{ij}[f_{ij}(\phi_{j,m} - 
\phi_{i,m})-f_{ij}(\phi_{j,0} - \phi_{i,0})].
\end{equation}
For sufficiently small signal strength, the phase dynamics of the oscillators 
are approximately those of the original. We can approximate 
$f_{ij}(x)=f_{ij}'(\Delta_{ji,0}) + \mathcal{O}(x^2)$ and let $\theta_{j,m}$ 
denote $\phi_{j,m}-\phi_{j,0}$. After a number of experiments, the network 
configuration converges to
\begin{equation} \label{eq:CH4_DJTheta}
\mathcal{D} = \hat{\mathcal{J}} \cdot \mathcal{\theta},
\end{equation}
where $\hat{J}_{ij}$ is the Laplacian matrix with
\begin{equation}
\hat{J}_{ij}=\left\{
          \begin{array}{ll}
           J_{ij}f_{ij}'(\Delta_{ji,0}), & \hbox{for $i\neq j$,} \\
           -\sum_{k,k\neq i} J_{ik}f_{ik}'(\Delta_{ki,0}), & \hbox{$i=j$.}
            \end{array}
         \right.
\end{equation}
Note that the phase and frequency difference matrices $\mathcal{\theta}$ 
and $\mathcal{D}$ are measurable. Thus, the network 
structure characterized by the matrix $\hat{\mathcal{J}}$ 
can be solved by $\hat{\mathcal{J}} = \mathcal{D}\cdot \mathcal{\theta}^{-1}$. 
The reconstruction method based 
on the response dynamics not only can reveal links among the nodes but also 
can provide a quantitative estimate of the interaction strengths represented in 
the matrix $\hat{\mathcal{J}}$. In general the number of experimental 
realizations required is $M=N$ to ensure a unique solution of the 
matrix $\hat{\mathcal{J}}$ to represent the network structure. 
However, most real networks are sparse 
in the sense that the degree of a node is usually much less than the 
network size, i.e., $K\ll N$. To obtain the matrix $\hat{\mathcal{J}}$ 
with the least number of links is effectively a constraint for reducing 
the required data amount through some optimization algorithms.
Specifically, constraint~(\ref{eq:CH4_DJTheta}) can be used to 
parameterize the family of admissible matrices through $(N-M)N$ parameters, 
$P_{ij}$, in terms of a singular value decomposition of 
$\mathcal{\theta}^{\rm T}=\mathcal{U}$ in a standard way, where the 
singular matrix $\mathcal{S}$ of dimension $M\times N$ contains the 
singular values on the diagonal. The coupling matrices can be reformulated 
to be $\hat{\mathcal{J}} = \mathcal{D}\cdot \mathcal{U} \cdot 
\tilde{\mathcal{S}} \cdot \mathcal{V}^{\rm T} + \mathcal{P} \cdot \mathcal{V}$,
where $P_{ij}$ is set to be 
zero for $j\leq M$, $\tilde{S}_{ij}=\delta_{ij}/\sigma_i$ if 
$\sigma_i>10^{-4}$ and $\tilde{S}_{ii}=0$ if $\sigma_i \leq 10^{-4}$. Finally, 
with respect to the parameter matrix $\mathcal{P}$, 
the incoming links of any node $i$ can be 
inferred by minimizing the 1-norm of the $i$th row vector of $\hat{J}$ as
\begin{equation}
\|\hat{J}_i \|_1 := \sum_{j=1;j\neq i}^{N} |\hat{j}_{ij}|.
\end{equation}
Numerical tests~\cite{TimC:2014} showed that the number $M$ of the 
required experimental realizations can be substantially smaller than $N$ to 
yield reasonable reconstruction results.

A detailed description of the response dynamics based approach to network
reconstruction can be found in Ref.~\cite{TimC:2014}. In general, the 
reconstruction method is applicable to networked systems whose behaviors
are dominated by the linearized dynamics about some stable state. An issue
concerns how the reconstruction method can be extended to networked systems 
characterized by more than one variable, i.e., systems beyond phase coupled
oscillators. Another issue is whether it would be possible to reconstruct 
a network from time series without any perturbation to the nodal dynamics. 
Addressing these issues is of both theoretical and practical importance. 

\subsection{Reconstructing complex networks via system clone} 
\label{subsec:CH4_Clone}

Exploiting synchronization between a driver and a response system through 
feedback control led to a method for reconstructing complex 
networks~\cite{YRK:2006}. The basic idea was to design a replica or a 
clone system that is sufficiently close to the original network without 
requiring knowledge about the network structure. From the clone system, the
connectivities and interactions among the nodes can be obtained directly, 
realizing the goal of network reconstruction.

The conventional method to deal with driver-response systems 
is to design proper feedback control to synchronize the state of the 
response system with that of the driver system. To achieve reconstruction,
the elements in the adjacency matrix characterizing the network topology 
of the clone system are treated as the variables to be synchronized with 
those of the original system. With an appropriate feedback control, the 
adjacency matrix of the clone system can be ``forced'' to converge to the 
unknown adjacency matrix in the original system due to synchronization. 
For this approach to be effective, the local dynamics of 
each node needs to be known. In addition, the local dynamical and 
the coupling functions need to be Lipschitz continuous~\cite{YRK:2006}.

More details of the synchronization approach can be described, as follows.
Consider a networked system described by a set of differential equations
\begin{equation}
\dot{\mathbf{x}}_i= \mathbf{f}(\mathbf{x}_i) + \sum_{j=1}^N a_{ij}\mathbf{h}_j
(\mathbf{x}_j),
\end{equation}
where $\mathbf{x}_i$ denotes the state of node $i$ $(i=1,\cdots, N)$, $a_{ij}$ 
is the $ij$th element of the adjacency matrix $\mathcal{A}$, 
$\mathbf{f}(\mathbf{x}_i)$ represents the local dynamics of node $i$, and 
$\mathbf{h}_j$ is the coupling function. In order to realize stable 
synchronization, the functions $\mathbf{f}_i$ and $\mathbf{h}_i$ are 
required to be Lipschitzian for all nodes. To design a clone system requires
that the functions $\mathbf{f}_i$ and $\mathbf{h}_i$ be known {\em a priori}. 
The clone system under feedback control can be written as 
\begin{eqnarray}
\dot{\mathbf{y}} & = & \mathbf{f}_i(\mathbf{y}_i)+
\sum_{j=1}^N b_{ij}\mathbf{h}_j(\mathbf{y}_j) + \mathbf{q}_i + \mathbf{u}_i, 
\nonumber \\
\dot{b}_{ij} & = & -\gamma_{ij} \mathbf{h}_j(\mathbf{y}_j)\cdot \mathbf{e}_i,
\end{eqnarray}
where $\gamma_{ij}$ are positive coefficients used to control the evolution 
of $b_{ij}$ so as to ``copy'' the dynamics of the original system, 
$\mathbf{q}_i$ represents the modeling errors, and 
$\mathbf{e}_i \equiv \mathbf{y}_i - \mathbf{x}_i$ denotes the difference 
between the states of the clone and the original systems. To ensure that the 
clone system can approach the original system, the following Lyapunov 
function can be exploited:
\begin{equation}
\dot{\Omega}=\sum_{i=1}^N e_i^2 +\sum_{i=1}^N \sum_{j=1}^N 
(1/\gamma_{ij})(b_{ij}-a_{ij})^2.
\end{equation}
The Lipschitzian and stability constraints give rise to a negative Lyapunov 
exponent and the feedback control as
\begin{equation} \label{eq:Ch4_feedback}
\mathbf{u}_i = -k_1 \mathbf{e}_i-\frac{1}{4\varepsilon_1}\delta^2\mathbf{e}_i,
\end{equation}
where $k_1>L_1 + NL_2$, $L_1$ and $L_2$ are the positive constants in the 
Lipschitzian constraint. It can be proven~\cite{YRK:2006} that feedback 
control in the form of Eq.~\eqref{eq:Ch4_feedback}  
in combination with the Lipschitzian constraint can guarantee that 
the Lyapunov exponent is zero or negative so that the clone 
system converges to the original system with small errors. Numerical tests
in \cite{YRK:2006} demonstrated the working of the synchronization-based 
reconstruction method. 

In an alternative approach~\cite{ST:2011}, it was argued that if the local 
dynamics of the nodes are available, a direct solution of the network 
topology is possible without the need of any clone system. The basic idea 
came from the fact that, if the local dynamics and the coupling functions 
are available, then the only unknown parameters are the coupling strengths 
associated with the adjacency matrix. With data collected at different times, 
a set of equations for the coupling strengths can be obtained, which can be
solved by using the standard Euclidean $L_2$-norm minimization. Numerical 
tests showed that this reconstruction method is effective for both transient 
and attracting dynamics~\cite{ST:2011}.

\subsection{Network reconstruction via phase-space linearization}
\label{subsec:CH4_PS_Linearization}

For nonlinear dynamical networks, there was a method~\cite{NS:2008}
based on chaotic time-series analysis, where time series were assumed
to be available from some or all nodes of the network. The nodal
dynamics were assumed to be described by autonomous systems with
a few coupling terms. That is, the network is sparse. The delay-coordinate
embedding method~\cite{Takens:1981,SYC:1991,PMNC:2007} was then used to
reconstruct the phase space of the underlying networked dynamical system.
The principal idea was to estimate the Jacobian matrix of the underlying
dynamics, which governs the evolutions of infinitesimal vectors in the
tangent space along a typical trajectory of the system. Mathematically,
the entries of the Jacobian matrix are mutual partial derivatives of the
dynamical variables on different nodes in the network. A statistically
significant entry in the matrix implies a connection between the two
nodes specified by the row and the column indices of that entry. Because
of the mathematical nature of the Jacobian matrix, i.e., it is meaningful
only for infinitesimal tangent vectors, linearization of the dynamics
in the neighborhoods of the reconstructed phase-space points is needed,
for which constrained optimization techniques~\cite{CDS:1998,CRT:2006a,
CRT:2006b,Candes:2006,Donoho:2006,Baraniuk:2007,CW:2008} were found to be
effective~\cite{NS:2008}. 

The approach~\cite{NS:2008} was based on using 
$L_1$-minimization in the phase space of a networked system to reconstruct 
the topology without knowledge of the self-dynamics of the nodes and without 
using any external perturbation to the state of nodes. In particular, one 
data point $x_*^t$ is chosen from the phase space at time $t$ and from 
the neighborhood of $N$ nearest data points from the time series. For 
node $i$ (the $i$th coordinate), the dynamics at $t$ around $x_*^t$ can 
be expressed by using a Taylor series expansion, yielding
\begin{equation}\label{eq:Taylor}
\mathbf{x}_i^{t+1}=\mathbf{F}_i(\mathbf{x}_*^t) + 
D\mathbf{F}_i(\mathbf{x}_*^t)\cdot (\mathbf{x}_i^t - \mathbf{x}_*^t) 
+ O(\|\mathbf{x}_i^t - \mathbf{x}_*^t \|^2),
\end{equation}
where the higher-order terms can be omitted when implementing the 
reconstruction. A key parameter affecting the reconstruction accuracy is
size of the neighborhood in the phase space. If the size is relatively large, 
more data points will be available but the equations will have larger 
errors. If the size is small, the amount of available data will be 
reduced. There is thus a trade-off for choosing the neighborhood size. 
Selecting a different time $t$, a set of equations in the form of 
Eq.~(\ref{eq:Taylor}) can be established for reconstruction. The direct 
neighbors of node $i$ (the $i$th column of the adjacency matrix $\mathcal{A}$) 
can be recovered by estimating the coefficients for $\mathbf{F}_i$ in the 
local neighborhoods in the phase space. The coefficients can be solved 
from the set of equations by using a standard $L_1$-minimization procedure. 
After the coefficients have been estimated, it is necessary to set a 
threshold to discern the neighbors of node $i$, where true and false 
positive rates can be used to quantify the reconstruction performance. 
Discrete dynamical systems were used to demonstrate the working of the
reconstruction method~\cite{NS:2008}.

\subsection{Reconstruction of oscillator networks based on noise induced 
dynamical correlation} \label{subsec:CH4_Noise}

The effect of noise on the dynamics of nonlinear and complex system has 
been intensively investigated in the field of nonlinear science. 
For example, the interplay between nonlinearity and stochasticity
can lead to interesting phenomenon such as stochastic 
resonance~\cite{BSV:1981,BPSV:1983,MW:1989,MPO:1994,GNCM:1997,GHJM:1998},
where a suitable amount of noise can counter-intuitively optimize the 
characteristics of the system output such as the signal-to-noise ratio. 
Previous studies also established the remarkable phenomenon of 
noise-induced frequency~\cite{SH:1989} or coherence 
resonance~\cite{PK:1997,LL:2001a,LL:2001b}. Specifically, when a nonlinear
oscillator is under stochastic driving, a dominant Fourier frequency in its
oscillations can emerge, resulting in a signal that can be much more temporarily
regular than that without noise~\cite{SH:1989,PK:1997,LL:2001a,LL:2001b}.
A closely related phenomenon is noise-induced collective oscillation
or stochastic resonance in the absence of an external periodic driving
in excitable dynamical systems~\cite{HDNH:1993}. In complex networks, 
there was a study of the effect of noise on the fluctuation of nodal 
states about the synchronization manifold~\cite{ZK:2006}, with a number
of scaling properties uncovered. 

The rich interplay between nonlinear dynamics and stochastic fluctuations 
pointed at the possibility that noise may be exploited for network 
reconstruction, leading to the development of a number of 
methods~\cite{WCHLH:2009,RWLL:2010,WRLL:2012}. In general, these methods
do not assume any {\em a priori} knowledge about the nodal dynamics and
there is no need to impose any external perturbation. Under the condition
that the influence of noise on the evolution of infinitesimal tangent 
vectors in the phase space of the underlying networked dynamical system
is dominant, it can be argued~\cite{RWLL:2010,WRLL:2012} that the dynamical 
correlation matrix that can be computed readily from the available nodal 
time series approximates the network adjacency matrix, fully unveiling 
the network topology. 

In a general sense, noise is beneficial for network reconstruction. Suppose 
that all nodes in an oscillator network are in a synchronous state. Without
external perturbation, the coupled oscillators behave as a single oscillator
so that the effective interactions among the oscillators vanish, rendering
impossible to extract the interaction pattern from measurements. However, 
noise can induce desynchronization so that the time series would contain 
information about the underlying interaction patterns. 

It was demonstrated~\cite{RWLL:2010,WRLL:2012} that noise can bridge dynamics 
and the network topology in that nodal interactions can be inferred from
the noise-induced correlations. Consider $N$ {\em nonidentical}
oscillators, each of which satisfies $\dot{\mathbf{x}}_i =
\mathbf{F}_i(\mathbf{x}_i)$ in the absence of coupling, where
$\mathbf{x}_i$ denotes the $d$-dimensional state variable of the
$i$th oscillator. Under noise, the dynamics of the whole
coupled-oscillator system can be expressed as:
\begin{equation} \label{eq:master}
\dot{\mathbf{x}}_i = \mathbf{F}_i(\mathbf{x}_i) - c \sum^N_{j=1}
L_{ij} \mathbf{H}(\mathbf{x}_j) + \eta_i,
\end{equation}
where $c$ is the coupling strength, $\mathbf{H}: \mathbb{R}^d
\rightarrow \mathbb{R}^d$ denotes the coupling function of
oscillators, $\eta_i$ is the noise term, $L_{ij} = -1$ if $j$
connects to $i$ (otherwise $0$) for $i \neq j$ and $L_{ii} =
-\sum_{j=1,j\neq i}^{N}L_{ij}$. Due to nonidentical oscillators
and noise, an invariant synchronization manifold does not exist.
Let $\mathbf{\bar{x}}_i$ be the counterpart of $\mathbf{x}_i$ in
the absence of noise, and assume a small perturbation $\xi_i$, we
can write $\mathbf{x}_i = \mathbf{\bar{x}}_i + \xi_i$.
Substituting this into Eq.~(\ref{eq:master}), we obtain:
\begin{equation} \label{eq:CH4_linearize}
\dot{\xi} = [ D\mathcal{\hat{F}}(\mathbf{\bar{x}}) - 
c\mathcal{\hat{L}} \otimes D\mathcal{\hat{H}}(\mathbf{\bar{x}}) ]
\cdot \xi + \eta,
\end{equation}
where $\xi = [\xi_1,\xi_2, \ldots , \xi_N ]^T$ denotes the
deviation vector, $\eta = [\eta_1,\eta_2,\ldots, \eta_N]^T$ is the
noise vector, $\mathcal{\hat{L}}$ names the Laplacian matrix of
elements $L_{ij}$ ($i,j=1,\ldots,N$), 
\begin{displaymath}
D\mathcal{\hat{F}}(\mathbf{\bar{x}}) = \mathrm{diag}
[D\mathcal{\hat{F}}_1(\mathbf{\bar{x}}_1),
D\mathcal{\hat{F}}_2(\mathbf{\bar{x}}_2),\cdots, 
D\mathcal{\hat{F}}_N(\mathbf{\bar{x}}_N)] 
\end{displaymath}
and $D\mathcal{\hat{F}}_i$ are $d\times d$ Jacobian matrices of 
$\mathbf{F}_i$, $\otimes$ denotes direct product, and 
$D\mathcal{\hat{H}}$ is the Jacobian matrix of the coupling function 
$\mathbf{H}$.

Let $\mathcal{\hat{C}}$ denote the dynamical correlation matrix of oscillators 
$\langle \xi \xi^T \rangle$, wherein $C_{ij} = \langle\xi_i \xi_j \rangle$ 
and $\langle \cdot \rangle$ is the time average. One obtains
\begin{equation} \label{eq:CH4_correlation}
0 =\langle d(\xi \xi^T)/dt\rangle = -\mathcal{\hat{A}} \cdot \mathcal{\hat{C}} 
- \mathcal{\hat{C}} \cdot \mathcal{\hat{A}}^T + \langle \eta \xi^T \rangle +
\langle \xi \eta^T \rangle,
\end{equation}
where $\mathcal{\hat{A}} = -D\mathcal{\hat{F}}(\mathbf{\bar{x}}) + c
\mathcal{\hat{L}} \otimes D\mathcal{\hat{H}}(\mathbf{\bar{x}})$.
$\xi(t)$ can be solved from Eq.~(\ref{eq:CH4_linearize}), yielding
the expression of $\langle \eta \xi^T \rangle$ and $\langle
\xi \eta^T \rangle$: 
\begin{displaymath}
\langle \xi \eta^T \rangle = \langle \eta \xi^T \rangle = \mathcal{\hat{D}}/2. 
\end{displaymath}
As a result, Eq.~(\ref{eq:CH4_correlation}) can be simplified to:
\begin{equation} \label{eq:CH4_general}
\mathcal{\hat{A}} \cdot \mathcal{\hat{C}} +
\mathcal{\hat{C}} \cdot \mathcal{\hat{A}}^T =
\mathcal{\hat{D}}.
\end{equation}
The general solution of $\mathcal{\hat{C}}$ can be written as 
\begin{displaymath}
\mbox{vec}(\mathcal{\hat{C}}) =\mbox{vec}(\mathcal{\hat{D}})/(\mathcal{\hat{I}}\otimes \mathcal{\hat{A}} + \mathcal{\hat{A}} \otimes \mathcal{\hat{I}}),
\end{displaymath}
where $\mbox{vec}(\mathcal{\hat{X}})$ is a vector containing all
columns of matrix $\mathcal{\hat{X}}$.

\begin{figure}
\centering
\includegraphics[width=0.8\linewidth]{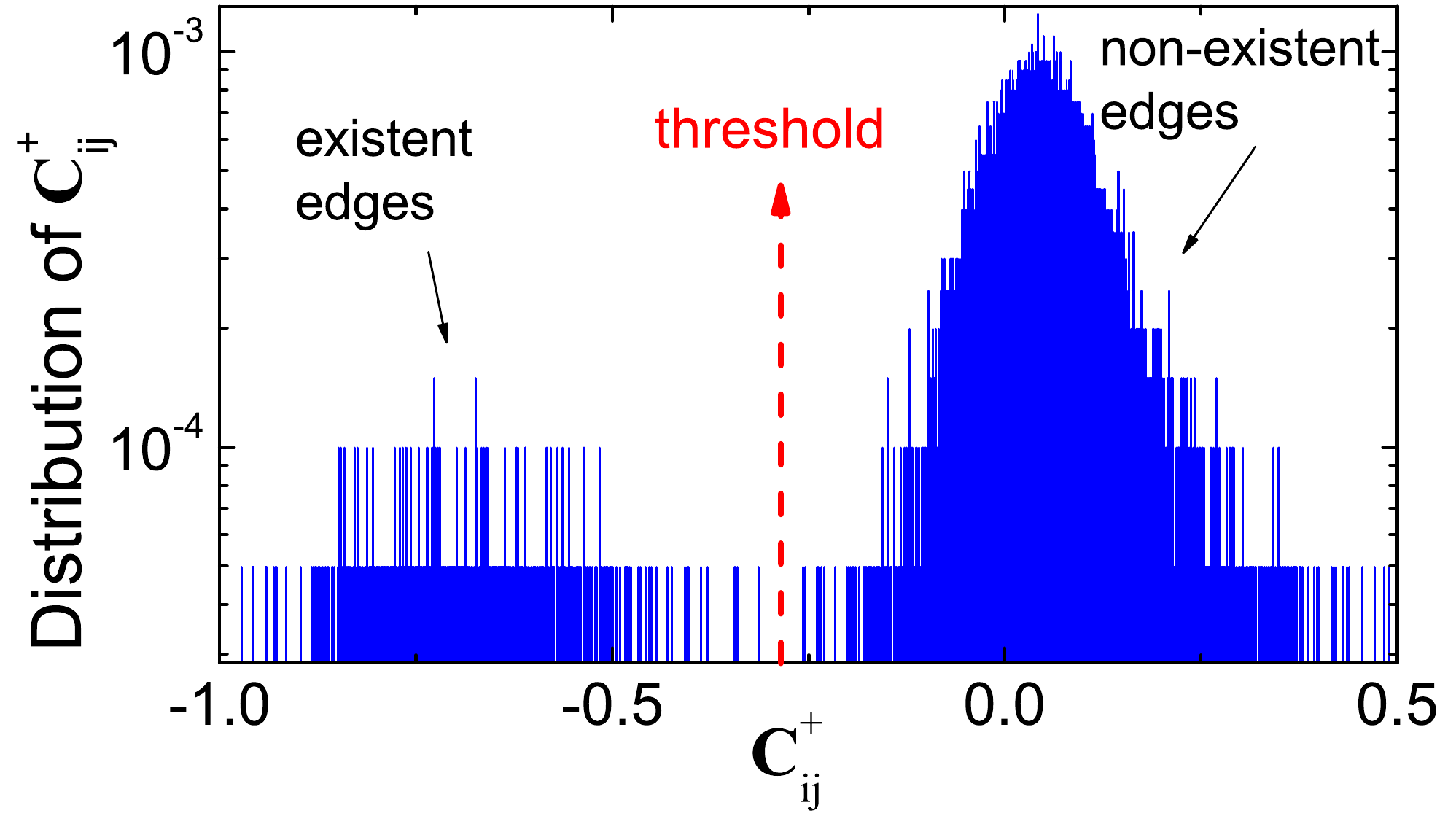}
\caption{\small {\bf Reconstruction of an oscillator network based on 
noise-induced dynamical correlation}. 
Illustration of the distribution of the element values of 
the pseudo inverse matrix of the dynamical correlation matrix 
$\mathcal{\hat{C}}$. The theoretical threshold is marked by red dashed arrow.}
\label{fig:CH4_distribution}
\end{figure}

For one-dimensional state variable and linear coupling, with Gaussian white 
noise $\mathcal{\hat{D}}=\sigma^2 \mathcal{\hat{I}}$, and negligible intrinsic 
dynamics $D\mathcal{\hat{F}}$, Eq.~(\ref{eq:CH4_general}) can be simplified
to 
\begin{displaymath}
\mathcal{\hat{L}}\cdot\mathcal{\hat{C}} + 
\mathcal{\hat{C}}\cdot\mathcal{\hat{L}}^T =
\sigma^2\mathcal{\hat{I}}/c. 
\end{displaymath}
For undirected networks, the solution of $\mathcal{\hat{C}}$ becomes
\begin{equation} \label{eq:CH4_undirect}
\mathcal{\hat{C}} = \frac{\sigma^2}{2c}\mathcal{\hat{L}}^\dag,
\end{equation}
where $\mathcal{\hat{L}}^\dag$ denotes 
the pseudo inverse of matrix $\mathcal{L}$. Since 
$\mathcal{\hat{C}}$ can be calculated from time series, and $\sigma$ and $c$ 
are constant, setting a threshold in the element values of matrix 
$\mathcal{\hat{C}}$ for identifying real 
nonzero elements can lead to full reconstruction of all the connections 
(Fig.~\ref{fig:CH4_distribution}). A key issue is how to determine 
the threshold.

\begin{figure*}
\centering
\includegraphics[width=\linewidth]{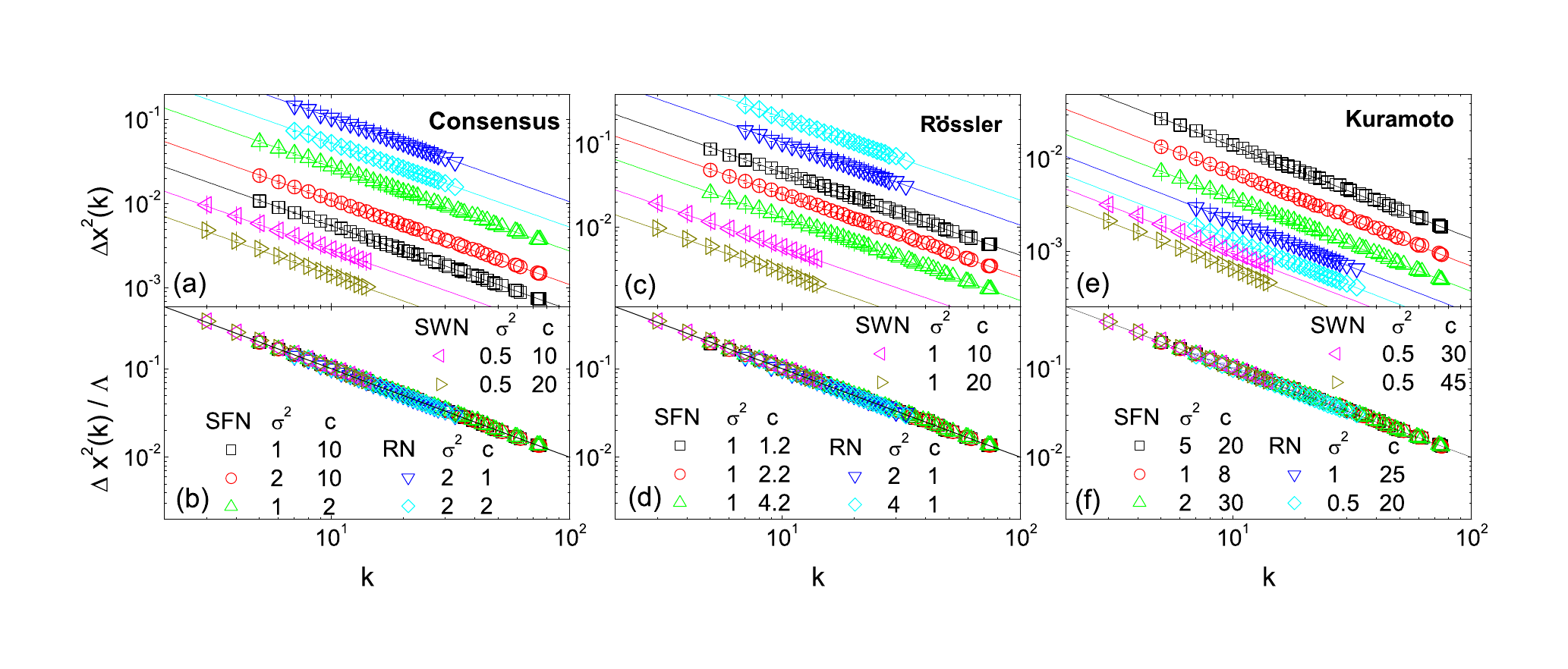}
\caption{\small {\bf Scaling of noise-induced fluctuations with nodal 
degree}. Average fluctuation $\Delta x^2(k)$ as a function of the
nodal degree $k$ for different values of noise variance $\sigma^2$
and coupling strength $c$ for scale-free, random and small-world
networks and three types of nodal dynamical processes 
[(a) consensus dynamics, (c) R\"ossler dynamics, and
(e) Kuramoto dynamics]. (b,d,f) Rescaled quantity $\Delta
x^2(k)/\Lambda$, where $\Lambda = \sigma^2(1+1/\langle k
\rangle)/(2c)$, versus $k$ for consensus, R\"ossler and Kuramoto
dynamics, respectively. Data points are from a single network
configuration and $\Delta x^2(k)$ is obtained by averaging over
all nodes of degree $k$ with error bars. The parallel lines in
(a), (c) and (e) are theoretical predictions, and the lines in (b), (d) and (f)
are the function $1/k$. Network size is 500. For the scale-free
network, the lowest degree is $k_{\min}=5$. For the random
network, the connection probability among nodes is 0.03. For the
small-world network, the average degree is 8 and the rewiring
probability is 0.1. The natural frequency $\omega_i$ in the
Kuramoto model is chosen independently from a prescribed
probability distribution $g(\omega)=3/4(1-\omega^2)$ for
$|\omega|\leq 1$ and $g(\omega) = 0$ otherwise. 
From Ref.~\cite{RWLL:2010} with permission.}
\label{fig:CH4_cii}
\end{figure*}

In Refs.~\cite{RWLL:2010,WRLL:2012}, it was argued that the threshold can 
be set based on the diagonal elements (autocorrelation) in $\mathcal{\hat{C}}$:
\begin{equation}
C_{ii} \simeq
\frac{\sigma^2}{2ck_i}\left(1+ \frac{1}{\langle k\rangle}\right).
\label{eq:CH4_local}
\end{equation}
The formula of $C_{ii}$ obtained from a second-order 
approximation is consistent with the finding of noise-induced 
algebraic scaling law in Ref.~\cite{WCHLH:2009}. 
Specifically, from Eq.~(\ref{eq:CH4_local}), let 
\begin{displaymath}
S \equiv \sum_{i=1}^{N} 1/C_{ii} = 2cl^2/[\sigma^2 (N+l)], 
\end{displaymath}
where $l=\sum_{i=1}^{N}k_i =N\langle k\rangle$ is twice the total number of 
links. The integral part of $l$ can be identified via 
\begin{displaymath}
l=(S\sigma^2 + \sqrt{S^2 \sigma^4 + 8cNS\sigma^2 })/4c. 
\end{displaymath}
The threshold $C_M^{\dag}$ (or $[\sigma^2/(2c)]C_M^{\dag}$) is chosen such that
$\sum^M_{m=1} \Phi(C_m^{\dag}) = l$, where $\Phi(C_m^{\dag})$ is the 
unnormalized distribution of $C_m^{\dag}$. Then the connection
matrix $\mathcal{L}$ can be obtained (see Fig.~\ref{fig:CH4_distribution}).
Figure~\ref{fig:CH4_cii} shows the scaling property of autocorrelation versus 
the node degree $k$ for different nodal dynamics and different types of 
networks, as predicted by Eq.~(\ref{eq:CH4_local}). The numerical results are 
in good agreement with the theoretical prediction.

The reconstruction method by virtue of noise was improved~\cite{CLL:2013}, 
based on reconstruction formula (\ref{eq:CH4_undirect}). It was argued that, 
to use the formula, the coupling strength $c$ and the noise variance 
$\sigma$ should be known {\em a priori}, a condition that may be difficult to
meet in realistic situations. A simple method to eliminate both $c$ and 
$\sigma$ without requiring any {\em a priori} parameters was then 
introduced~\cite{CLL:2013}. Specifically, note that
\begin{eqnarray}\label{eq:CH4_eliminate}
\frac{\sigma^2}{2c}C_{ii}^+ &=& k_i, \nonumber \\
\frac{\sigma^2}{2c}C_{ij}^+ &=& -A_{ij}, \hbox{   } i\neq j,
\end{eqnarray}
where $\sigma$ and $c$ can be simultaneously eliminated by the ratio 
$r_{ij}$ of the diagonal and off-diagonal elements of $\mathcal{\hat{C}}^+$ 
in Eq.~(\ref{eq:CH4_eliminate}):
\begin{equation}
r_{ij}=\left\{
         \begin{array}{ll}
           0, & \hbox{$i$ and $j$ are disconnected}, \\
           -\frac{1}{k_i}, & \hbox{$i$ and $j$ are connected}.
         \end{array}
       \right.
\end{equation}
A threshold is still necessary to fully separate links and zero elements 
in the adjacency matrix. A heuristic method was developed~\cite{CLL:2013} 
for determining the threshold. While based on the same principle, the 
improved reconstruction method appears indeed more practical.

An alternative method to reconstruct the network topology of a dynamical 
system contaminated by white noise was developed~\cite{ZZNMWH:2015} through 
measurement of both the nodal state ${\bf x}$ and its velocity $\dot{{\bf x}}$,
with the following formula: 
\begin{equation}
\mathcal{\hat{A}}= \mathcal{\hat{B}} \cdot \mathcal{\hat{C}}^{-1},
\end{equation}
where $\mathcal{\hat{C}} = \langle \mathbf{x} \mathbf{x}^{\rm T} \rangle$ is 
the state-state correlation matrix, and 
$\mathcal{\hat{B}} = \langle \dot{{\bf x}} {\bf x}^{\rm T} \rangle$ is the 
velocity-state correlation matrix. An interesting feature is that, by 
measuring the velocity, the white noise variance is not required 
to be known and the formula applies (in principle) to any strength of noise. 
However, measuring the instantaneous velocity of the state variables can 
be difficult and the errors would affect the reconstruction performance
dramatically. There is in fact a trade-off between the generality 
of the reconstruction method and the measurement accuracy. Nevertheless, 
the work~\cite{ZZNMWH:2015} is theoretically valuable, and it was validated 
by using a linear system and a nonlinear cell cycle dynamics model. In 
the nonlinear model, there are both active and 
inhibitive links, corresponding to positive 
and negative elements in the adjacency matrix. It was demonstrated that both 
classes of links can be successfully reconstructed~\cite{ZZNMWH:2015}.

\subsection{Reverse engineering of complex systems} \label{subsec:CH4_RE}

\paragraph*{Automated reverse engineering of nonlinear dynamical systems.}
Reconstructing nonlinear biological and chemical systems is of great 
importance. An automated reverse engineering approach was developed 
to solve the problem by using partitioning, automated probing and 
snipping~\cite{BongardLipson:2007}. Firstly, partitioning allows the 
algorithm to characterize the variables in a nonlinear system separately by 
decoupling the interdependent variables. It was argued that Bayesian networks 
cannot model mutual dependencies among the variables, a common pattern 
in biological and other regulation networks with feedback. It was also 
articulated that the partitioning procedure is able to reveal the underlying 
structure of the system to a higher degree as compared with alternative 
methods. Secondly, instead of passively accepting data to model a system, 
automated probing uses automatically synthesized models to create a test 
criterion to eliminate unsuitable models. Because the test cannot be 
analytically derived, candidate tests are optimized to maximize the 
agreement between data and model predictions. Thirdly, snipping 
automatically simplifies and optimizes models. For all models obtained by 
automated probing, one calculates their prediction errors against system 
data, perturbs the existent models randomly, evaluate the newly created 
models, and compare the performance of the modified models with that 
of the corresponding original models. In this way, models are constantly 
evolving, with those with better performance replacing the inferior ones. This 
process tends to yield more accurate and simpler models. The process is 
akin to a genetic algorithm, but in the former, models evolve and are 
constantly improved, whereas in the latter, some adaptive functions evolve.

Four synthetic systems and two physical systems were 
used~\cite{BongardLipson:2007} to test the automated reverse engineering 
approach. The method was demonstrated to be robust against noise and  
scalable to interdependent and nonlinear systems with many variables. The
limitations of the method were discussed as well. Especially, the method is 
restricted to systems in which all variables are observable. In addition, 
discrete time series as input to the method can lead to inconsistence with
the synthesized models. Without including any fuzzy effect, the process 
may not yield explicit models. Potential solutions to these problems were 
suggested~\cite{BongardLipson:2007}. 

\paragraph*{Constructing minimal models for complex system dynamics.}
In Ref.~\cite{BLB:2015}, the authors introduced a method of reverse 
engineering to recover a class of complex networked system described by 
ordinary differential equations of the form
\begin{equation}\label{eq:CH4_mini_Main}
\frac{dx_i}{dt}= M_0(x_i(t))+ \sum_{j=1}^{N} A_{ij} M_1(x_i(t))M_2(x_j(t)),
\end{equation}
where the adjacency matrix $\mathcal{A}$ defines the interacting components,
$M_0(x_i(t))$ describes the self-dynamics of each node, 
$M_1(x_i(t))M_2(x_j(t))$ characterizes the interaction between nodes $i$ 
and $j$. Equation~\eqref{eq:CH4_mini_Main} was argued to be suitable for 
many social, technological and social systems.

The system dynamics is uniquely characterized by three independent 
functions and the aim is to construct the model in the form 
\begin{equation}
{\bf m}=(M_0(x),M_1(x), M_2(x)),
\end{equation}
corresponding to a point in the model space $\mathbb{M}$. The goal was 
to develop a general method to infer the subspace $\mathbb{M}(\mathcal{X})$ 
by relying on minimal {\em a priori} knowledge of the structure of $M_0(x)$, 
$M_1(x)$, and $M_2(x)$. The key lies in using the system's response to 
external perturbations, a common technique used in biological experiments 
such as genetic perturbation, which is feasible for technological and 
social networked systems as well. The link between the observed system 
response and the leading terms of ${\bf m}$ was established, enabling 
the formulation of $\mathcal{X}$ into a dynamical equation. Contrary to 
traditional reverse engineering, the approach~\cite{BLB:2015} gives 
the boundaries of $\mathbb{M}(\mathcal{X})$ rather than a specific model 
${\bf m}$.

To infer ${\bf m}$, the three functions were expressed~\cite{BLB:2015} in 
terms of a Hahn series~\cite{SSigmund:1995}
\begin{eqnarray} \label{eq:CH4_Hahn}
M_0(x) & = & \sum_{n=0}^{\infty} A_n(x_0 - x)^{\Pi_0(n)} \\
M_1(x) & = & \sum_{n=0}^{\infty} B_n(x_0 - x)^{\Pi_1(n)} \nonumber \\
M_2(x) & = & \sum_{n=0}^{\infty} C_n(x_0 - x)^{\Pi_2(n)}. \nonumber
\end{eqnarray}
The challenge was to discern the coefficients $A_n$, $B_n$ and $C_n$ for 
uncovering the functional form~(\ref{eq:CH4_Hahn}), which can be overcome 
by exploiting the system's response to external perturbation.

\begin{figure}
\centering
\includegraphics[width=\linewidth]{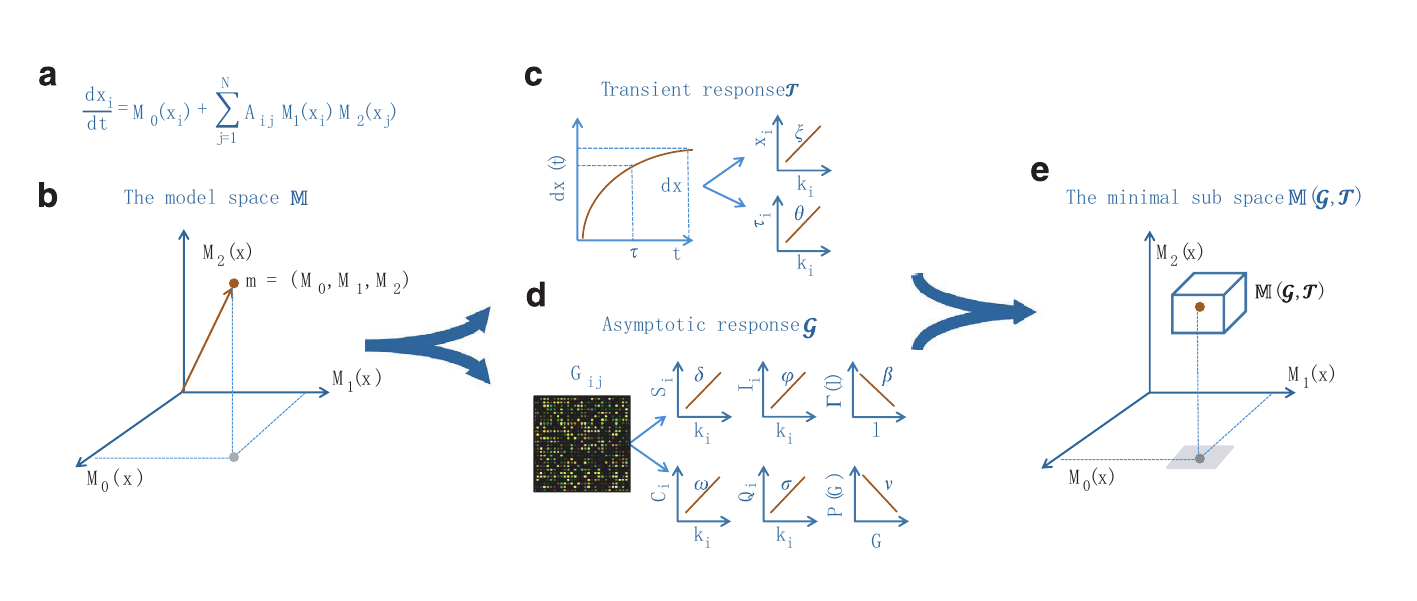}
\caption{\small {\bf Constructing a minimal model for complex dynamics}. 
(a) The unknown nonlinear equation of a dynamical system, which 
is to be reconstructed from empirical observations. (b) The dynamic model 
${\bf m}$ is composed of three functions: $M_0(x)$, $M_1(x)$ and $M_2(x)$, 
corresponding to a point in the model space, $\mathbb{M}$. (c,d) The 
system's response to perturbations is used to extract a set of observable 
functions directly linked to ${\bf m}$. (c) The transient response is 
characterized by the steady-state $\xi_i$ and the relaxation time $\tau_i$, 
from which the exponents $x$ and $y$ can be extracted. (d) The asymptotic 
response for measuring six additional functions, providing $\delta$, 
$\varphi$, $\beta$, $\omega$, $\sigma$ and $\nu$. (e) The subspace in the model 
space. From Ref.~\cite{BLB:2015} with permission.}
\label{fig:CH4_mini_space}
\end{figure}

Under perturbation, the temporal dynamics is characterized 
by the time-dependent relaxation from the original steady state, $x_i$, 
to the perturbed steady-state $x_i(t \rightarrow \infty) =x_i + dx_i$. 
System~(\ref{eq:CH4_mini_Main}) can then be linearized about the steady 
state. After relaxation, the system's new state is characterized by a 
response matrix. Based on the permanent perturbation, ${\bf m}$ can be 
estimated via
\begin{eqnarray}
M_0(x) &\sim& \frac{1}{(R(x))^2} \int \left[(R(x))^{2+\theta} 
+ \mathcal{O}((R(x))^{\mathbf{\phi}+(2+\theta)}) \right]dx \nonumber \\
M_1(x) &\sim& M_0(x)R(x) \nonumber \\
M_2(x) &\sim& \left\{
            \begin{array}{ll}
              (R(x))^{\delta+1-\varphi} + 
\mathcal{O}( (R(x)^{\mathbf{\phi} + (\delta +1 -\varphi)})), & \beta =0 \\
              y_0-(R(x))^\beta + 
\mathcal{O}((R(x))^{\mathbf{\phi} + \beta} ), & \beta >0
            \end{array}
          \right.
\end{eqnarray}
where
\begin{eqnarray}
R(x)\sim\left\{
          \begin{array}{ll}
            x^{-\frac{1}{\xi}} + 
\mathcal{O}(x^{{\bf \phi}(-\frac{1}{\xi})}) , & \delta=0 \\
            (x_0-x)^{\frac{1}{\delta}} + 
\mathcal{O}((x_0-x)^{{\bf \phi}(\frac{1}{\delta})}), & \delta >0
          \end{array}
        \right.
\end{eqnarray}
and $x_0$ and $y_0$ are arbitrary constants.
A set of parameters need to be determined, which exhibit certain scaling 
properties. Specifically, the relaxation time scales with $\theta$, i.e., 
$\tau_i \sim k_i^\theta$, the steady-state activity either scales with 
$\xi$ as $x_i \sim k_i^\xi$ or has the relation $x_i \sim k_i^\xi$, 
node $i$'s impact on neighbors obeys $I_i \sim k_i^{\varphi}$; and
$i$'s stability against perturbation in its vicinity leads to 
$S_i \sim k_i^{\delta}$ (see Fig.~\ref{fig:CH4_mini_space}).

The determination of the parameters gives only the leading terms of 
$M_0(x)$, $M_1(x)$ and $M_2(x)$. Additional terms can be used 
to better model complex systems. The reconstructed model is called 
a minimal model because it captures the essential features of the underlying 
mechanism without any additional constraint. The subspace given by the 
reconstruction method is robust to parameter selection. Models subject to 
the subspace can produce consistent data with observations~\cite{BLB:2015}.

\clearpage

\section{Inference approaches to reconstruction of biological networks}
\label{sec:CH5_Bio_Net}

High-throughput technologies such as microarrays and RNA sequencing produce 
a large amount of experimental data, making genome-scale inference of 
transcriptional gene regulation possible. The reconstruction of biological 
interactions among genes is of paramount importance to biological sciences
at different levels. A variety of approaches aiming to reconstruct gene 
co-expression networks or regulation networks have been developed.

In general, there are three types of experimental gene data: gene 
co-expression data, gene knockout data, and transcriptional factors.
Gene interaction networks can be classified into two categories: 
(1) co-expression networks, in which the nodes represent genes and the 
edges represent the degree of similarity in the expression profiles of 
the genes, and (2) transcription-regulatory networks, in which the nodes 
represent either transcription factors or target genes, and edges 
characterize the causal regulatory relationships. 

Reconstructing neuronal networks and brain functional networks from 
observable data has also been an active area of research. Typical 
experimental data include blood oxygen level dependent (BOLD) signals, 
electroencephalography (EEG) and stereoelectroencephalography (SEEG) data, 
magnetoencephalography (MEG) data, spike data, calcium image data, etc. 
To uncover the interaction structure among the neurons or distinct brain 
domains from the signals, a number of reconstruction approaches have been 
developed and utilized in neuroscience.

\subsection{Correlation based methods} \label{subsec:CH5_correlation}

Correlation based methods are widely used for inferring the associations
between two variables.

\subsubsection{Value based methods}

{\em Pearson's correlation coefficient} measuring the strength of the 
linear relationship between two random variables is widely used in many 
fields~\cite{Pearson_1895}. The correlation is defined as
\begin{equation}
{\rm corr}=\frac{\sum^n_{i=1}(x_i - \bar{x})(y_i-\bar{y})}{(n-1)S_xS_y},
\end{equation}
where $\bar{x}$ and $\bar{y}$ are the sample means and $S_x$ and $S_y$ 
are the standard deviations of $x$ and $y$, respectively. The Pearson's 
correlation assumes that data is normally distributed and is sensitive 
to outliers.

{\em Distance Covariance} (dCov) provides a nonparametric test to examine 
the statistical dependence of two variables~\cite{szekely2009brownian}. For 
some given pairs of measurement $(x_i, y_i), i=1, 2, \dots, n$ for 
variables $X$ and $Y$, let $\mathcal{A}$ denote the pairwise 
Euclidean distance matrix of $X$ with $a_{ij}=|x_i - x_j|$ and $\mathcal{B}$ 
be the corresponding matrix for $Y$, 
where $|.|$ denote the Euclidean norm. Define the doubly centered distance 
matrix $\mathcal{A}^{\rm c}$ whose elements are given by
$a^c_{ij} = a_{ij}-\bar{a}_{i.} - \bar{a}_{.j}+\bar{a}_{..}$, 
where $\bar{a}_{i.}$ is $i^{th}$ row mean, $\bar{a}_{.j}$ is the $j^{th}$ 
column mean, and $\bar{a}_{..}$ is the grand mean of $\mathcal{A}$. 
Centered distance matrix $\mathcal{B}^{\rm c}$ 
can be defined similarly. The squared sample distance 
covariance is defined to be the arithmetic average of the products of 
$\mathcal{A}^{\rm c}$ and $\mathcal{B}^{\rm c}$:
\begin{equation}
{\rm dcov}^2(X,Y)=\frac{1}{n^2}\sum_{ij}a^c_{ij}b^c_{ij}.
\end{equation}

The {\em Theil-Sen estimator}, proposed in 
Refs.~\cite{wilcox1998note,peng2008consistency}, is defined as
\begin{equation}
\hat{\beta}_1 = {\rm median} \{m_{ij}=\frac{y_i- y_j}{x_i - x_j}: x_i \ne x_j, 1\le i \le j \le n\},
\end{equation}
where the median can characterize the relationship between the two variables.
This estimate is robust, unbiased, and less sensitive to outliers. 

{\em Partial correlation and information theory} 
(PCIT)~\cite{Fuente_2004,Reverter_2008} extracts all possible interaction 
triangles and applies Data Processing Inequality (DPI) to filter indirect 
interactions using partial correlation. The partial correlation coefficient 
$c^{\rm p}_{ij}$ between two genes $i$ and $j$  within an interaction 
triangle $(i,j, k)$ is defined as
\begin{equation}
c^p_{ij}=\frac{{\rm corr}(x_i, x_j)-{\rm corr}(x_i, x_k){\rm corr}(x_j, x_k)}
{\sqrt{(1-{\rm corr}(x_i, x_k))^2(1-{\rm corr}(x_j, x_k))^2}},
\end{equation}
where ${\rm corr}(.,.)$ is Pearson's correlation coefficient.

\subsubsection{Rank based methods}

Compared with value based correlation measures, rank based correlation 
measures are more robust and insensitive to outliers. The 
{\em Spearman and Kendall measures} are the most commonly used. In particular, 
Spearman's correlation is simply the Pearson's correlation coefficient 
incorporated with ranked expression~\cite{nie2011tf}, and Kendall's $\tau$ 
coefficient~\cite{kendall1938new} is defined as
\begin{equation}
\tau(x_i, x_j) = \frac{{\rm con}(x^r_i, x^r_j)-{\rm dis}(x^r_i, x^r_j)}
{\frac{1}{2}n(n-1)},
\end{equation}
where $x^r_i$ and $x^r_j$ are the ranked expression profiles of genes $i$ 
and $j$, ${\rm con}(.,.)$ and ${\rm dis}(.,.)$ represent the numbers of 
concordant and disconcordant pairs, respectively.

{\em Inner Composition Alignment} (ICA)~\cite{HKKN:2011} was proposed to 
infer directed networks from short time series by extending the Kendall's 
$\tau$ measure. Given time series $x^l$ and $x^k$ of length $n$ from  
subsystems $l$ and $k$ over the same time intervals, let $\pi^l $ be the 
permutation that arranges $x^l$ in a nondecreasing order. The series 
$g^{k,l}=x^k(\pi^l)$ is the reordering of the time series $x^k$ with 
respect to $\pi^k$. The ICA is formulated as
\begin{equation}
\tau^{l \rightarrow k}= 1 - \frac{\sum^{n-2}_{i=1}\sum^{n-1}_{j=i+1}w_{ij}
\Theta[(g^{k,l}_{j+1}-g^{k,l}_i)(g^{k,l}_i - g^{k,l}_j)]}
{\frac{1}{2}(n-1)(n-2)},
\end{equation}
where $w_{ij}$ denotes the weight between points $i$ and $j$, and $\Theta[x]$ 
is the Heaviside step function. Compared with Kendall's $\tau$ measure, 
ICA can infer direct interactions and eliminate indirect interaction by using 
the partial version of ICA.

{\em Hoeffding's D coefficient} is a rank-based nonparametric measure of 
association~\cite{fujita2009comparing}. The statistic coefficient $D$ is 
defined as
\begin{equation}
D = \frac{(n-2)(n-3)D_1+D_2-2(n-2)D_3}{n(n-1)(n-2)(n-3)(n-4)},
\end{equation}
where
\begin{eqnarray}
D_1&=&\sum^n_{i=1}Q_i(Q_i-1),\\
D_2&=&\sum^n_{i=1}(R_i-1)(R_i-2)(S_i-1)(S_i-2),\\
D_3&=&\sum^n_{i=1}(R_i-2)(S_i-2)Q_i,
\end{eqnarray}
$R_i$ is the rank of $X_i$, $S_i$ is the rank of $Y_i$, and the bivariate 
rank $Q_i$ is the number of both $X$ and $Y$ values less than the 
$i^{th}$ point.

It was pointed out~\cite{Wang_2014} that real gene interactions may change as 
the intrinsic cellular state varies or may exist only under a specific 
condition. That is, for a long time series of the co-expression data, 
perhaps only part 
of the data are meaningful for revealing interactions. In this regard, two 
new co-expression measures based on the matching patterns of local expression 
ranks were proposed. Specifically, when dealing with time-course data, the 
measure $W_1$ is defined as
\begin{equation}
W_1 = \sum^{n-k+1}_{i=1}I(\phi(x_i,\dots,x_{i+k-1})=\phi(y_i,\dots,y_{i+k-1}))
+I(\phi(x_i,\dots,x_{i+k-1})=\phi(-y_i,\dots,-y_{i+k-1})),
\end{equation}
where $I(.)$ is an indicator function, and $\phi$ is the rank function that 
returns the indices of the elements after they have been sorted in an 
increasing order, and $W_1$ counts the number of continuous subsequences of 
length $k$ with matching and reverse rank patterns. For non-time-series data, 
where the order is not meaningful, a more general measure, $W_2$ can be 
defined:
\begin{equation}
W_2 = \sum_{1\le i_1<\dots < i_k \le n}I(\phi(x_{i_1},\dots,x_{i_k})
=\phi(y_{i_1},\dots,y_{i_k}))+I(\phi(x_{i_1},\dots,x_{i_k})
=\phi(-y_{i_1},\dots,-y_{i_k})).
\end{equation}

\subsection{Causality based measures}

{\it Wiener-Granger Causality} (WGC), pioneered by 
Wiener~\cite{wiener1956theory} and Granger~\cite{Granger:1969}, 
is a commonly used measure to infer the causal influence between 
two variables, e.g., the expression time series of 
genes and the spiking neural time series. The basic idea of WGC is 
straightforward. Given two time series $\{X_t\}$ and $\{Y_t\}$, if using the 
history of both $X$ and $Y$ is more successful to predict $X_{t+1}$ than 
exclusively using the history of $X$, $Y$ is said to be G-cause X. The 
idea of WGC is similar to that of the transfer entropy, but WGC uses some 
correlation measure instead of mutual information. WGC is widely used in 
measuring the functional connectivity among subdomains of brain based on 
EEG, MEG, SEEG data~\cite{bressler2011wiener}. Fundamentally,  
Granger test is a linear method operated on the hypothesis that the underlying 
system can be described as a multivariate stochastic process. Thus, in 
principle, there is no guarantee that the method would be effective for 
nonlinear systems, in spite of efforts to extend the methodology to
strongly coupled systems~\cite{CRFD:2004,AMS:2004,MPS:2008}.

In the traditional Granger framework, measurement noise is generally
detrimental in the sense that, as its amplitude is increased the
value of the detected causal influence measure decreases monotonically,
leading to spurious detection outcomes~\cite{HNG:2014}.
The transfer entropy framework is applicable~\cite{Schreiber:2000} to both
linear and nonlinear systems, but often the required data amount is
prohibitively large. In the special case of Gaussian dynamical variables,
the two methods, one of the autoregressive nature (Granger test) and another
based on information theoretic concepts (transfer entropy, to be discussed
below), are in fact equivalent to each other~\cite{BBS:2009}. An alternative
information theoretic measure, the causation entropy, was also
proposed~\cite{SCB:2014,CSB:2015,STB:2015}.

{\em Convergent cross mapping framework} (CCM) is based on delay coordinate 
embedding, the paradigm of nonlinear time series 
analysis~\cite{Takens:1981,SYC:1991,KS:book,DS:2011}. The CCM method
can deal with both linear and nonlinear systems with small data sets,
and it has been applied to data from different contexts, such as EEG
data~\cite{Mcbride:2014}, FMRI~\cite{WWDN:2014}, fishery
data~\cite{HSNKLLSW:2014}, economic data~\cite{HF:2014}, and cerebral
auto-regulation data~\cite{HALC:2014}. It was also found that properly
applied noise can enhance the CCM performance in inferring causal 
relations~\cite{JHHLL:2016}.

The nonlinear dynamics based CCM method was proposed~\cite{SMYHDFM:2012} to 
detect and quantify causal influence between
a pair of dynamical variables through the corresponding time series.
The starting point is to reconstruct a phase space, for each variable,
based on the delay-coordinate embedding method~\cite{Takens:1981}.
Specifically, for time series $x(t)$, the reconstructed vector is
$X(t)=[x(t),x(t-\tau),...,x(t-(E_x-1)\tau)]$, where $\tau$ is the
delay time and $E_x$ is the embedding dimension. For variable $y$,
a similar vector can be constructed in the $E_y$ dimensional space.
Let $\textbf{M}_\textbf{X}$ and $\textbf{M}_\textbf{Y}$ denote the
attractor manifolds in the $E_x$- and $E_y$-dimensional space, respectively.
If $x$ and $y$ are dynamically coupled, there is a mapping relation between
$\textbf{M}_\textbf{X}$ and $\textbf{M}_\textbf{Y}$. The CCM method
measures how well the local neighborhoods in $\textbf{M}_\textbf{X}$
correspond to those in $\textbf{M}_\textbf{Y}$. In particular, the
cross-mapping estimate of a given $Y(t)$, denoted as
$\hat{Y}(t)|\textbf{M}_\textbf{X}$, is based on a simplex
projection~\cite{SM:1990,Sugihara:1994} that is essentially a
\emph{nearest-neighbor algorithm} involving $E+1$ nearest neighbors of
$X(t)$ in $\textbf{M}_\textbf{X}$. (Note that $E+1$ is the minimum number of
points required for a bounding simplex in the $E$-dimensional space.) The
time indices of the $E+1$ nearest neighbors are denoted as
$t_1, t_2 , ..., t_{E+1}$ in the order of distances to $X(t)$ from the
nearest to the farthest, i.e., point $X(t_1)$ is the nearest-neighboring
point of $X(t)$ in $\textbf{M}_\textbf{X}$. These time indices are used to
identify the points (putative neighborhoods) in $\textbf{M}_\textbf{Y}$,
namely, to find the points at the corresponding instants:
$Y(t_1)$, $Y(t_2)$, $...$, and $Y(t_{E+1})$, which are used to estimate
$\hat{Y}(t)$ through the weighted average
\begin{equation}
    \hat{Y}(t)|\textbf{M}_\textbf{X} = \sum_{i=1}^{E+1} w_i(t)\cdot Y(t_i),\\
\end{equation}
where
\begin{equation}
    w_i(t) = \mu_i(t)/\sum_j \mu_j(t)\\
\end{equation}
is the weight of the vector $Y(t_i)$,
\begin{equation}
    \mu_i(t) =\exp\{-d[X(t),X(t_i)]/d[X(t),X(t_1)]\},\\
\end{equation}
and $d[X(t), X(t_i)]$ is the Euclidean distance between the two vector
points $X(t)$ and $X(t_i)$ in $\textbf{M}_\textbf{X}$. An estimated
time series $\hat{y}(t)$ can then be obtained from
$\hat{Y}(t)|\textbf{M}_\textbf{X}$. Likewise, the cross mapping from $Y$
to $X$ can be defined analogously so that the time series of $x(t)$ can
be predicted from the cross-mapping estimate
$\hat{X}(t)|\textbf{M}_\textbf{Y}$.

The correlation coefficient between the original time series $y(t)$ and
the predicted time series $\hat{y}(t)$ from $\textbf{M}_\textbf{X}$,
denoted as ${\rho}_{Y|\textbf{M}_\textbf{X}}$, is a measure of CCM causal
influence from $y$ to $x$. Larger value of ${\rho}_{Y|\textbf{M}_\textbf{X}}$
implies that $y$ is a stronger cause of $x$, while
${\rho}_{Y|\textbf{M}_\textbf{X}}\leq 0$ indicates that $y$ has
no influence on $x$. The relative strength of causal influence can be defined as
$R={\rho}_{X|\textbf{M}_\textbf{Y}}-{\rho}_{Y|\textbf{M}_\textbf{X}}$,
which is a quantitative measure of the casual relationship between $x$
and $y$. A positive value of $R$ indicates that $x$ is the CCM cause of $y$.

\subsection{Information-theoretic based methods}

\subsubsection{Mutual information}

{\em Mutual Information} (MI)~\cite{mackay2003information} is used widely 
to quantify the pairwise mutual dependency between two variables. For two 
variables $X$ and $Y$, the mutual information is defined in term of
\begin{equation}
{\bf MI}(X;Y)=\sum_{x\in X}\sum_{y\in Y} p(x,y)\log 
\left (\frac{p(x,y)}{p(x)p(y)} \right),
\end{equation}
where $p(x,y)$ is the joint probability distribution functions of $X$ and $Y$, 
and $p(x)$ and $p(y)$ are the marginal probability distribution function of 
$X$ and $Y$, respectively. Note that MI is nonnegative and symmetric: 
${\rm MI}(X;Y)={\rm MI}(Y;X)$. The larger the value of MI, the more highly 
the two variable are correlated. The key to applying the mutual information for 
quantifying associations in continuous data is to estimate the probability 
distribution from the data. A difficult and open issue is how to obtain 
unbiased mutual information value. 

There are four commonly used MI-based network reconstruction methods: 
relevance network approach (RELNET)~\cite{Butte_2000,Butte_2000a}, context 
likelihood of relatedness (CLR)~\cite{Faith_2007}, maximum relevance/minimum 
redundancy feature selection (MRNET)~\cite{Meyer_2007,Meyer_2010}, and the 
algorithm for accurate reconstruction of cellular networks 
(ARACNE)~\cite{Margolin_2006,Zoppoli_2010}. 

In RELNET, a threshold $\tau$ is 
used to distinguish actual links from null connections. However, 
this method has a limitation, i.e., the indirect 
relationship may lead to high mutual information values and failures to 
distinguish direct from indirect relationships.

The CLR method integrates z-score to measure the significance of the 
calculated MI. For node $i$, the mean $\mu_i$ and the standard deviation 
$\delta_i$ of the empirical distribution of the mutual information 
${\rm MI}(x_i; x_k), (k=1,\dots, n)$ are calculated. The z-score 
$z_{ij}$ of ${\rm MI}(x_i;x_j)$ is defined as
\begin{equation}
z_{ij}=\max \left (0, \frac{MI(x_i;x_j)-\mu_i}{\delta_i} \right).
\end{equation}
The value of $z_{ji}$ can be defined analogously. A combined score, which 
quantifies the relatedness between $i$ and $j$, is expressed as 
$\hat{z}_{ij} = \sqrt{z_{ij}^2 + z_{ji}^2}$.

The MRNET method can be used to infer a network by repeating the maximum 
relevance, minimum redundancy (MRMR)~\cite{MRMR1,MRMR} feature selection 
method for all variables. The key to the MRMR algorithm lies in selecting 
a set of variables with a high value of the mutual information with the target 
variable (maximum relevance) but meanwhile the selected variables are 
mutually maximally independent (minimum redundancy). The aim of this 
method is to associate direct interactions with high rank values, but associate 
indirect interactions with low rank values.

The MRMR algorithm is essentially a greedy algorithm. Firstly, the MRMR 
selects the variable $x_i$ with the highest MI value regarding the target $y$. 
Next, given a set of selected variables, a new variable $x_k$ is chosen 
to maximize the value of $s^y_k = u^y_k - r^y_k$, where $u^y_k$ is the 
relevance term and $r^y_k$ is the redundancy term. More precisely, one has
\begin{eqnarray}
u^y_k & = & MI(x_k; y), \\
r^y_k & = & \frac{1}{|S|}\sum_{x_l \in S}MI(x_k; x_l).
\end{eqnarray}
By setting each gene to be the target, one can calculate all the values 
of $s^{x_i}_{x_j}$, and the relatedness value between any pair of $x_i$ 
and $x_j$ is the maximum of $(s^{x_i}_{x_j}, s^{x_j}_{x_i})$. A modification 
of MRNET can be found in Ref.~\cite{Meyer_2010}.

ARACNE is the extension of RELNET. In the ARACEN, Data Processing 
Inequality (DPI) \cite{cover2012elements}, a well-known information theoretic 
property, is used to overcome the limitation in the RELNET. Specifically, 
the DPI stipulates that if variables $x_i$ and $x_j$ interact only through a 
third variable $x_k$, one has 
\begin{equation}
{\rm MI}(x_i; x_j)\le \min(MI(x_i, x_k), MI(x_j, x_k)),
\end{equation}
which is used to eliminate indirect interactions. ARACNE starts from a 
network in which the value of MI for each edge is larger than $\tau$. The 
weakest edge in each triplet, for example the edge between $i$ and $j$, is 
regarded as an indirect interaction and is removed if
\begin{equation}
{\rm MI}(x_i; x_j)\le \min({\rm MI}(x_i, x_k), {\rm MI}(x_j, x_k))-\epsilon ,
\end{equation}
where $\epsilon$ is a tolerance parameter.

\subsubsection{Maximal information coefficient}

{\em Maximal Information Coefficient} (MIC) is a recently proposed 
information based method, belonging to the larger class of maximal 
information-based, non-parametric exploration statistics~\cite{Reshef_2011}. 
The basic idea of MIC is that, if a relationship exists between two 
variables, a grid can be drawn on the scatterplot of the two variables 
to partition the data so that the relationship between the two variables 
in the two-dimensional diagram can be uncovered.

The MIC of a given pair of data $D$ is calculated, as follows. First, 
one partitions the x-values of $D$ into $x$ bins and the y-values into 
$y$ bins according to a uniform partition grid G(x,y). One then 
calculates the mutual information with respect to this partition, 
which is denoted as $I(D|_G)$. For fixed $D$ with a different grid $G$, 
even with the same $x$ bins and $y$ bins, the value of the mutual 
information will be different. A characteristic matrix $\mathcal{M}(D)$ can 
be defined as
\begin{equation}
\mathcal{M}(D_{x,y})=\frac{I^*(D, x, y)}{\log\ \min(x,y)},
\end{equation}
where $I^*(D, x, y)=\max \{I(D|_G)\}$ is the maximum over all grids $G$ with 
$x$ and $y$ bins. The value of MIC of the variable pair $D$ is given by
\begin{equation}
{\rm MIC}(D)= \max \{M(D)_{x,y}\}.
\end{equation}
There were claims~\cite{Kinney_2014,Simon_2014} that the MIC does not 
outperform MI, suggesting that the mutual information may be a more 
natural and practical way to quantify statistical associations.

\subsubsection{Maximum entropy}

One way to model the spiking activity of a population of neurons is 
through the classic Ising model~\cite{Cocco_2009,Schneidman_2006}, in
terms of the following maximum entropy distribution:
\begin{equation}
P(\sigma_1,\sigma_2,\dots,\sigma_N)=\frac{1}{Z}\exp{\left[\sum_i{h_i\sigma_i} 
+\frac{1}{2}\sum_{i\ne j}J_{ij}\sigma_i\sigma_j \right]},
\end{equation}
where $\sigma_i=\pm 1$ is the spiking activity of cell $i$ and the 
normalized factor is the partition function $Z$ in statistical physics. 
The effective coupling strengths $J_{ij}$ are then chosen so that the 
averages $(\langle \sigma_i \rangle,\langle \sigma_i \sigma_j \rangle)$ 
in this distribution agree with the results from numerical or actual 
experiments.

\subsection{Bayesian network}

The Bayesian networks (BNs)~\cite{yeung2005bayesian, Ferrazzi_2007} are 
a commonly used probabilistic graphical model, represented as a direct 
acyclic graph (DAG), in which nodes represents a set of random variables 
and the direct links signify their conditional dependencies. For example, in 
the case of inferring a gene regulatory network, a Bayesian network could 
represent the probabilistic relationships between transcription factors 
and their target genes. The BN method is associated with, however, high 
computational cost (especially for large networks). Unlike the alternative
pairwise association estimation methods discussed above, the BN method 
requires that all possible DAGs be exhaustively or heuristically searched,
scored, and kept either as a best-scoring network or a network constructed 
by averaging the scores of all the networks. In order to optimize the 
posterior probabilities in BN, a variety of heuristic search algorithms 
were developed, e.g., simulated annealing, max-min parent and children 
algorithm~\cite{Tsamardinos_2003}, Markov blanket 
algorithm~\cite{Aliferis_2010}, Markov boundary induction 
algorithm~\cite{statnikov2010analysis}. The optimization algorithms 
integrated into BN notwithstanding, the method is practically applicable 
but only to biological networks of relatively small size.

\subsection{Regression and resampling}

Interring gene regulatory networks can be treated as a feature selection 
problem. The expression level of a target gene can be predicted by its 
direct regulate transcription factors. It is often assumed that the links among
the genes are sparse. For example, the lasso~\cite{tibshirani1996regression} 
is a widely used regression method for network inference, and it is a 
shrinkage and selection method for linear regression. Specifically, 
it minimizes the usual sum of squared errors, with a bound on the $L_1$ norm 
of the coefficients, which is used to regularize the model. Given the 
expression data of genes, in a steady state experiment, the feature 
selection can be formulated as
\begin{equation}
\min \frac{1}{2n} ||y - Xw||^2_2 + \lambda ||w||_1,
\end{equation}
where $\lambda$ is an adjustable parameter for controlling the sparsity 
of the coefficients. A direct use of Lasso to infer networks has two 
shortcomings: (a) it is known to be an unstable procedure in terms of 
the selected features, and (b) it does not provide confidence scores for 
the selected features. As a result, a stability selection procedure is 
often integrated into Lasso~\cite{Haury_2012}. One first resamples the 
data into several sub-data based on bootstrap, then apples the Lasso 
to solve these sub-data regression problems, and aggregates the final 
score for each feature to select more confident features.

Beside the steady-state data, time series data can also be tackled using 
the Lasso in the sense that the current expression level of the 
transcription factors is the predictors of the change in the expression of 
the target gene. A group Lasso method using both steady-state and time 
series data was proposed~\cite{yuan2006model}, in which the pair 
coefficients of a single transcription factor across both steady-state 
and time-series data are either both zero or both non-zero.

The gene regulation can also be modeled with nonlinear models, e.g., 
polynomial regression models~\cite{Li_2008, Song_2012} and sigmoid 
functions~\cite{Maetschke_2014}.

\subsection{Supervised and semi-supervised methods}

Supervised and semi-supervised methods treat reconstruction as 
a classification problem. A number of methods were developed to accomplish 
this goal, with some works showing that the performance of the supervised 
methods is better than that of the 
unsupervised methods~\cite{Huynh-Thu_2010}. Generally,
prior to applying a supervised method, two types of training data 
sets are required: the expression profile of each gene and a list of prior 
information about whether the known transcriptional factors and genes are
regulated. 

{\em Support Vector Machine} (SVM) is a prevailing supervised classification 
method that has been successfully used in inferring gene regulatory networks. 
The basic idea of SVM is to find an apparent gap that divide the data points of 
the separate categories as widely as possible. By integrating the 
kernel function, SVM can be extended to nonlinear classification.
{\em Gene Network Inference with Ensemble of Trees} (GENIE), similar to MRNET, 
identifies the best subset of regulator genes with random forest and 
extra-trees for regression and feature selection instead of MI and MRMR.
The method of {\em Supervised inference of regulatory networks} 
(SIRENE)~\cite{Mordelet_2008} uses SVM as a classifier to learn the decision 
boundary given the training dataset, and then generates the labels 
(whether links exist or do not) for the prediction dataset. Through a 
semi-supervised method, the unlabeled data are also included into the 
training data~\cite{Cerulo_2010}.

\subsection{Transfer and joint entropies}

{\em Transfer Entropy} (TE), as an information theoretic method, is 
often used for reconstructing neuron networks~\cite{lungarella2006mapping,
buehlmann2010optimal,honey2007network}. Compared with the mutual 
information, TE can be employed to reveal the causal relationships from the 
historical time series of two variables. Given the time series $\{x_t\}$ 
and $\{y_t\}$ for $x$ and $y$ respectively, TE can be expressed as:
\begin{equation}
{\rm TE}_{y->x}=H(x_t|x_{t-1}:x_{t-m}) - H(x_t|y_{t-1}:y_{t-l}, 
x_{t-1}:x_{t-m}),
\end{equation}
where $H(x)$ is the Shannon entropy of $x$, and $l,m$ are the length of 
the historical time series. In the case of inferring neuronal networks, $x_t$ 
can be the number of spikes in a specific time window, and $l$ and $m$ 
are often set to be 1. 
{\em Joint Entropy} (JE)~\cite{Garofalo_2009}. In calculating the MI or 
correlations for neural networks, the number of spikes in each time window 
is used, but the temporal patterns between spikes are ignored. For JE, the 
cross-inter-spike-intervals (cISI), defined as ${\rm cISI} = t_y - t_x$ 
between two neurons, is used. The JE is calculated based on cISI as
\begin{equation}
{\rm JE}(X,Y) = H({\rm cISI}),
\end{equation}
where $H$ is the entropy of cISI. In Ref~\cite{Garofalo_2009}, it was shown 
that TE and JE perform better than other methods. In Ref.~\cite{Ito_2011}, 
the authors demonstrated that, by integrating high-order historical time 
series ($l > 1$ and $m>1$) and multiple time delays, TE can be improved 
markedly.

\subsection{Distinguishing between direct and indirect interactions}

Based on a motif analysis, it was found~\cite{Marbach_2010} that three 
types of generic and systematic errors exist for inferring networks: 
(a) fan-out error, (b) fan-in error, and (c) cascade error. The cascade 
error originates from the tendency for incorrectly predicted ``shortcuts'' 
to cascades, also known as an error to misinterpret indirect links as 
direct links. For example, if nodes 1 and 2, and nodes 2 and $3$ in the 
true (direct) network are strongly dependent upon each other, 
then high correlations will 
also be visible between nodes 1 and $3$ in the observed (direct and 
indirect) network. Such errors are always present in the pairwise 
correlation, mutual information or other similarity metrics between a 
pair of nodes.

It was proposed~\cite{Feizi_2013} that a deconvolution method 
can be used to distinguish direct dependencies in networks. In particular,
the weights of an observed network $\mathcal{G}_{obs}$ to be inferred by 
the correlation or mutual information can be modeled as a sum of both 
direct weights $\mathcal{G}_{dir}$ in the real network and 
indirect weights $\mathcal{G}^2_{dir}$, 
$\mathcal{G}^3_{dir}$, etc. due to indirect paths: 
\begin{equation}
\mathcal{G}_{obs}=\mathcal{G}_{dir} + \mathcal{G}^2_{dir} + 
\mathcal{G}^3_{dir}+\dots =\mathcal{G}_{dir}\cdot (\mathcal{I} -
\mathcal{G}_{dir})^{-1}.
\end{equation}
The direct network can then be obtained as
\begin{equation}
\mathcal{G}_{dir}=\mathcal{G}_{obs}\cdot (\mathcal{I}+\mathcal{G}_{obs})^{-1}
\end{equation}
Through a matrix similarity transformation, one obtains 
$\mathcal{G}_{obs}=\mathcal{U} \cdot \mathcal{S}_{obs} \cdot 
\mathcal{U}^{-1}$, where $\mathcal{S}_{obs}$ is a diagonal matrix 
whose elements are $\lambda_{obs}$, which holds similarly for 
$\mathcal{G}_{dir}$: $\mathcal{G}_{dir}=\mathcal{U} \cdot 
\mathcal{S}_{dir} \cdot \mathcal{U}^{-1}$, and 
$\lambda_{dir}=\lambda_{obs}/(1+\lambda_{obs})$. The direct network 
can then be inferred.

The problem of extracting local response from global response to 
perturbation was first addressed using a modular response analysis 
(MRA)~\cite{Kholodenko_2002}. Since then many works have 
appeared~\cite{stark2003top,kholodenko2007untangling,
kholodenko2012computational}). A local response coefficient $r_{ij}$ 
is defined as
\begin{equation}
r_{ij}=\lim_{\Delta x_j}\left (  \frac{\partial \ln{x_i}}
{\partial \ln{x_j}}\right), i\ne j,
\end{equation}
which quantifies the sensitivity of module $i$ to the change in module 
$j$ with the states of other modules unchanged. However, in a real 
situation, an external intervention to perturb a parameter $p_j$ 
intrinsic to module $j$ can cause a local change in $x_j$ and then 
propagates to the whole system. After the network has relaxed into a 
new steady state, the global response coefficient, an often measured 
quantity in practice, can be obtained as 
$R_{i p_j}=d\ln{x_i}/dp_j$. The relation 
between the local and global responses can be established.

The idea of extracting local from global responses using statistical 
similarity measure based methods and a global silence method were
articulated and developed~\cite{Barzel_2013}. Given the observed global 
response matrix $\mathcal{G}$, measured by the correlation or the 
mutual information, the local response matrix $\mathcal{S}$ can be 
obtained as
\begin{equation}
\mathcal{S} =[\mathcal{G} - \mathcal{I} + \mathcal{D} \cdot 
(\mathcal{G}-\mathcal{I})\cdot\mathcal{G}]\cdot\mathcal{G}^{-1},
\end{equation}
where $\mathcal{I}$ is the identity matrix and the function $D(\mathcal{M})$ 
sets the off-diagonal elements of matrix $\mathcal{M}$ to be zero. 
Through this method, the 
indirect links become silenced so that the direct and indirect interactions 
can be distinguished.

\clearpage

\section{Discussions and future perspectives}
\label{sec:CH6_Discussions}

Nearly two decades of intense research in complex networks have resulted
in a large body of knowledge about network structures and their effects 
on various dynamical processes on networks. The types of processes that 
have been investigated include synchronization~\cite{LHCS:2000,GH:2000,
JJ:2001,BP:2002,WC:2002a,WC:2002b,HCK:2002,NMLH:2003,BBH:2004,BHLN:2005,
CHAHB:2005,DHM:2005,ZK:2006a,ZK:2006b,PLGK:2006,HPLYY:2006,WHLL:2007,
GWLL:2008}, virus spreading~\cite{PSV:2001,EK:2002,WMMD:2005,CBBV:2006,
HPL:2006,GJMP:2008,WGHB:2009,MA:2010,BV:2011}, traffic flow~\cite{ADG:2001,
EGM:2004,ZLPY:2005,WWYXZ:2006,MGLM:2008,TZ:2011,YWXLW:2011,MB:2012},
and cascading failures~\cite{ML:2002,ZPL:2004,ZPLY:2005,GC:2007,Gleeson:2008,
HLC:2008,YWLC:2009,WYL:2010,BPPSH:2010,PBH:2010,PBH:2011,LWLW:2012}.
A typical approach in the field is to 
implement a particular dynamical process of interest on networks whose 
connecting topologies are completely specified. Often, real-world networks 
such as the Internet, the power grids, transportation networks, and various 
biological and social networks are used as examples to demonstrate the 
relevance of the dynamical phenomena found from model networks. While this 
line of research is necessary for discovering and understanding various 
fundamental phenomena in complex networks, the ``inverse'' or ``reverse
engineering'' problem of predicting network structure and dynamics from 
data is also extremely important. The basic hypothesis underlying the 
inverse problem is that the detailed structure of the network and the nodal
dynamics are totally unknown, but only a limited set of signals or time
series measured from the network is available. The question is whether
the intrinsic structure of the network and the dynamical processes can be 
inferred solely from the set of measured time series. Compared with the 
``direct'' network dynamics problem, there has been less effort in the 
inverse problem. Data-based reverse engineering of complex networks, with 
great application potential, is important not only for advancing network 
science and engineering at a fundamental level but also for meeting the 
need to address an array of applications where large-scale, complex networks 
arise. In this Review, we attempt to provide a review, as comprehensive as 
possible, of the recent advances in the network inverse problem. 

A focus of this Review is on exploiting compressive sensing for addressing
various inverse problems in complex networks (Secs.~\ref{sec:CS_NDSI} and
\ref{sec:CH3_CS_CN_Reconstruction}). The basic principle of compressive sensing 
is to reconstruct a signal from sparse observations, which are obtained 
through linear projections of the original signal on a few predetermined
vectors. Since the requirements for observations can be relaxed considerably
as compared with those associated with conventional signal reconstruction 
schemes, compressive sensing has received much recent attention and it is 
becoming a powerful technique to obtain high-fidelity signal for applications 
where sufficient observations are not available. 

The key idea rendering possible exploiting compressive sensing for reverse 
engineering of complex network structures and dynamics lies in the fact that
any nonlinear system equation or coupling function can be approximated by 
a power-series expansion. The task of prediction becomes then that of 
estimating all the coefficients in the power-series representations of the 
vector fields governing the nodal dynamics and the interactions. Since the 
underlying vector fields are unknown, the power series can contain 
high-order terms. The number of coefficients to be estimated can therefore 
be quite large. Conventional wisdom would count this as a difficult 
problem since a large amount of data is required and the computations 
involved can be extremely demanding. However, compressive sensing is ideally 
suited for this task. In this Review we have described in detail
how compressive sensing can be used for uncovering the full topology and 
nodal dynamical processes of complex, nonlinear oscillator networks,  
for revealing the interaction patterns in social networks hosting 
evolutionary game dynamics, for detecting hidden nodes, for identifying
chaotic elements in neuronal networks, for reconstructing complex geospatial
networks and nodal positioning, and for mapping out spreading dynamics on
complex networks based on binary data.
   
In addition to the compressive sensing based paradigm, we have also 
reviewed a number of alternative methods for reconstruction of complex 
networks. These include methods based on response to external driving
signals, system clone (synchronization), phase-space linearization, 
noise induced dynamical correlation, and reverse engineering of complex 
systems. Representative methodologies to reconstruct biological networks
have also been briefly reviewed: methods based on correlation, causality,
information-theoretic measures, Bayesian inference, regression and 
resampling, supervision and semi-supervision, transfer and joint 
entropies, and  distinguishing between direct and indirect interactions.     

Data-based reconstruction of complex networks belongs to the broad field 
of nonlinear and complex systems identification, prediction, and control.
There are many outstanding issues and problems. In the following we list 
several open problems which, in our opinion, are at the forefront of this
area of research. 
 
\subsection{Localization of diffusion sources in complex networks}
\label{subsec:Ch6_localization}

Dynamical processes taking place in complex networks are ubiquitous in
nature and in engineering systems, examples of which include disease
or epidemic spreading in the human society~\cite{NNK:2009,HNC:2013}, 
virus invasion in computer networks~\cite{BBCS:2005,WGHB:2009}, and 
rumor propagation in online social networks~\cite{LC:2011}. 
Once an epidemic emerges, it is often of
great interest to be able to identify its source within the network
accurately and quickly so that proper control strategies can be devised
to contain or even to eliminate the spreading process. In general,
various types of spreading dynamics can be regarded as diffusion
processes in complex networks, and it is of fundamental interest to be
able to locate the {\em sources of diffusion}. A straightforward,
brute-force search for the sources requires accessibility of global
information about the dynamical states of the network. However, for large
networks, a practical challenge is that our ability to obtain and process
global information can often be quite limited, making brute-force search
impractical with undesired or even disastrous consequences. For example,
the standard breadth-first search algorithm for finding the shortest paths,
when implemented in online social networks, can induce information
explosion even for a small number of searching steps~\cite{BKMRRST:2000}.
Recently, in order to locate the source of the outbreak of Ebola virus in
Africa, five medical practitioners lost their lives~\cite{GGASPKJMFD:2014}.
All these call for the development of efficient methodologies to locate
diffusion sources based only on limited, practically available information
{\em without the need of acquiring global information about the dynamical
states of the entire network}.

There have been pioneering efforts in addressing the source localization
problem in complex networks, such as those based on the maximum likelihood
estimation~\cite{SZ:2010,LTL:2013}, belief propagation~\cite{ABDLZ:2014}, 
the phenomena of hidden geometry of contagion~\cite{BH:2013}, particle 
diffusion and coloration~\cite{ZY:2013}, and inverse 
spreading~\cite{SCFDWS:2015}. In addition, some approaches have
been developed for identifying super spreaders that promote spreading
processes stemming from sources~\cite{KGHLMSM:2010,PMAZM:2014,MM:2015}. 
In spite of these efforts,achieving accurate source localization from a 
small amount of observation remains challenging. A systematic framework 
dealing with the localization of diffusion sources for arbitrary network 
structures and interaction strength has been missing.

A potential approach~\cite{HHLW:2016} 
to addressing the problem of network source localization is to investigate the 
fundamental issue of {\em locatability}: given a complex network and limited 
(sparse) observation, are diffusion sources locatable? Answering this question
can then lead to a potential solution to the practical and challenging problem 
of actual localization: given a network, can a minimum set of nodes be 
identified which produce sufficient observation so that sources at arbitrary 
locations in the network can actually be located? A two-step solution
strategy was recently suggested~\cite{HHLW:2016}. Firstly, one develops a 
minimum output analysis to identify the minimum number of messenger/sensor 
nodes, denoted as $N_\text{m}$, to fully locate any number of sources in an 
efficient way. The ratio of $N_\text{m}$ to the network size $N$, 
$n_\text{m}\equiv N_\text{m}/N$, thus characterizes the source locatability 
of the network in the sense that networks requiring smaller values of 
$n_\text{m}$ are deemed to have a stronger locatability of sources. The 
minimum output analysis can be carried out by taking advantage of the dual 
relation between the controllability theory~\cite{YZDWL:2013} and the
canonical observability theory~\cite{Kalman:1959}. Secondly, given
$N_\text{m}$ messenger nodes, one can formulate the source localization problem
as a sparse signal reconstruction problem, which can be solved by using
compressive sensing. The basic properties of compressive sensing allow one to
accurately locate sources from a small amount of measurement from the 
messenger nodes, much less than that required by the conventional observability 
theory. Testing these ideas using variety of model and real-world networks, 
the authors found that the connection density and degree distribution play a 
significant role in source locatability, and sources in a homogeneous and 
denser network are more readily to be located. A striking and counterintuitive 
finding~\cite{HHLW:2016} was that, for an undirected network with one connected 
component and random link weights, a single messenger node is sufficient to 
locate any number of sources.

Theoretically, the combination of the minimum output analysis (derived from
the controllability and observability theories for complex networks) and
the compressive sensing based localization method 
constitutes a general framework for
locating diffusion sources in complex networks. It represents a
powerful paradigm to quantify the source locatability of a
network exactly and to actually locate the sources efficiently and accurately.
Because of the compressive sensing based methodology, 
the framework is robust against
noise, paving the way to practical implementation in a noisy environment.
The framework also provides significant insights into the open problem of 
developing source localization methods for time variant complex networks 
hosting nonlinear diffusion processes.

\subsection{Data based reconstruction of complex networks with
binary-state dynamics}

Complex networked systems consisting of units with binary-state
dynamics are common in nature, technology, and society~\cite{barrat2008}.
In such a system, each unit can be in one of the two possible states, e.g., 
being active or inactive in neuronal and gene regulatory 
networks~\cite{kumar2010}, cooperation or defection in
networks hosting evolutionary game dynamics~\cite{SF:2007}, being
susceptible or infected in epidemic spreading on social and technological
networks~\cite{pastor2015}, two competing opinions in social
communities~\cite{shao2009}, etc. The interactions among the units
are complex and a state change can be triggered either
deterministically (e.g., depending on the states of their neighbors)
or randomly. Indeed, deterministic and stochastic state changes
can account for a variety of emergent phenomena, such as the outbreak
of epidemic spreading~\cite{granell2013}, cooperation among selfish
individuals~\cite{santos2005}, oscillations in biological
systems~\cite{koseska2013}, power blackout~\cite{buldyrev2010},
financial crisis~\cite{galbiati2013}, and phase transitions in natural
systems~\cite{balcan2011}. A variety of models were introduced
to gain insights into binary-state dynamics on complex
networks~\cite{Newman:book}, such as the voter models
for competition of two opinions~\cite{voter}, stochastic propagation
models for epidemic spreading~\cite{sis}, models of rumor diffusion
and adoption of new technologies~\cite{castellano2009}, cascading
failure models~\cite{bashan2013}, Ising spin models for ferromagnetic
phase transition~\cite{ising}, and evolutionary games for cooperation
and altruism~\cite{santos2008}. A general theoretical approach to dealing
with networks hosting binary state dynamics was developed~\cite{gleeson2013} 
based on the pair approximation and the master equations, providing a 
good understanding of the effect of the network structure on the emergent 
phenomena.

Compressive sensing can be exploited to address the inverse problem of 
binary-state dynamics on complex networks, i.e., the problem of reconstructing 
the network structure and binary dynamics from data. Of particular relevance
to this problem is spreading dynamics on complex networks, where the
available data are binary: a node is either infected or healthy.
In such cases, a recent work~\cite{SWFDL:2014} demonstrated that
the propagation network structure can be reconstructed
and the sources of spreading can be detected by exploiting compressive
sensing (Sec.~\ref{subsec:CS_SPN}). 
However, for binary state network dynamics, a general
reconstruction framework has been missing. The
problem of reconstructing complex networks with binary-state dynamics
is extremely challenging, for the following reasons. (i) The switching
probability of a node depends on the states of its neighbors according
to a variety of functions for different systems, which can be linear,
nonlinear, piecewise, or stochastic. If the function that governs the
switching probability is unknown, difficulties may arise
in obtaining a solution of the reconstruction problem. (ii) Structural
information is often hidden in the binary states of the nodes in an
unknown manner and the dimension of the solution space can be extremely
high, rendering impractical (computationally prohibitive) brute-force
enumeration of all possible network configurations. (iii) The presence
of measurement noise, missing data, and stochastic effects in the
switching probability make the reconstruction task even more challenging,
calling for the development of effective methods that are robust
against internal and external random effects.

A potential approach~\cite{LWLG:2016} to developing a general and robust 
framework to reconstruct complex networks based solely on the binary
states of the nodes without any knowledge about the switching functions 
is based on the idea of linearizing the 
switching functions from binary data. This allows one
to convert the network reconstruction into a sparse signal reconstruction 
problem for local structures associated with each node. Exploiting the 
natural sparsity of complex networks, one can employ the lasso~\cite{lasso}, 
an L$_1$ constrained fitting method for statistics and data mining, to 
identify the neighbors of each node in the network from sparse binary data 
contaminated by noise. The underlying mechanism that justifies the linearization
procedure by conducting tests using a number of linear, nonlinear and
piecewise binary-state dynamics can be established on a large number of 
model and real complex networks~\cite{LWLG:2016}. Universally high 
reconstruction accuracy was found even for small data amount with noise. 
Because of its high accuracy, efficiency and robustness against noise and 
missing data, the framework is promising as a general solution to the inverse
problem of network reconstruction from binary-state time series. 
The data-based linearization method can be useful for dealing with general 
nonlinear systems with a wide range of applications.

\subsection{Universal structural estimator and dynamics approximator for
complex networks}

Is it possible to develop a universal and completely data-driven framework
for extracting network topology and identifying the dynamical processes,
without the need to know {\em a priori} the specific types of network
dynamics? An answer to this question would be of significant value not
only to complexity science and engineering but also to modern data
science where the goal is to unearth the hidden structural information
and to predict the future evolution of the system. A partial solution 
to this problem emerged recently, where the concept of {\em universal 
structural estimator and dynamics approximator} for complex networked 
systems was proposed~\cite{CL:2016}, and it was demonstrated that such a 
framework or ``machine'' can indeed be developed for a large number of 
distinct types of network dynamical processes. While it remains an open
issue to develop a rigorous mathematical framework for the universal 
machine, the preliminary work~\cite{CL:2016} can be regarded as an 
initial attempt towards the development of a universal framework for 
network reconstruction and dynamics prediction.

The key principle is the following. In spite of the dramatic difference in 
the types of dynamics in terms of, e.g., the interaction setting and state 
updating rules, there are two common features shared by many dynamical 
processes on complex networks: (1) they are stochastic, first-order Markovian 
processes, i.e., only the current states of the systems determine their 
immediate future; and (2) the nodal interactions are local. The two features 
are characteristic of a \emph{Markov network} (or a \emph{Markov random
field})~\cite{bishop:2006book,RN:2009book}. In particular, a Markov network
is an undirected and weighted probabilistic graphical model that is effective
at determining the complex probabilistic interdependencies in situations
where the directionality in the interaction between connected nodes cannot
be naturally assigned, in contrast to the directed Bayesian
networks~\cite{bishop:2006book,RN:2009book}. A Markov network has two 
types of parameters: a node bias parameter that controls its preference of state
choice, and a weight parameter characterizing the interaction strength
of each undirected link. The joint probability distribution of the state
variables $\mathbf{X} = (x_1, x_2, \ldots, x_N)^{\mathrm{T}}$
is given by $P(\mathbf{X}) = \prod_{C}\phi(\mathbf{X}_C)/\sum_{\mathbf{X}}
\prod_{C}\phi(\mathbf{X}_C)$, where $\phi(\mathbf{X}_C)$ is the potential
function for a well-connected network clique $C$, and the summation in the
denominator is over all possible system state $\mathbf{X}$. If this joint
probability distribution is available, literally all conditional probability
interdependencies can be obtained. The way to define a clique and to
determine its potential function plays a key role in the Markov network's
representation power of modeling the interdependencies within the
system. In Ref.~\cite{CL:2016}, the possibility was pursued of modeling
the conditional probability interdependence of a variety of dynamical
processes on complex networked systems via a binary Ising Markov network
with its potential function in the form of the Boltzmann factor, $\exp(-E)$,
where $E$ is the energy determined by the local states and their
interactions along with the network parameters (the link weights and node
biases) in a \emph{log-linear} form~\cite{AHS:1985}. This is effectively
a sparse Boltzmann machine~\cite{AHS:1985} adopted to complex network
topologies without hidden units. (Note that hidden units can play a
critical role in ordinary Boltzmann machines~\cite{AHS:1985}). A temporal 
evolutionary mechanism was introduced~\cite{CL:2016} as a persistent sampling 
process for such a machine based on the conditional probabilities obtained 
via the joint probability, and generate a Markov chain of persistently sampled
state configurations to form state transition time series for each node.
The model was named a \emph{sparse dynamical Boltzmann machine} 
(SDBM)~\cite{CL:2016}.

In dynamical processes on complex networks, such as epidemic
spreading or evolutionary game dynamics, the state of each
node at the next time step is determined by the probability conditioned
on the current states of its neighbors (and its own state in some cases).
There is freedom to manipulate the conditional probabilities
that dictate the system behavior in the immediate future 
through change in the parameters such as weights and biases. A basic question
is then, for an SDBM, is it possible to properly choose these parameters
so that the conditional probabilities produced are identical or nearly
identical to those of a typical dynamical process with each given observed
system state configuration? If the answer is affirmative, the SDBM can
serve as a dynamics approximator of the original system, and the
approximated conditional probabilities possess predictive power for the
system state at the next time step. When such an SDBM is found for
many types of dynamical process on complex networks, it effectively serves
as a universal dynamics approximator. Moreover, if the detailed statistical
properties of the state configurations can be reproduced in the long time
limit, that is, if the time series generated by this SDBM are statistically
identical or nearly identical to those from the original system, 
the SDBM is a generative model of the observed data (in the language of
machine learning), which is potentially capable of long term prediction.

When an approximator exists for each type of dynamics on a
complex network, the topology of the SDBM is nothing but that of the
original network, providing a simultaneous solution to the problem of
network structure reconstruction. Previous works on the inverse static
or kinetic Ising problems led to methods of reconstruction for Ising
dynamics through maximization of the data 
likelihood (or pseudo-likelihood) function
via the gradient descent approaches~\cite{T:1998,CM:2011,AE:2012,NB:2012,
R:2012,RH:2011,MJ:2011,SBD:2011,Z:2012,ZAAHR:2013,BLHSB:2013,DR:2014}. 
Instead of adopting these approaches, compressive sensing can be exploited.
By incorporating a K-means clustering algorithm into the sparse solution 
obtained from compressive sensing, it was demonstrated~\cite{CL:2016} that 
nearly perfect reconstruction of the complex network topology can be achieved. 
In particular, using $14$ different types of dynamical processes on complex 
networks, it was found that, if the time series data generated by these 
dynamical processes are assumed to be from its equivalent SDBMs, the universal 
reconstruction framework is capable of recovering the underlining network 
structure of each original dynamics with almost zero error. This represents 
solid and concrete evidence that SDBM is capable of serving as a universal 
structural estimator for complex networks. In addition to being able to 
precisely reconstruct the network topologies, the SDBM also allows the link 
weights and the nodal biases to be calculated with high accuracy. In fact, the
the universal SDBM is fully automated and does not require any subjective 
parameter choice~\cite{CL:2016}.
 
\subsection{Controlling nonlinear and complex dynamical networks}

An ultimate goal of systems identification and prediction is to control.
The coupling between nonlinear dynamics and complex network structure presents 
tremendous challenges to our ability to formulate effective control methodologies. 
In spite of the rapid development of network science and engineering toward 
understanding, analyzing and predicting the dynamics of large complex network 
systems in the past fifteen years, the problem of controlling nonlinear 
dynamical networks has remained largely open.

The field of controlling chaos in nonlinear dynamical systems has been
active for more than two decades since the seminal work of Ott, Grebogi,
and Yorke~\cite{OGY:1990}. The basic idea was that chaos, while
signifying random or irregular behavior, possesses an intrinsically
sensitive dependence on initial conditions that can be exploited for
controlling the system using only small perturbation. This feature,
in combination with the fact that a chaotic system possesses an infinite
set of unstable periodic orbits, each leading to different system
performance, implies that a chaotic system can be stabilized about
some desired state with optimal performance. Controlling chaos has since 
been studied extensively and examples of successful experimental implementation 
abound in physical, chemical, biological, and engineering 
systems~\cite{BGLMM:2000}. The vast literature on controlling chaos, however, 
has been mostly limited to low-dimensional systems, systems that possess 
one or a very few unstable directions (i.e., one or a very few positive 
Lyapunov exponents~\cite{GL:1997}). Complex networks with nonlinear dynamics
are generally high dimensional, rendering inapplicable existing methodologies 
of chaos control.

In the past several years, a framework for determining the \emph{linear}
controllability of network based on the traditional control and graph theories
emerged~\cite{LH:2007,LCWX:2008,RJME:2009,LSB:2011,WNLG:2011a,NA:2012,
YRLLL:2012,NV:2012,YZDWL:2013,LSBA:2013,MDB:2014,RR:2014,Wuchty:2014,
WBSS:2015,YTBSLB:2015,CWWL:2015}, leading to a quantitative understanding of the
effect of network structure on its controllability. In particular, a
structural controllability framework was proposed~\cite{LSB:2011}, revealing
that the ability to steer a complex network toward any desired state, as
measured by the minimum number of driver nodes, is determined by the set
of maximum matching, which is the maximum set of links that do not share
starting or ending nodes. A main result was that the number of driver
nodes required for full control is determined by the network's degree
distribution~\cite{LSB:2011}. The framework was established for weighted
and directed networks. An alternative framework, the exact controllability
framework, was subsequently formulated~\cite{YZDWL:2013}, which
was based on the principle of maximum multiplicity to identify the
minimum set of driver nodes required to achieve full control of networks
with arbitrary structures and link weight distributions. Generally, a
limitation of such rigorous mathematical frameworks of controllability is
that the nodal dynamical processes must be assumed to be \emph{linear}.
For nonlinear dynamical networks, to establish a mathematical
controllability framework similar to those based on the classic Kalman's
rank condition~\cite{Kalman:1963,Lin:1974,Luenberger:book} for linear
networks is an unrealistically broad objective.
Traditionally, controllability for nonlinear control can be
formulated based on Lie brackets~\cite{SL:book}, but to extend the
abstract framework to complex networks may be difficult. A recent
work extended the linear controllability and observability
theory to nonlinear networks with symmetry~\cite{WBSS:2015}. In spite of
the previous works, at the present there is no known general framework
for controlling nonlinear dynamics on complex networks.

Due to the high dimensionality of nonlinear dynamical networks and the
rich variety of behaviors that they can exhibit, it may be
prohibitively difficult to develop a control framework that is
universally applicable to different kinds of network dynamics.
In particular, the classic definition of linear controllability,
i.e., a network system is controllable if it can be driven from an arbitrary
initial state to an arbitrary final state in finite time, is generally
not applicable to nonlinear dynamical networks. Instead,
controlling collective dynamical behaviors may be more pertinent and
realistic~\cite{WC:2002,LWC:2004,SBGC:2007,CZL:2014}.
Our viewpoint is that, for nonlinear dynamical networks, control
strategies may need to be specific and system-dependent. Recently,
a control framework for systems exhibiting 
multistability was formulated~\cite{WSHWWGL:2016}. A defining 
characteristic of such systems is that, for a realistic parameter setting, 
there are multiple coexisting attractors in the phase 
space~\cite{GMOY:1983,MGOY:1985,FG:1997,FG:2003,LT:book,NYLDG:2013}.
The goal is to drive the system from one attractor to another using 
physically meaningful, temporary and finite parameter perturbation, assuming 
that the system is likely to evolve into an undesired state (attractor) or 
the system is already in such a state and one wishes to implement control to 
bring the system out of the undesired state and steer it into a desired one. It
should be noted that dynamical systems with multistability are ubiquitous in 
the real world ranging from biological and ecological to physical
systems~\cite{May:1977,Alley:2003,Chase:2003,SPD:2005,BM:2005,HGME:2007,
WZXW:2011}.

In biology, nonlinear dynamical networks with multiple attractors have
been employed to understand fundamental phenomena such as cancer
mechanisms~\cite{Huang:2013}, cell fate differentiation~\cite{HEYI:2005,
SGLE:2006,FK:2012,LZW:2014}, and cell cycle control~\cite{BT:2004,
YTWNJY:2011,RKGGBGRT:2013}. For example, Boolean network models were used to
study gene evolution~\cite{KPST:2004}, attractor number variation with
asynchronous stochastic updating~\cite{GD:2005}, gene expression in
the state space~\cite{HEYI:2005}, and organism system growth rate
improvement~\cite{MGAB:2008}. Another approach is to abstract key regulation
genetic networks~\cite{MTELT:2009,Faucon:2014} (or motifs) from all
associated interactions, and to employ synthetic biology to modify,
control and finally understand the biological mechanisms within these
complicated systems~\cite{YTWNJY:2011,SGLE:2006}. An earlier application
of this approach led to a good understanding of the ubiquitous phenomenon
of bistability in biological systems~\cite{Gardner:2000}, where there are
typically limit cycle attractors and, during cell cycle control, noise can
trigger a differentiation process by driving the system from a limit circle
to another steady state attractor~\cite{SGLE:2006}. In general,
there are two candidate mechanisms for transition or switching between
different attractors~\cite{FK:2012}: through signals transmitted within
cells and through noise, which were demonstrated recently using synthetic
genetic circuits~\cite{WSLELW:2013,WMX:2014}. More recently, a detailed
numerical study was carried out of how signal-induced bifurcations in a
tri-stable genetic circuit can lead to transitions among different cell
types~\cite{LZW:2014}.

The control and controllability framework for nonlinear dynamical networks 
with multistability can be formulated~\cite{WSHWWGL:2016} based on the concept 
of {\em attractor networks}~\cite{Lai:2014}. An attractor network is defined 
in the phase space of the underlying nonlinear system, in which each node 
represents an attractor and a directed edge from one node to another indicates 
that the system can be driven from the former to the latter using experimentally
feasible, temporary, and finite parameter changes. A well connected
attractor network implies a strong feasibility that the system can be
controlled to reach a desired attractor. The connectivity of the
attractor network can then be used to characterize the controllability
of the nonlinear dynamical network. More specifically, for a given pair of 
attractors, the relative weight of the shortest path is the number of 
accessible control parameters whose adjustments can lead to the attractor 
transition as specified by the path. Gene regulatory networks (GRNs) were 
used~\cite{WSHWWGL:2016} to demonstrate the {\em practicality}
of the control framework, which includes low-dimensional, experimentally
realizable synthetic gene circuits and a realistic T-cell cancer network
of 60 nodes. A finding was that noise can facilitate control by reducing the 
required amplitude of the control signal. In fact, the development of the 
nonlinear control framework~\cite{WSHWWGL:2016} was based entirely on
physical considerations, rendering feasible experimental validation.

The framework can be adopted to controlling nonlinear dynamical networks
other than the GRNs. For example, for the Northern European power grid
network recently studied by Menck et al.~\cite{MHKS:2014}, a rewiring method
was proposed and demonstrated to be able to enhance the system stability
through the addition of extra transmission lines. For a power grid network,
the synchronous states are desired while other states, e.g., limit cycles,
detrimental. Treating the link density (or number) as a tunable parameter,
the minimum transfer capacity required for extra lines to realize the
control can be estimated through our method. Another example is Boolean
networks with discrete dynamics, for which a perturbation method was 
proposed based on modifying the update rules to rescue
the system from the undesired states~\cite{CA:2014}. 
An attractor network can be constructed based on perturbation to multiple
parameters to drive the system out of the undesired, damaged states
toward a normal (desired) state. For biological systems, an epigenetic 
state network (ESN) approach was proposed~\cite{WSZWTX:2014} to
analyze the transitions among different phenotypic processes. In an ESN,
nodes represent attractors and edges represent pathways between a pair of
attractors. By construction, different parameter values would result
in a different ESN. This should be contrasted to an attractor network,
in which nodes are attractors but 
edges are directed and represent controllable paths (through parameter 
perturbation) to drive the system from one attractor to another.

\clearpage

\section*{Acknowledgement}
We thank Dr.~R.~Yang (formerly at ASU), Dr.~R.-Q.~Su (formerly at ASU),
and Mr. Zhesi Shen for their contributions to a number of original papers 
on which this Review is partly based. This work was supported by ARO under 
Grant No.~W911NF-14-1-0504. W.-X. Wang was also supported by NSFC under
Grants No.~61573064 and No.~61074116, as well as by the Fundamental 
Research Funds for the Central Universities, Beijing Nova Programme.


\end{document}